\documentclass[12pt,a4paper]{report}

\usepackage{amssymb}
\usepackage{array}
\usepackage{amsmath}
\usepackage{amsthm}
\usepackage{graphicx}
\usepackage[dvips]{epsfig}
\usepackage[usenames,dvipsnames]{color}
\usepackage{hyperref}
\usepackage[utf8x]{inputenc}
\usepackage[left=1.0in,right=1.0in,top=1.0in,bottom=1.0in]{geometry}

\hypersetup{
    colorlinks=true,         
    linkcolor=blue,          
    citecolor=red,        
    urlcolor=Violet            
}

\newcommand{\order}[1]{\ensuremath{\mathcal{O}(#1)}}
\newcommand{\diby}[2]{\ensuremath{\frac{\delta #1}{\delta #2}}}

\newcommand{\pb}[2]{\ensuremath{\lf\{#1,#2 \rt\}}}

\newcommand{\mean}[1]{\ensuremath{\lf\langle #1 \rt\rangle }}
\newtheorem{proposition}{Proposition}
\newtheorem{defi}{Definition}
\newtheorem{theorem}{Theorem}
\newcommand{\R}{\mathbb{R}}

\let\oldmarginpar\marginpar
\renewcommand\marginpar[1]{\oldmarginpar{\color{red}\raggedright\footnotesize #1}}

\def\lf {\ensuremath{\left}}
\def\rt {\ensuremath{\right}}
\def\ra {\ensuremath{\rightarrow}}

\def\hg {\ensuremath{\mathcal H_\text{gl}}}
\def\ts {\ensuremath{T_\phi \frac{S}{\sqrt g}}}

\def\wts {\ensuremath{\widetilde{T_\phi S}}}
\def\be{\begin{equation}}
\def\ee{\end{equation}}
\def\bea{\begin{eqnarray}}
\def\eea{\end{eqnarray}}
\def\nn{\nonumber}

\newcommand{\hn}[1]{\ensuremath{\mathcal H_{(#1)}}}
\newcommand{\wn}[1]{\ensuremath{\omega_{(#1)}}}
\newcommand{\sn}[1]{\ensuremath{S_{(#1)}}}
\newcommand{\ncmd}{\newcommand}
\ncmd{\ad}{\mbox{ad}}
\ncmd{\Ad}{\mbox{Ad}}
\ncmd{\diff}{\mbox{Diff}(M)}
\ncmd{\End}{\mbox{End}}
\ncmd{\Exp}{\mbox{Exp}}
\ncmd{\HH}{\mbox{H}}
\ncmd{\V}{\mbox{V}}
\ncmd{\riem}{\mbox{Riem}(M)}
\ncmd{\Aut}{\mbox{Aut}}
\ncmd{\Id}{\mbox{\tiny Id}}
\ncmd{\I}{\mathcal{I}}
\ncmd{\Ker}{\mbox{Ker}}
\ncmd{\M}{\mathcal{M}}
\ncmd{\DD}{\mathcal{D}}
\ncmd{\super}{\mathcal{S}}
\ncmd{\Lg}{\mathfrak{g}}
\ncmd{\Q}{\mathcal{Q}}
\def\hg {\ensuremath{\mathcal H_\text{gl}}}
\def\ts {\ensuremath{T_\phi \frac{S}{\sqrt g}}}
\newtheorem{prop}{Proposition}
\newtheorem{lem}{Lemma}
\newtheorem{cor}{Corollary}
\newtheorem{theo}{Theorem}

\title{{\bf{\huge{THE DYNAMICS OF SHAPES}}}}

\author{Henrique de A. Gomes\\~
\\
\small{under supervision of Prof. John W. Barrett.}\\
~\\
\\
Thesis submitted for the title of Doctor of Philosophy\\
 at the  University of Nottingham}

\begin{document}

\maketitle
\begin{abstract}This thesis consists of two parts, connected by one central
theme: the dynamics of the ``shape of space". To give the reader some inkling of
what we mean by ``shape of space", consider the fact that the shape of a
triangle is given solely by its three internal angles; its position and size in
ambient space are irrelevant for this ultimately intrinsic description.
Analogously, the shape of a 3-dimensional space is given by a metric up to
coordinate and conformal changes. Considerations of a relational nature strongly
support the development of such dynamical theories of shape. The first part of
the thesis concerns the construction of a theory of gravity dynamically
equivalent to general relativity (GR) in 3+1 form (ADM). What is special about
this theory is that it does not possess foliation invariance, as does ADM. It
replaces that ``symmetry" by another: local conformal invariance. In so doing it
more accurately reflects a theory of the ``shape of space", giving us reason to
call it  \emph{shape dynamics} (SD). Being a very recent development, the
consequences of this radical change of perspective on gravity are still largely
unexplored. In the first part we will try to present some of the highlights of
results so far, and indicate what we can and cannot do with shape dynamics.
Because this is a young, rapidly moving field, we have necessarily left out some
interesting new results which are not yet in print and were developed alongside
the writing of the thesis.  The second part of the thesis will develop a gauge
theory for ``shape of space"--theories. To be more precise, if one admits that
the physically relevant observables are given by shape, our descriptions of
Nature carry a lot of redundancy, namely absolute local size and absolute
spatial position. This redundancy is related to the action of the
infinite-dimensional conformal and diffeomorphism groups on the geometry of
space. We will show that the action of these groups can be put into a language
of infinite-dimensional gauge theory, taking place in the configuration space of
3+1 gravity. In this context gauge connections acquire new and interesting
meanings, and can be used as ``relational tools''.  \end{abstract}

\section*{\centering Acknowledgements}

 I vividly remember being on the beach, at quite a young age, trying to understand my father's explanation of Newton's law of action and reaction. I would like to dedicate this thesis to him, for first getting me interested in physics and thereafter never failing to give
me his full support or sharing his enthusiasm. I would like to thank first and foremost my mother, whose words of wisdom always put me back on track whenever I, for one reason or another, lost motivation. I would like to thank Julian Barbour for his interest and immense support for my work, and for reading and extensively commenting a first draft of this thesis. In the great tradition of Einstein, Bohr, Poincar\'e, and others, his unwavering focus on foundational principles is an example of what physics has to gain from Natural Philosophy.  He is largely responsible for inspiring us to give shape dynamics its due attention.  Julian so eloquently argues for shape dynamics that even his son, Boris Barbour -- who is not a physicist -- was recruited to draw the figures contained in this text, and which I ``borrowed". My thanks goes out to Boris as well. My immense gratitude also of course goes out to my good friends Tim and Joy (and Nate and Ruthie) Koslowski for being very gracious hosts to me more than one time. I feel very grateful for having found a collaborator so knowledgeable and yet  with whom I can communicate so effortlessly. I would also like to thank Sean Gryb, without whom the ``perfect storm" leading to Shape Dynamics would not have been complete. Our heated discussions have enhanced my understanding of shape dynamics a very great deal. My supervisor John Barrett, must also figure prominently in this list. My admiration for his extremely honest approach to physics (and to being a physicist) inspires me to no end. John was somehow able to make sure I was on the right track for my degree while still allowing me to follow my own interests. This is a balance I had never thought possible.

\paragraph{}{\it Thank you all.}

\vfill
\begin{flushright}

In quantum gravity, {\it we are all in the gutter, but some of us are looking at the stars.}\\ -- Popular saying, pre-cognitively adapted by Oscar Wilde.

\end{flushright}

\tableofcontents
\chapter{Introductory remarks}

\section{A tale of two theories}
Probably one of the most regurgitated quotes of theoretical physics is Minkowski's 1908 address at the 80th Assembly of German Natural Scientists and Physicians:
\begin{quote}
Henceforth space by itself, and time by itself, are doomed to fade away into mere shadows, and only a kind of union of the two will preserve an independent reality.\end{quote}
By the time Minkowski pronounced the now famous words, the experimental and
theoretical bases for relativity were on solid ground. The experimental absence
of the ether had been explained away, and the 4-dimensional unification of
electricity and magnetism was one of the theoretical triumphs of the early
20th century. The foundation was laid for one of the great edifices of modern
physics: general relativity.

In a different part of the world of physics, simultaneously with these advances,
the apparently completely different field of thermodynamics and heat emission
had given birth to the quantum. The infant came into the scene dissipating the
second of Lord Kelvin's famous ``clouds" over physics:  the then experimentally
disproved law of black-body radiation.\footnote{Both Rayleigh-Jeans'and Wien's,
for different ends of the spectrum.} The newborn was destined for greatness, and
it did not disappoint.  Under the teenage guise of quantum mechanics, and its
later adult incarnation, quantum field theory, it dominated much of modern
theoretical and experimental physics.

Both fields grew up side by side basking in glory after glory, with relativity
perhaps reaching its maturity earlier than quantum mechanics, being put into its
present form already in 1916. After the teenage years, under the auspices of
Schr\"odinger, Klein, Gordon, and most prominently Dirac, the two met and the
encounter evolved quantum mechanics into quantum field theory, arguably the most
successful theory ever developed. On the other hand, general relativity remained
largely unmoved by quantum mechanics.

As it stands however, quantum field theory is not the final word in this tale. It incorporates at best a sterile version of general relativity, one in which quantum fields are not allowed to feed back into the geometry of space-time. At worst, it still requires a structure that allows one to separate space-time into space and time, thereby foiling Minkowski's grandiose prediction.\footnote{In other words, unless the background metric has a global time-like Killing vector, one cannot canonically  define a vacuum or ground state, as the concept of a vacuum is not invariant under diffeomorphisms. In general, under a diffeomorphism,  the mode decomposition of the transformed eigenfunctions  will contain negative frequencies even if they were positive before the transformation.} In spite of valiant attempts by many physicists over the past 80 years, it remains true that the two theories are not completely on talking terms.

Our objective in the present thesis  is to give an alternative view of gravity,
one which breaks away from space-time -- indeed breaks space-time -- into space
and time. This might seem at first like a step back. But as we will see, many
other conceptual challenges are resolved by the approach advertised in this
thesis. In so doing we hope to remove  the most glaring point of conceptual
disagreement between our two protagonists: the different notions of
\emph{time} each one clings to. Or, being a little more conservative, we hope to
at least smooth out the issue to the point where a compromise can be reached
and gravitational phenomena be made more sympathetic to quantum mechanics.

\section{The problem of time}
It is no secret that time plays very different roles in quantum
field theory  and general relativity. In the former
time is part of an absolute framework with respect to which dynamical operators
(or states) are defined, but in the latter everything is dynamical. This
mismatch is the source of many great
difficulties encountered in the attempts to create an overarching
unified framework of quantum gravity \cite{Anderson:2010xm}.

In the canonical formulation, more amenable to a standard quantum theoretical
treatment,  space-time is essentially `sliced-up', and Einstein's equations
are described as an
evolution of the geometry of spatial slices through time. In effect, one
attempts to revert to separate notions of space and time as much as possible to
be able to apply the Hamiltonian analysis. It is in this formulation that we can
see most
clearly some of the problems that relativistic ``time" creates in the quantization of gravity.  The ADM formulation of the Einstein equations
\cite{Arnowitt:1962hi} leads directly to constraints. These constraints are such that they are associated with ``symmetries" of the system, symmetries whose action generate certain transformations of the physical description of the Universe.

One set of constraints, known collectively as the momentum constraint, is associated with foliation-preserving 3-diffeomorphisms. In other words, its action preserves the ``slicing", and thus the separation of space-time into space and time remains intact. This action has a well defined group  representation on phase
space, which simplifies its treatment considerably. Only a single initial configuration of the Universe is needed to obtain the resulting final configuration under the action of the symmetry. Moreover this is valid for the finite (as opposed to infinitesimal) action of the group, a property which  does not hold for the remaining constraints, as we will see. These characteristics make it fairly simple to quotient out the symmetry associated to the momentum constraint, eliminating its related
unphysical degrees of freedom. The resulting quotient space,
called \emph{superspace}, parametrizes initial data obeying the constraint, and
is the proper physical arena that eliminates the redundancy generated by
that symmetry.

\begin{figure}\label{fig:foliations}
 \begin{center}
 \includegraphics[width=11cm]{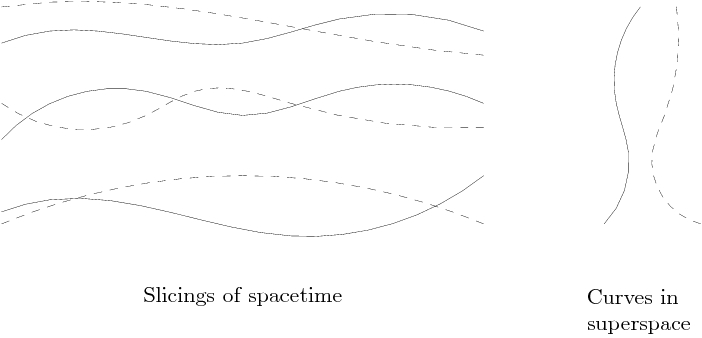}

\caption{Because there is no distinguished definition of simultaneity in
general relativity, a spacetime can be sliced in many different
ways. This slicing, or foliation, freedom leads to many different
representations of the spacetime by curves in superspace. Two slicings and
corresponding curves in superspace are shown.}

 \end{center}

\end{figure}

The other set of constraints, which we denote by
$S(x)$, is called \emph{the Hamiltonian constraint}, and it generates
evolution of the spatial variables. Already at the classical level a severe problem immediately arises. Because
there is an $S(x)$
constraint at each space point, generating evolution independently, the time evolution is ``many fingered'', which
means that the spatial slices can be made to evolve at arbitrarily chosen
different rates at different points. In contrast to the action of
3-diffeomorphisms, this ``symmetry" changes the original decomposition of space-time into space and time. By generating
different foliations of space-time, it yields curves in phase space that bear no
simple relationship to each other (see figure \ref{fig:foliations}). Unlike the
momentum constraints, it does not, by itself, have a group action on phase
space, and one cannot straightforwardly quotient phase space. Dirac, when
speaking about the difference between the constraints (in the setting of quantum
theory) phrased it in the following simple way \cite{Dirac:CMC_fixing}: ``Thus
we have the situation that we cannot specify the initial state for a problem
without solving the equations of motion. The formalism is thus not suitable to
dealing with practical problems." And therein lies the problem: to quotient out
the symmetry and obtain the physical configuration space, one must basically
solve the equations of motion.

 \section{A tale of two parts}

\subsubsection{A theory of space \emph{and} time}

Our objective in the first part of the thesis will be to develop a theory of
gravity that is indeed a theory of space \emph{and} time.  Furthermore, it has a
different symmetry group than general relativity and carries a ``proper" group
action on the gravitational variables. As we put it, in shape dynamics (SD) we
are in the business of symmetry ``trading". It will be a theory of gravity in
the sense that  the solutions of a
particular gauge fixing of general relativity are equivalent to the solutions of
a particular gauge fixing of the new symmetry in the present theory. The new
symmetry  is that of 3-dimensional conformal
transformations,\footnote{Transformations that either preserve the total volume
in the case that the Universe is closed and without boundary, or respect given
boundary conditions if the Universe is spatially asymptotically flat. } i.e.,
transformations that change local scale.  
These transformations act truly as a
symmetry group in the phase space of general relativity. Their simple linear
action allows us to easily track the effect of the symmetry transformations and
eliminate their associated redundancy (i.e.,  we can quotient by their action).

We thus obtain the space of physical configurations called \emph{conformal
superspace}, or, more lyrically, \emph{shape space}. It is the natural setting for a description of the Universe which 
relies solely on the ``shape of space". In such a description, spatial angles take the forefront, while local size is relegated to a quantity measured only with respect to an arbitrary local scale.

In this new formulation, we replace evolution generated by infinitely
many local Hamiltonian constraints by evolution generated by a unique single
global constraint $\mathcal{H}_0$. Time evolves rigidly, in step everywhere.  We believe this
squarely addresses issues related to the problem of time, and offers real hope of a
definitive solution to it. This part will be based on the papers: \cite{Gomes2011a,Gomes2011,Gomes2011b, Gomes:2011dc}.

\subsubsection{A geometrical gauge-theory setting.}

 Since in Shape Dynamics we have a ``proper" group action on phase space, the question of how far one can pursue the gauge-theoretic scenario of usual gauge theories, such as electrodynamics, immediately becomes relevant. In the second part of this thesis, we will describe such a gauge setting, with connection forms, gauge choices  and the like,  for both the
3-diffeomorphism group and the 3-conformal group as actions on Riem.
Both the actions of the groups and their algebras are perfectly well-defined and
correspond exactly to their finite-dimensional counterparts. This part is
entirely based on paper \cite{deA.Gomes:2008uv}.

    This thesis is formed from  two separate but deeply related subjects: the
construction of a theory of gravity embodying a different symmetry principle,
called \emph{shape dynamics} (SD), and the construction of a gauge theory for
the configuration space of general relativity (GR).
\section{Notation and other warnings.}

\subsection{Other warnings}

We try to pursue our proofs to the point where only subtle technical
functional-analytic matters, such as domains in Frech\'et spaces etc, start to
appear. Even though we do not present full mathematical proofs to the bitter end
taking these issues into account, we give strong plausibility arguments of why
they should go through without a hitch.

This thesis is a merger of two independent but very much interrelated lines of
research. One is the construction of the theory of shape dynamics, which takes a
leading role and is very prominent in the present work. The other is the working
out of specific geometric gauge theoretic structures in the configuration space
of GR. The former is a self-contained theory, with a robust conceptual
background. The latter is a conceptual framework, or the development of a set of
tools, that can be applied in the future to, among other things,  the theory of
shape dynamics itself.

We have chosen in this thesis to keep the background material for the two
sections separate. It seemed  more pedagogic to distinguish the technical
background necessary for each part, and only after both edifices have been
constructed to bridge them, although of course by then the reader will already
see a clear connection between the two parts. This approach has its drawbacks;
it is not always possible to keep the two parts completely  separate. In
particular, we have three cases where one section must borrow from the other, or
vice-versa. One of these instances happens when we explain Barbour's
best-matching ideas, and find that the ideal way to explain them is to use
almost exclusively the technical material present in Part II, defining gauge
structures in Riem, the space of 3-metrics  (chapter \ref{sec:Riem}). A second
instance is that we use the Fredholm alternative when studying the
asymptotically flat case of shape dynamics in chapter \ref{chapter:SD_AF}.
Lastly, we give an intuitive geometrical picture of shape dynamics, and we must
also mention the use of a section for the conformal bundle, presented also in
chapter \ref{sec:Riem}. For this item, we had no option but to include it in the
first part of the thesis. We could have as easily  included the first and last
items in Part II of the thesis and made reference to part I,  or vice-versa,
but although they  might be less displaced technically if placed in Part II,
we find them  conceptually better situated in Part I.

\subsection{Notation}
We explicitly mention only a few of the items individually, those that may be more confusing to the reader without further explanations.

\subsubsection{Concerning the 3-metric $g$.}
One important difference between the usual notation and the one utilized in this thesis has to be mentioned here at the beginning. Since we will focus mostly on the 3+1 picture of gravity, we will use $g_{ab}$ to denote the \emph{3-dimensional Riemannian metric}, and not the four-dimensional Lorentz one.
 Whenever we write $\sqrt
g$ we mean of course the square root of the determinant of the metric. But we
will also use $g$ for the determinant itself, or for the metric as an argument
of some function(al), such as $f(g,\pi)$. This does \emph{not} mean that $f$ is
a function of the determinant of the metric and the trace of the momenta, but of
the full metric and full momenta. The distinction should be clear from the
context. At some points, when it is convenient to use index-free tensor
notation, we will adopt a boldface $\mathbf{g}$ for the metric tensor.

Throughout the paper  semi-colon denotes covariant differentiation, and we
will, when it is convenient, use abstract index notation
(parentheses denote symmetrization of indices, and square brackets
anti-symmetrization). Also, again when it is convenient, we shall use $\nabla_a$
to denote the intrinsic Levi-Civita covariant derivative related to the
3-metric, and $D_a$ the one related to the 4-dimensional one.

The one parameter family of natural metrics on the tangent space to Riem (the
configuration space of all 3-metrics) is taken to be given by
\cite{Giulini:1993ct}:
 \be\label{supermetric} \mathcal{G}_\beta(u,v)_g=\int_M
G^{abcd}_\beta u_{ab}v_{cd} d\mu_g,\ee where, for tangent vectors $u,v\in
T_g(\mbox{Riem})$, the \emph{generalized DeWitt metric} is defined as
   \be\label{deWitt}
G_{\beta}^{abcd}:=g^{ac}g^{bd}-\beta g^{ab}g^{cd}
   \ee
{with inverse}
\be G^{\beta}_{abcd}:=g^{ac}g^{bd}-\lambda g^{ab}g^{cd},\ee where by inverse we
mean
$G_{\beta}^{abnm}G^{\beta}_{cdnm}=\delta^a_c\delta^b_d$. The relation between $\beta$ and $\lambda$ is that $\lambda=\frac{\beta}{3\beta-1} $. The usual
 DeWitt metric is $G_1$.
  We briefly note that the DeWitt metric is usually taken to be
  $(\sqrt{g}/2)(g^{ac}g^{bd}+g^{ad}g^{bc}-2g^{ab}g^{cd})$, but if we are only dealing with
   symmetric two-valence tensors, its action amounts to the one we have used,
apart from
   the $\sqrt{g}$ factor, which we input on the volume form.

  \subsubsection{Functional dependence,  brackets, function spaces. }
  We employ square brackets for functional dependence, as $F[g]$ for example,
and Kucha\v r's notation for mixed functional and local dependence, $F[g,x)$ for a functional of $g$ that yields a local function. Sometimes, when there is more than one functional dependence and still local dependence, we separate the functional arguments by commas and the local dependence by semi-commas $F[g,\pi:~x)$.

  We will use bras and kets both for the mean of quantities, such as $\mean
{f}$. This is not to be confused to its use when separated by a comma, for a given contraction between
dual vector spaces and the vector spaces themselves, such as $\mean {v, w}$.

Another non-standard notation we will be employing is that of $f\equiv g$, for
some given function $f:\Gamma\rightarrow C^\infty(M)$. This is meant to signify
that $f(x)=g(x)$ strongly, i.e., over all of phase space $\Gamma$ and for every
$x\in M$.

The space of smooth functions over the manifold $M$ will be denoted by $C^\infty(M)$. The space of smooth sections over a given vector bundle $E$ will be given by $\Gamma^\infty(E)$.

\subsubsection{Conformal transformations}
The acronym vpct signifies volume-preserving-conformal-transformation and we shall employ it widely. The calligraphic
$\mathcal{T}_\phi$ is the notation for the conformal transformation map, and should not be confused with $T_xM$, meaning the tangent space to $M$ at $x\in M$, nor with $T_xf:T_xM\rightarrow T_{f(x)}N$ which is the tangent map to $f:M\rightarrow N$ at $x$. It should also not be confused when we denote the general linking theories in Chapter \ref{chapter:linking_theory} by $T_L$.

\part{Shape Dynamics}
\chapter{Introducing Shape Dynamics}

We will now introduce the main subject of this thesis: the theory of shape
dynamics (SD). The first aim of this chapter is to provide the reader with an
outline of the theory, that is, its  motivations and main results so far. It
will serve the purpose of pointing north, and the rest of Part I will guide us
there and hopefully, once we have arrived, indicate some interesting directions
to explore.

\section{Technical Background}\label{sec:technical_back}

Before actually presenting an introduction to the main subject of this thesis, we have to  present some technical baggage without which it makes little sense. In other words, this section will be an \emph{introduction to the introduction} of shape dynamics. The theory itself will be constructed in the next few chapters (chapter \ref{chapter:linking_theory} through \ref{chapter:SD_AF}).

 We will start by giving a streamlined view of constrained dynamics, which suffices for our purposes. The main result which we wish to present in the first section is that for systems without a ``true Hamiltonian" a complete description of the dynamics can be made by separating first and second class constraints (section \ref{sec:1stClass_vs_2nd}) and strongly solving the second class constraints. We also give a more geometric view of the whole Dirac analysis, including that for systems possessing a  ``true Hamiltonian".

In \ref{sec:ADM}, we  then present the ADM 3+1 decomposition,
which is the starting point of almost all canonical approaches to gravity,
finishing with the ADM constraints and the Dirac algebra in section
\ref{sec:const.alg_ADM}. After introducing these theoretical constructs, we will
be able to discuss work that led to the construction of shape dynamics, such as
York's method for solving the initial value problem of general relativity and
Barbour et al's first principles derivation of those equations.
\subsection{Constrained dynamics}

\subsubsection{Lagrangian dynamics}
In this thesis we are mainly concerned with a dynamical formulation of physical systems. That means we will focus on how such systems develop through time, a view in some aspects different from the usual 4-dimensional covariant field theory.

In the Lagrangian formulation of mechanics, one is given a \emph{Lagrangian} $ L(q^\alpha(t), \dot q^\alpha(t))$, where  $q^\alpha$ are the coordinates of the system, $\dot q^\alpha$ are their time derivatives, and $\alpha$ is an index that parametrizes them. For example, for a single particle in $\R^3$, $\alpha$ runs from 1 to 3.\footnote{ For two particles in $\R^3$ for example, it is convenient to subdivide the 6 values of $\alpha$ into two subsets of three, parametrized by the particle to which they belong: $\alpha=(\beta,i)~|~\beta\in\{1,2,3\},i\in\{1,2\}$.} As the coordinates $q^\alpha$ describe every possible configuration of the system, the space of coordinates parametrize what is called the configuration space of the system, $Q$.

With the Lagrangian, one forms an \emph{action functional}
\be S[q^\alpha(\gamma)]=\int_{t_1}^{t_2} L(q^\alpha(\gamma(t)),\dot q^\alpha(\gamma(t))) dt
\ee
by integrating the Lagrangian over a given path $\gamma:[0,1]\to Q$.
Now, upon variation and integration by parts, assuming that the system is fixed
at both initial and final configurations, one obtains from the least action
principle $\delta S=0$ the Euler--Lagrange equations:
\be \label{equ:Euler-Lagrange}\frac{d}{dt}\frac{\delta L}{\delta\dot q^\alpha}=\frac{\delta L}{\delta q^\alpha}
.\ee
 We chose to use the notation $\diby{F}{q^\alpha}$ as opposed to $\frac{\partial
F}{\partial q^\alpha} $  because this generalizes the partial derivatives
directly to functional derivatives in the infinite dimensional case, just as the
sum indicated by repeated indices generalizes to integrals.

If we use the chain rule for the $\frac{d}{dt}$ derivative, we get from \eqref{equ:Euler-Lagrange}:
\be\label{equ:accel_lagrange} \ddot q^{\beta}\frac{\delta^2 L}{\delta\dot q^{\beta}\delta\dot q^\alpha}+\dot q^{\beta}\frac{\delta^2 L}{\delta q^{\beta}\delta\dot q^\alpha} =\frac{\delta L}{\delta q^\alpha}
.\ee
From this it becomes clear that the accelerations are uniquely determined by the positions and velocities if and only if the matrix $M_{\alpha\beta}:=\frac{\delta^2 L}{\delta\dot q^{\beta}\delta\dot q^\alpha} $ is invertible. If it isn't, our system possesses some kind of redundancy in its description, and this is indicative of {\it gauge symmetries}.

 Although the Euler--Lagrange equations derived from variation of the Lagrangian
completely describe the dynamics of the system, it is a rather cumbersome ordeal
to obtain directly from them information about redundancy in the description of
the system. The more suitable method to unravel such information is to use the
Hamiltonian formalism.

\subsection{Hamiltonian dynamics}\label{sec:hamiltonian_dyn}

 For Hamiltonian dynamics, we seek to perform a change of variables
$(q^\alpha,\dot q^\alpha)\to (q^\alpha,p_\alpha)$, where
\be\label{equ:def_can_momenta}p_\alpha:=\frac{\partial L}{\partial\dot
q^{\alpha}}.
\ee
In other words, one defines the action of the \emph{Legendre transform} as
$\mbox{LT}:TQ\to T^*Q$. Here $TQ$ denotes the tangent space to the configuration
space. This is the space parametrized by the doubles $(\dot q^\alpha,
q^\alpha)$, where $\dot q^\alpha$ denotes an element of the tangent space to
$q^\alpha$, i.e., the tangent of a curve at the point $q^\alpha$. Thus we say
that $\dot q\in T_qQ$. The space $T^*Q$ is obtained by replacing the tangent
space at each point by the \emph{cotangent} space, i.e., by the vector space
consisting of linear functionals on $T_qQ$, for each point $q$.\footnote{These
are all simple examples of vector bundles over $Q$, the trivializing charts
being induced by the tangent map of the original charts of $Q$.} The momenta
$p_\alpha$ are \emph{not} elements of the tangent space, but of the cotangent
space. That means that we can define their action on the tangent vectors $\dot
q^\alpha$ without the need of an inner product. In fact, we \emph{have} to
define these elements of $T_q^*Q$  by the way in which they act on elements of
the tangent space.

  Thus we characterize the map $\mbox{LT}_q(v)\in T_q^*Q$  by defining it to act
on $w\in T_qQ$ as
\be\label{equ:momentum_map} \langle \mbox{LT}_q(v),
w\rangle=\frac{d}{dt}L(q,v+tw)
\ee
As an example, let us set $L(q,v)= \frac{m}{2}F(q)g^{ab}v_av_b+V(q)$. Then
$$\frac{d}{dt}L(q,v+tw)=mF(q)g^{ab}v_aw_b,
$$ and we can see that the linear functional, at $v$, is just given by $mF(q)g^{ab}v_a$, which is indeed an element of the cotangent space, and parametrizes the momenta with the position and velocity  vectors.

As it happens however, under usual assumptions the map $\mbox{LT}$ might not be injective nor surjective. In particular the full $T^*Q$ might not be accessible to the dynamical system. However, this is far from being a
disadvantage of the Hamiltonian approach. Quite the contrary, a dynamical system
may possess some redundancy in its description -- one  is in fact
``over-parametrizing" it -- and this property of the Hamiltonian approach is a
warning sign that the system has this feature.  Let us see
how this is related, in the Lagrangian approach, to the unique determination of
the accelerations from  the velocities and positions.

The condition for the map $\mbox{LT}$  to be (at least locally) an isomorphism
is that the block diagonal matrix
\be \left[\begin{array}{rl}\mbox{Id}&0\\
0&\frac{\delta p_\alpha}{\delta \dot q^\beta}
\end{array}\right]
\ee  be invertible. Since one of the blocks contains the identity Id (this is
just the matrix $\diby {q^\alpha}{q^\beta} $),  we arrive at the same condition
imposed from equation \eqref{equ:accel_lagrange} that such a situation reflects
the fact that
\be \frac{\delta^2 L}{\delta\dot q^{\beta}\delta\dot q^\alpha}=\frac{\delta p_\alpha}{\delta \dot q^\beta}=M_{\alpha\beta}
\ee
has to be invertible.
The constraints on $T^*Q$ that we get usually form submanifolds of $T^*Q$ (this
relies on our assumption that the rank of $M_{\alpha\beta}$ is constant)) and
can thus be put in the form of functionals of $T^*Q$, let us say
$\chi^I(q,p)=0$, which implicitly define the said manifolds (see the
\emph{regular value theorem} \ref{theo:regular_value}).

The regularity assumption of constant rank and the further assumption that the
constraints are \emph{irreducible}\footnote{We assume that all the $\chi^I$ are
linearly independent, i.e., that the one forms $d\chi^I$ are linearly
independent, i.e., that we have an irreducible set of constraints. If they were
not, we would have to choose a basis for the constraints.}, implies again from
theorem \ref{theo:regular_value} that the $\chi^I$ form  a complete coordinate
system in phase space for \emph{the complement of} the constraint surface. For
example, if we had a 4-dimensional phase space with two sets of irreducible
constraints, theorem \ref{theo:regular_value} guarantees that we can find
coordinates in phase space, $x,y,z,w$, such that $x,y$ parametrize the
constraint surface and $z,w$ are given by the two constraint functions. This has
the strong consequence that a vector $X$ is tangent to the constraint surface
\emph{if and only if} $X[\chi^I]=0$. The fact that $X$ is tangent to the
constraint surface already implies the ``if" part, since $\chi^I=0$ on the
entire constraint surface it doesn't change along $X$. The ``only if'' part is a
result of writing everything in a coordinate system and on the fact that indeed
we have a complete coordinate system as mentioned above.

Such constraints, arrived at from the sole definition of the momenta,  are called \emph{primary}, in allusion to the fact that the equations of motion need not be used to derive them. In Lagrangian variables, these relations are merely identities, as we will see in practice in section \ref{sec:constructionPrinciple}. It follows  that the inverse transformation from the momenta to the
 velocities, even when we restrict ourselves to the constraint surface, is
multi-valued. Given a point in phase space that fulfills the constraints, the
``inverse image" through $LT$ is not unique, and in order to render it
single-valued,  and thereby indicate the location of the velocities $\dot q$  on
the inverse manifold, one needs to introduce extra parameters in at least the
same number as there are primary constraints. These
parameters will appear as Lagrange multipliers in the Hamiltonian formulation.

Thus we can see that we do not fully characterize dynamical redundancy solely by
restricting ourselves to the constraint surface. Dynamical redundancy, or
symmetry, is present in the Lagrangian characterization of the system as
much as in the Hamiltonian. The transform  $\mbox{LT}$ being injective
signifies that a restriction to the constraint surface (the image of
$\mbox{LT}$) still ``includes"  the full Lagrangian characterization, symmetries
and all.  Part of the power of the Hamiltonian formulation is exactly that it
gives us a starting point to study redundancy in the description of dynamical
systems, so let us get to it.

The canonical Hamiltonian is defined as
\be\label{equ:def:can_Hamiltonian}H_0:=p^\alpha\dot q_{\alpha}-L
.\ee
If we compute the variation of \eqref{equ:def:can_Hamiltonian}, we get
\begin{eqnarray} \delta H_0&=&\delta p^\alpha\dot q_{\alpha}+p^\alpha\delta\dot q_{\alpha}- \delta\dot q^\alpha\frac{\delta L}{\delta\dot q^\alpha}- \delta q^\alpha\frac{\delta L}{\delta q^\alpha}\nonumber\\
&=& \delta p^\alpha\dot q_{\alpha}- \delta q^\alpha\frac{\delta L}{\delta q^\alpha}.\label{equ:deltaHamiltonian}
\end{eqnarray}
 As the total variation depends only on the variation of $p$ and $q$, this means that  $\dot q^\alpha$ enters $H_0$ only in the precise combination that gives $p_\alpha$, and thus the non-trivial dependence of the Hamiltonian can be set to be just $H(q,p(q,\dot q))$.

Since the system can only access the surfaces in $T^*Q$ defined by $\chi^I(q,p)=0$, and the Hamiltonian is a function of $(q,p)$, one would be inclined to conclude that we may arbitrarily extend the Hamiltonian in $T^*Q$ out of the surface:
\be\label{equ:def_H_1} H_1=H_0+\rho_I\chi^I(q,p)\approx H_0,\ee
where we introduced the notation $\approx$ to mean \emph{weak equality}; i.e.,
equalities that are valid only over the constraint surface. Here we are summing
over the $I$ index, and $\rho_I$ is an arbitrary coefficient. These extra
parameters could then be seen as coordinates on $TQ$ that determine the exact
position (a ``height") over the inverse images of the momenta.
However, such a conclusion would be hasty. The issue, which will be explained
better in  section \ref{sec:geometric_Pb} below, is that the dynamics does not
depend on the value of the Hamiltonian itself, but on its \emph{flow}, or
gradient. In fact, it is true that we can amend the Hamiltonian in such a way
(for first class constraints), but for a different reason than the one stated
above.

From \eqref{equ:deltaHamiltonian} and \eqref{equ:def_H_1} we get
\begin{eqnarray}\label{equ:eoms_hamiltonian1}\dot q^\alpha&=& \frac{\delta H_0}{\delta p_\alpha}+\rho _I\frac{\delta\chi^I}{\delta p_\alpha}\\
- \frac{\delta L}{\delta q^\alpha}=\dot p_\alpha&=& \frac{\delta H_0}{\delta
q^\alpha}+\rho _I\frac{\delta\chi^I}{\delta
q^\alpha}\label{equ:eoms_hamiltonian2}.\end{eqnarray}
The equations of motion \eqref{equ:eoms_hamiltonian1}-\eqref{equ:eoms_hamiltonian2} can be derived from the variation of the Legendre transform of the action with generating function $H$:
\be \delta\int dt \left[\dot q^\alpha p_\alpha- H-\rho_I\chi^I(q,p) \right]=0
\ee
subject to the boundary conditions that the variations vanish at the endpoints.
If we are able to explicitly solve the constraints, i.e., if we can impose the
conditions $\chi^I(q,p)\equiv 0$, then we can use the simpler variational
principle subject to the conditions
\be \delta\int dt \left[\dot q^\alpha p_\alpha- H \right]=0.
\ee

\subsection{Poisson brackets and symplectic flows}\label{sec:1stClass_vs_2nd}

It is through \eqref{equ:eoms_hamiltonian1}-\eqref{equ:eoms_hamiltonian2} that we choose to introduce Poisson brackets into the dynamical analysis, as these equations generalize by the chain rule to arbitrary functionals of the dynamical variables $(q,p)$ to
\be\label{equ:def:Poisson_brackets}\dot
F[q,p]=\diby{F}{q^\alpha}\diby{H}{p_\alpha}-\diby{F}{p_\alpha}\diby{H}{q^\alpha}
=:\{F,H\},
\ee
where a sum over the index $\alpha$ is understood, and the usual notation for
Poisson brackets $\{\cdot,\cdot\}$ was introduced. One can immediately see the
value of Poisson brackets for evolution through a Hamiltonian, but they can be
generalized beyond that, to signify the evolution of any constraint under the
action of another.  This is completely necessary when one starts talking about
symmetries, as one would like to know if one or another constraint is invariant
under its action.

\subsubsection{First- and second-class constraints}

 A more geometric picture of the workings of both first- and second-class
constraints will be given below. For now we give a more pragmatic approach to
the classifications of constraints. A set of constraints $\chi^I$ will  be
called \emph{first class} if their Poisson bracket vanishes weakly on the
constraint surface, i.e. $C^{IJ}=\{\chi^I, \chi^J\}=a_K\chi^K\approx 0$. In
contrast, a set will be called \emph{second class} if $C^{IJ}$ does not vanish
on the constraint surface. A given constraint $\chi^1$ will be said to be first
class with respect to this set if $C^{1J}\approx0$.

 It can easily be seen that for a set of $N$ second-class constraints for
which the $N\times N$ matrix $C^{IJ}$ is not of maximal rank on the constraint
surface, i.e., $\det( C^{IJ})\approx 0$, there exists at least one linear
combination of the constraints that is first class with respect to all of the
rest. By definition, there exists a vector (or an N-tuple) $a_I$ for which
$a_IC^{IJ}=0$. Then clearly $a_I\chi^I$ is still first class. By iterating this
procedure we arrive at a set of purely first-class constraints $\Psi^i$, and
purely second-class ones $\chi^I$. The Poisson bracket matrix is then given by
 \begin{equation}
 \left(
   \begin{array}{cc}
     \{\Psi^i,\Psi^j\}&\{\Psi^i,\chi^I\}\\
     \{\chi^J,\Psi^j\}&\{\chi^J,\chi^I\}
   \end{array}
 \right) \approx
 \left(
   \begin{array}{cc}
     0&0\\
     0&C^{IJ}
   \end{array}
 \right),
\end{equation}
where $C^{IJ}$ is invertible.

Second-class constraints cannot be interpreted as gauge generators,
or, even indeed  as generators of any transformation that is  physically significant. Because it does not preserve the constraints its symplectic flow will take us out of the allowed surfaces for dynamics. So what does one do with second class constraints? We use the invertibility of the matrix of purely second class constraints to define a projection of the dynamics into the constraint surface. That is, we define the Dirac bracket:
\be\label{equ:def:Dirac_bracket}\{\cdot,\cdot\}_{\mbox{\tiny DB}}:=\{\cdot,\cdot\}-\{\cdot, \chi^I\}{C^{-1}}_{IJ}\{\chi^J,\cdot\}
\ee
As can easily be checked,  $\{\chi^I,\cdot\}_{\mbox{\tiny DB}}=0$. I.e. the symplectic flow (defined on Section \ref{sec:geometric_Pb}) of any of the second class constraints automatically vanishes with this bracket. The Dirac brackets effectively  \emph{project} the dynamics to the constraint surface and thus reduce the degrees of freedom of the theory (again, see Section \ref{sec:geometric_Pb}). This obligatory projection of the dynamics implies that the constraints are imposed \emph{strongly}: the constraints should be taken to be zero everywhere (since we are forcefully projecting dynamics to the surface where they are zero). In relatively simple cases one or more pair of conjugate variables can be found such that the purely second class constraints can be solved for them in terms of the other variables. Considering such variables as coordinates in phase space, these values define surfaces in $T^*Q$. We can then completely project dynamics to the surface thus defined by completely eliminating said variables (using the second class constraints as definitions) and reverting to the usual Poisson brackets. In this simple case we eliminate the degrees of freedom of the system that refer to these constraints and move our analysis to the projected surface. In other words,  using the second-class constraints equations as \emph{definitions} we reduce our phase space, and hence our Poisson bracket, to the remaining variables only, expressing all quantities in terms of the remaining variables.  If we cannot find a way to express all second class constraints in such a manner the
dynamics must be formulated using the Dirac bracket, whereby one keeps the second class constraint in their implicit form and all variables are retained.

As a matter of fact, one of the main aspects of the present work is based exactly on what is described here: separate a first class constraint from the purely second class ones and then solve the latter for a pair of conjugate variables.

\subsubsection{Gauge fixings}

In the absence of a true Hamiltonian, i.e., a Hamiltonian not entirely made up
of constraints,\footnote{As a matter of fact, recent work shows that Dirac's
conjecture, namely that all primary first-class constraints generate gauge
symmetries, holds only in the absence of a time labeling. It does not need
to hold for a constraint that generates time reparametrization,  any more than
it does for  a true Hamiltonian \cite{barbour_foster:dirac_thm}.} the  presence
of primary first-class constraints is  associated with gauge symmetry. The
associated gauge freedom indicates that there is more than one set of
canonical variables that corresponds to a given physical state. In practice it is sometimes desirable to eliminate this freedom by imposing further
restrictions on the canonical variables. This should eliminate part (partial
gauge fixing) or all of the arbitrariness in the choice of canonical variables
representing the same physical states.  The inclusion of such extra conditions
in the formalism is permissible
 because they only
remove unobservable elements of the system  and do not impinge on the gauge-invariant properties.

For a certain extra (imposed) constraint $G(p,q)=0$ to be considered a gauge fixing, we must demand two properties of gauge transformations as related to the fixing:
\begin{itemize}
\item {\bf Existence:} The particular choice of gauge that the condition
$G(p,q)=0$ imposes has to be reachable from any point on the constraint surface
through a gauge transformation that this condition purports to fix, i.e., there
has to \emph{exist} a gauge transformation that fixes the gauge to satisfy
$G(p,q)=0$.
\item{\bf Uniqueness:} There must be \emph{only one gauge transformation} that fixes the variables to satisfy the gauge-fixing condition $G(p,q)=0$.
\end{itemize}
These conditions can be similarly formulated in the language of fiber bundles by the concept of a \emph{section} (see section \ref{sec:Riem}).

\subsection{Geometric interpretations and the case of a true
Hamiltonian}\label{sec:geometric_Pb}

\subsubsection{Interlude: geometric interpretation}

We will try not to give too technical an account of  the introduction of
symplectic geometry, but aim to give merely a pedestrian approach to the
meaning  of Poisson brackets of general functions on phase space.
If one looks closely at equation \eqref{equ:def:Poisson_brackets}, one can see that indeed it is a derivation, i.e. it obeys Leibiniz's  rule:
$$\{f,gh\}=g\{f,h\}+\{f,g\}h,
$$ which indicates that we can see the linear operator $\{f,\cdot\}$ as a kind
of vector field in phase space. We then generalize what was done for the
Hamiltonian and define the \emph{symplectic flow} of a given phase-space
function $f$ as ${v}_f:=\{f,\cdot\}$. It will act on other functions as a
directional derivative ${v}_f[h]$ and measure how much $h$ changes in the
direction of ${v}_f$. That is, it measures how other phase-space functions
change  under ``evolution" through the action of the corresponding phase-space
function $f$. Now, as we have already mentioned in section
\ref{sec:hamiltonian_dyn}, the
phase-space function $f$ implicitly
defines a surface through the regular value theorem \ref{theo:regular_value}
provided certain regularity assumptions. In the presence of a metric, one would
usually say that the differential one-form $df$ is ``perpendicular" to the
surface $f^{-1}(0)$, because its dual vector field $df^\sharp$ is defined as
$X[f]=df(X)=:g(df^\sharp,X)$, which obviously vanishes for  any vector field
tangent to $f^{-1}(0)$. In the case of symplectic geometry, one does not define
the analogous operation through the use of a metric but a symplectic
two-form, usually denoted by $\omega$. Explicitly,
\be\label{equ:symplectic_form1} \omega({v}_f,\cdot):=df
\ee and furthermore
\be\label{equ:symplectic_form2} \omega({v}_f,{v}_h)=\{f, h\}.
\ee

Now, just as a vector field can be tangential to a given manifold, so can
symplectic flows.  Suppose then that a surface $\mathcal{N}$ in phase space is
given by the intersection of regular manifolds defined by the inverse values of
the functions $\chi^I$, i.e., $\mathcal{N}:=\{(q,p)~|~\chi^I(q,p)=0 ~\forall I
\}$.  Then $\mathcal{N}$ will be said to be \emph{first class} if: for all
phase-space functions $f$ such that $f$ vanishes on $N$, i.e. $df(X)=0$ for all
$X\in T\mathcal{N}$, then ${v}_f[\chi^I](p,q)=0$ for all $I$ and $(p,q)\in N$.
The statement is equivalent to the much simpler statement  that
$\{f,\chi^I\}=a_J\chi^J$, since this will indeed be zero whenever we are on the
surface. The geometric translation is indeed very simple: all symplectic flows
${v}_f$ of functions $f$ that vanish on the surface $\mathcal{N}$ are tangent to
the surface.

By contrast, we can define a second-class manifold (or set of regular functions
$\chi^I$) if all  symplectic flows (of functions that vanish on the surface)
take us out of the surface (i.e. are not tangent to it).

As anticipated in section \ref{sec:hamiltonian_dyn}, we now explain with more
completeness why we must add arbitrary summands of constraints to the
Hamiltonian function. As we saw, by the regularity assumptions, a  phase-space
vector $X$ is tangent to a first-class constraint surface if and only if
$X[\chi^I]=0$ for all the $\chi^I$ making up the said surface. Thus by the above
\eqref{equ:symplectic_form1}, for any such vector
$$\omega({v}_\chi, X)=0,
$$
meaning that for the pull-back (or, let us say, the projection) of the
symplectic form $\omega$ to the constraint surface, $\tilde\omega$, the
directions given by ${v}_\chi$ are degenerate, and dynamical flows are not
uniquely defined, i.e., $\tilde\omega({v}_\chi,\cdot)=0$. Thus, as their
dynamical effect is not felt over the constraint surface, we can arbitrarily add
summands of $\chi^I$ to the definition of the Hamiltonian function
\eqref{equ:def_H_1} without further consequence.

 \subsubsection{The case of a true Hamiltonian.}

If we have a Hamiltonian that consists of ``pure constraints", as happens in GR,
then after separating the constraints into pure second- and first-class ones,
and solving for the second-class ones (and thereby setting them strongly to
zero), we are done, as we are left with only first-class constraints. Thus  all
smearings (Lagrange multipliers) in the total Hamiltonian would propagate all
the constraints, making the dynamical system consistent. But if we have a true
Hamiltonian, let us call it $H_0$, we have more work to do.

 As we will not be dealing with this result directly,  we but briefly remark on
the Dirac procedure, using the geometric interpretation presented above. What
geometrically happens when we are obliged to add constraints to the theory? Let
us start with, say, the initial primary first-class constraints $\chi^I_1$.
We are restricting the domain of the dynamics, as we said before, to a
subsurface of the total phase space, ${(\chi^I_1)}^{-1}(0)$,  and we must add
the new constraints to the Hamiltonian with some Lagrange multipliers,
$\rho_I\chi_1^I$, as they will have no observable effect on the dynamics.

Let us call this initial surface $\mathcal{J}_1$. We have to check if the
Hamiltonian propagates the corresponding constraints. In the geometric picture,
this means we have to find a subsurface $\mathcal{J}_2\subset\mathcal{J}_1$ to
which the symplectic flow of the Hamiltonian is tangent. The problem
is that the flow ${v}_H$ might be tangent to  $\mathcal{J}_1$ \emph{only} at the
points $\mathcal{J}_2\subset \mathcal{J}_1$, i.e., we can only say that the
Hamiltonian vector field is contained in the tangent space
$T_{\mathcal{J}_2}\mathcal{J}_1:=\{v\in T_q\mathcal{J}_1~|~q\in
\mathcal{J}_2\}$. It does not need to be tangent to the entirety of
$\mathcal{J}_1$. Of course, ${v}_H$ is contained in the ``full" tangent space to
$\mathcal{J}_1$ at those points, and need not be tangent to the subspace
$\mathcal{J}_2$ itself (i.e., contained in $T\mathcal{J}_2$). But now we cannot
restrict dynamics to $\mathcal{J}_2$ because the Hamiltonian flow will take us
out of that surface again. And thus we must find a subsurface of $\mathcal{J}_2$
to which the Hamiltonian flow is tangent, i.e., ${v}_H\in
T_{\mathcal{J}_3}\mathcal{J}_2$.

We keep on doing this until we finally reach a $\mathcal{J}_k$ over which the
symplectic flow ${v}_H$ is tangent to the \emph{entire} surface. At the end of
the algorithm, we will be left with a surface (a set of
constraints $\chi^I$) over which the extended Hamiltonian
$H_0+\rho_I\chi^I(q,p)$ is completely tangent. Then the dynamics is  said to be consistent.

\subsection{ADM 3+1 split}\label{sec:ADM}

General relativity in its original formulation is very elegant and powerful, describing the physics of space-time simply as 4-dimensional Lorentzian geometry. While this is indeed a very simple  framework, we human beings do not directly observe space-time, but instead we notice an evolution, or change, of space. Therefore, under some circumstances, it is very useful to have a more direct translation between experience and theory by formulating what is called a 3+1 description of general relativity.

\subsubsection{Gauss--Codazzi relations}

To arrive at the so-called 3+1 description, we have to assume that space-time,
$(M,^4g)$, where $M$ is a four-dimensional manifold and $^4g$ is a Lorentzian
metric on it, is diffeomorphic to the direct product $\R\times\Sigma$, where
$\Sigma$ is a 3-dimensional manifold representing space and $t\in\R$ represents
time. Such space-times are called {\it globally hyperbolic} and exist if and
only if  the primary condition that allows us to split space and time is
satisfied. Namely, we have to assume \emph{causality}; that no closed time-like
curves\footnote{Curves $\gamma:S^1\to M$ such that $^4g(\gamma',\gamma')<0$.}
exist \cite{Bernal:2006xf}. Of course, a particular slicing of space-time will
still be a matter of choice, {\it not} considered in the standard presentation
of GR to be something intrinsic to the world.\footnote{The main aim of this
thesis however is to convince the reader that indeed a natural splitting does
exist.} A choice of such a slicing is equivalent to a choice of a regular
function $f:M\to \R$ (in this case, this is equivalent to saying that the
gradient of $f$ is not zero anywhere) for which $\partial^\mu t$ is time-like.

\subsubsection{Adapted coordinates, shift and lapse.}

 As we have assumed that the time function is regular, the regular values of $f$ form 3-dimensional manifolds, which we call $\Sigma(t_0)=f^{-1}(t_0)$.
   Using the submersion theorem, we can always find a local coordinate system
$\{x^{\bar\mu}\}$ over the open set $U$,  where, for $p\in U$,
$f(p)=f(x^0(p),\dots ,x^4(p))=x^0(p)$, and we use barred variables only when we
feel we need to emphasize that we are in an adapted coordinate system.

   The one-form $df$ is then given by $dx^0$ and the intrinsic coordinates of
each hypersurface are given by $x^1,x^2,x^3$. Thus  the vectors
$\partial_{a}:=\frac{\partial}{\partial x^{a}}$ span the tangent space to each
hypersurface, where we used latin indices to denote the adapted spatial
coordinates.  To make coordinate independence more transparent, we can express
the components of these vector fields in terms of a general basis
$\{y^{\alpha}\}$ as $e_{\bar\nu}$:
   \be\label{equ:projection_vecs} \frac{\partial}{\partial x^{\bar\nu}}=\frac{\partial y^{\alpha}}{\partial x^{\bar\nu}}\frac{\partial }{y^{\alpha}}=:
e_{\bar\nu}^{\alpha}\partial_{\alpha},\ee
 where  $e_{\bar\nu}^{\alpha}$ can be interpreted as the components of the
vector field $\partial_{\bar\mu}$ (i.e., as the vector field itself). As
$\partial_a$ is tangent to the hypersurfaces and $\partial_\alpha$ is a general
coordinate vector field, $e_{a}^{\alpha}$ can alternatively be seen to act as a
projection onto the hypersurface $\Sigma$ \cite{Poisson:2003nc}.

     Let $v^\mu$ be tangent to $\Sigma$. As the value of $f=x^0$ is constant
over each surface,  $v[f]=v^\mu \partial_\mu f=0$ by definition. In the
adapted coordinates, this is just saying $v^0=0$. In this subsection, we will
try to keep both notations, $f$ and $x^0$, side by side, so that the reader does
not forget that it is actually an arbitrary function that is defining the
hypersurfaces. Let us pause to note the important geometric fact that in order
to define the foliation we need only a regular function $f$, which does not
require the aid of coordinate systems. However, when we define the curves
parametrized by $x_0$, all other coordinates being held constant, we have made
an arbitrary choice of coordinates and endowed our description with extra
structure. Since $df=dx^0$ is defined independently of this structure, it is not
necessary that the vectors tangent to the \emph{chosen} coordinate curves $x^0$,
$\partial_0:=\frac{\partial}{\partial x^0}$ have much to do with the previously
existing one-form $df$. Let us see what this implies.

      The superscipt $\flat$ usually denotes dualization of a one-form to a
vector field by use of the metric. We adjoint a $4$ to it, to make clear that we
are using the full four-metric. Then, using the notation $^4\flat$ to mean the
metric dual to the one-form $dx^0$, we have the vector field (\emph{not} written
in components):
     \be\label{equ:metric_decomp}
(dx_0)^{^4\flat}:={^4g}(dx^0,\cdot)={^4g}^{\mu\nu}\partial_\mu f\otimes
\partial_\nu=\partial^\nu f\otimes \partial_\nu={^4g}^{0\nu}\partial _\nu.\ee
     Thus the vector field with components $\partial^\nu f$, or ${^4g}^{0\nu}$ in adapted coordinates, is a (un-normalized) normal to the hypersurfaces. We call $n^\mu$ the unit normal to $\Sigma$, to which $\partial^\mu f$ is parallel. It is  straightforward to find the metric induced norm squared of $\partial^\mu f$, in adapted coordinates, using \eqref{equ:metric_decomp}:
     \be\label{equ:time_proj_N} ||(dx_0)^{^4\flat}||^2=||(dx_0)||^2=-{^4g}^{00}.
     \ee

      Of course, for a general 4-metric  the metric dual to the one-form $dx^0$
is {\emph not} equal to $\partial_0$, which is the algebraic dual to $dx^0$ and
tangent vector to the $x^i=\text{const}$ curves. By \eqref{equ:metric_decomp},
we can tell that this is the case if ${^4g}^{0\nu}=\delta^{0\nu}$.
      Thus we cannot say that the vectors tangent to the $x^0$ coordinates are orthogonal to the  hypersurfaces $\Sigma$. We decompose $\partial_0$ into its components parallel to the hypersurface, $N^\mu$, and orthogonal to it $Nn^\mu$. In short
   \be\label{equ:shift_def} \partial_0=Nn+\vec N.\ee
Since $dx^0(\partial_0)=1$ we get from \eqref{equ:time_proj_N} that
\be\label{equ:def:N} -N^2={^4g}^{00},
\ee
and we can simply define $\vec N$ through $\partial_0-Nn$.

\begin{figure}\label{fig:3+1decomposition}
 \begin{center}
 \includegraphics[width=11cm]{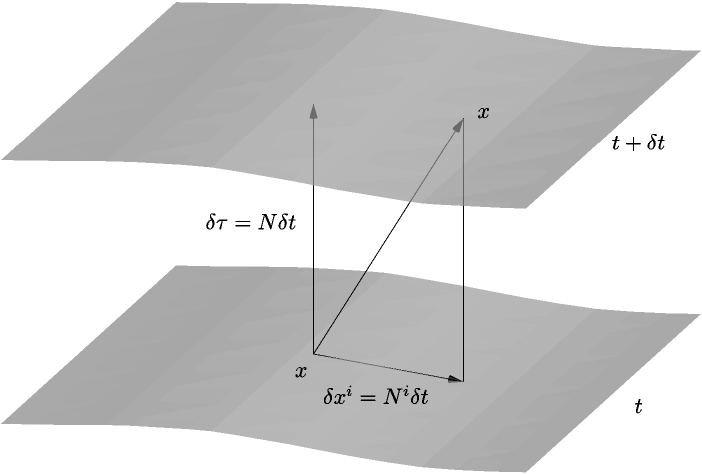}

\caption{The $3+1$ decomposition of spacetime.}

 \end{center}

\end{figure}

Alternatively, using the projectors $e_{a}^{\alpha}$, and  abbreviating $\partial_0$ by the vector $t^\alpha$:
\be\label{equ:shift_def2}t^\alpha=Nn^\alpha+N^a e_{a}^{\alpha}
\ee See figure \ref{fig:3+1decomposition}.

   The four metric $^4g$ induces a metric on $\Sigma$, which is just its
restriction to vectors tangential to $\Sigma$. We call this induced metric $g$
and can straightforwardly check that
   \be\label{equ:intrinsic_metric}g_{\mu\nu}={^4g}_{\mu\nu}+n_\mu n_\nu
   \ee is indeed the induced metric. Using \eqref{equ:intrinsic_metric}, we
write the orthogonal projection operator onto $\Sigma$ as
$g_\mu^\nu=\delta_\mu^\nu+n_\mu n^\nu$. It is easy to check that $g_\mu^\nu
n^\mu=0$ and  $g_\mu^\nu g^\mu_\rho=g^\nu_\rho$. Alternatively, we can use the
components $e_a^\alpha:=\partial_a^\alpha$ as defined in
\eqref{equ:projection_vecs} to project indices. Thus, in intrinsic $\Sigma$
coordinates
\be\label{equ:intrinsic_metric2} g_{ab}=e_a^\alpha e_b^\beta
({^4g})_{\alpha\beta}.
\ee
 Both definitions have their advantages and disadvantages.

\subsubsection{Extrinsic curvature}

 The extrinsic curvature is a two-form, given by the tangential component of the covariant derivative of the normal vector. Quite a mouthful, so let us write it out explicitly in coordinate-free notation:
 \be\label{equ:extrinsic_curvature}K(u,v):=^4g(D_un,v)=-^4g(n,{D}_uv)=^4g({D}_vn,u)
 \ee
where we define the Levi-Civita covariant derivative associated with $^4g$ as
$D={^4\nabla}$. In the next to last equality, we used the metric compatibility
of the connection and orthogonality of $u$ and $v$ and, in the last, we noticed
that by Frobenius theorem the commutator $[u,v]$ is tangent to $\Sigma$, which
enabled us to write $-^4g(n,{D}_uv)=-^4g(n,{D}_vu)$.
Since $n$ is normalized, we can also straightforwardly check that
$K(n,\cdot)=0$; thus, although it depends on the normal $n$, which is not
intrinsic to $\Sigma$,  we can write the extrinsic curvature with indices in
$\Sigma$, as $K_{ab}$. We will denote the trace $g^{ab}K_{ab}=K$.

We can split the covariant derivative for vector fields on $\Sigma$ into normal
and  parallel components, defining the intrinsic covariant derivative to
$\Sigma$ as
\be {D}_uv= K(u,v)n+\nabla_uv.
\ee
In this way the definition of $\nabla$ is intrinsic on $\Sigma$; fact, given
that the original covariant derivative is the Levi-Civita one for $(M,^4g)$ --
that it is metric preserving and torsion-free --  it can be shown that $\nabla$
is the Levi-Civita one for $(\Sigma, g)$.

\subsubsection{Gauss--Codazzi relation}
Using this decomposition, we can rewrite the 4-dimensional Ricci scalar ${^4R}$,
and thus the Einstein--Hilbert Lagrangian density, in terms of the intrinsic
geometry of $\Sigma(t)$ and $K^{ab}$:
\be\label{equ:Gauss_codazzi0}
^4R=(R+K^{ab}K_{ab}-K^2)-2({n^\alpha}_{;\beta}n^\beta-n^\alpha
{n^\beta}_{;\beta})_{;\alpha}.
\ee
The final thing we must do is express $\sqrt{^4 g}$ in terms of our present set of dynamical variables. The expression for the determinant gives us, since $N$ is precisely the projection of $\partial_0$ along the normal, $$^4g^{00}=\frac{\text{cofactor} ({^4g}_{00})}{g}=-N^2$$
we get that $\sqrt{-{^4g}}=N\sqrt g$.
Up to boundary terms,  we now have
\be\label{equ:Gauss_codazzi}\int_M d^4 x R\sqrt
{-{^4g}}=\int_{t_1}^{t_2}dt\int_{\Sigma_t}d^3x(R+K^{ab}K_{ab}-K^2)N\sqrt g, \ee
where $R$ is the intrinsic 3-dimensional Ricci scalar of $g$.
\subsection{Constraint algebra for ADM.}\label{sec:const.alg_ADM}
Now, to find the Hamiltonian, we must express \eqref{equ:Gauss_codazzi} in terms
of the metric velocities $\dot g_{ab}=\mathcal{L}_{\mathbf{ t}} g_{ab}$. By
definition $\mathcal{L}_{\partial_{\bar\mu}}\partial_{\bar\nu}=0$. In
particular, using components as in \eqref{equ:projection_vecs},
$$ \mathcal{L}_te^\alpha_a=0.
$$
  Using this last equation in \eqref{equ:intrinsic_metric2}, we have
\be \mathcal{L}_t g_{ab}= e_a^\alpha e_b^\beta
\mathcal{L}_t({^4g})_{\alpha\beta}=e_a^\alpha e_b^\beta(D_\alpha t_\beta+D_\beta
t_\alpha).
\ee Using \eqref{equ:shift_def2}, we have
\begin{eqnarray*}D_\alpha t_\beta+D_\beta t_\alpha &=& D_\beta(Nn_\alpha+N_\alpha)+D_\alpha(Nn_\beta+N_\beta)\\
&=&2n_{(\alpha} N_{,\beta)}+2ND_{(\beta} n_{\alpha)}+2D_{(\alpha}N_{\beta)}.
\end{eqnarray*} Upon projection
\be\label{equ:dot g} \dot g_{ab}=2NK_{ab}+2N_{(a;b)},
\ee or, to put it the other way around,
\be\label{equ:def:extrinsic_curv} K_{ab}=\frac{1}{2N}(\dot g_{ab}-2N_{(a;b)}).
\ee

We first rewrite the integrand of \eqref{equ:Gauss_codazzi} as
\be (R+K^{ab}K_{ab}-K^2)N\sqrt g= (R+G^{abcd}K_{ab}K_{cd})N\sqrt g,
\ee where $G^{abcd}=g^{ac}g^{bd}-g^{ab}g^{cd}$ is the DeWitt supermetric.  We get
\be\label{equ:def:metric_momenta}\pi^{ab}=G^{abcd}K_{cd}\sqrt g
\ee
Now, from \eqref{equ:Gauss_codazzi} and using \eqref{equ:dot g}, we get
\begin{eqnarray*} \pi^{ab}\dot g_{ab}&=&
G^{abcd}K_{ab}(2NK_{cd}+2N_{(c;d)})\sqrt g.
\end{eqnarray*}
Since
$$N G^{abcd}K_{ab}K_{cd}\sqrt g=N\frac{G_{abcd}\pi^{ab}\pi^{cd}}{\sqrt g},
$$
where $G_{abcd}=g_{ac}g_{bd}-\frac{1}{2}g^{ab}g^{cd},$ we finally get
\be H=\int_{t_1}^{t_2}dt\int_{\Sigma_t}d^3x \left(
(\frac{G_{abcd}\pi^{ab}\pi^{cd}}{\sqrt g}-R\sqrt g)N-2\pi^{ab}N_{(a;b)}\right).
\ee
Note the important fact that, being a vector field, the shift is originally written as $N^a$, and thus we should write
 $$ -2\pi^{ab}N_{(a;b)}= -2\pi^{ab}N^c_{;b}g_{ca}.
 $$We can now consider the constraints. We first have $\pi_N=0$,  which by the Hamilton equations means $\dot\pi_N=\frac{\delta H}{\delta N}=0$. In turn, this enforces the \emph{ scalar constraint}
\be\label{equ:scalar constraint}S(x):= \frac{G_{abcd}\pi^{ab}\pi^{cd}}{\sqrt g}(x)-R(x)\sqrt g(x)=0.
\ee
Similarly, we obtain the vectorial momentum  constraint
 \be\label{equ:momentum constraint0}H_a:= g_{ca}{\pi^{cb}}_{;b} =0,
\ee  which is many times written in the equivalent form
\be\label{equ:momentum constraint} H^a:={\pi^{ab}}_{;a}=0
\ee Both of these can be rewritten using the extrinsic curvature:
\begin{eqnarray}R-K^{ab}K_{ab}-K^2&=&0\label{equ:scalar constraint2}\\
(K^{ab}-g^{ab}K)_{;a}&=&0.\label{equ:momentum constraint2}
\end{eqnarray}

This is the point of departure for a constraint analysis of the 3+1 formulation of general relativity. We get the constraint algebra calculated in \eqref{equ:Dirac_ctraint_alg}:
\begin{eqnarray}
\{S(N_1),S(N_2)\}&=&g^{ab}H_b(N_1\nabla_a N_2-N_2\nabla_a N_1)\label{equ:ADM_alg:SS}\\
\{S(N),H^a(\xi_a)\}&=&-S(\mathcal{L}_\xi N)\\
\{H^a(\xi_a),H^b(\eta_b)\}&=& H^a([\xi,\eta]_a)\end{eqnarray}
where we use the notation for smearing $S(N)=\int d^3 x N(x)S(x)$ and \\ $H^a(\xi_a)= \int d^3 x H^a(x)\xi_a(x) $, $N\in C^\infty(M)$ and $\xi_a\in \Gamma^\infty(TM)$ is a smooth vector field.

Note that \eqref{equ:ADM_alg:SS} involves the infamous ``structure functions" $g^{ab}$  when we use the correct form of the momentum constraint \eqref{equ:momentum constraint0}.\footnote{Note that $\nabla_a N=dN$, which as a one-form, does not involve the metric at all. }It is the appearance of the metric in the original form of the momentum constraint that flushes out the appearance of the  ``structure functions", as opposed to structure constants  in the Dirac algebra.

\section{A brief history of 3D conformal transformations in standard general
relativity}

To orient the reader on how the present work originated, we give here, in some
detail, an account of previous work that led up to it. We start with the
attempt by Weyl to introduce some notion of relativity of size into the
structure of general relativity, an enterprize very close in spirit to our own
motivations. We then discuss how first Lichnerowicz and then York successfully
developed the 3-dimensional conformal tools to solve the initial-value problem
of GR for almost all initial data.

\subsection{The Weyl connection}

In 1918, H. Weyl had a happy thought \cite{weyl:conformal_1918}. If, when
generalizing Euclidean geometry to Riemannian geometry, we need extra
information to characterize parallel directions at different points, shouldn't
we worry about how to characterize ``parallel" (or equal) lengths? The
assumption of equal lengths comes from one of the elements of the definition of
the Levi-Civita connection; namely, that it preserves the metric tensor:
$$ Z[ \mathbf{g}(X,Y)]=\mathbf{g}(\nabla_Z X,Y)+\mathbf{g}(X,\nabla_Z Y).
$$
Weyl's idea was to include in the definition of the connection a one-form
$\theta$ such that
$$ Z[ \mathbf{g}(X,Y)]=\mathbf{g}(\nabla_Z X,Y)+\mathbf{g}(X,\nabla_Z Y)+\theta(Z) \mathbf{g}(X,Y).
$$
In other words, it would no longer be true that $\nabla g=0$, but $\nabla g=g\otimes \theta$. In this way, the extra term says that even when we parallel translate the direction of a vector there is also an infinitesimal change in its length, given by its initial length times the value of the one form $\theta$ in the given direction. This he hoped to be connected to the electromagnetic $U(1)$ connection $A^\mu= \theta^\mu$. However, as Einstein  soon pointed out, if $\theta$ was non-zero, the lengths of objects would be path-dependent, something not observed in Nature.

Our construction bears strong similarities to Weyl's initial attempts,
especially as regards this question: how do we compare lengths at distinct
points? According to relationalist principles, we in fact cannot. One way to get
around Einstein's criticism would be to limit the  Weyl potential $\theta^\mu$
to be given by $\theta^\mu=\partial^\mu \phi$ for some scalar function $\phi$.
In this way we would have an integrable connection and lengths would still be
``relative" but would not depend on the path taken. Unfortunately this solution
ceases to be interesting for incorporating electromagnetism, because it
obligates the curvature tensor $F^{\mu\nu}$ to be zero.  Nonetheless, we are not
interested in using conformal transformations for the coupling of
electromagnetism, and Weyl's  enquiries were  an important stimulus for the
further work on the meaning of relative size that has culminated in the work
presented in this thesis.

\subsection{Lichnerowicz and York's contribution to the initial value problem.}\label{sec:LY}

Conformal transformations are here defined as those transformations that change the local spatial scale. A priori they  have nothing to do with the passage of time and therefore appear to have nothing in common with the scalar Hamiltonian constraint.
Yet, in a study begun through purely mathematical considerations, the great relativist James York came to quite a revolutionary conclusion: we can adjust the local scale so as to find appropriate initial data for GR, i.e. data that solve the scalar and the momentum constraints.  He began his 1973 paper \cite{York:york_method_prl} on the conformal approach to the initial value problem by stating:
\begin{quote}An increasing amount of evidence shows that the true dynamical degrees of freedom of the gravitational field can be identified directly with the conformally invariant geometry of three--dimensional spacelike hypersurfaces embedded in spacetime.[...] the configuration space that emerges is not superspace (the space of Riemannian three--geometries) but `conformal superspace'[the space of which each point is a conformal equivalence class of Riemannian three--geometries]$\times$[the real line](ie, the time, $T$).''\end{quote}
 Perhaps a more careful choice of words would have been ``An increasing amount of evidence \emph{suggests}", as, although it was indeed shown by York
   that one could \emph{construct initial data} for GR using three-dimensional conformally invariant initial data, a conformally invariant version of general relativity -  with its own sets of conformally invariant constraints and evolution equations - was not developed. The use of the conformal factor was input by hand to aid in the solvability of an equation. It did not involve any sort of canonical analysis and thus did not contain stronger statements about the dynamical system as a whole.

The first important step towards solving the initial value problem for GR, given by \eqref{equ:scalar constraint} and \eqref{equ:momentum constraint}, was taken by Lichnerowicz \cite{Lichnerowicz}. He did so by realizing that if $K^{ab}$ is \emph{traceless}, then the \eqref{equ:momentum constraint2} means $K^{ab}$ must also be divergenceless, or \emph {transverse}.

Now, transverse traceless (TT) tensors are equivariant with respect to conformal transformations. That is, if $A^{ab}$ is a $TT$ tensor with respect to $g$, conformal transformations  act on $A_{ab}$ in such a way that the conformally transformed $A^{ab}$ is TT with respect to the transformed metric. For more information on this, see section \ref{sec:York_splitting}. There we also show that if $g_{ab}$ transforms\footnote{Note that Lichnerowicz and York did not use the exponentiated action of the conformal group, which differs from our treatment. Because of this they had to deal with other questions, such as positivity of the conformal factor.  } as ${4\phi}g_{ab} $ then $A_{ab}$ must transform as ${-2\phi}A_{ab}$. A short explanation for this conformal weighing is that, besides the usual ${4\phi}$ factor, the ${6\phi}$ coming from the density $\sqrt g$ must be compensated for.

Alternatively, in the language of inner products of metric velocities in Riem (see Chapter \ref{sec:Riem}), to maintain the conformal invariance of the superspace inner product, one must demand that the lapse have the conformal weight given in definition \eqref{defi:conf_lapse}, section \ref{sec:conformal_diff_group}. Then straightforwardly \eqref{equ:dot g} yields the appropriate weight.

A more straightforward procedure is to not use the extrinsic curvature formulation, but the momentum one. Then, calling $\sigma^{ab}$ the traceless part of $\pi^{ab}$ we get the weighting:
$$\sigma^{ab}\rightarrow e^{-4\phi}\sigma^{ab}
$$ which matches the conformal weight associated with the momenta in the rest of this work. We shall call $\sigma_{TT}^{ab}$ a choice of transverse  $\sigma^{ab}$.

Now since a conformal change in the TT tensor will still satisfy the momentum constraint \eqref{equ:momentum constraint2}, one can choose an arbitrary one and try to solve for it the modified scalar constraint, given from \eqref{equ:scalar constraint} as
 \be\label{equ:modified S} -\bar R+\frac{\sigma_{TT}^{ab}\sigma^{TT}_{ab}}{g}=0
 \ee
 where $\bar R$ is the conformally transformed Ricci scalar obtained from \eqref{equ:deltaR}:
\be\label{equ:RicciConf}\bar R=R[\phi^4 g]=-8\phi^{-5}\nabla_g^2 \phi+ R[g]\phi^{-4} .\ee It is sometimes useful to rewrite this as:
\be\label{equ:RicciConf2}R[\phi^4 g]=\phi^{-5}(-8\nabla_g^2 \phi+ R[g])\phi.
\ee
The scalar constraint thus becomes:
\be\label{equ:Lich}8\nabla^2 \phi-R\phi+\frac{\sigma_{TT}^{ab}\sigma^{TT}_{ab}}{g}\phi^{-7}=0
\ee
Of course,  \eqref{equ:modified S} only makes sense if we can find a conformal transformation that makes $\bar R$ positive everywhere. Such metrics are said to be in the positive Yamabe class, and this imposes a restriction on initial data to belong to this class.

In 1970, James York contributed to the program by adding a constant trace term to the TT momenta $\pi^{ab}=\sigma^{ab}_{TT}+\frac{1}{3}c\sqrt {g}g^{ab}$, where $c$ is a spatial constant. With this simple addition,  the scalar constraint as an equation for the conformal factor becomes:
\be\label{equ:LY} 8\nabla^2 \phi-R\phi+\frac{\sigma_{TT}^{ab}\sigma^{TT}_{ab}}{g}\phi^{-7}-\frac{2}{3}\phi^5c^2=0.
\ee
As long as $c\neq 0$, this places no restriction on the scalar curvature ab initio. In \cite{Niall_73}, York and O\'Murchadha, using Leray--Schauder degree theory, showed that the specific form of the polynomial in $\phi$, $R\phi+\frac{\sigma_{TT}^{ab}\sigma^{TT}_{ab}}{g}\phi^{-7}-\frac{2}{3}\phi^5c^2$, implies that, as long as $\pi^{ab}\not\equiv 0$, equation \eqref{equ:LY} always possesses a unique solution. We will not go into details of the proof, as it is involved and requires too much background material. The important point is that  the initial value problem was shown to be solvable for any choice of metric, TT tensor, and non-zero constant $c$. The initial data that are constructed have constant trace of the extrinsic curvature and are thus
called a constant--mean--curvature (CMC) solution to the constraints. From the physics point of view, this method, which did not arise in any way from canonical analysis, has been regarded as a felicitous ``device" for solving the initial--value problem, which distances the LY method from Shape Dynamics.  It did not yield, as stressed in the beginning of the section, a conformally invariant theory. We will have further comments on the mathematical similarities once we have presented Shape Dynamics in its full form, at the end of chapter
\ref{sec:SD_CMC}.

\subsection{Dirac's fixing of the foliation.}\label{sec:Dirac}
In 1958, Dirac \cite{Dirac:CMC_fixing} saw the need to fix the foliation of GR for the Hamiltonian framework, as a step for quantization.  In effect (as we later discovered), he essentially describes the steps we take in chapter \ref{chapter:SD_AF} after enforcing the gauge fixing. Indeed, the gauge-fixing that he attempts to work with, and deems the most natural, is given by $\pi=0$.

Let us briefly review the main steps. After basically (re)constructing the 3+1 decomposition and the constraints, defining the Dirac bracket, and pinpointing foliation invariance as the main obstacle to quantizing GR, all in under 4 pages, Dirac recognized that a more powerful approach to quantizing GR would be to fix the gauge of the scalar constraint, after which it would no longer be required as an operator equation on a wave-functional. He then proceeded to show  how one would go about doing that.

In the general setting of dynamical systems, he abstractly described the method contained in section \ref{sec:1stClass_vs_2nd} and introduced a gauge fixing in a manner which we now describe.  Suppose there  are initially $\chi^m$ first class constraints.  Introduce $Y^n, ~n=1\cdots N$ gauge-fixing constraints, all second class with respect to the initial constraints. Thus there are now $2N$ second class constraints, and   we must separate the first and second class sets of constraints. As ``there is no room for second class constraints in the quantum theory",  we must either use the Dirac bracket, or completely solve for the second class ones. The two procedures are equivalent. Suppose that $N$ of the second class constraints are of the form $p^n=0~,~n=1\cdots N$, where $p^n$ is the  momentum conjugate to $q_n$. That means that the remaining second--class constraints must contain all of the coordinates  $q_n~,~n=1\cdots N$ in a linearly independent manner, otherwise there would be at least one $p^n$ which would  still be first--class.\footnote{We basically use the reciprocal argument when we come to equation \eqref{equ: matrix_1st_vs_2nd}.} This means it might  in principle be possible to solve the remaining second class constraints for $q_n$, i.e.:
\be\label{equ:dirac's_solution} q_n=f_n(q_{N+1},\cdots,p_{N+1},\cdots)
\ee
Then using equation \eqref{equ:dirac's_solution} and $p^n=0$ we can completely eliminate these variables from the system (and obviate the need for a Dirac bracket), as they play no effective role.  We will not get to grips with exactly how Dirac effectively performed this fixing with $\pi=0$ in this section, but leave it for section \ref{chapter:SD_AF}, where a better description and direct comparison  are more natural.

We  pause to mention that this procedure basically outlines what we will do with general relativity to arrive at SD. The major departure from  Dirac is that we will introduce extra degrees of freedom, and thus our equation analogous to $p^n=0$ will be given by $\pi_\phi(x)=0$, and a good part of our efforts will be devoted to proving that we can indeed solve the remaining second class constraints for $\phi$ as a functional of $({g},{\pi})$, culminating in Theorem \ref{prop}.

Dirac's procedure, unlike ours, is not coordinate independent, but if put on a firmer grounding (in other aspects as well) might well have culminated in our results for asymptotically flat SD, of chapter \ref{chapter:SD_AF}.
However the manner in which we perform our trading explicitly maintains conformal symmetry to very significant advantage. Nonetheless, however unwittingly, Dirac's attempt implicitly used conformal methods applied to the quantization of gravity.

\section{Barbour et al's work. }\label{sec:Barbour}

The most influential of all previous work on conformal methods however came from Barbour et al, in work contained in various papers: \cite{Barbour94, barbour-2002-19, barbour:bm_review} and especially \cite{Barbou:CS_plus_V}. We will try to keep the account of this beautiful body of work  to a minimum. We do this in the interest of brevity, but more importantly,  this work is more eloquently described than what we would be able to achieve here in various sources (for the most updated, complete and masterfully written account, see \cite{Barbour2011}, whose reading we strongly encourage). We would fear misrepresenting the area in any attempt to be complete.

\subsection{Poincar\'e's principle.}

\begin{figure}\label{fig:triangle_shape_space}
 \begin{center}
 \includegraphics[width=11cm]{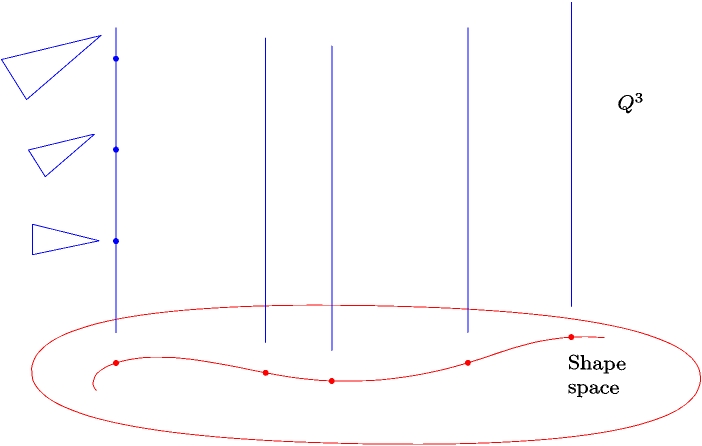}

\caption{Shape space for 3 particles. The points on a vertical line correspond
to the
different representations in Euclidean space of a shape of the triangle formed by the particles. The effects of rotation and scaling are shown.}

 \end{center}

\end{figure}

To introduce some of the main ideas, let us consider a Newtonian system of $N$ particles. Although it seems completely
transparent when expressed in an inertial frame of reference, from the relational point of view the dynamics are not determined uniquely from initial interparticle separations and their rates of change. One also needs as extra data the angular momentum of the entire system, which is not encoded in such data. For a relationalist, this is disturbing, as was already noted by Poincar\'e. This discrepancy led Barbour to formulate what he called the Poincar\'e principle. We will let Barbour explain this concept in his own words:
\begin{quote}Poincar\'e,
writing as a philosopher deeply committed to relationalism, found the need for
them [the extra data] repugnant [Poincar\'e 1902, Poincar\'e 1905]. But, in
the face
of the manifest presence of angular momentum in the solar system, he resigned
himself to the fact that there is more to dynamics than, literally, meets the
eye.[...]{Poincar\'e's penetrating analysis [...] only takes into account the role of angular momentum in the `failure' of
Newtonian dynamics when expressed in relational quantities. Despite its
precision and clarity, it has been almost totally ignored in the discussion of
the absolute vs relative debate in dynamics.}[...]
For some reason, Poincar\'e did not consider Mach's suggestion
[Mach 1883] that the
universe in its totality might somehow determine the structure
of the dynamics observed locally. Indeed, the universe exhibits evidence for
angular momentum in innumerable localized systems but none overall. This
suggests that, regarded as a
closed dynamical system, it has no angular momentum and meets the
\emph{Poincar\'e principle}: [...] a
point and a tangent vector  in the universe's
shape space determine its evolution. \end{quote}
Now of course we are faced with the question; what exactly is shape space? To define it requires some degree of arbitrariness. For example in the case of the $N$ particles, we could deem shape space to be given by a $3N-6$ dimensional space of Euclidean coordinates ($3N$) minus translations and rotations ($6$), which do not change the inter-particle separations. Or, if we are more radically relationalist, we can also argue that, having no absolute ruler, we can only compare distances, and so one of the distances serves as unity, giving us $3N-7$ dimensions to this space. According to Barbour, this choice is the only possible in a complete relationalist setting. We will try to translate these concepts to geometrodynamics soon.
We illustrate the concept of shape space as a quotient of configuration space for the case of a triangle (or just $N=3$) in figure \ref{fig:triangle_shape_space}.
A quotient space, in pedestrian language, is a space $X$ obtained from some other space $Y$ by considering certain elements of $Y$ to be equivalent. By using this concept we will get rid of such extraneous structure. It is to  have a theory existing in shape space that the tool of best-matching was devised.

\subsection{Best matching.}

One of the keys to understanding Barbour's ideas is to try to define motion itself in a relationalist setting, incorporating Poincar\'e's principle.  How do we know a given object has moved from one place to another? Well in the relationalist approach, we can only compare its relative position with respect to some other objects serving as a reference system. One initial attempt might be to say an object has moved if relative to other fixed objects it has different coordinates. But suppose we lived in a swarm of bees, how would we go about defining movements? Barbour explains the problem in the following excerpt, and gives us a hint of the solution  \cite{Barbour2011}:
\begin{quote}
We can now see that there are two very different ways of interpreting general
relativity. In the standard picture, spacetime is assumed from the beginning
and it must locally have precisely the structure of Minkowski space. From the
structural point of view, this is almost identical to an amalgam of Newton's
absolute space and time. This near identity is reflected in the essential
identity locally of Newton's first law and Einstein's geodesic law for the
motion
of an idealized point particle. In both cases, it must move in a straight line
at a uniform speed. As I already mentioned, this very rigid initial structure is
barely changed by Einstein's theory in its standard form. In Wheeler's aphorism
 ``Space tells matter how to move, matter tells space how to
bend.'' But what we find at the heart of this picture is Newton's first law
barely changed. No explanation for the law of inertia is given: it is a --
one is tempted to say \emph{the} -- first principle of the theory. The
wonderful structure of Einstein's theory as he constructed it rests upon it as a
pedestal. I hope that the reader will at least see that there is another way of
looking at the law of inertia: it is not the point of departure but the
destination reached after a journey that takes into account all possible ways in
which the configuration of the universe could change.
\end{quote}

The answer Barbour came up with is called \emph{best matching}. To explain it in a simpler setting than geometrodynamics first, let us consider a system of 3 particles in Euclidean space $\R^n$. Any three particles form a triangle $P_1=(A^a,B^a,C^a)$ at any given moment $t$, and another triangle $P_2=(A^a+\delta A^a,B^a+\delta B^a, C^a+\delta C^a)$ at $t+\delta t$.  To define the infinitesimal motion happening during an infinitesimal interval of time, we have to first find out what would \emph{not} constitute a motion, i.e., what is deemed to leave the physical configuration fundamentally untouched. In the example of the triangles, we could say that any rotation leaves the physical configuration untouched $(A^a,B^a,C^a)\mapsto (\Lambda^a_bA^b,\Lambda^a_bB^b,\Lambda^a_bC^b) $. Or we could be more radical, arguing that since all we can ever really measure are ratios of distances, we should also include dilatations in this group, $(A^a,B^a,C^a)\mapsto (\lambda A^a,\lambda B^a,\lambda C^a) $. Remember this is a world in which \emph{only} the 3 particles exist.

\begin{figure}\label{fig:bm_triangle}
 \begin{center}
 \includegraphics[width=11cm]{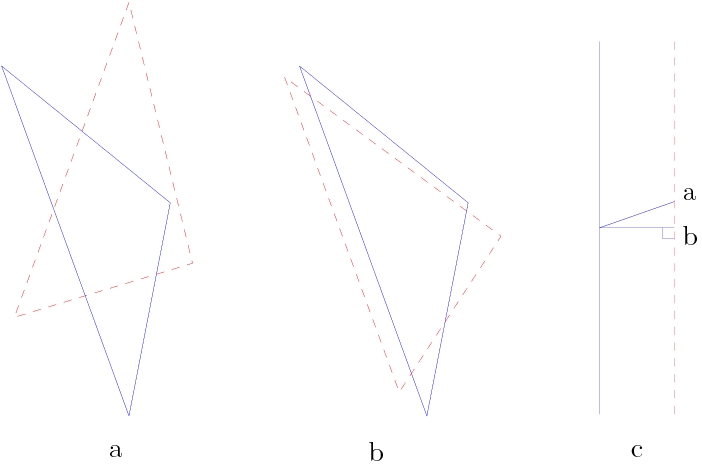}

\caption{$a$) An arbitrary placing of the dashed triangle
relative to the solid triangle; b) the best-matched placing ; $c$) this part already hints at the fiber bundle description. The two
positions of the triangle configurations on their group orbits in $Q^N$. The
connecting velocity is orthogonal with respect to the supermetric on $Q^N$. Best matching brings the centers of mass to coincidence
and reduces the net rotation to zero.}

 \end{center}

\end{figure}

 The abstract group that one chooses to characterize this ``no real change" of configurations of a given system  has been recently termed the \emph{geometrical group} \cite{3-man}. Let us call it $G$ for now.  Having chosen the geometrical group $G$, we have to define an
 \emph{``infinitesimal distance functional" between configurations} , $d\mathcal{S}$. This ``distance" does not, a priori, have to be invariant under the action of the geometric group,\footnote{This fact is what allows best-matching to generalize gauge theory \cite{3-man}. In fact calling it a distance functional is also not entirely accurate since for many cases considered in the literature it does not satisfy the basic postulates required of a norm.}  Suppose we have two descriptions, or configurations, of triangles  $P_1$ and $P_2$, lying in $Q^3$ but over different shapes in the quotient space (see figure \ref{fig:triangle_shape_space}).   We know that $P_2$ is in fact equivalent to all other descriptions related to it by the geometrical group. What best matching does is to select the description of that final state that is closest to $P_1$. Formally,
 $$ P_2':=g'\cdot P_2 ~ ~\mbox{where}~~g'~|~\inf_{g\in G}\mathcal{D}(P_1-g\cdot P_2)
 $$
 Let us give now the pedestrian approach to the above description: we have two triangles, $P_1$ and $P_2$, and we move the second one however we like without breaking it (i.e. obeying the geometric group) until it is ``most similar" (i.e. until it minimizes the distance) to the first one. That is, until it is best matched. See figure \ref{fig:bm_triangle} for an illustration of best-matching in the way described here.

 As noticed by the author (\cite{Private} and later put into \cite{deA.Gomes:2008uv}), if the whole construction can be put in configuration space, with the geometrical group giving an orbit foliation, the notion of shape space becomes analogous to that of a base space in a principal fiber bundle given by the whole of configuration space (see section \ref{sec:Riem}). Then it might be that the whole approach can be put into the  same geometrical terms as usual gauge theory. Thus the Poincar\'e principle would be equivalent to saying that one has to build a theory on the quotient space, i.e. the base space of the fiber bundle. And then, if this is attainable, best matching becomes very reminiscent of the description of the way a connection form works in gauge theory. It will give a notion of ``parallel transport" of coordinates.
 Both the Poincar\'e principle and best-matching can thus be geometrically incorporated in a principal fiber bundle setting, as we shall explain in the second part of this thesis.

 As mentioned in the caption to figure \ref{fig:bm_triangle}, one can prove that the best-matched velocity, when induced by a metric in configuration space, always implies that the corrected velocity of the triangle be orthogonal to the orbits, with respect to said metric. This will be explored more thoroughly in Part II. The statement that best-matching brings the centers of mass to coincidence and brings the net rotation to zero can be seen as particular cases of  \eqref{equ:bm_cond} below.

 \subsection{Best matching in pure geometrodynamics}\label{sec:BM_pure}

 Unfortunately, this section requires some of the introductory material contained in chapter \ref{sec:Riem}, but we shall attempt to make it as self-contained as possible.

  For the implementation of best matching in ``pure" geometrodynamics, we first choose the geometrical group to be  $G=\mbox{Diff}(M)$, the group of 3-diffeomorphisms of the manifold $M$ and the distance functional to be given by  $I[g,\dot g]$, some action functional
on the configuration space
 $$\mbox{Riem}(M):=\{g_{ab}~|~g_{ab}~\mbox{is a 3-dimensional Riemannian metric}\}
 $$
 dependent on the metric and metric velocities. To effect the best matching (in the Lagrangian setting), we first transform all the dynamical variables $g_{ab}\mapsto f\cdot g_{ab}$, where $f\in \mbox{Diff}(M)$, and the $\cdot$ represents a given action of the diffeomorphisms (which we explain in Part II). This implies a certain transformation for $\dot g_{ab}$ and so forth, which we will not make explicit here, since they are discussed extensively in chapter \ref{sec:Riem}. In any case, this is important in that it will imply a certain transformation property for $I[g,\dot g]$, which now becomes a functional $I[g,\dot g,f,X]$ dependent on $f$ and its infinitesimal generating vector field $X$.

Given an initial metric $(M,g)$ and
another, $(M,g+\delta g)$, infinitesimally close-by, best matching is equivalent to finding the  infinitesimal diffeomorphim
(change of coordinates) $X'$ that makes the norm of $\delta g$ with respect to  $I[g,\delta g,f,X]$
an extremum. In this manner, Barbour argues that subsequent instants
in time should have points in their copies of $M$ identified such that the metrics are ``as close as possible".

As it happens, for any action taken to be the integral of a density, the diffeomorphism  parameter $f$ does not appear. We can thus restrict to the case where the dependence of $I$ is given solely by  $I[g,\dot g,X]$. Looking for a geodesic principle in configuration space that incorporated the arguments above, Barbour argued initially for an action of the type:
\be\label{Jacobi}I[g,\dot g,X]=\int_M {FG_\lambda(\dot g-L_X g,\dot g-L_X g)}d\mu_g\ee
where $G_\lambda$ is the generalized DeWitt supermetric, and $F$ is a positive functional of $g$. As argued in \cite{Barbour94}, for a truly geodesic, timeless picture, where the only notion of time is that given by change, one should also demand that this ``distance" functional be reparametrization invariant.

For the usual geodesic principle, one usually looks towards minimizing some function of the type
\be\label{equ:geod_princ}\int_\gamma dt\sqrt{ \gamma'\cdot\gamma'}
\ee for curves $\gamma$, and some inner product $\cdot$.
This suggests taking a global square root:
\be\label{Jacobi_global}I[g,\dot g,X]=\sqrt{\int_M FG(\delta g-L_X g,\delta g-L_X g)d\mu_g}\ee
 (for $F$ an undetermined conformal factor to the supermetric) which should be taken to be extremal with respect to $X$. This would at least heuristically define  a geodesic principle in superspace (see chapter \ref{sec:Riem} for the definition of superspace, and chapter \ref{chapter:connection_forms} for the mathematical difficulties inherent in trying to define an induced metric in superspace). In the relationalist approach of Barbour this is seen as highly desirable \cite{Barbou:CS_plus_V}.

 \subsubsection{BSW form of gravity.}
 Encouragingly,
 the Einstein-Hilbert action, in the alternative, $3+1$ lapse-eliminated BSW
 formulation \cite{BSW},
 is of the type \eqref{Jacobi_global} for $F={R}$ (the three scalar curvature) and $\lambda=1$, i.e. $G_\lambda=G_{\mbox{\tiny DW}}$, but only if we take a \emph{local} square root:
 \be\label{Jacobi_local}I[g,\dot g,X]=\int_M\sqrt{RG(\delta g-L_X g,\delta g-L_X g)}d\mu_g\ee
 This however forfeits the geodesic picture, as we no longer have a variational principle for a functional of the form \eqref{equ:geod_princ}.

 Let us say a few words about the above action \eqref{Jacobi_local}, called the BSW action. It is obtained from the ADM action by regarding the metric and metric velocities as fundamental variables,
 i.e. by regarding $k_{ab}:=\dot g_{ab}-L_Xg_{ab}=NK_{ab}$, where $K_{ab}$ is given in \eqref{equ:def:extrinsic_curv} as one of the fundamental variables. Then instead of the primary scalar constraint \eqref{equ:scalar constraint}, one gets that the terms involving the lapse $N$ appear as:
 $  \frac{G^{abcd}k_{ab}k_{cd}}{N}-NR.
 $
  Upon  varying and solving the action with respect to the lapse, one obtains
   $N=\sqrt{\frac{T}{R}}$, where $T=G^{abcd}k_{ab}k_{cd}$.  See section \ref{sec:prop_const_Lagrange} for more on the propagation of constraints in the BSW action. As it should, this Lagrangian point of view transforms the scalar constraint into an identity.

   In other words, for these choices of $F$ and $G$, and a local square root, the action \eqref{Jacobi} gives GR in
  the BSW form, which could arguably
be said to be a Jacobian timeless form for GR \cite{Lanczos}. \footnote{We note however that in this case $N$ is not a lapse
potential, in accordance in definition \ref{lapse potential}.}

The local square root is the source of many (so far) insurmountable mathematical difficulties \cite{Giulini:1993ct} in trying to formulate the theory as either a geodesic theory in superspace or a gauge theory with a metric-induced connection, as in Part II of this thesis (the two problems are interconnected). However, as we will not need it in the following sections, and in the interest of brevity we limit its discussion to what has been just said. In the author's opinion, there is still no  first--principles justification for the local square-root, or at least not as convincingly as there exists for the other structures present in Barbour's relational construction.

\subsection{Introduction of the conformal group and  emergence of CMC by best matching}

Let us go back to  the case of the triangle (see figure \ref{fig:bm_triangle}). Irrespectively of the form of the distance functional we assume, we can always use the best-matching algorithm. In fact, if one defines the distance functional simply as any given action $S[g]$ in configuration space, one has an interesting consequence. To see this, suppose the group $G$ acts on congiguration space, $\lambda\cdot q$ where $\lambda$ designates a general element of $G$. Then after making the substitutions in the action, both $\lambda$ and $\dot\lambda$ appear in the action. As we show below, the statement then  that the action will be extremized for infinitesimal variations along the orbit  translates to:
\be\label{equ:bm_cond} \diby{L}{\dot \lambda}=\pi_\lambda=0
\ee
When one does this for the diffeomorphism group in geometrodynamics, either for the ADM or the BSW action, one automatically recovers the momentum constraint \eqref{equ:momentum constraint}, whereas for the full conformal group one recovers the \emph{maximal slicing} constraint $\pi=0$. Albeit straightforward, we will not show this now, as it is contained in its general form in chapter \ref{chapter:linking_theory}, and in its particulars, in chapters \ref{chapter:SD_CMC}, \ref{chapter:SD_AF} and \ref{chapter:connection_forms}.

Let us quickly present an alternate view however, which is not presented in the main text and which can be shown to be equivalent to \eqref{equ:bm_cond} by an application of the chain rule.  Recall first of all equation \eqref{equ:momentum_map}. Given $v,w\in T_qQ$, the Legendre map will give us momenta defined by:
\be\label{equ:momentum_map_geom}\langle \mbox{LT}_q(v),
w\rangle= \frac{d}{dt}L(q,v+tw)
\ee
Thus, In accordance with  the best-matching ansatz, we would like  the action to
be an extremal with respect to all possible infinitesimal velocity displacements
along the orbit of the group: \be\label{equ:bm_cond_geom}\langle \mbox{LT}_q(v),
w\rangle=\frac{d}{dt}L(q,v+tw)=0~~\mbox{ for all}~~ w=T\lambda(u)\ee
where $u\in \mathfrak{g}$, the Lie-algebra of $G$. This means that we require all the conjugate momenta to annihilate the tangent space to the orbits (represented here by $w$). The rhs of \eqref{equ:bm_cond_geom} implies
$$ {\diby{L}{w}}_{(q,v)}=0
$$
which implies \eqref{equ:bm_cond}.

As a more concrete example, in the case of configuration space  given by Riem, and thus the configurations $q$ being given by three-metrics $g$ and $G$ being the group of 3-diffemorphisms, we get,  upon contraction of the metric conjugate momenta \eqref{equ:def:metric_momenta} with a tangent vector to the diffeomorphims orbits, assumed for now given by $w=X_{(a;b)}$, the equation \eqref{equ:momentum constraint}.  Or for that matter, upon contraction \footnote{Contraction here of course includes integration over $M$.} with an element of tangent space to the conformal orbit, of the form $\phi g_{ab}$, we get $\pi=0$, and finally, with an element of the tangent space to the volume-preserving-conformal orbit (see section \ref{sec:tangent_space_vpct}) we get $\pi-\mean{\pi}\sqrt g$.

In \cite{Barbou:CS_plus_V} Barbour et al, after reaching some obstacles in their attempt to apply full conformal best-matching to GR \cite{Anderson:2002ey},\footnote{Upon careful scrutiny one can see that these difficulties are related to the non-invertibility of the corresponding lapse fixing operator. See proposition \ref{prop:meanN_0}.} applied best-matching with respect to the group of volume-preserving-conformal transformations. After quite a bit of algebra, they correctly derived the new momentum constraint $\pi_\phi= \pi-\mean{\pi}\sqrt g=0$, along with the transformed scalar constraint -- which had the form of \eqref{equ:LY} -- and the momentum constraint. Of course this is one constraint too many, and implies that there is a gauge fixing of the scalar constraint. They correctly derived the conditions that this implies on the lapse function (which is roughly our equation \eqref{lapse}). But as the scalar constraint is gauge fixed, the interpretation of \eqref{equ:LY} can be only that of an initial value equation, and not of a constraint imposed at every point.\footnote{There is also the question of over-imposing the scalar constraint, since they depart from the Lagrangian formulation, and fix the lapse as in the BSW approach \eqref{Jacobi_local}, turning the scalar constraint into an identiy already satisfied by what they term as the momenta. This will be more thoroughly discussed in section \ref{chapter:SD_CMC}.} Thus they conclude with
\begin{quote}
Our new principles enable us to derive Hamiltonian GR, the prescription for solving its initial-value problem, and the
condition for maintaining the CMC condition in a single package.
\end{quote}
We will discuss some of the factors that distance this earlier work from Shape Dynamics in chapter \ref{chapter:SD_CMC}. However, needless to say, this work was \emph{the} foundation stone of the whole program of Shape Dynamics.

\section{Results and directions for Shape Dynamics.}

\subsection{Brief statement of results.}

 We have found a theory of gravity with two physical degrees of freedom that possesses local scale invariance. Only in a certain conformal gauge it is identical to ADM in constant mean curvature gauge. The total Shape Dynamics Hamiltonian is given by
\begin{equation}\label{equ:prelim_H_SD}
  H_{\text{SD}}=\alpha\mathcal{H}_{\mbox{\tiny{gl}}}+\int_\Sigma d^3 x \big(\rho(x) 4(\pi(x)-\mean{\pi}\sqrt g)+\xi^a(x)H_a(x)\big)
\end{equation}
in the ADM phase space $\Gamma$ parametrized by the usual coordinates $(g,\pi)$, where $\alpha\in\R$, $\rho(x)\in C^\infty(M)$ is an arbitrary Lagrange multiplier function,  and $\mathcal{H}_{\mbox{\tiny{gl}}}[g,\pi]$ is our unique global Hamiltonian, which is a \emph{non-local} functional of $g_{ab},\pi^{ab}$ which does \emph{not} depend on the point $x\in M$. Shape Dynamics possesses the local first class constraints
\begin{equation}
4(\pi(x)-\mean{\pi}\sqrt g)~,~{H}^a.
\end{equation} where $\mean{f}$ is  the global mean of the function $f$ over the 3-manifold $M$. These are the generators of volume-preserving conformal transformations and spatial diffeomorphisms, respectively.
The non-zero part of the constraint algebra is given solely by:
\begin{eqnarray}
\{H^a(\eta_a),H^b(\xi_b)\}&=&H^a([\vec\xi,\vec\eta]_a)\nonumber\\
\{H^a(\xi_a),\pi(\rho))\}&=&\pi\mathcal{L}_\xi(\rho)\nonumber
\end{eqnarray}
For the asymptotically flat case we have a similar result, where the conformal generator is given by $\pi(x)$ only, and its Lagrange multiplier respects certain asymptotic boundary conditions.

We have furthermore found how we can extend the treatment that leads to \eqref{equ:prelim_H_SD} to the electromagnetic, massive and massless scalar fields (see chapter \ref{chapter:coupling}), with their usual Hamiltonians. 
We also show here that different approximation schemes are available for the global Hamiltonian, and in this thesis we perform a large volume expansion for it (chapter \ref{sec:HJ}), obtaining the first three terms. We use this expansion to find the Hamilton--Jacobi version of the global Hamiltonian, a first step towards quantization. We have found that this bears strong resemblance to certain holographic dualities between gravity and traditional conformal field theories \cite{Boer2000}.

 Shape Dynamics provides the theory that fulfills the requirements of a complete theory of the gravitational field on conformal superspace.
Our results  justify York's intuitive remarks regarding the configuration space of gravity: conformal superspace is not the reduced configuration space of general relativity but that of  Shape Dynamics.
Shape Dynamics also meets Barbour's relational arguments for a truly relational theory of the Universe, encapsulated by the aphorism: ``size and motion are relative, and time is given by change".

It is also true, although unseen by us at the time of its conception, that SD is the completion and formalization of Dirac's 1958 paper \cite{Dirac:CMC_fixing}.  Made explicit and put into context however, it gains significance way beyond that of a mere ``fixation of the coordinates", providing truly an alternative description of gravity.

The local constraints are all linear in momenta, being easily implementable in configuration space. The true gravitational degrees of freedom are easily found. The constraint algebra is very simple, making it possible that the attempts at the quantization of gravity that encounter the obstacle posed by structure functions being present in the algebra of constraints (as opposed to structure constants) might be more successful in Shape Dynamics.

\chapter{Linking theory for Shape Dynamics}\label{chapter:linking_theory}

This chapter is based on the paper \cite{Gomes2011}. It is aimed at introducing the general mechanism behind Shape Dynamics. If the reader feels it has become too abstract we recommend following the explicit example given in the next chapter.
\section{General trading of symmetries}\label{sec:general_gauge}

Before introducing Shape Dynamics (SD) per se, we will present in this short chapter a general mechanism that  relates equivalent gauge theories. The central concept in this mechanism is that of a \emph{linking gauge theory}. Roughly speaking, a linking gauge theory between two gauge theories  $\mathcal{A},\mathcal{B}$ with first class constraints $A$ and $B$, respectively, both existing on the phase space parametrized by $(q,p)$, is a gauge theory possessing additional fields and corresponding additional first class constraints. Such a gauge theory qualifies as a linking gauge theory if it yields theories $\mathcal{A}$ and $\mathcal{B}$ under two distinct gauge fixings of the additional fields. We show that whenever there is a linking gauge theory that links two gauge theories then these two gauge theories are equivalent.

The method by which we proceed can be said to be closely analogous to the Stuckelberg mechanism \cite{Stueckelberg}, whereby one adds a fictitious field to a given system in order to reveal some hidden properties it might possess. This is the main aim of our actual usage of the linking theory in GR, through a fictitious addition of a "conformal" field, we reveal a hidden conformal invariance present in ADM. 

\subsection{Linking Theories}

A gauge theory can be denoted by data $T=(\Gamma,\{.,.\},\{\chi_i\}_{i\in \mathcal I},\{\Psi_j\}_{j\in \mathcal J})$, where $\Gamma$ denotes the phase space carrying the Poisson structure $\{.,.\}$, the set $\{\chi_i\}_{i\in\mathcal I}$ denotes first class constraints and the set $\{\Psi_j\}_{j\in\mathcal J}$ denotes second class constraints. We shall from the start restrict the  study to a class of theories with no explicit Hamiltonian (it can be included in the set of first class constraints as the constraint $H-\varepsilon$ that enforces energy conservation) and no second class constraints. The initial value problem of $T$ is given by finding the space $\mathcal C=\{x\in \Gamma:\chi_i(x)=0\forall i\in\mathcal I\}$ and the canonical equations of motion are given by the Hamilton vector fields $v_{H(\lambda_i)}$ defined through the action on smooth phase space functions $f$ as
\begin{equation}
 v_H(f) = \{f,\sum_{i\in\mathcal I} \lambda_i \chi_i\},
\end{equation}
where the $\lambda_i$ are arbitrary Lagrange multipliers. Furthermore, one is able to impose (partial) gauge-fixing conditions $\{\sigma_i\}_{i\in \mathcal I^0}$, such that (some of) the Lagrange multipliers $\lambda_i$ are determined by the condition that $v_H$ is tangent to $\mathcal C_{\mbox{gf}}=\mathcal C \cap \{x\in \Gamma:\sigma_i(x)=0~\forall i\in\mathcal I^0\}$. Hence, gauge-fixing conditions turn (some of) the first class constraints into second class constraints and transforms the initial value problem into a gauge-fixed initial-value problem $\mathcal C_{\mbox{gf}}$.

There can exist a nontrivial physical equivalence between gauge theories, based on the observation that physical quantities are gauge-invariant. To be precise, we call two gauge theories $ {T}_1, {T}_2$ {\bf  equivalent}, if there is a (partial) gauge-fixing $\Sigma_1=\{\sigma^1_i=0\}_{i\in \mathcal I^0_1}$ of $ {T}_1$ and another partial gauge fixing $\Sigma_2=\{\sigma^2_i=0\}_{i\in \mathcal I^0_2}$ of $ {T}_2$, such that the initial value problems $C^1_{\mbox{gf}}=C^2_{\mbox{gf}}$ and the (partially) gauge-fixed Hamilton-vector fields coincide.

Let us define a \emph{ general linking gauge theory} $L=({T}_L,\Sigma_1,\Sigma_2)$, where
$${T}_L=(\Gamma_{\mbox{\tiny Ex}},\{.,.\},\{\chi_i\}_{i\in\mathcal I})$$
is a gauge theory as described before and $\Sigma_1=\{\sigma^1_k\}_{k\in \mathcal K}$ and $\Sigma_2=\{\sigma^2_l\}_{l\in\mathcal L}$ are two sets of partial gauge-fixing conditions such that $\Sigma_1 \cup \Sigma_2$ is a (partial) gauge-fixing condition for $\mathcal {T}_L$ and we assume that we can split the set $\mathcal X=\{\chi_i\}_{i\in \mathcal I}$ of first class constraints into three independent subsets: $\mathcal X_1,\mathcal X_2$ and $\mathcal X_0$, where $\mathcal X_1$ is gauge fixed by $\Sigma_1$, $\mathcal X_2$ is gauge fixed by $\Sigma_2$ and $\mathcal X_0$ is not gauge fixed by either $\Sigma_1$ or $\Sigma_2$.

Given a linking gauge theory, we can construct two equivalent gauge theories:
\begin{eqnarray} {T}_1&=&(\Gamma_{\mbox{\tiny Ex}},\{.,.\},\mathcal X_0\cup \mathcal X_2,\{\rho_j\}_{j\in \mathcal J}\cup\Sigma_1\cup\mathcal X_1)\\
   {T}_2&=&(\Gamma_{\mbox{\tiny Ex}},\{.,.\},\mathcal X_0\cup \mathcal X_1,\{\rho_j\}_{j\in \mathcal J}\cup\Sigma_2\cup\mathcal X_2)\end{eqnarray}
  These are equivalent gauge theories, because both can be gauge-fixed to
   $$(\Gamma_{\mbox{\tiny Ex}},\{.,.\},\mathcal X_0,\{\rho_j\}_{j\in \mathcal J}\cup\Sigma_1\cup\Sigma_2\cup\mathcal X_1\cup\mathcal X_2).$$

This construction becomes nontrivial if we construct the Dirac-bracket and reduced phase space for $ {T}_L^1$ and $ {T}_L^2$. In particular, in the important case where the phase space $\Gamma_{\mbox{\tiny Ex}}$ is a direct product of $\Gamma$ with another phase space $\tilde \Gamma$, which we assume to be coordinatized by a canonically conjugate pair $\{\phi_i,\pi_\phi^i\}_{i\in\mathcal I}$ for simplicity. Moreover, let us assume a {\bf special} set of first class constraints $\chi_1~,~\chi_2$, which are equivalent  to (define the same constraint surface as) the constraints:
\begin{equation}\label{equ:special-constraints}
 \begin{array}{rcl}
  &\tilde\chi_1:=\phi_i - f_i \approx & 0\\
   &\tilde\chi_2:= \pi_\phi^i - g_i \approx & 0,
 \end{array}
\end{equation}
where $f_i, g_i$ are functions on $\Gamma$ for all $i\in \mathcal I$.\footnote{The functions $f_i$ and $g_i$  obey certain conditions so that $\tilde\chi_1$ and $\tilde\chi_2$ are equivalent to the first class constraints $\chi_1$ and $\chi_2$. These conditions do not play any role since we will start already from  explicitly first class $\chi_1~,~\chi_2$. See last paragraph of this subsection. } By equivalence between $\chi_1~,~\chi_2$ and $\tilde\chi_1~,~\tilde\chi_2$ we mean solely that the constraints $\chi_1~,~\chi_2$ can be solved for certain values of $\phi_i$ and $\pi_\phi^i$.

 Moreover, we assume special gauge-fixing conditions
\begin{equation}\label{equ:special-gaugefixing}
\Sigma_1= \phi_i= 0,\,\,\,\,\Sigma_2=\pi_\phi^i= 0
\end{equation}
for all $i\in \mathcal I$. Note that $\Sigma_1$ completely fixes the gauge of $\chi_2 $ to zero and vice-versa. The special form of the constraints and gauge fixing conditions allows us to perform the phase space reduction explicitly. That is, we  use the equations \eqref{equ:special-constraints} as definitions and completely eliminate the variables $\phi_i$ and $\pi_\phi^i$ from the system together with the second class constraints,  reverting to the usual Poisson bracket in the reduced phase space, as mentioned in Section \ref{sec:Dirac}.

 To see this is indeed equivalent to using the Dirac bracket, consider functions $F_r$ on $\Gamma_{\mbox{\tiny Ex}}$ that are independent of $\{\phi_i,\pi_\phi^i\}_{I\in\mathcal I}$, which are in one-to-one correspondence with functions on $\Gamma$, and we construct their Dirac bracket $\{.,.\}_D$ for the gauge-fixing $\phi_i\approx 0$:
\begin{equation}\label{Dirac bracket}
 \{F_1,F_2\}_{\mbox{\tiny{DB}}}=\{F_1,F_2\}+\{F_1,\phi_i\}\{\pi_\phi^i-g^i,F_2\}-\{F_1,\pi_\phi^i-g^i\}\{\phi_i,F_2\}=\{F_1,F_2\},
\end{equation}
where Einstein summation over $i$ is assumed, and we used the facts that $\{\phi_i,g^j\}=0$ and that $\phi, \pi_\phi$ are canonically conjugate. The Dirac bracket thus reduces to the Poisson bracket on the reduced phase space $\Gamma \subset \Gamma_{\mbox{\tiny Ex}}$ and as $\chi_2$ is completely gauge-fixed to zero,  the remaining first class constraints are
\begin{equation}
 f_i \approx 0 \,\,\textrm{ for all }i\in \mathcal I.
\end{equation}
Performing the analogous phase space reduction for the gauge-fixing condition $\pi_\phi^i\approx 0$, we arrive at
\begin{proposition}\label{prop:equivalence}
  Given a gauge theory on a phase space $\Gamma_{\mbox{\tiny Ex}}=\Gamma\times \tilde \Gamma$ with special first class constraints which are equivalent to constraints of the form (\ref{equ:special-constraints}), and special gauge fixing conditions of the form (\ref{equ:special-gaugefixing}) then $\mathcal {T}_1=(\Gamma,\{.,.\},\{f_i\}_{i\in\mathcal I}\cup \mathcal X_0)$ and $\mathcal {T}_2=(\Gamma,\{.,.\},\{g_i\}_{i\in\mathcal I}\cup \mathcal X_0)$ are equivalent gauge theories.
\end{proposition}
Note that this proposition only assumes that the constraints can be formally written in the form (\ref{equ:special-constraints}). However, any set of constraints that can in principle be solved for $\phi_i$ and $\pi^i_\phi$ on the respective gauge fixing surface as in (\ref{equ:special-constraints}) suffices for the construction of the phase space reduction.

\subsection{A Construction Principle for Linking Theories}\label{sec:constructionPrinciple}

We will now give a simple construction principle for special linking theories linking a known gauge theory to a desired gauge theory with different symmetry. For this purpose we consider elementary degrees of freedom $q_i$ whose dynamics is governed by an action $S[q]:=\int dt L(\dot q,q)$. If the dynamical system is consistent, then the Legendre transform to the canonical system will yield first and second class constraints, however we ignore second class constraints in this subsection (as they will not appear in such a fashion in the model we will be studying) and assume we have a conjugate pair $(q_i,p^i)$ of canonical degrees of freedom that coordinatize our phase space $\Gamma$ and  a purely first class system of constraints:
\begin{equation}\label{equ:originalConstraints}
 \chi_\rho(q,p)\approx 0.
\end{equation}
In the Lagrangian picture, we extend our configuration space to include auxiliary degrees of freedom $\phi_\alpha$, but still with the Lagrangian $L(\dot q,q)$, which implies that the Legendre transform yields a phase space with an additional canonically conjugate pair $(\phi_\alpha,\pi^\alpha)$ and with additional first class constraints
\begin{equation}\label{equ:originalAdditionalConstraints}
 C^\alpha=\pi^\alpha=\diby{L}{\dot\phi_\alpha}\approx 0,
\end{equation}
whose Poisson-brackets with the original constraints (\ref{equ:originalConstraints}), as well as with any $f(p,q)$, and among themselves, vanish strongly by construction. Let us now apply a point transformation
\begin{equation}\label{equ:pointTransfromaion}
  \mathcal {T}_\phi : q_i\to Q_i(q,\phi)
\end{equation}
parametrized by the auxiliary degrees of freedom $\phi_\alpha$, such that $Q_i(q,0)=q_i$, which reverts the system to the original Lagrangian. This transformation is a canonical transformation generated by the generating functional
\begin{equation}\label{equ:generatingFunctional}
 F=Q_i(q,\phi)P^i+\phi_\alpha \Pi^\alpha.
\end{equation}
Using the shorthand $M^i_j=\frac{\partial Q_j}{\partial q_i}=\frac{\partial \dot Q_j}{\partial \dot q_i}$ as well as $R^\alpha_j:=\frac{\partial Q_j}{\partial \phi_\alpha}=\frac{\partial \dot Q_j}{\partial \dot\phi_\alpha}$ we can denote the canonical transformation from $(q_i,p^i,\phi_\alpha,\pi^\alpha)$ to $(Q_i,P^i,\Phi_\alpha,\Pi^\alpha)$ generated by (\ref{equ:generatingFunctional}) in the compact form
\begin{equation}\label{equ:canonicalTransformation}
  \begin{array}{rcccl}
   q_i&\to&Q_i&=&Q_i(q,\phi)\\
   p^i&\to&P^i&=&\left(M^{-1}\right)^i_j p^j\\
   \phi_\alpha&\to&\Phi_\alpha&=&\phi_\alpha\\
   \pi^\alpha&\to&\Pi^\alpha&=&\pi^\alpha-R^\alpha_j \left(M^{-1}\right)^j_k p^k.
  \end{array}
\end{equation}
{Alternatively, we can obtain these formulae from the transformed Lagrangian, eg.,
\be\label{equ:canonicalTransformationLagrange}p^i=\frac{\partial L}{\partial \dot q_i}=\frac{\partial L}{\partial \dot Q_j}\frac{\partial Q_j}{\partial q_i}=P^jM_j^i,\ee as $Q$ is the only variable dependent on $q$. In the same way
\be\pi^\beta=\frac{\partial L}{\partial \dot \Phi_\alpha}\frac{\partial\Phi_\alpha}{\partial \phi_\beta}+\frac{\partial L}{\partial \dot Q_j}\frac{\partial Q_j}{\partial \phi_\beta} \ee} yields the respective equation in \eqref{equ:canonicalTransformation}.

Let us now consider the system of canonically transformed constraints (\ref{equ:originalConstraints}) and (\ref{equ:originalAdditionalConstraints}):
\begin{equation}\label{equ:transformedConstraints}
 \begin{array}{rcl}
   \chi_\rho(p,q) &\to& \chi_\rho\left(Q(q,\phi),P(p,q,\phi)\right)\approx 0\\
   C^\alpha &\to& \pi^\alpha-R^\alpha_j \left(M^{-1}\right)^j_k p^k\approx 0,
  \end{array}
\end{equation}
which is of course still first class. Notice that the previously (almost) trivial constraints $C^\alpha$ now take a quite nontrivial form.  To construct a special linking theory, we now assume that we can split the constraints $\chi_\rho\left(Q(q,\phi),P(p,q,\phi)\right)$ into two sets $\chi^1_\alpha(q,p,\phi)$ and $\chi^2_\mu(q,p,\phi)$, where the first set can be solved for $\phi_\alpha$ and the second (weakly) Poisson commutes with $\pi^\alpha$. So we can write the constraints (\ref{equ:transformedConstraints}) equivalently as
\begin{equation}\label{equ:constraintsInLinkingForm}
 \begin{array}{rcl}
   0&\approx& \phi_\alpha -\phi^0_\alpha(q,p)\\
   0&\approx& \chi^2_\mu(q,p,\phi)\\
   0&\approx& \pi^\alpha-R^\alpha_j \left(M^{-1}\right)^j_k p^k,
 \end{array}
\end{equation}
which is of the form needed for a special linking theory. We can thus impose the two sets of gauge fixing conditions
\begin{equation}\label{equ:gaugeFixingConditions}
  \pi^\alpha=0 \,\,\textrm{ and }\,\, \phi_\alpha=0,
\end{equation}
which gauge-fix the first and (respectively) last line of (\ref{equ:constraintsInLinkingForm}).

The gauge fixing conditions $\phi_\alpha=0$ can be worked out easily. Since $\phi_\alpha$ commutes with the two first lines of \eqref{equ:constraintsInLinkingForm}, it imposes no further conditions on these. Hence the union of these two sets of constraints is still $\chi_\rho\left(Q(q,\phi),P(p,q,\phi)\right)$. Now
$$\{\phi_\alpha, \pi^\beta-R^\beta_j \left(M^{-1}\right)^j_k p^k\}=\delta_\alpha^\beta.
$$
Thus it completely gauge fixes the constraint $C^\beta=\pi^\beta-R^\beta_j \left(M^{-1}\right)^j_k p^k$, whose Lagrange-multiplier  is constrained to vanish for propagation. In other words, we set the two (sets of) second class constraints $\phi_\alpha$ and $C^\beta$ strongly to zero, using these equations as definitions for $\phi_\alpha$ and $\pi^\alpha$.
 One can thus perform the phase space reduction by setting $(\phi_\alpha,\pi^\alpha)=(0,R^\alpha_j \left(M^{-1}\right)^j_k p^k)$. Since $\pi_\alpha$ appears nowhere but in $C^\alpha$, which is used solely as a definition of $\pi_\alpha$ itself, phase space reduction reduces the constraints $\chi_\rho\left(Q(q,\phi),P(p,q,\phi)\right)$ to $\chi_\rho\left(q,p\right)$ and reverts us to the original gauge theory.

Let us now examine the gauge fixing conditions $\pi^\alpha=0$. They clearly do not (weakly) Poisson commute with the constraints $\chi^1_\alpha$, because these can be written as $\phi_\alpha -\phi^0_\alpha(q,p)$. In fact, the assumption that the remaining constraints are not gauge fixed by $\pi^\alpha=0$ is redundant. Given constraints of the form of \eqref{equ:constraintsInLinkingForm} we can prove that  $\pi^\alpha=0$ does not gauge fix any further constraint(s). For this we assume that there is a subset $\sigma_\rho$ of the constraints $\chi^2_\mu$ and $C^\alpha$ that is gauge-fixed, which implies that the matrix
\begin{equation}\label{equ: matrix_1st_vs_2nd}
 \left(
   \begin{array}{ccc}
     \{\chi^1,\chi^1\}&\{\chi^1,\pi\}&\{\chi^1,\sigma\}\\
     \{\pi,\chi^1\}&\{\pi,\pi\}&\{\pi,\sigma\}\\
     \{\sigma,\chi^1\}&\{\sigma,\pi\}&\{\sigma,\sigma\}
   \end{array}
 \right) \approx
 \left(
   \begin{array}{ccc}
     0&A&0\\
     -A&0&b\\
     0&-b&0
   \end{array}
 \right)
\end{equation}
is invertible. The block containing $A$ is invertible by assumption and hence the determinant of the entire matrix vanishes, since the conjugate block vanishes identically. This lies in contradiction with the assumption, which stated that the matrix was invertible. This was to be expected, because we can write one of the constraints as $\phi-f(g,\pi)$, it makes its Poisson Bracket with the gauge fixing automatically invertible and thus completely exhausts the gauge fixing.

It follows that $\pi^\alpha=0$ can just gauge fix the $\chi^1_\alpha$.  Thus, to perform the phase space reduction we trivialize the constraints $\chi^1_\alpha$ and set $(\phi_\alpha,\pi^\alpha)=(\phi_\alpha^0,0)$. On the reduced phase space this then gives the first class constraints
\begin{equation}\label{equ:tradedConstraints}
 \begin{array}{rcl}
   0&\approx&\chi^2_\mu\left(q,p,\phi^0(q,p)\right)\\
   0&\approx&\left(R^\alpha_j \left(M^{-1}\right)^j_k\right)\left(q,p,\phi^0(q,p)\right) p^k=:D^\alpha,
 \end{array}
\end{equation}
and thus we have effectively traded the constraints $\chi_\alpha^1$ for $D^\alpha$.

In summary we have shown the following proposition
\begin{proposition}\label{prop:constructionPrinciple}
 Given a dynamical system with first class constraints $\chi_\mu$ and a point transformation $q_i\to Q_i(q,\phi)$ parametrized by auxiliary degrees of freedom $\phi_\alpha$ such that a subset $\chi^1_\alpha$ of the constraints can be solved for $\phi_\alpha$ as a function of $(q,p)$ after applying the canonical transformation that implements the point transformation, then the above construction provides a linking theory that provides equivalence with a theory in which the $\chi^1_\alpha$ are replaced by the constraints $D^\alpha$ as defined in equation (\ref{equ:tradedConstraints}).
\end{proposition}

Conversely, the phase space reduction of a linking theory can be  viewed as an embedding of the equivalent gauge theories in  the linking theory. In this picture one has two embeddings $i_{orig.}$ and $i_{dual}$ that embed the original resp. dual gauge theory in the linking theory by
\begin{equation}
 \begin{array}{rcccl}
   i_{orig.}&:&(q,p)&\mapsto&(q,p,0,\pi_\phi^0)\\
   i_{dual}&:&(q,p)&\mapsto&(q,p,\phi^0,0),
 \end{array}
\end{equation}
where we wrote $\pi_\phi^0=D^\alpha$ to bring out the similarity between the two embedding. It should be noted however that in spite of the apparent similarity, there is an important formal asymmetry. Namely, whereas setting $\phi=\phi_0$ will influence the reduced dual system \eqref{equ:tradedConstraints}, as  $\pi_\phi$ does not figure anywhere in the original system, setting its value to some given function $D^\alpha$ does not influence the reduced system (see paragraph following the paragraph of equation \eqref{equ:constraintsInLinkingForm}).

\subsection{Diagram of trading}

We started by showing how an equivalence of gauge theories follows from the existence of a linking gauge theory on an extended phase space $\Gamma\times \tilde \Gamma$.  One can sketch the construction of a pair of equivalent gauge theories A and B on a reduced phase space $\Gamma$ as follows
\begin{equation}\label{diagram:trading}
  \begin{array}{ccccc}
     &\textrm{partial gauge fixing}&&\textrm{partial gauge fixing}&\\
     \textrm{theory A}&\longleftarrow&\textrm{linking theory}&\longrightarrow&\textrm{theory B}\\
      \textrm{on }\Gamma \times \tilde\Gamma& \phi_I=0&\textrm{ on }\Gamma \times \tilde\Gamma&\pi_\phi^I=0& \textrm{on }\Gamma \times \tilde\Gamma\\
     \downarrow &&&& \downarrow\\
     \textrm{ reduced } &&&& \textrm{ reduced }\\
     \textrm{ theory A} &&&& \textrm{ theory B}\\
     \textrm{on }\Gamma &&&& \textrm{on }\Gamma\\
     &\longrightarrow&\textrm{Dictionary}&\longleftarrow&\\
     &&\textrm{ on }\Gamma_{red},&&
  \end{array}
\end{equation}
where $\phi_I$ and $\pi_\phi^I$ is a canonical pair coordinatizing $\tilde \Gamma$ and the \lq\lq Dictionary\rq\rq is a further gauge fixing of the two equivalent theories such that the two theories coincide. The dictionary can be used to easily identify trajectories of the equivalent theories with one another.

To summarize the procedure in plain words and notation: we start with a theory $A$ which possesses some explicit symmetry $a$, but which we suspect might possess some hidden symmetry $b$. We then artificially introduce a field parametrizing symmetry $b$ into the system $A$, producing a system $LT_A$ even more redundantly parametrized. The outcome of the introduction of $b$ however, is not the presence of the symmetry $b$, but of some other symmetry $c$. In some very restricted set of cases, upon very particular gauge fixings of $LT_A$ (namely setting either $\phi_b$ or its conjugate momenta to zero), one may obtain a system which explicitly possesses symmetry $b$.  
 
 We stress that in most cases, the procedure outlined above is not applicable, i.e. one obviously cannot trade any two given symmetries. If for example the theory already possesses the symmetry we are trying to trade the two gauge fixings coincide and there is no gain in the procedure. In the majority of cases though, what will happen is that the system produced through the gauge fixing will entail a tower of constraints, rendering it inconsistent. 

\chapter{Trading GR for SD: compact closed $\Sigma$ case}\label{chapter:SD_CMC}
\subsubsection{Summary of this chapter}
In this section we will apply the method of the linking theory presented in the previous chapter to a specific extension of ADM gravity for compact closed space. After extending ADM to include certain scalar fields representing conformal transformations and their conjugate fields, a linking theory in the sense of the diagram \eqref{diagram:trading} will be presented. One gauge fixing of the linking theory will then eliminate conformal freedom but retain lapse freedom, resulting in ADM, and another will fix the lapse freedom but retain conformal freedom, resulting in Shape Dynamics. The most difficult technical steps will be taken in section  \ref{sec:SD_CMC}, where we will split the constraints into first and second class and show that the second class constraints can be uniquely solved for the extra variables. This chapter is based in \cite{Gomes2011a, Gomes2011}, but contains more detailed (and somewhat different) proofs of the main propositions.

\section{Construction of the Linking Theory}

We start with the equivalent of \eqref{equ:originalConstraints} and denote the usual ADM constraints as
\begin{equation}
 \begin{array}{rcl}
   S&=&\frac{\pi^{ab}\pi_{ab}-\frac{1}{2}\pi^2}{\sqrt g}-\sqrt g R\\
   H^a(\xi_a)&=&\int d^3 xg_{ab}\mathcal{L}_\xi\pi^{ab}.
  \end{array}
  \end{equation}
  where for ease of manipulation we wrote the smeared version of the diffeomorphism constraint.

Following the method described in the previous chapter, we will embed the original system into an extended phase space. This extended phase space is chosen to include the auxiliary variables $(\phi,\pi_\phi)$. The scalar function $\phi\in C^\infty(M)$ will parametrize conformal transformations of the system. We denote the space of conformal transformations by $\mathcal{C}$.
 The nontrivial canonical Poisson brackets are
\begin{equation}\label{equ:canonicalPbs}
  \begin{array}{rcl}
    \{g_{ab}(x),\pi^{cd}(y)\}&=&\delta^{(cd)}_{ab}\delta(x,y)\\
    \{\phi(x),\pi_\phi(y)\}&=&\delta(x,y).
  \end{array}
\end{equation}
The extended phase space for these fields is now given by:
$$(g_{ij},\pi^{ij}, \phi,\pi_\phi)\in\Gamma_{\mbox{\tiny {Ex}}}:=\Gamma_{\mbox{\tiny {Grav}}} \times \Gamma_{{\mbox{\tiny {Conf}}}}$$
with the additional constraint analogous to \eqref{equ:originalAdditionalConstraints}:
\begin{equation}\pi_\phi\approx 0
\end{equation}

 We will explore the space of {\it volume-preserving conformal transformations} acting as canonical transformations in $\Gamma_{\mbox{\tiny {Ex}}}$. Given any conformal transformation $\phi$ we can define a surjection into a volume-preserving one $\phi\mapsto\hat\phi$ as follows:
 \be\label{equ:vpct_def}\hat \phi(x):=\phi(x)-\frac 1 6 \ln\langle e^{6\phi}\rangle_g\ee where we use the mean $$\langle f\rangle_g:=\frac 1 V \int d^3x\sqrt{|g|} f(x)$$ and 3-volume $V_g:=\int d^3x\sqrt{g}$. We will abuse notation and extend the use of the \emph{mean} $\mean{\cdot}$ to densities by dividing out by the appropriate power of $\sqrt g$. For example, instead of writing $\mean{\pi/\sqrt g}$, we will just redefine the mean for this scalar density as:
 $$\langle \pi\rangle_g:=\frac 1 V \int d^3x\pi
 $$

 We denote the space of volume-preserving conformal transformations as $\mathcal{C}/\mathcal{V}$, and it is redundantly parametrized by $\mathcal{C}$. One can check that the redundancy, i.e the equivalence relation, is given by $\hat\phi\equiv\hat\phi'$ if and only if $\phi'=\phi+c$ where $c$ is a spatial constant. We also note that
 $$ \{\int \pi_\phi(x) d^3x,\hat\phi\}=0
 $$ This can be derived from \eqref{equ:tangent_hatphi}, with $f=1$. Thus we have
 $$\{f(g,\pi,\hat\phi,\pi_\phi),\int \pi_\phi(x)d^3x\}=0.
 $$

  As we will see, this deviation from the simplest case of  unconstrained conformal transformations is necessary in order to have some combination of the linking theory scalar constraint $\{ \mathcal {T}_{\phi}S(x), x\in\Sigma\}$ that is not fixed by (remains first class wrt) the condition $\pi_\phi=0$. This will allow shape dynamics have a unique non-zero global Hamiltonian and be matched to ADM in something other than the frozen lapse regime.

Following \eqref{equ:generatingFunctional}, we construct the generating function \begin{equation}\label{equ:generatingFunctional2}F_\phi:=\int_\Sigma d^3x \left(g_{ab}(x)e^{4\hat\phi(x)}\Pi^{ab}(x)+\phi(x)\Pi_\phi\right),\end{equation} where capitals denote the transformed variables. We find the canonical transformation analogous to \eqref{equ:canonicalTransformation} operating in extended phase space:
\begin{equation}\label{equ:canonicalTransformation_grav}
 \begin{array}{rcl}
   g_{ab}(x)&\to& \mathcal {T}_{\phi} g_{ab}(x):=e^{4\hat\phi(x)}g_{ab}(x)\\
   \pi^{ab}(x)&\to&\mathcal {T}_{\phi} \pi^{ab}(x):=e^{-4\hat\phi(x)}\left(\pi^{ab}(x) -\frac{g^{ab}}{3}\sqrt {g}\langle \pi\rangle (1-e^{6\hat\phi})\right)\\
      \phi(x)&\to&\mathcal {T}_{\phi} \phi(x):=\phi(x)\\
   \pi_\phi(x)&\to&\mathcal {T}_{\phi} \pi_\phi(x):=\pi_\phi(x)-4(\pi(x)-\mean{\pi}\sqrt g).
 \end{array}
\end{equation}

Again, although slightly more inconvenient, one can find essentially the same set of transformations from the first line of \eqref{equ:canonicalTransformation_grav} through a Lagrangian analysis, as in \eqref{equ:canonicalTransformationLagrange}. We explicitly check that this transformation is indeed canonical in section \ref{sec:canonical_transfs_app}.

The three sets of constraints that we now have are  the transformed scalar and diffeomorphism constraints of GR  as well as the transform of $\pi_\phi$,
\begin{equation}\label{equ:transformedConstraintsSD}
 \begin{array}{rcl}
  \mathcal {T}_\phi S&=&\mathcal {T}_\phi(\frac{\pi^{ab}\pi_{ab}-\frac{1}{2}\pi^2}{\sqrt g}-\sqrt g R)\\
  \mathcal {T}_\phi H^a&=&\mathcal {T}_\phi(\nabla_b\pi^{ab})\\
  \mathcal{Q}&=&\pi_\phi-4(\pi-\mean{\pi}\sqrt g)
 \end{array}
\end{equation}
where we have used the shorthand $\pi^{ab}g_{ab}=\pi$ and the notation $\mathcal{Q}$ to denote the analogous constraint of  \eqref{equ:transformedConstraints} in Chapter \ref{chapter:linking_theory}, where we used the notation $C^\alpha$. We write down here explicitly the form of the transformed scalar constraint, but note that it will not be explicitly used until section \ref{sec:HJ}.
\begin{multline}\label{equ:transformedScalar}
 \ts = \frac{1}{g e^{12\hat\phi}} \lf( \pi^{ab}\pi_{ab} - \frac{\pi^2}{2} - \frac{\mean{\pi}}{6} (1 -e^{6\hat\phi})^2 g + \frac{\mean{\pi}}{3} \pi (1 - e^{6\hat\phi}) \sqrt g \rt) \\+ 2\Lambda - \frac{R-8(|\nabla\phi|^2+\nabla^2\phi)}{e^{4\hat\phi}}.
\end{multline}
The transformed diffeomorphism constraint will be worked out below, in \eqref{equ:transf_diffeo}.

Any functional of the original phase space variables, when transformed by $\mathcal{T}_\phi$, strongly commutes with $\mathcal{Q}$, as
\be\label{equ:0Pb}0=\{f(g_{ab},\pi^{cd}),\pi_\phi(x)\}=\mathcal{T}_\phi\{f(g_{ab},\pi^{cd}),\pi_\phi(x)\}=\{\mathcal{T}_\phi f(g_{ab},\pi^{cd}),\pi_\phi(x)-4(\pi-\mean{\pi}\sqrt g)\}.\ee
Thus it is clear that on the original phase space, $\pi-\mean{\pi}\sqrt g$ generates infinitesimal volume-preserving conformal transformations. This can be seen more clearly if we use the fact that $\pi-\mean{\pi}\sqrt g$ is invariant under the transformation \eqref{equ:canonicalTransformation_grav} to rewrite \eqref{equ:0Pb} as:
$$\{T_\phi f(g_{ab},\pi^{cd}),\pi_\phi(x)\}_{\phi=0}=T_\phi\{ f(g_{ab},\pi^{cd}),4(\pi-\mean{\pi}\sqrt g)\}_{\phi=0}=\{ f(g_{ab},\pi^{cd}),4(\pi-\mean{\pi}\sqrt g)\}$$
which can be checked explicitly as well.\footnote{ One can fairly easily visualize how this comes about by just considering how $\pi(x)$ generates infinitesimal conformal transformation. That is $\{g_{ab}(x),\pi(\rho)\}=\rho(x) g_{ab}(x)$ and $\{\pi^{ab}(x),\pi(\rho)\}=-\rho(x) \pi^{ab}(x)$.}

The diffeomorphism constraint in the linking theory, $\mathcal {T}_\phi H^a(\xi_a)$,  can be explicitly calculated as follows (in smeared density form):
        \begin{eqnarray}
        \xi_aT_{\phi}H^a &=& (\pi^{ab}-\frac{g^{ab}}{3}\sqrt {g}\langle \pi\rangle (1-e^{6\hat\phi}))\mathcal{L}_\xi g_{ab}+4(\pi-\langle\pi\rangle\sqrt g(1-e^{6\hat\phi}))\mathcal{L}_\xi\phi\nonumber\\
        &=& \pi^{ab}\mathcal{L}_\xi g_{ab}-\frac{2}{3}\sqrt {g}\langle \pi\rangle (1-e^{6\hat\phi})\xi^a_{~;a}+\pi_\phi\mathcal{L}_\xi\phi+4e^{6\hat\phi}\sqrt {g}\langle \pi\rangle \mathcal{L}_\xi\phi\nonumber\\
        &\dot=& \pi^{ab}\mathcal{L}_\xi g_{ab}+\pi_\phi\mathcal{L}_\xi\phi\label{equ:transf_diffeo}
        \end{eqnarray}
         where we have used integration by parts and the fact that the constraint $\mathcal{Q}$ vanishes on the image of $\mathcal {T}_\phi$ strongly. Thus the constraint (in smeared density form) $\pi^{ab}\mathcal{L}_\xi g_{ab}+\pi_\phi\mathcal{L}_\xi\phi$ explicitly generates diffeomorphisms in extended phase space.

We define the total Hamiltonian
\be\label{Hamiltonian} H_{\mbox{\tiny {Total}}}=\int d^3x [N(x)\mathcal {T}_\phi S(x)+\xi^a(x)\mathcal {T}_\phi H_a(x)+\rho(x)\mathcal{Q}(x)]
\ee
We do not explicitly topologize phase space for now and only later assume that we can turn it into a Banach space compatible with the Poisson bracket.
This completely defines the linking $\mathcal {T}_L$ as contained in the previous section. We define the linking theory as the gauge theory defined in this section together with the two sets of gauge-fixing conditions and constraint sets
\begin{eqnarray}
\mbox{Constraints}&:& \mathcal X_1=\mathcal{Q}\,\,\,\textrm{ and }\,\,\, \mathcal X_2=\phi-\phi_0\,\,\,\textrm{ and }\,\,\, \mathcal X_0=\mathcal {T}_\phi H^a\cup \langle N_0 \mathcal {T}_{\phi}S\rangle\nonumber\\
\mbox{Gauge fixing}&:&\Sigma_1=\{\pi_\phi(x)=0\}_{x\in\Sigma} \,\,\,\textrm{ and }\,\,\, \Sigma_2=\{\phi(x)=0\}_{x\in\Sigma}\label{gauge conditions},\end{eqnarray}
 where $\phi_0$ and $N_0$ will be specified shortly in a way that ensures that $\phi-\phi_0$ combined with $\langle N_0 \mathcal {T}_{\phi_0}S\rangle$ is equivalent to $\mathcal {T}_\phi S(x)$ on the surface $\pi_\phi\equiv0$.

\section{Recovering General Relativity for compact closed manifolds}

The only non-vanishing Poisson bracket of the gauge fixing condition $\phi(x)=0$ with the constraints of the linking theory is
\begin{equation}
 \{\phi(x),\Q(\rho)\}=\rho(x),
\end{equation}
which determines the Lagrange-multiplier $\rho(x)=0$, and effectively eliminates $\pi_\phi$ from the theory. We can thus perform the phase space reduction by setting the two second class constraints strongly to zero:
\begin{eqnarray}
\phi(x)&\equiv&0\\
\pi_\phi(x)&\equiv& 4(\pi-\mean{\pi}\sqrt g)(x)\end{eqnarray}
This eliminates both of the extra conjugate variables, and makes the constraint $\Q(0)$  empty. Moreover, for phase space functions independent of $\phi,\pi_\phi$ one finds in the same way  as in \eqref{Dirac bracket} that the Dirac-bracket coincides with the canonical Poisson bracket. Since $\pi_\phi$ does not appear anywhere else but in equation $\Q =0$, which is now seen as its definition,  the constraints on the reduced phase space are
\begin{equation}
S(x)~~\mbox{and}~~ H^a(x)
\end{equation}
The resulting gauge theory is thus ADM gravity.

\section{Recovering Shape Dynamics for compact closed manifolds.}\label{sec:SD_CMC}
Our main aim in this subsection will be to prove that part of the scalar constraints can be written in the form $\phi-\phi_0(g,\pi)\approx 0$  on the gauge-fixing surface $\pi_\phi\equiv 0$, in a particular way that will be useful for us, and then use the results section \ref{sec:general_gauge}.

The only weakly non-vanishing Poisson-bracket of the gauge-fixing condition $\pi_\phi(x)=0$ with the constraints of the linking theory is
\be\label{equ:Pb_pi_phi,pi}\{\mathcal {T}_\phi S(N),\pi_\phi(x)\}=4\mathcal {T}_\phi\{S(N),\pi(x)-\mean{\pi}\sqrt g(x)\},\ee
which leads to (calculated in the appendix, see \eqref{equ:PbSpi}):
\begin{equation}\label{lapse}4\{S(N),\pi(x)-\mean{\pi}\sqrt g(x)\}=8(\nabla^2-\frac{1}{4\sqrt g}\langle\pi\rangle\pi-R)N(x)-8\langle \Delta N\rangle-6S(x)N(x).\end{equation}
   In \eqref{lapse} $\Delta$ is the differential operator appearing in the first term:
\begin{equation}
\label{Delta}\Delta:=\nabla^2-\frac{1}{4\sqrt g}\langle\pi\rangle\pi-R.
\end{equation}
This is the important operator of the theory for compact closed manifolds, and we will have to make small detour to study some of its properties. We leave this diversion to later, namely to Proposition \ref{prop:meanN_0}. For now let us assume the end result of the proposition: $\Delta$ has a unique fundamental solution (i.e. it is an invertible operator).

First of all, $\Delta$ is an elliptic linear second order differential operator on a compact manifold. Merely from ellipticity, a fundamental solution (or Green's function) always exists, in the sense that there exists a distribution $G_y(x)$ such that
$$ \Delta G_y(x)=\delta(x,y).
$$ However for our purposes it is not sufficient that the fundamental solution exist; it needs also to be unique, which is what we for now assume, and later on, in Proposition \ref{prop:meanN_0}, proceed to prove.

 If this is so, then if $\Delta N=0$ the only solution is $N\equiv 0$. Thus to solve \eqref{lapse} non-trivially, all we require is that $\Delta N=c$, where $c$ is any non-zero spatial constant. This fact allows us to escape integral-differential equations and just stick to very simple partial differential ones.
 We then adjust this constant so that $\mean{N_0}=1$ for our lapse smearing. To be able to do this we must prove that there exists a $c'$ such that the solution obeys $\mean{N_0}\neq0$. This is done in Proposition \ref{prop:meanN_02}. By scaling $c'$ appropriately to get $\mean{N_0}=1$, we fix the ambiguity and get the unique kernel $N_0$ such that:
 \be \label{N_0}\begin{array}{rcl}
\Delta N_0[g,\pi,x)-\mean{\Delta N_0[g,\pi,x)}&=&0\\
\mean{N_0}&=&1\end{array}\ee
for each $(g,\pi)$.

Thus, by the canonical transformation properties and \eqref{equ:Pb_pi_phi,pi}, the solution to
$$\{\mathcal {T}_\phi S[N],\pi_\phi(x)\}\approx0$$ is $ N_0[\mathcal {T}_\phi g,\mathcal {T}_\phi\pi,x)$, which always  exists uniquely provided $\pi^{ab}(x)\not\equiv 0$.
We thus have  \emph{one} linear combination among the infinitely many $\mathcal {T}_\phi S(x)$ constraints that remains first class with respect to all the other constraints and is also not gauge fixed by $\pi_\phi=0$. This constraint no longer has a spatial index, as it carries an integration. We denote this global constraint by
\be\label{H_gf}
H_{\mbox{\tiny gl}}:=\mathcal {T}_\phi S(N_0).
\ee
Now, we do not fix the lapse gauge to be given by $N_0$, but we separate the constraints into a first class part, given by
$$
\mbox{\bf First class:}~~~\{~~H_{\mbox{\tiny gl}}, \{\mathcal{Q}(x),x\in\Sigma\}, \{\mathcal {T}_\phi H^a(x),x\in\Sigma\}~~\}
$$
 and a purely second class part, given by
$$
  \mbox{\bf Second class:}~~~~\{~~\{\widetilde{\mathcal {T}_\phi S}(x):=\mathcal {T}_\phi S(x)-H_{\mbox{\tiny gl}}\sqrt{g},x\in\Sigma\}, \{\pi_\phi(x),x\in\Sigma\}~~\}.
$$
We will discuss the affirmation that $\widetilde{\mathcal {T}_\phi S}$ is indeed purely second class in the next subsection.

\subsection{Constraint Surface for Shape dynamics in compact closed manifolds.}\label{sec:Const.Surf.SD}

Now we show that the constraint $\widetilde{\mathcal {T}_\phi S}$ is equivalent to a constraint of the form $\phi-\phi_0(\Gamma_{\mbox{\tiny Grav}})$, the form necessary for the workings of proposition \ref{prop:equivalence} already anticipated in (\ref{gauge conditions}).

 We have
 \begin{equation}{ \mathcal{T}S}(x):\Gamma\times T^*(\mathcal{C}/\mathcal{V})\rightarrow C^\infty(M),\end{equation}  Since this map does not depend on $\pi_\phi$, we can fix $\pi_\phi(x)=f(x)$. Then
  \begin{equation} { {\mathcal {T}}S}(x)_{\pi_\phi=f(x)}:\Gamma\times \mathcal{C}/\mathcal{V}\rightarrow C^\infty(M).
\end{equation}
where we note that in fact ${\mathcal {T}_\phi}S(x)$ depends solely on $\mathcal{C}/\mathcal{V}$. Furthermore, everything said here using $C^\infty(M)$ as the domain can (and should) be extended to the square-integrable\footnote{We assume that, for all practical purposes, we can carry on as if all our spaces were Banach. The fact that they are not does not impose great obstacles for our approach, since we can use the constructions of section \ref{sec:Riem} (Sobolev lemma and the such) to regularize the domains.\label{footn}} domain \cite{EinsteinCentenary-FM}.

Consider the linear operator:
   $$\delta_{\mathcal{C}}{\mathcal {T}_\phi}S(g_0,\pi_0,\pi_\phi^0)_{|\phi=0}: {T}_0(\mathcal{C}/\mathcal{V})\rightarrow C^\infty(M).
   $$ where $T_xN$ denotes the tangent space at $x\in N$, and, as in usual partial derivatives, one holds the coordinates $(g,\pi,\pi_\phi)$ fixed. We will omit from now on the ``initial" point $(g_0,\pi_0,\pi_\phi^0)$ where we take the derivative.  One can explicitly check (see \eqref{equ:tangent_hatphi} in the appendix) that the tangent space to $\mathcal{C}/\mathcal{V}$ at $0$, ${T}_0(\mathcal{C}/\mathcal{V})$,  is given by smooth functions of the form $[f]:=f(x)-\mean f$, i.e. it is the linear version of the surjective `~$\hat~$~' map \eqref{equ:vpct_def}, which redundantly parametrizes the elements of $\mathcal{C}/\mathcal{V}$ by elements of $C^\infty(M)$.

The tangent map is given by:
\begin{equation}\label{iso2} \delta_{\mathcal{C}}{\mathcal {T}_\phi}S_{|\phi=0}:=\frac{\delta {{\mathcal {T}_\phi}S(x)}}{\delta\phi(y)}_{|\phi=0}=\{{\mathcal {T}_\phi}S(x),\pi_\phi(y)\}_{|\phi=0}
=\Delta(x)\delta(x,y)-\frac{\Delta(x)}{V}\end{equation}
 where $\Delta$ is given by \eqref{Delta}. Contraction of \eqref{iso2} with $N(x)$ yields  $\Delta N-\langle \Delta N\rangle$. Here we have denoted the derivative in the second coordinate, the one parametrized by $\phi$, by a subscript $\mathcal{C}$. We note that contraction in the $x$ variable requires us to use the adjoint $\delta_{\mathcal{C}}{\mathcal {T}_\phi}S^*_{|\phi=0}$ (see \eqref{Equ:pi_phi_derivative}), and as this is \emph{not} a self-adjoint operator the distinction is important. Thus
 $$ (\delta_{\mathcal{C}} \mathcal {T}_\phi S)^*\cdot N= \Delta N-\langle \Delta N\rangle
 $$

 If we then use \eqref{Equ:pi_phi_derivative}, we get from
   \begin{multline}\label{equ:def_N_0}\{\mathcal {T}_\phi S(N_0),\pi_\phi(\rho)\}=\langle (\delta_{\mathcal{C}} \mathcal {T}_\phi S)\cdot\rho,N_0\rangle_{C^\infty(M)}+\langle  \mathcal {T}_\phi S,\delta_{\mathcal{C}}(N_0)\cdot\rho\rangle_{C^\infty(M)}\\
   \approx\langle (\delta_{\mathcal{C}} \mathcal {T}_\phi S)\cdot\rho,N_0\rangle_{C^\infty(M)}=\langle (\delta_{\mathcal{C}} \mathcal {T}_\phi S)^*\cdot N_0,\rho\rangle_{C^\infty(M)}=0
 \end{multline}
 for all $\rho\in \mathcal {T}_0(\mathcal{C}/\mathcal{V})$ which means that under this inner product $\text{Im}(\delta_{\mathcal{C}} \mathcal {T}_\phi S) $ is perpendicular to $N_0$. \footnote{In fact, we should take $N$ to be a test function over $M$ in the square-integrable domain and not necessarily in $C^\infty(M)$. This affects none of our arguments (see \cite{EinsteinCentenary-FM}).\label{footn2} } By uniqueness, $N_0$ generates the whole annihilator of $\text{Im}(\delta_{\mathcal{C}} \mathcal {T}_\phi S) $.

 {Let  $\mbox{Ker}(W)$ denote the annihilator of the subspace $W\subset V$.  There exists an isomorphism between $\mbox{Ker}(W)$ and the dual to $V/W$ \cite{Rudin}. Now put $V=C^\infty(M)$ and $W=N_0$. This, together with the fact that the space generated by $N_0$ is a closed linear subspace of the dual, tells us that
 \be C^\infty(M)/N_0\simeq \text{Im}(\delta_{\mathcal{C}} \mathcal {T}_\phi S)\ee which is what we'll need.} Thus from now on assume $C^\infty(M)/N_0\simeq \text{Im}(\delta_{\mathcal{C}} \mathcal {T}_\phi S)$.

As $N_0$ is a closed one-dimensional linear subspace of $C^\infty(M)$, the tangent space to $C^\infty(M)/N_0$ is isomorphic to $C^\infty(M)/N_0$ at each point. Now we construct a modification of $\mathcal {T}_\phi S(x)$ such that it has the same tangent map but  its range must be such that:
  \begin{equation} \widetilde{\mathcal {T}_\phi S(x)}_{\pi_\phi=f(x)}:\Gamma\times \mathcal{C}/\mathcal{V}\rightarrow \text{Im}(\delta_{\mathcal{C}} \mathcal {T}_\phi S).
\end{equation}
 As predicted, this modified map is given by $\widetilde{\mathcal {T}_\phi S(x)}= \mathcal {T}_\phi S(x)-H_{\mbox{\tiny gl}}\sqrt{g}$, since as one can readily check the tangent map indeed stays the same and:
 $$\langle \widetilde{\mathcal {T}_\phi S}, N_0\rangle=\widetilde{\mathcal {T}_\phi S}(N_0)=\mathcal {T}_\phi S(N_0)-H_{\mbox{\tiny gl}}=0
 $$
where we must use the fact that we chose $\int N_0\sqrt g d^3x=1$ (otherwise we would have a numerical factor between the two elements of $\wts$).

For a heuristic explanation of what we are doing,  in the language of linear algebra, smearing functions can be viewed as a choice of a linear combination of the (continuously infinite) set of constraints. It follows from \eqref{H_gf} that \eqref{iso} has a kernel: the linear combination given by the smearing $N_0$.  We thus take a set of constraints that is linearly independent of \eqref{H_gf} given by $\widetilde {\mathcal {T}}_\phi S={\mathcal {T}}_\phi S-H_{\mbox{\tiny gl}}\sqrt{g}$.

We have \emph{not} yet proven that $\delta_{\mathcal{C}} \widetilde{\mathcal {T}}_\phi S $ is a topological linear isomorphism. We have shown that it is a surjective linear map, but we must still prove injectivity. We must still show that if  for some $\rho_0\in \mathcal{C}\simeq C^\infty(M)$, \be\label{equ:injectivity}\langle (\delta_{\mathcal{C}} \widetilde{\mathcal {T}_\phi S})^*\cdot N,\rho_0\rangle_{\mathcal{C}}=0\ee for all $N\in C^\infty(M)$ then $[\rho_0]=0$, i.e. $\rho(x)=\mean{\rho}$, which would mean we have a zero kernel of the linear map $\delta_{\mathcal{C}} \widetilde{\mathcal {T}_\phi S}$.

The differential operator $\Delta$ is invertible, possessing a Green's function. Thus for any function $f$ there exists some $N_f$ for which $\Delta N_f=f$. Since $(\delta_{\mathcal{C}} {\mathcal {T}}_\phi S)^*\cdot N=\Delta N-\mean{\Delta N}$, if \eqref{equ:injectivity} holds, we must have  $\langle f-\mean{f},\rho_0\rangle=0$ for any  function $f$.

Suppose then that $\rho_0(y)\neq 0$ for some   $y\in M$. Let us take $f_y(x)=\delta(x,y)\rho_0(x)$ (i.e. we take the point source of the field $\rho_0$).\footnote{We note that $f$ is here a locally integrable function, and not a smooth one. We could have done everything in this section appropriately in this setting (see \cite{Yamabe} and \cite{Gilberg} for the appropriate versions of the theorems used), but as this would introduce too many complications to this already involved construction we decided to leave it out.} Then $f_y(x)-\mean f_y=\delta(x,y)\rho(x)-\frac{\rho(y)}{V}$ and
$$\langle f_y(x)-\mean{f_y},\rho_0\rangle=\rho_0^2(y)-\rho_0(y)\mean{\rho_0}=0$$
which means $\rho_0(y)=\mean {\rho_0}$ and thus $[\rho_0]=0$. By the canonical transformation properties of $\mathcal{T}_\phi$, one can extend this construction to arbitrary $\phi$. We have thus proven
\begin{proposition}\label{prop2}
The linear map given by $\delta_{\mathcal{C}}\widetilde{\mathcal {T}_\phi S}(x):T_0(\mathcal{C}/\mathcal{V})\rightarrow \text{Im}(\delta_{\mathcal{C}} \mathcal {T}_\phi S)\simeq C^\infty(M)/N_0$
where $\widetilde{\mathcal {T}_\phi S}(x)=\mathcal {T}_\phi S(x)-H_{\mbox{\tiny gl}}\sqrt{g} $, is a toplinear isomorphism for all $(\phi, g,\pi)$ provided $\pi^{ab}\not\equiv 0$.
  \end{proposition}
  We have shown that it is a linear continuous bijection, and hence a topological linear isomorphism \cite{Lang}. $\square$.

   Thus not only can we form the Dirac bracket using $\{\widetilde{{\mathcal {T}}_\phi H}(x),\pi_\phi(y)\}^{-1}$,  but we can now use the implicit function theorem for Banach spaces for the function $\wts(x)_{\pi_\phi=f(x)}:\Gamma\times \mathcal{C}/\mathcal{V}\rightarrow \text{Im}(\delta_{\mathcal{C}} \mathcal {T}_\phi S)\simeq C^\infty(M)/N_0 $ to assert (with the caveat of footnote \ref{footn}) that
\begin{theorem}\label{prop}
 There exists a unique $\hat\phi_0:\bar\Gamma\rightarrow \mathcal{C}/\mathcal{V}$, where $\bar\Gamma$ is the restriction of phase space to $\pi^{ab}(x)\not\equiv 0$,  such that \end{theorem}
 $$ (\widetilde{{\mathcal {T}}_\phi S})^{-1}(0)=\{(g_{ij},\pi^{ij},
 \hat\phi_0[g_{ij},\pi^{ij}],\pi_\phi)~|~(g_{ij},\pi^{ij})\in\Gamma_{\mbox{\tiny {Grav}}}\}.
$$ In other words, we can find the solution to $\widetilde{{\mathcal {T}}_\phi S}(g,\pi,\phi,\pi_\phi)=0$ for all $(g,\pi,\pi_\phi)$, $\pi^{ab}\neq 0$, by setting $\phi=\phi_0$.
$\square$.

\subsection{Constructing the theory on the constraint surface}

We now have a surface in $\Gamma_{\mbox{\tiny {Ex}}}$, defined by $\pi_\phi=0$ and $\phi=\phi_0$, on which $\widetilde{T\mathcal{H}}=0$, and whose intrinsic coordinates are $g_{ij},\pi^{ij}$.
Furthermore, the Dirac bracket on the surface exists,  and  on the constraint surface we now have the symplectic structure
\begin{equation}
  \label{reduced Pb}\{\cdot,\cdot\}_{|{\mbox\tiny{reduced}}}:={\{\cdot,\cdot\}^{\Gamma_{\mbox{\tiny {Ex}}}}_{\mbox{\tiny{DB}}}}
    =\{\cdot_{|\phi=\phi_0,\pi_\phi=0},\cdot_{|\phi=\phi_0,\pi_\phi=0}\}.
\end{equation}Equivalently, for phase space functions independent of $\phi, \pi_\phi$, analogously to \eqref{Dirac bracket}:
\begin{multline} {\{F_1(x),F_2(y)\}^{\Gamma_{\mbox{\tiny {Ex}}}}_{\mbox{\tiny{DB}}}}_{|\phi=\phi_0,\pi_\phi=0}=\\
\{F_1(x),F_2(y)\}+\{F_1(x),(\phi-\phi_0)(x')\}\{\pi_\phi(x'),F_2(y)\}-\{F_1(x),\pi_\phi(x')\}\{(\phi_0-\phi)(x'),F_2(y)\}
\\ {=\{F_1(x),F_2(y)\}}\end{multline}
where the repeated variable $x'$ is integrated over. As a last corollary  of the use of the Dirac bracket we have:
\begin{cor}
For any phase space functional $f(g,\pi)$, the transformed functional $T_{\phi_0[g,\pi]}f(g,\pi)$ is volume-preserving-conformally invariant (vpct-invariant). \end{cor}
To prove this, we must merely use \eqref{equ:Pb_pi_phi,pi}:
\be\label{equ:vpct_equivariance}\{\mathcal {T}_\phi f(y),\pi_\phi(x)\}_{\mbox{\tiny{DB}}}=4\{\mathcal {T}_\phi f(y),\pi(x)-\mean{\pi}\sqrt g(x)\}_{\mbox{\tiny{DB}}
}=0=4\{\mathcal {T}_{\phi_0} f(y),\pi(x)-\mean{\pi}\sqrt g(x)\}
\ee

One can immediately see from \eqref{reduced Pb} that the first class constraints ${4(\pi-\mean\pi\sqrt g)},\mathcal{T}_\phi H^a$ and $ \langle T N_0S\rangle$ remain first class. Alternatively, for the diffeomorphism constraint, we can directly observe from \eqref{equ:transf_diffeo} that setting $\pi_\phi=0$ effects ${\mathcal {T}}_\phi H^a\rightarrow H^a$, yielding the usual diffeomorphism constraint.  We thus find the total Shape Dynamics Hamiltonian
\begin{equation}\label{equ:total_hamiltonian_SD}
  H_{\text{SD }}=\mathcal N\langle {\mathcal {T}}_{\phi_0} N_0S\rangle+\int_\Sigma d^3 x \left(\rho(x) 4(\pi(x)-\mean\pi\sqrt g)+\xi^a(x)H_a(x)\right)
\end{equation}
in the ADM phase space $\Gamma$ with the first class constraints
\begin{equation}
  \langle {\mathcal {T}}_{\phi_0}N_0S\rangle~,~{D:=4(\pi-\mean\pi\sqrt g)}~, ~{H}^a.
\end{equation} We have thus effectively fixed the gauge $N=N_0$ at the surface $\phi=\phi_0$. We have lost the freedom to fix the lapse, but retained the freedom to choose the conformal Lagrange multiplier $\rho$.

The non-zero part of the constraint algebra is given by:
\begin{eqnarray}
\{H^a(\eta_a),H^b(\xi_b)\}&=&H^a([\vec\xi,\vec\eta]_a)\nonumber\\
\label{equ:constraintAlgebra}
\{H^a(\xi_a),D(\rho))\}&=&D(\mathcal{L}_\xi\rho)
\end{eqnarray}
which substantially simplifies the algebra of constraints of gravity if compared to the ADM constraint algebra \eqref{equ:Dirac_ctraint_alg}.  Now we have two subalgebras, and the commutator of the two does not contain any structure functions.

\subsubsection{Further fixing of the gauge}

As an explicit check to see whether we indeed have the same theory, we can further gauge fix both ADM and Shape Dynamics to a system which possesses exactly the same gauge fixed Hamiltonian.
To do this, we merely input further gauge fixings: $S(x)=0$ in Shape Dynamics and $D=0$ in ADM. On the Shape Dynamics side, we have that restriction to the gauge fixing surface implies that we are over $(g,\pi)$ for which  $\phi_0(g,\pi)=0$. On the GR side, we have CMC slicing, which requires that $N=N_0$, and thus we arrive explicitly at GR in CMC gauge from both sides and have thus verified that the trajectories of the two theories are the same.

\subsubsection{Closing remarks on the procedure.}
 Let us briefly summarize some of the key questions that might arise from our presentation of SD. 
 First of all, it should be noted that imposing the gauge fixing $S=0$ in the shape dynamics side will only fix a volume-preserving conformal transformation. The argument is the following: for it to be a gauge fixing, we should be able to take any initial data $(g,\pi)$ to one that satisfies the gauge-fixing condition. We know that there exists a vpct such that we can bring any $(g,\pi)$ to one that satisfies $S=\mean{S}$, i.e. that brings the scalar constraint to a constant value. But the global constraint $H_{\mbox{\tiny gl}}$ demands that this constant be zero. So the gauge fixing $S=0$ is (in Shape Dynamics) weakly equal to the constraint $S-\mean{S}$.

What we mean by this is that gauge fixing the SD constraint does not solve the entire LY equation \eqref{equ:LY}, but the LY equation with an inhomogeneous term. This permits the rest of the scalar constraint to be accounted for by our global Hamiltonian, whose action indeed can and does change the volume.

In SD, the actual gauge fixing is, analogously to $\pi-\mean{\pi}=0$,  $S-\mean{S}=0$.  The gauge fixing $S-\mean{S}$ completely gauge fixes the vpct constraint, because the Dirac bracket is invertible between these two conditions. So we can always find a unique vpct factor so that $S-\mean{S}=0$, and this is of course also $\phi_0(g,\pi)$. But in the gauge fixation of SD,  we get the complete fixation of the Lagrange multiplier (or the velocity of the vpct facto) $\rho=0$, due to the invertibility of the bracket. 

\subsection{Properties of $N_0$}\label{sec:lapse}

We now show that indeed we have the invertibility properties that we need from the operator $\Delta$ and that $N_0$ is unique and such that $\mean{N_0}=1$. First of all, $\Delta$ is an \emph{elliptic, linear, self-adjoint second order differential operator on a compact manifold}. The  setting for the analysis of its properties could not be more convenient.  Ellipticity and linearity already guarantee a  fundamental solution (or Green's function) \cite{Gilberg}, in the sense that there exists a distribution $G_y(x)$ such that
$$ \Delta G_y(x)=\delta(x,y).
$$ However for our purposes it is not sufficient that the fundamental solution exist; it also needs to be unique,  which we  required  to solve for $\phi_0$ using the implicit function theorem. Thus we need to show that for our purposes no non-zero homogeneous solution exists, i.e., $\Delta N=0\Rightarrow N\equiv0$.

First, one should note that we require solvability of \eqref{lapse} only on the surface $\pi_\phi=0$, which reduces $\mathcal Q$ to $D=\pi-\langle\pi\rangle\sqrt g$.
Now, let us rewrite the operator $\nabla^2-\frac{1}{4}\langle\pi\rangle\pi-R$ in the form of \eqref{equ:PbSpi2} already using $D$:
\be\label{equ:positive_linearity} \Delta=\nabla^2-\frac{1}{4}\langle\pi\rangle^2-\frac{\pi^{ab}\pi_{ab}}{g}+\frac{1}{2}\mean{\pi}^2=
\nabla^2+\frac{1}{4}\langle\pi\rangle^2-\frac{\pi^{ab}\pi_{ab}}{g}
\ee
Now we further split $\pi^{ab}=\bar\sigma^{ab}\sqrt g+\frac{1}{3}g^{ab}\pi$ using its traceless part $\sigma^{ab}=\bar\sigma^{ab}\sqrt g$. Then
$$ -\frac{\pi^{ab}\pi_{ab}}{g}=-\bar\sigma^{ab}\bar\sigma_{ab}-\frac{1}{3g}\pi^2
$$
Substituting this back in \eqref{equ:positive_linearity} gives
\be\label{equ:new_Delta}\Delta=\nabla^2-\frac{1}{12}\langle\pi\rangle^2-\bar\sigma^{ab}\bar\sigma_{ab}.
\ee
Thus our operator $\Delta$ can be written as $\Delta=\nabla^2-f[g,\pi;x)$ where
  $f[g,\pi;x):=\bar\sigma^{ab}\bar\sigma_{ab}+\frac{1}{12}\mean{\pi}^2\geq 0$.

  This already implies that the only homogeneous solution $\Delta N=0$ is $N(x)=0$ \cite{Gilberg}. To see this in a simple way, suppose that there exists an $x$ such that $N(x)<0$. Then since $\Sigma$ is compact, it follows that $N$ attains a minimum, let us say at $x_0$. Then
 $$0<\nabla^2N(x_0)=f(x_0) N(x_0)<0,$$
 a contradiction. The same reasoning applies for $N(x)>0$.  Thus we are guaranteed not only existence but uniqueness of Green's functions for this case. For a complete proof taking into account the appropriate domains as square integrable functions, etc, see Theorem 2.8 in \cite{Yamabe}.
We thus have proven
 \begin{prop}\label{prop:meanN_0} If $\pi^{ab}(x)\not\equiv 0$,\footnote{Of course, if $\pi^{ab}=0$ in vacuum, then by the scalar constraint  the scalar curvature also vanishes, which all but trivializes the range of such static solutions. } the operator $\Delta$ appearing in \eqref{equ:Pb_pi_phi,pi} has a unique Green's function associated with it.
 \end{prop}
Using the fact that $\mathcal {T}_\phi$ is a canonical transformation, this also means that we can find a unique solution away from $\phi=0$ whenever $\mathcal {T}_\phi \pi^{ab}(x)\not \equiv 0$. In its turn, this would require that
$$ \pi^{ab}(x) -\frac{g^{ab}(x)}{3}\sqrt {g(x)}\langle \pi\rangle (1-e^{6\hat\phi}(x))=0
$$ which upon contraction with $g_{ab}$ and integrating yields $\mean{\pi}=0$, which means again $\pi^{ab}(x)=0$.

  As stated in section \ref{sec:SD_CMC}, we can choose the spatial constant $\Delta N'=c'$ such that $N'(x)>0$ for every $x$. Suppose then that $c'<0$ (here the primes have nothing to do with derivatives). From the above argument it follows that if $N'$ were negative anywhere, then at the minimum $x_0$, we would have $N'(x_0)<0$ and $\nabla^2 N'\geq0$. Thus $\Delta N'>0$ everywhere, which would imply $c'>0$; a contradiction. Thus $N'$ is non-negative. By the linearity in $N$, we can now  scale the constant $c'\rightarrow c$ so that $\mean{N}=1$.
 We have
 \begin{prop}\label{prop:meanN_02} There exists a unique constant $c<0$ such that the solution $\Delta N=c$ implies $N\geq 0$ and $\mean{N}=1$.
 \end{prop}

\subsubsection{Satisfaction of the exceptional requirements for the functioning of our mechanism. }

We briefly pause here to call attention to the fact that had we attempted to use not volume-preserving-conformal-transformations (vpct's), but full conformal ones, the corresponding operator one ends up with instead of \eqref{Delta} is given by
$$ (\nabla^2+R)N=0
$$which not only does not possess the same good invertibility properties, but even when invertible implies a frozen lapse. This fact, alluded to in the introductory section \ref{sec:LY}, is one of the reasons why it is of such fundamental importance that we use the volume-preserving ansatz. It allows a good definition of the theory everywhere on phase space (except for a trivial subset) and it allows the system to evolve, i.e. time to progress.

Furthermore, for our usage of the implicit function theorem and overall workings of the theory, we utilized many times and in diverse situations the fact that the gauge fixing $\pi_\phi$ is conjugate to a simple Lie transformation. These threefold requirements; $i)$ that it be a partial gauge fixing of $S(x)$ (owing to the good properties of the operator \eqref{equ:new_Delta}), $ii)$ that it still allows for time-evolution (owing to its leftover global constraint), and $iii)$ that it generates a symmetry (owing to the fact that indeed $D$ generates a first class constraint and true Lie group, see \ref{sec:group_structure}) are each extremely non-trivial demands. For example, a given gauge-fixing might leave an infinite amount of scalar constraints unfixed, which would be an unlikely advantage in the description of gravity. Or it might leave none, which would not leave room in the theory for time evolution. Alternatively, it might have led to an infinite chain of  constraints,  leading to an inconsistent system. Moreover, it might have been impossible to eliminate the extra variables simultaneously with solving the second class constraints. Moreover, we might have been able to exchange all but one of the scalar constraint with some other constraint, but this constraint could have been such that no real gain in simplicity would have been gained, as for instance if it was not linear in the momenta.  These  are the main factors that obstruct the construction of such symmetry trading in general.

\section{Construction of the section in $T^*\mbox{Riem}$}\label{sec:constructionTM}

 In section \ref{sec:Const.Surf.SD} we focused our attention on the operator $\delta_{\mathcal{C}}\mathcal{T}S$. There our primary objective was to use the implicit function theorem to solve the second class constraints and eliminate the additional variable $\phi$. Now that we have established the properties of $\phi_0$ in proposition \ref{prop},  we consider the whole construction in $T^*\mbox{Riem}$. After the elimination of the auxiliary variables $(\phi,\pi_\phi)$, what we have is that there exists a unique functional $\phi_0:T^*\mbox{Riem}\rightarrow \mathcal{C}/\mathcal{V}$ such that
 \be \label{equ:section_def}\frac{S({\mathcal T}_{\phi_0[g,\pi]}g_{ab},{\mathcal T}_{\phi_0[g,\pi]}\pi^{ab})}{\sqrt g}(x)=\frac{1}{V}\int d^3y S({\mathcal T}_{\phi_0[g,\pi]}g_{ab},{\mathcal T}_{\phi_0[g,\pi]}\pi^{ab})(y)N_0[{\mathcal T}_{\phi_0[g,\pi]}g_{ab},{\mathcal T}_{\phi_0[g,\pi]}\pi^{ab}](y)
 \ee
where we have established the properties of $N_0$ in propositions \ref{prop:meanN_0} and \ref{prop:meanN_02}. We now investigate the slightly peripheral question of whether $\phi_0$ yields a section of a bundle.

In section \ref{sec:York_splitting} we shall see that $T^*\mbox{Riem}$ indeed forms a bundle under conformal transformations as a gauge group. Furthermore, we can apply the extended version of theorem \ref{theo:slice} (see theorem 1.6 in \cite{FiMa77}) to the conformal group. The extension to which we refer also guarantees the existence of a section for $\mathcal{C}/\mathcal{V}$ whenever $\mathcal{V}$ forms a normal subgroup of $\mathcal{C}$.  Volume-preserving conformal transformations are particular instances of conformal transformations, but they do not form a group. Since they depend on the metric $g$, we only have the more primitive notion of a groupoid.\footnote{We have already established that the projection of $\mathcal{C}$ under the ``hat" map \eqref{equ:vpct_def} acts as a subgroup for each $g$ in section \ref{sec:group_structure}.} Nonetheless, as the map \eqref{equ:vpct_def} induces the equivalence relation in $\mathcal{C}$: $\phi_1\sim\phi_2\Leftrightarrow \phi_1=a\phi_2$, we have a quotient by the normal subgroup (since the group is abelian) of constant functions:  $\mathcal{C}/\R$.  Since the action of the group and the groupoid are both smooth, uniqueness and existence of $\hat\phi_0$ implies that this defines a ``section" also for the volume-preserving conformal transformations.  In \cite{Michorbook} it is shown more explicitly how one can extend the notion of a principal bundle to the smooth action of groupoids, we will not get into the technicalities and from now on assume we have the existence of a ``section" under the volume-preserving conformal transformations.

 So is $\phi_0[g,\pi]$ a ``section"? We note that we already have existence and uniqueness of an element of each orbit $\mathcal{O}_{(g,\pi)}$ for each $(g,\pi)$. Leaving aside the technical issue of whether one can define a section for a groupoid (see above), we should prove that under volume-preserving conformal transformations the ``section" stays the same, i.e.:
\be\label{equ:inv_section} ({\mathcal T}_{\phi_0[g,\pi]}g_{ab},{\mathcal T}_{\phi_0[g,\pi]}\pi^{ab})=({\mathcal T}_{\phi_0[\mathcal{T}_\lambda g,\mathcal{T}_\lambda\pi]}\mathcal{T}_\lambda g_{ab},{\mathcal T}_{\phi_0[\mathcal{T}_\lambda g,\mathcal{T}_\lambda \pi]}\mathcal{T}_\lambda\pi^{ab})
\ee
This would imply that we can see the section as a function from the quotient space $T^*\mbox{Riem}/(\mathcal{C}/\mathcal{V})$ to $T^*\mbox{Riem}$ which intersects orbits once (uniqueness) and transversely (existence, which implies that every orbit is intercepted by the section).
Furthermore, from \eqref{equ:vpct_equivariance},
\be\label{equ:gen_conf_invariance} \{\mathcal {T}_{\phi_0} f(y),\pi(x)-\mean{\pi}\sqrt g(x)\}=0
\ee
for any phase space functional $f[g,\pi](x)$, in particular for the canonical variables. Since  $\pi(x)-\mean{\pi}\sqrt g(x)$ generate vpct's, we indeed have \eqref{equ:inv_section}. To see this immediately from \eqref{equ:gen_conf_invariance}, let $\lambda(x)_t$ be a one-parameter family of Lagrange multipliers for the vpct symmetry. Then
$$ \frac{d}{dt}_{|t=0}{\mathcal T}_{\phi_0[\mathcal{T}_{\lambda_t} g, \mathcal{T}_{\lambda_t} \pi]}\mathcal{T}_{\lambda_t}g_{ab}=\int d^3 x\lambda'(x)\{\mathcal {T}_{\phi_0} g_{ab}(y),\pi(x)-\mean{\pi}\sqrt g(x)\}=0
$$
where $\lambda'(x)=\frac{d}{dt}_{|t=0}\lambda(x)_t$.

\subsubsection{Pictorial representation.}

\begin{figure}[h!]
\begin{center}
\includegraphics[width=0.45\textwidth]{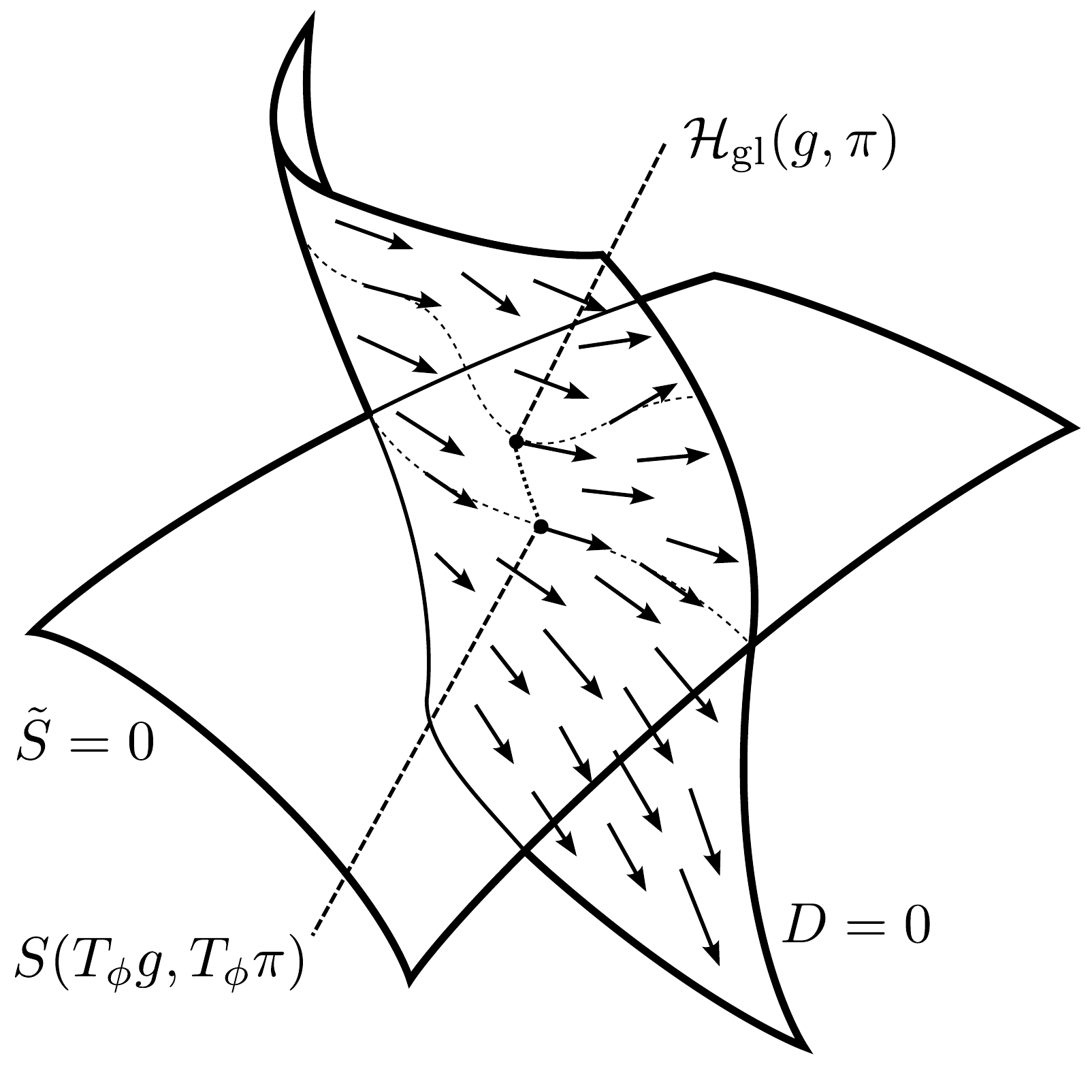}
\caption{The definition of $\hg$.  The Hamilton vector field $\hg$ is defined by the value of $S[g,\pi]$ on the surface $\tilde S = 0$: $\hg[g,\pi]:=S[T_{\phi_0}g,T_{\phi_0}\pi]$. We  show the constraint surface $D= 0$ intersecting $\tilde S = 0$ along the gauge fixing surface and illustrate the vector flow of  $\hg[g,\pi]$ along $D=0$.   }\label{fig:sd}
\end{center}
\end{figure}

This enables us to use the picture shown in figure \ref{fig:sd}.

 This provides us with an illustration of the relation between GR to SD. Both theories are defined by constraint surfaces on the ADM phase space. There is a subset, $\tilde S$, of $S$, defined as
 \be\label{equ:def:tilde_S}\tilde S[g,\pi,x):=S[g,\pi,x)-S(N_0)[g,\pi]\sqrt g(x).
 \ee
 We then have a manifold\footnote{One could attempt to use the fact that our surface, defined by \eqref{equ:section_def}, could now also be defined by the regular value theorem and with the aid of proposition \ref{prop:regularValue_S}. However one must remember here that in general the rhs of \eqref{equ:section_def} is a spatial constant that depends on the point $[g,\pi]$. So we want to regard the inverse value of the whole real line $\R$. One would have to use the transversality theorem to prove that $S^{-1}(\R)$ is a manifold. This amounts to proving that the composite map ${pr}\circ\delta_{(g^0,\pi^0)} S:T_{(g^0,\pi^0)}(T^*\mbox{Riem})\rightarrow C^{\infty}(M)/{\R}$ (where we have used the abbreviation $(g^0,\pi^0)=({\mathcal T}_{\phi_0[g,\pi]}g_{ab},{\mathcal T}_{\phi_0[g,\pi]}\pi^{ab})$, and $pr:C^{\infty}(M)\rightarrow C^{\infty}(M)/\R$  is the projection), is surjective and its kernel splits. We will not attempt to prove this in this thesis, as we already have an alternative proof.} defined by:
 \be \tilde S^{-1}(0)=({\mathcal T}_{\phi_0[g,\pi;x)}g_{ab}(x),{\mathcal T}_{\phi_0[g,\pi;x)}\pi^{ab}(x)).
 \ee
 That is, for each $(g,\pi)$ there is a single element in the vpct-class of $(g,\pi)$ that belongs to $\tilde S^{-1}(0)$. It is a section of $T^*$Riem under the group of vpct transformations (it intersects once each orbit). Thus by exploring the symmetry of $D$ we can solve all but one linear combination of the scalar constraints,  the global scalar constraint: $\hg[g,\pi]=S[\mathcal{T}_{\phi^0}g,\mathcal{T}_{\phi^0}\pi]$.

\subsection{Possibility of bypassing the linking theory construction}

 The crucial result in the intrinsic $T^*\mbox{Riem}$ view is that there exists a partial gauge fixing ($D=4(\pi-\mean{\pi}\sqrt g)$ ) of the scalar constraint that is also a generator of symmetry. This relies heavily on the fact is that one can invert the Poisson bracket $\{D,\tilde S\}$ for a certain linear combination of the scalar constraints: $\tilde S$. Furthermore, this gauge fixing leaves just one independent constraint $\hg$. But could we have done everything straight off in $T^*\mbox{Riem}$, without the introduction of the extended phase space $(\phi,\pi_\phi)$?

 Here the reader must be careful. For indeed we could have found $\tilde S$ directly from the picture in $T^*\mbox{Riem}$. But to find $\hg$ we exploited the conformal transformations, eventually finding a functional $\phi_0$ which solved for $\tilde S$ and thus yielded our global Hamiltonian $\hg$. At the end of the process, we got rid of the extra variables together with the second class constraints, expressing everything independently of the gauge fixing function and its conjugate ($\pi_\phi$ and $\phi$ respectively).

  We could attempt to forego the extended phase space and Linking Theory in the construction, by just considering $D$ as a gauge-fixing in $T^*\mbox{Riem}$. But the way would have to have been a little more roundabout. To follow the Dirac analysis properly, we now make explicit the steps we would have to take, writing in parentheses the ones we already took in the construction of SD.

  Step 1: impose $D(x)=0$ as a gauge-fixing in ADM (impose $\pi_\phi=0$). Step 2: we would first have to explicitly find the variable canonically conjugate to $D(x)$, let us call it $q_D$ (this is just $\phi$ in our case). Step 3: then we would have to find variables that Poisson commuted with both $D(x)$ and its canonical conjugate, $q_D$. These would already be vpct invariant variables (in our case these were just $(g,\pi)$). Step 4: find the purely second class combination of the constraints with respect to $D=0$. This is given by $\tilde S$. (in our case this part is exactly analogous, ours being given by $\widetilde T_\phi S$). Step 5:  Finally, show  that indeed we could solve all but one of the original scalar constraints $S(x)$ for $q_D$ as a function of the remaining variables (this is our $\phi^0[g,\pi,x)$, given in Proposition \ref{prop}).

    As all the quantities would be expressed in terms of variables that Poisson commuted with $D(x)$ we would automatically have a vpct invariant theory without going into extended phase space. In the author's opinion, this is roughly what Dirac had in mind in \cite{Dirac:CMC_fixing}, albeit solely from a gauge fixing point of view, i.e. not involving a ``conformal transformation" conceptual background, and also limiting the analysis to asymptotically flat space. We will discuss this a bit further at the end of chapter \ref{chapter:SD_AF}.

    Thus we come to the conclusion that without all the baggage presented in this chapter, in particular the use of the linking theory and extended phase space, the picture presented in figure \ref{fig:sd} is of very limited value. It does not in itself present how the dynamics are made consistent. Nonetheless it is a powerful pictorial representation of the end result.
    
\section{Comparisons with earlier work}    

Now that we have explicitly constructed the theory of Shape Dynamics (SD) we will present it against the backdrop of two its three main sources, leaving the comparison with Dirac to the next chapter.  

\subsection{Comparison with earlier work: Barbour et al}\label{sec:SD_vs_CS+V}
In \cite{Barbou:CS_plus_V} Barbour et al implemented best matching with respect to volume-preserving-conformal transformations (vpcts). By doing so they were able to derive the constraint $\pi-\mean{\pi}\sqrt g=0$ and came indeed close to arriving at the same theory we present in this thesis (SD). The main issue that blind-sighted them was that they used the Lagragian BSW formalism \eqref{Jacobi_local}. This means they never sought to distinguish between the first- and second-class sets of constraints, but were solely concerned with propagating $D$. Thus they indeed found the transformed version of equation \ref{lapse}, but could not  possibly have found our global scalar constraint \eqref{equ:total_hamiltonian_SD}. Furthermore, the Lagrangian BSW formalism gives the scalar constraint as an identity (see section \ref{sec:Barbour}) so that there is no place for the trading of constraints we have in Shape Dynamics. There is also some confusion over the fact that they attempt to interpret the transformed constraint $\mathcal{T}_\phi S(x)=0$ as an equation for $\phi$. Even if we are willing to grant this, more importantly the extra variables $(\phi,\pi_\phi)$ are never seen as an extension of the phase space, so that the initial degrees of freedom are not extended and no extra constraint gained. The solvability of equation $\mathcal{T}_\phi S(x)=0$, which is the LY equation, is then interpreted as \emph{defining a physical scale for each metric}. Of course if this is the case there is no room for admitting scale-invariance, as in SD.

In summary, \cite{Barbou:CS_plus_V} constituted a very important initial step towards SD, but fell short of finding a theory that contained vpct-invariance and a global scalar constraint.

\subsection{Comparison with earlier work: York}

We have already stressed in section \ref{sec:LY} the fact that, although it did not arise from canonical analysis of any sort,  the York conformal method must be physically conceived as just that: a method for solving the initial value problem of general relativity.
From the purely mathematical point of view, the form of the defining equation for Shape Dynamics is not too far removed from the Lichnerowicz-York equation \eqref{equ:LY}. The main difference in the form of the equations is that our principal result, contained in Theorem \ref{prop}, requires the solution of a non-homogeneous version of \eqref{equ:LY}, i.e. containing a 0-th order term in $\phi$. It is not yet clear whether we can apply the same arguments presented in \cite{Niall_73} to such an equation, which is why we resort to the implicit function theorem. We furthermore remark that it was not possible in the case of the paper \cite{Niall_73} to apply such methods to the solvability of \eqref{equ:LY} for two main reasons. First, the LY equation \eqref{equ:LY} in \cite{Niall_73}  did not arise from a gauge-fixing. This gauge fixing has to satisfy the fundamental requirement of also being a symmetry generator for the implicit function theorem to aid us in solving for the second class constraint. Secondly, even if one did try to use the generator of pure conformal transformations (without the volume-preserving condition of Shape Dynamics), the corresponding version of the linear operator given in \eqref{equ:new_Delta} would not be invertible everywhere on phase space.

\chapter{Trading GR for SD: asymptotically flat case}\label{chapter:SD_AF}

Let us now apply proposition \ref{prop:equivalence} and the construction leading to proposition \ref{prop:constructionPrinciple} to General Relativity to extend the results of section \ref{sec:SD_CMC} to asymptotically flat Cauchy surfaces. One of the leading differences is that it now makes no sense to talk about ``volume-preserving" as the volume is infinite. To recall, the volume-preserving condition is what allowed us to find the purely second class part of the scalar constraint and invert the Poisson bracket formed between itself and  the gauge fixing. A similar role will be played by an asymptotic fall-off condition on the conformal factor, as we will see.

\section{Constructing the Linking Theory}

To construct the linking gauge theory on a Cauchy-surface $\Sigma=\mathbb R^3$, we must first properly define the appropriate setting. We fix a Euclidean global chart (with radial coordinate $r$) and impose asymptotically flat boundary conditions. We implement this through the fall-off conditions of the 3-metric $g_{ab}$, its conjugate momentum density $\pi^{ab}$, the lapse $N$ and shift $N^a$ in the limit $r\to \infty$:
\begin{equation}\label{equ:boundary-cond}
 \begin{array}{rclcrcl}
  g_{ab}&\to&\delta_{ab}+\mathcal O(r^{-1}),&&\pi^{ab}&\to&\mathcal O(r^{-2}),\\
  N &\to&1+\mathcal O(r^{-1}),&&N^a&\to&\mathcal O(r^{-1}).
 \end{array}
\end{equation}
We call $\mathcal{C}$ the space of functions on $\Sigma$ with the fall-off rate ascribed to $N$. We should note that these conditions are not of utmost importance in what follows, it is only the fall-off conditions on $\phi$ that we ascribe below that is of relevance.  

We start with the equivalent of \eqref{equ:originalConstraints} and denote the usual ADM constraints as
\begin{equation}
 \begin{array}{rcl}
   S&=&\frac{\pi^{ab}\pi_{ab}-\frac{1}{2}\pi^2}{\sqrt g}-\sqrt g R\\
   H^a&=&\nabla_b\pi^{ab}.
  \end{array}
  \end{equation}

As before, we now embed the original system in an extended phase space that includes the auxiliary variables $(\phi,\pi_\phi)$. In accordance with the boundary conditions we assume the scalar $\phi$ falls off as
\begin{equation}
 e^{4\phi}\to 1+\mathcal O(r^{-1})
\end{equation}
for $r\to\infty$ and that its conjugate momentum density $\pi_\phi$ falls off sufficiently fast at $r\to\infty$. We call the space of such $\phi$ $\mathcal{C}_r$. The nontrivial canonical Poisson brackets are still given by \eqref{equ:canonicalPbs}. Again have an extended phase space with $(\phi,\pi_\phi)$,
with the additional constraint \eqref{equ:originalAdditionalConstraints}
$\pi_\phi\approx 0$ determined by our embedding.

Following \eqref{equ:generatingFunctional}, we construct the generating function \begin{equation}\label{equ:generatingFunctional2}F_\phi:=\int_\Sigma d^3x \left(g_{ab}(x)e^{4\phi(x)}\Pi^{ab}(x)+\phi(x)\Pi_\phi\right),\end{equation} where capitals denote the transformed variables. Note the lack of hatted variables in this case. We find the canonical transformation analogous to \eqref{equ:canonicalTransformation}:
\begin{equation}
 \begin{array}{rcl}
   g_{ab}(x)&\to& T_\phi g_{ab}(x):=e^{4\phi(x)}g_{ab}(x)\\
   \pi^{ab}(x)&\to&T_\phi \pi^{ab}(x):=e^{-4\phi(x)}\pi^{ab}(x)\\
      \phi(x)&\to&T_\phi \phi(x):=\phi(x)\\
   \pi_\phi(x)&\to&T_\phi \pi_\phi(x):=\pi_\phi(x)-4\pi(x)
 \end{array}
\end{equation}
and again subsequently use these transformed variables to construct three sets of constraints: the transformed scalar and diffeomorphism constraint of GR  as well as the transform of $\pi_\phi$,
\begin{equation}
 \begin{array}{rcl}
  \mathcal{Q}&=&\pi_\phi-4\pi.
 \end{array}
\end{equation}

Using a scalar Lagrange-multiplier $\rho$, which is required to fall off as $\mathcal O(r^{-1})$ as $r\to\infty$, we define the total Hamiltonian\footnote{We should for general purposes add a regularizing boundary term
to the total Hamiltonian, as it diverges in the present form. However since this does not impinge on either the equations of motion nor on the constraints, we omit it in order to avoid cluttering  the paper.}
\be\label{Hamiltonian} H_{\mbox{\tiny {Total}}}=\int d^3x [N(x)T_\phi S(x)+\xi^a(x)T_\phi H_a(x)+\rho(x)\mathcal{Q}(x)]
\ee

This completely defines the linking $T_L$ as contained in section \ref{sec:constructionPrinciple} in an analogous fashion to the compact closed case treated in the previous section (see \eqref{gauge conditions}).

\section{Recovering General Relativity in the asymptotically flat. }

Again the only nonvanishing Poisson-bracket of the gauge fixing condition $\phi(x)=0$ with the constraints of the linking theory is
\begin{equation}
 \{\phi(x),Q(\rho)\}=\rho(x),
\end{equation}
and everything follows in the same way: we can follow through from \eqref{gauge conditions}, and, in the language of Proposition \ref{prop:equivalence}, arrive at $\rho \equiv 0$ and $f_i\approx 0$ equivalent to $S(x)\approx 0$. We have lost the freedom to fix $\rho$, but retained the freedom to fix the lapse.

\section{Recovering Shape Dynamics in the asymptotically flat case}
Our main aim in this section will be to prove that part of the scalar constraints can again be written again in the form $\phi-\phi_0(g,\pi)\approx 0$ on the gauge-fixing surface $\pi_\phi\equiv 0$.

The only weakly non-vanishing Poisson-bracket of the gauge-fixing condition $\pi_\phi(x)=0$ with the constraints of the linking theory is $\{T_\phi S(N),\pi_\phi(x)\}=4T_\phi\{S(N),\pi(x)\}$, which leads to
\begin{equation}
 \{ S(N),\pi(x)\}=
 2(\nabla^2N-NR)\sqrt g-\frac 3 2 NS\approx 2\sqrt g(\nabla^2-R)N
\end{equation}
The differential operator
\begin{equation}
\label{Delta}\Delta=\nabla^2-R
\end{equation}
is an  elliptic, second order, self-adjoint operator,  invertible \emph{for the given boundary conditions}. So for the  boundary conditions given in \eqref{equ:boundary-cond}, we have the unique kernel \be\label{N_0}N_0[g,\pi]\not\equiv 0\ee
Thus, by the canonical transformation properties, we have a unique solution for
$\{T_\phi S[N],\pi_\phi(x)\}=0$,  $T_\phi N_0$.

Again, as in \eqref{H_gf}, we denote the  one linear combination, among the infinitely many $T_\phi S(x)$ constraints, that remains first class with respect to all the other constraints by
\be
H_{\mbox{\tiny gl}}:=\mathcal{T}_\phi S(N_0).
\ee
Again, we separate the constraints into a first class part, given by
$$
\mbox{\bf First class:}~~~\{~~H_{\mbox{\tiny gl}}, \{\mathcal{Q}(x),x\in\Sigma\}, \{{\mathcal T}_\phi H^a(x),x\in\Sigma\}~~\}
$$
 and a purely second class part, given by
$$
  \mbox{\bf Second class:}~~~~\{~~\{\widetilde{\mathcal{T}_\phi S}(x):=\mathcal{T}_\phi S(x)-H_{\mbox{\tiny gl}}\sqrt{g},x\in\Sigma\}, \{\pi_\phi(x),x\in\Sigma\}~~\}.
$$

\subsection{Constraint Surface for Shape Dynamics}

Now we again show that, even for the asymptotically flat case, the constraint $\widetilde{{\mathcal T}_\phi S}$ is equivalent to a constraint of the form $\phi-\phi_0(\Gamma_{\mbox{\tiny Grav}})$.

 We have that
 \begin{equation}{ \mathcal{T}_\phi S}(x):\Gamma\times T^*(\mathcal{C}_r)\rightarrow C^\infty(M),\end{equation}  Since these equations do not depend on $\pi_\phi$, we can fix $\pi_\phi(x)=f(x)$. Then
  \begin{equation} { {{\mathcal T}_\phi}S}(x)_{\pi_\phi=f(x)}:\Gamma\times \mathcal{C}_r\rightarrow C^\infty(M).
\end{equation}
 Thus
   $$\delta_{\mathcal{C}}{{\mathcal T}_\phi}S_{|\phi=0}:\mathcal{C}_r\rightarrow C^\infty(M).
   $$
    We thus take the set of constraints that is linearly independent of \eqref{H_gf}, which is given by $\widetilde {\mathcal T}_\phi S$. Clearly $\widetilde {\mathcal T}_\phi S(N_0)=0$, which means indeed it lives in the dual space of the quotient of $\mathcal{C}$ by $N_0$.  Effectively, we must subtract from any $N\in\mathcal{C}$ the function $ N_0$ given by \eqref{N_0}.

   Consider the linear self adjoint elliptic operator we presently have:
\begin{equation}\label{iso} \delta_{\mathcal{C}}{{\mathcal T}_\phi}S_{|\phi=0}:=\frac{\delta {{{\mathcal T}_\phi}S(x)}}{\delta\phi(y)}_{|\phi=0}=\{{{\mathcal T}_\phi}H(x),\pi_\phi(y)\}_{|\phi=0}=\Delta(x)\delta(x,y)
\end{equation}
If we then use \eqref{Equ:pi_phi_derivative}, we again have
   \begin{equation}\{\mathcal {T}_\phi S(N),\pi_\phi(\rho)\}\approx\langle (\delta_{\mathcal{C}} \mathcal {T}_\phi S)\cdot\rho,N\rangle=\langle (\delta_{\mathcal{C}} \mathcal {T}_\phi S)^*\cdot N,\rho\rangle
 \end{equation}
 Then from the Fredholm alternative (Theorem \ref{theo:Fredholm}) and self-adjointness
 \be\label{equ:splitting2}C^\infty(M)\simeq  \text{Im}(\delta_{\mathcal{C}} \mathcal {T}_\phi S) \oplus \text{Ker}(\delta_{\mathcal{C}} \mathcal {T}_\phi S)^*= \text{Im}(\delta_{\mathcal{C}} \mathcal {T}_\phi S) \oplus \text{Ker}(\delta_{\mathcal{C}} \mathcal {T}_\phi S)
 \ee
 The splitting is then given by
\be \text{Im}(\delta_{\mathcal{C}} \mathcal {T}_\phi S) \oplus \text{Ker}(\delta_{\mathcal{C}} \mathcal {T}_\phi S)=\text{Im}(\delta_{\mathcal{C}} \mathcal {T}_\phi S) \oplus N_0
\ee

   Thus  the appropriate versions of both Proposition \ref{prop2} and Theorem \ref{prop} work in the asymptotically flat case.
 It is important that there is a non-zero homogeneous solution to the $\Delta$ operator, so we are left with a first class component of ${\mathcal T}_\phi S$. This is the reason for using not full conformal transformations in the compact case, but only those that preserve the total spatial volume. The analogous restriction arises from the fall-off conditions in the present case.

The construction of the theory on the constraint surface and the further fixing of the gauge therefore proceeds in the same manner as was shown in the constant mean curvature case. Of course  instead of having constant mean curvature slicing, we now have the maximal slicing $\pi=0$.

\subsection{Comparison with Dirac's work}\label{sec:comparison:Dirac} As mentioned already in section \ref{sec:Dirac}, and followed up in section \ref{sec:constructionTM}, in 1958 Dirac  already anticipated much of the constructions present in shape dynamics in the asymptotically flat case. We pick up from the end of section \ref{sec:Dirac}. After setting $\pi(x)=0$ as a gauge-fixing condition, he noticed that it will be second class only with respect to the scalar constraint $S(x)$.  Then he finds that there exists \emph{a} variable canonically conjugate to $\pi(x)$, namely $\ln (g^{1/3})$, where $(g^{1/3})$ is a tensor density of weight $2/3$ (his equation 28). He then changed variables to consider the metrics with unit determinant $\tilde g_{rs}$ (a coordinate-dependent statement), and the tensor density of weight $2/3$: $\tilde \pi^{rs}:=(\pi^{rs}-\frac{1}{3}\pi g^{rs})g^{1/3}$, both of which have zero Poisson bracket with $\pi(x)$ and its conjugate variable. Then he assumed that one can solve the scalar constraint by fixing the value of the variable $\ln (g^{1/3})$ as a function of the tilded variables. After one does this, one has completely eliminated $\pi(x)$, its conjugate and the scalar constraint. Although Dirac never mentions it, since the remaining variables all commute with $\pi(x)$ we have a conformally invariant theory. Of course, as we are taking Poisson brackets between tensor densities whose density does not add up to 1, these statements are all coordinate-dependent, which is why he later mentions that one would have to also fix the spatial coordinates.

This indeed bears a strong resemblance to the work presented in this section, and indeed with the method we use for the Hamilton--Jacobi equation (section \ref{sec:HJ}). Even if we disregard the spatial coordinate-dependence of the procedure Dirac outlined, there are still several disparities with what we did. One, as we mentioned at the end of section \ref{sec:constructionTM}, is that Dirac heuristically went through the entire procedure from a gauge fixing point of view, and not as a symmetry trading, one of the main selling points of SD which he never considered or mentioned. Furthermore, Dirac falls into a bit of contradiction, as he retains a non-zero global (which he calls ``main")  Hamiltonian (his equation 32). But without boundary considerations (which he did not seem to have for these particular equations), solving the scalar constraint (his equation 30) would set the ``main" Hamiltonian (his equation 17) also to zero. Lastly, of course, he never showed whether one can indeed solve the scalar constraint in terms of the variable canonically conjugate to $\pi(x)$ (the specific case of \eqref{equ:dirac's_solution}). Had he done so, we would have also anticipated the work of York et al \cite{Niall_73} (see section \ref{sec:LY}).

Concluding, although there are several enticing hints and interesting directions, that original paper falls reasonably short of actually defining Shape Dynamics in the asymptotically flat case.

\section{Lagrangian Picture}\label{sec:Lagrangian}

Let us consider the Lagrangian of the linking gauge theory in an attempt to relate the local degrees of Shape Dynamics with those of General Relativity. The local degrees of freedom of standard General Relativity are given by the ADM-decomposition of a 4-metric, i.e. a 3-metric, shift vector field and lapse field, while shape dynamics is a local theory of a 3-metric, a shift vector field, the conformal field $\phi$ and the conformal Lagrange-multiplier $\rho$. While the 3-metric and shift vector field are naturally identified, one needs the Euler--Lagrange equations to investigate further.

Using $D(\xi)=\int d^3x H^a(x)\xi_a(x)$, $C(\rho)=\int d^3x \mathcal Q(x) \rho(x)$ and the supermetric $G_{abcd}=g_{ac}g_{bd}-\frac 1 2 g_{ab}g_{cd}$ we can write the action for the linking theory in canonical form as
\begin{equation} \label{equ:linking-action}
 \begin{array}{rcl}
  S&=&\int dt\left(d^3x \left(\dot g_{ab}\pi^{ab}+\dot\phi \pi_\phi\right) -\left({\mathcal {T}}_\phi S[N]+D[\xi]+C[\rho]\right)\right)\\
   &=&\int dtd^3 x\left( \frac{1}{4N}G^{abcd}(\dot g_{ab}-\mathcal{L}_\xi g_{ab}-\rho g_{ab})(\dot g_{cd}-\mathcal{L}_\xi g_{cd}-\rho g_{cd})+N {\mathcal {T}}_\phi R\right),
 \end{array}
\end{equation}
where we used the equations of motion
\begin{equation}
 \begin{array}{rcl}
   \dot g_{ab}&=&2N \pi^{cd}G_{abcd}+\mathcal L_\xi g_{ab}+\rho g_{ab}\\
   \dot \phi&=&-\rho-\mathcal{L}_\xi\phi
 \end{array}
\end{equation}
to eliminate the momenta. Coming purely from the  Lagrangian one could now think that it would be possible to find an equation that relates the lapse and the conformal Lagrange multiplier. This basically would mean we could relate the local speed of time to local speed of scale. To see that this is not  possible, we consider the  general construction principle for linking theories as explained in section \ref{sec:constructionPrinciple}.  We start with the Gauss-Codazzi split of the Einstein-Hilbert action
\begin{equation}
 S=\int dt d^3 x\sqrt{|g|}\left(\frac 1{4N}\left(\dot g_{ab}-(\mathcal L_\xi g)_{ab}\right)G^{abcd}\left(\dot g_{cd}-(\mathcal L_\xi g)_{cd}\right)+N R[g]\right)
\end{equation}
and use the transfromation $g_{ab}(x)\to {\mathcal {T}}_\phi g_{ab}(x):= e^{4\phi(x)}g_{ab}(x)$, which again yields the action for the linking theory, i.e. the second line of (\ref{equ:linking-action}), with $-\rho$ replaced by $\dot \phi$ as it should be
\begin{equation}
  S=\int dtd^3 x\left( \frac{1}{4N}G^{abcd}(\dot g_{ab}-\mathcal{L}_\xi g_{ab}+\dot\phi g_{ab})(\dot g_{cd}-\mathcal{L}_\xi g_{cd}+\dot\phi g_{cd})+N {\mathcal T}_\phi R\right).
\end{equation}
It is quite easy to show that the constraint $Q$ comes out as a primary constraint from the Legendre transform of this Lagrangian, as it was to be expected from section \ref{sec:constructionPrinciple}. However, there is no relationship possible between the Lagrange multiplier $\rho$ of the conformal constraint and the lapse $N$, neither in the linking theory (where both Lagrange multipliers are free), nor in Shape Dynamics (where the lapse is fixed but $\rho$ is free). One could now argue that one can write a relationship between $N$ and $\phi, \dot\phi$ by imposing the constraint $\pi_\phi=0$. But this choice, as we showed in the previous section, fixes $\phi=\phi_0, \dot\phi=\dot\phi_0$. We thus find for this case that the Lagrangian for Shape Dynamics is written as the second line of equation (\ref{equ:linking-action}) with $\phi=\phi_0$ and $-\rho=\dot \phi_0$, which admits no dynamical relation anymore.

\chapter{Causal Structure and coupling to different fields}\label{chapter:coupling}
One of the outstanding features of Shape Dynamics (SD), is that it contain one single global Hamiltonian constraint which generates evolution. As this theory no longer possesses many fingered time, or Lorentz invariance for that matter, it becomes a crucial concern of the program to establish its causal structure. The natural way to do this, which is what we pursue in this section,is to study propagation of a scalar field.

\section{General criteria for coupling.}

When we try to couple different fields to gravity, we will have to face basically one question: how does one scale fields? What exponent shall we choose in  $\psi\rightarrow e^{n\hat\phi}\psi$? This is an important issue because if the scaling is not correct we could encounter two difficult obstructions.

 The first obstruction is that if we are dealing with a field that possesses some kind of gauge symmetry it might not be possible to find a constraint $\mathcal{Q}$, as in  \eqref{equ:transformedConstraintsSD}, that is first class with respect to the gauge constraint. The second is that the conserved charge (which we call $D$) implicit inside $\mathcal{Q}$  that defines the foliation  might depend on the field. In this case, there might be an even worse consequence if the field possesses some sort of gauge symmetry (like electromagnetism). For then the charge could turn out to depend on the gauge potential. This could be the case even if we can find a field-dependent $D$ that is first class with respect to the gauge generator of the field. Indeed with any other choice of scaling than the one chosen in the text, this is what happens with the electromagnetic field, where the generator of the gauge symmetry is the Gauss constraint. We will not show it, but it is possible to make a scaling choice different from the one we make in the following section, but still such that the Gauss constraint is propagated. But what happens then is that the charge is $U(1)$ gauge dependent.

 So any of the couplings that we choose should  pass these two hurdles. As it turns out, the solution exists only when we require the fields to have trivial scaling with respect to the conformal factor,  $\psi\rightarrow \psi$.

\subsection{Coupling to the scalar field.}\label{sec:coupling:scalar}
The most natural way to approach the coupling of a scalar field is not to try to develop it directly in SD, but to make use of the linking theory to facilitate its introduction. We shall concentrate on the case of the closed spatial manifold without boundary.

Let us define the fields we shall be working with. The usual Hamiltonian density for a scalar field $\psi$  is given by
\be\label{equ:Hamiltonian_scalar} H_\psi =\frac{\pi_\psi^2}{\sqrt g}+g^{ab}\nabla_a\psi\nabla\psi_b\sqrt g
\ee
where $\pi_\psi$ is the momentum conjugate  to the scalar field. The original gravitational constraints amended by the constraints arising from the coupling to the scalar field can be written as
\begin{equation}
 \begin{array}{rcl}
   S&=&\frac{\pi^{ab}\pi_{ab}-\frac{1}{2}\pi^2+\pi_\psi^2}{\sqrt g}-\sqrt g (R-g^{ab}\nabla_a\psi\nabla\psi_b)\\
   H^a(\xi_a)&=&\int d^3 x(g_{ab}\mathcal{L}_\xi\pi^{ab}+\psi \mathcal{L}_\xi\pi_\psi)
  \end{array}
  \end{equation}
where for ease of manipulation we wrote the smeared version of the diffeomorphism constraint.

We now embed the original system in an extended phase space including the auxiliary variables $(\hat\phi,\pi_{\hat\phi})$ in the same way.
The new nontrivial canonical Poisson bracket is
\begin{equation}\label{equ:canonical_relations}
  \begin{array}{rcl}
    \{\psi(x),\pi_\psi(y)\}&=&\delta(x,y).
  \end{array}
\end{equation}
The extended phase space for these fields is now:
$$(g_{ij},\pi^{ij}, \psi, \pi_\psi,\phi,\pi_\phi)\in\Gamma_{\mbox{\tiny {Ex}}}:=\Gamma_{\mbox{\tiny {Grav}}}\times \Gamma_{\mbox{\tiny {Scalar}}}\times \Gamma_{{\mbox{\tiny {Conf}}}}$$
with the additional constraint :
\begin{equation}\pi_\phi\approx 0
\end{equation}

We construct the generating function \begin{equation}F_\phi:=\int_\Sigma d^3x( g_{ab}(x)e^{4\hat\phi(x)}\Pi^{ab}(x)+\phi\Pi_\phi+\psi\Pi_\psi)\end{equation} and find the previous canonical transformations and two new ones:
\begin{equation}
 \begin{array}{rcl}
      \psi(x)&\to&{\mathcal {T}}_\phi \psi(x)=\psi(x)\\
   \pi_\psi(x)&\to&{\mathcal {T}}_\phi \pi_\psi(x):=\Pi_\psi=\pi_\psi
 \end{array}
\end{equation}
We again use the transformed variables to construct three sets of constraints: the transformed scalar- and diffeomorphism-constraint of GR  as well as the transform of $\pi_\phi$,
\begin{equation}
 {\mathcal {T}}_\phi S;~ {\mathcal {T}}_\phi H^a ;~
  \mathcal{Q}:=\pi_\phi-4(\pi-\langle \pi\rangle\sqrt g)
\end{equation}
It can again be shown that the $\mathcal{Q}$ constraint restricts the functions in $\Gamma_{\mbox{\tiny {Ex}}}$ to be in a one to one relation with the functions on the embedding of $\Gamma_{\mbox{\tiny {Grav}}}\times \Gamma_{\mbox{\tiny {Scalar}}}$ independent of $\pi_\phi$ (equivalently  $\mathcal{Q}$ holds on the image of ${\mathcal {T}}_\phi$ as applied to functions dependent solely on the original phase space coordinates).  The fact that
 \be {\mathcal {T}}_\phi H^a(\xi_a)\approx\int d^3 x(g_{ab}\mathcal{L}_\xi\pi^{ab}+\psi \mathcal{L}_\xi\pi_\psi+\phi \mathcal{L}_\xi\pi_\phi)
 \ee can also be explicitly computed using $\mathcal{Q}$. We will refrain from doing these calculations as they do not differ from the vacuum case.

 The linking theory gravitational Hamiltonian is:
\be H_{\mbox{\tiny {Total}}}=\int d^3x [N(x){\mathcal {T}}_\phi S(x)+\xi^a(x){\mathcal {T}}_\phi H_a(x)+\rho(x)\mathcal{Q}(x)]
\ee Now we use the gauge fixing $\pi_\phi=0$, and find that it weakly commutes with all constraints except for ${\mathcal {T}}_\phi S$.
We get from equation \eqref{equ:PbSpiScalar}, the modification of the $\Delta$ operator \eqref{Delta}:
\be \Delta_{\mbox{\tiny scalar}}= (\nabla^2-\frac{\pi\langle\pi\rangle}{4\sqrt g}-R+g^{ab}\nabla_a\psi\nabla_b\psi)
\ee
To show that this has the desired properties, we must again show that the linear term $ -\frac{\pi\langle\pi\rangle}{4\sqrt g}-R+g^{ab}\nabla_a\psi\nabla_b\psi$ is non-positive. To do so, we use the scalar constraint to get the equation in the form of \eqref{equ:PbSpi2}:
\be\Delta_{\mbox{\tiny scalar}}= -\frac{G_{abcd}\pi^{ab}\pi^{cd}-\pi_\psi^2}{\sqrt g}-\frac{1}{4}\pi\mean{\pi}+\sqrt g\nabla^2
\ee
which allows us to use the same decomposition as in \eqref{equ:new_Delta}. Since the extra term $-\pi_\psi^2$ is negative, we still have uniqueness and existence of the solution $N_0[g,\pi,\pi_\psi]$.

We then define the gauge-sfixed part of ${\mathcal {T}}_\phi S$ as $\widetilde {{\mathcal {T}}_\phi S}={\mathcal {T}}_\phi S-{\mathcal {T}}_\phi S(N_0)$. It can be shown that $\widetilde {{\mathcal {T}}_\phi S}$ can be written as $\phi-\phi_0(g,\pi,\psi,\pi_\psi)$ in the same way as was done in \eqref{prop}. This constraint exhausts the gauge fixing $\pi_\phi=0$ (they have invertible Poisson bracket), and as second class constraint can be set strongly to zero alongside $\pi_\phi$ to eliminate the extra variables and allow us to use the usual Poisson bracket instead of the Dirac ones.

The two outstanding features of the coupling are that the constraint $\mathcal{Q}$ does not depend on the scalar field $\psi$ and that one can still uniquely solve the lapse fixing equation for a functional $N_0[g,\pi,\pi_\psi]$ such that $\langle N_0\rangle=1$. We thus have well defined shape dynamics coupled to a scalar field given by the first class constraints:
\be\label{equ:scalar_field_constraints} \langle {\mathcal {T}}_{\phi_0} S N_0\rangle;~~\{ H^a(x) ,~ x\in \Sigma\};~~\{D(x):=4(\pi(x)-\langle \pi\rangle\sqrt g(x)),~x\in\Sigma\}
\ee where we have used that with the reduction ${\mathcal {T}}_\phi H^a\rightarrow H^a$.

\subsubsection{Adding a cosmological constant}
We now show that our proof only works for a certain range of the cosmological constant. If the new (modified) scalar constraint is given by adding $\Lambda\sqrt g$, it is trivial to see that this will contribute with a term $\frac{3}{2}\Lambda N\sqrt g$ to the Poisson bracket $\{S(N),\pi\}$. Thus to complete the $-\frac{3}{2}S(x)$ term, we must add and subtract $3\Lambda\sqrt g$ getting the modified version of  \eqref{Delta}:
\be \Delta_{\Lambda}= (\nabla^2-\frac{\pi\langle\pi\rangle}{4\sqrt g}-R+\frac{3}{2}\Lambda)
\ee But if we now use the scalar constraint to get the equation in the form of \eqref{equ:PbSpi2}, we obtain:
\be \Delta_{\Lambda}=-\frac{G_{abcd}\pi^{ab}\pi^{cd}}{\sqrt g}-\frac{1}{4}\pi\mean{\pi}+\sqrt g(\frac{\Lambda}{2}+\nabla^2)
\ee Using the same techniques as before (section \ref{sec:lapse}), we only have guaranteed uniqueness and existence for $\Lambda\leq 2\bar\sigma^{ab}\bar\sigma_{ab}+\frac{1}{6}\mean{\pi}^2$. We note that for asymptotic de Sitter, $\Lambda=3\leq 6= 2\bar\sigma^{ab}\bar\sigma_{ab}+\frac{1}{6}\mean{\pi}^2$. It is still interesting that such a bound exists. Adding a massive potential term to the scalar field in the previous section, of the form $\psi^2\sqrt g$, puts the same bound on the density of such a field. The set of conditions in which a field initially respecting this bound will evolve to one that does not is still under study.

\subsection{Coupling to the electromagnetic field.}\label{sec:coupling:EM}

In coupling the electromagnetic constraints we have one more ingredient than in the scalar field case. Namely, now we must also remember to include the Gauss constraint.

The Hamiltonian density for electromagnetism is
\be H_{\mbox{{\tiny EM}}} =-A_{[a,b]}A_{[c,d]}(x)g^{ac}(x)g^{bd}(x)\sqrt{g}(x)+
\frac{E^a(x)E^b(x)g_{ab}(x)}{\sqrt g}(x)\ee
where $E^a$ is the vector density canonically conjugate to $A_a$ (and not a vector field). The constraints are
\begin{equation}
 \begin{array}{rcl}
   S&=&\frac{\pi^{ab}\pi_{ab}-\frac{1}{2}\pi^2}{\sqrt g}-\sqrt g (R+H_{\mbox{{\tiny EM}}})\\
   H^a(\xi_a)&=&\int d^3 x(g_{ab}\mathcal{L}_\xi\pi^{ab}+A_a \mathcal{L}_\xi E^a)\\
   G&=&\nabla_a \bar E^a
  \end{array}
  \end{equation}
  where $\bar E^a\sqrt g =E^a$ defines the electric vector field $\bar E^a$. We then replace the last equation of \eqref{equ:canonical_relations} by
  $$     \{A_{a}(x),E^{c}(y)\}=\delta^{c}_{a}\delta(x,y)\\
  $$
  The generating functional is
  \begin{equation}\label{equ:generatingFunctional_EM}F_\phi:=\int_\Sigma d^3x( g_{ab}(x)e^{4\hat\phi(x)}\Pi^{ab}(x)+\phi\Pi_\phi+A_a\mathcal{E}^a ),\end{equation}The new transformations are
  \begin{equation}
  \begin{array}{rcl}
  A_a(x)&\to&{\mathcal {T}}_\phi A_a(x)=A_a(x)\\
   E^a(x)&\to&{\mathcal {T}}_\phi E^a(x):=\mathcal {E}^a(x)=E^a(x)
 \end{array}
\end{equation}
The new constraints are
\be\label{equ:EM_constraints} \langle {\mathcal {T}}_\phi S N_0\rangle~;~\{{\mathcal {T}}_\phi H^a(x) ,~ x\in \Sigma\}~;~\{D(x):=4(\pi(x)-\langle \pi\rangle\sqrt g(x)),~x\in\Sigma\}~;~\{e^{-6\hat\phi(x)}G(x),~x\in\Sigma\}
\ee
  Here we used
  $$ \nabla_a \bar E^a=\frac{1}{ \sqrt g}\partial_a \sqrt g\bar E^a=\frac{1}{\sqrt g}\partial_a E^a
  $$

Now, we reproduce \eqref{equ:new_Delta} for electromagnetism.  We get from equation \eqref{equ:PbSpiEM}, the modified $\Delta$ operator of \eqref{Delta}:
\be \Delta_{\mbox{\tiny EM}}= (\nabla^2-\frac{\pi\langle\pi\rangle}{4\sqrt g}-R+\frac{H_{\mbox{\tiny EM}}}{2\sqrt g})
\ee
To show that this has the desired properties, we must again show that the linear term $ -\frac{\pi\langle\pi\rangle}{4\sqrt g}-R+\frac{H_{\mbox{\tiny EM}}}{2\sqrt g}$ is non-positive. To do so, we use the scalar constraint to get the equation in the form of \eqref{equ:PbSpi2}:
\be 2(-\frac{G_{abcd}\pi^{ab}\pi^{cd}}{\sqrt g}-\frac{1}{2}H_{\mbox{\tiny EM}}-\frac{1}{4}\pi\mean{\pi}+\sqrt g\nabla^2)
\ee
which allows us to use the same decomposition as in \eqref{equ:new_Delta}. Since the extra term $-\frac{1}{2}H_{\mbox{\tiny EM}}$ is negative, we still have uniqueness and existence of the solution $N_0[g,\pi,A,E]$.

In the same way as with the scalar field, we now have well defined shape dynamics coupled to vacuum electromagnetism, given by the first class constraints:
\be\label{equ:scalar_field_constraints} \langle {\mathcal {T}}_{\phi_0} S N_0\rangle;~~\{ H^a(x) ,~ x\in \Sigma\};~~\{D(x):=4(\pi(x)-\langle \pi\rangle\sqrt g(x)),~x\in\Sigma\};~~\{G(x),~x\in\Sigma\}
\ee
\section{Emergence of the causal structure.}\label{sec:causal_reconstruction}

General Relativity is a theory of the spacetime metric, but the physical interpretation of this metric arises through a clock and rod model. Terms like
light-cone put the operational meaning of geometry to the forefront. Shape
Dynamics does not immediately provide a spacetime metric, but a spacetime
interpretation of Shape Dynamics comes from a clock and rod model in the
same way as in General Relativity. The simplest one of these is a multiplet of
free scalar fields, which we will consider in this section.
For this we assume that the field strength $\psi^i(x,t)$ of the $i$ components of a
scalar multiplet and conjugate momentum density $\pi_\psi^i(x,t)$ can be prepared at
every point $x\in\Sigma$ and initial time $t$, and that both are measurable at later times.
Moreover, we assume that the fields can be prepared as test fields, i.e. the field
strength and momentum density is small enough, so that the back-reaction on gravity
can be  neglected. To recover the spacetime metric at a point $(x_0,t_0)$ we first
consider the equations of motion for test fields with Hamiltonian \eqref{equ:Hamiltonian_scalar}:
$$  H_\psi =\frac{\pi_\psi^2}{\sqrt g}+g^{ab}\nabla_a\psi\nabla\psi_b\sqrt g
,$$ which are:
\be\label{equ:matter_prop} \begin{array}{rcl}\{S(N), \psi\}&=&2\frac{2N\pi_\psi}{\sqrt g}\\
\{S(N),\pi_\psi\}&=&\sqrt {g} g^{ab}\nabla_a\nabla_b\psi\end{array}
.\ee
We now prepare the first six components of the scalar multiplet around a given point $x_0^i=0$ (in some chart) as
\be \psi_{(ab)}(x)=\psi_{(ab)}(x_0)+\delta_a^i\delta_b^jx_ix_j+  \order{x^3}
\ee
This determines through \eqref{equ:matter_prop} the metric at point $x_0$ up to $\sqrt g$.  The initial velocity of the field is prepared such that
\be \dot\psi(x_0)=1
\ee
which translates into $\pi_\psi=\sqrt g$ and thus we get from \eqref{equ:matter_prop} the lapse $N(x_0)$.
We thus can recover the ADM-decomposition of the metric by inverting this system for the components of the ADM metric.

\subsection{Testing different actions with the reconstruction of the metric.}

The shape dynamics Hamiltonian, given by $\hg$, is a complicated beast, given implicitly by an inverse elliptic operator. Suppose however that we would like, from first principles, to come up with a Hamiltonian that still obeys the symmetries of SD and that furthermore matches general relativity to some given approximation, what would be a reasonable form? In answer to this question, let us once again consider the smeared form of the scalar constraint:
\be \int d^3 xN\left(\frac{\pi^{ab}\pi_{ab}-\frac{1}{2}\pi^2}{\sqrt g}-\sqrt g (R-2\Lambda)\right)
\ee
The most straightforward attempt to make something  that is volume-preserving-conformally invariant out of this is to have a gauge fixing: $g_{ab}\mapsto e^{\lambda[g,x)}g_{ab}$ where $\lambda:\mbox{Riem}\to \mathcal{C}/\mathcal{V}$ is a section of the fiber bundle $\mbox{Riem}$ with gauge group  $\mathcal{C}/\mathcal{V}$. The most natural one, for which the gauge fixing properties are known to work is the so  called Yamabe gauge \cite{Yamabe}. For this gauge, the scalar curvature is a constant:
\be\label{Yamabe} R(e^{\lambda[g,x)}g)(x)=R_0
\ee

To make it diffeomorphism invariant, we integrate over each term with the constant smearing $N=1$. Since our gauge freedom does not involve the volume, in the end we get the following Hamiltonian, separated by the powers of volume involved in the conformal weight of each term:
\be\label{equ:H_CFT}
   H_{3} := \lf( 2\Lambda - \frac 3 2 \mean{\pi}^2 \rt) - \frac{\mean{R[e^{4\lambda[g,x)}g]}}{V^{2/3}}
           + \frac 1{V^2} \mean{\frac{\bar \sigma^{ab} \bar\sigma_{ab}}{e^{6\lambda[g,x)}g}}
\ee
where $ \bar \sigma^{ab} = \lf({V}\rt)^{\frac{2}{3}}\lf(\pi^{ab} - \frac 1 3 \mean{\pi} g^{ab} \sqrt{g} \rt)$ is the traceless part of $\pi^{ab}$, and $\lambda[g,x)$ is the Yamabe functional, chosen so that we keep vpct invariance of each term. There are of course other choices of vpct invariant actions.
Interestingly, upon making a formal volume expansion of the SD Hamiltonian we get $H_{3}$ as the first three terms appearing (see section \ref{sec:HJ}), which is why we labeled this Hamiltonian as $H_3$. This guarantees that at least to the allowed order in volume $H_{3}$ agrees dynamically with general relativity.

In the same way, we can use the construction principle for the Hamiltonian of the gravitational field coupled to a scalar field presented in the previous section.   For an interesting comparison with general relativity, we can now use the reconstruction of space-time presented in section \ref{sec:causal_reconstruction}, replacing $\hg$ with $H_{3}$.

\chapter{Volume expansions and Hamilton-Jacobi approach.}\label{sec:HJ}
In this section, we give an explicit perturbative construction of SD in a large-volume expansion.  This was our first attempt at an approximation scheme to the global Hamiltonian of SD. More fruitful schemes are being carried out at present, but will not be contained in this thesis.

The comparison with earlier work in the classical HJ approach to GR \cite{Freidel2008, Boer2000} brings to light at least one great practical advantage of SD over GR: the ability to implement \emph{all} local constraints of SD, as in SD we are able to work out all \emph{local} degrees of freedom due to the linearity in the momenta of the constraints. Remarkably, this construction provides a rigorous classical correspondence between gravity at very large volumes and  conformal field theory (CFT).

The SD Hamiltonian can be explicitly constructed by solving an elliptic differential equation whose coefficients are local phase space functions. The nonlocality of the solution of this differential equation introduces nonlocality into SD. However, we will show that, with some caveats,  it is possible to construct explicit solutions in a large volume expansion at least to third order.

This analysis may be of interest for the semiclassical dS/CFT-correspondence. The AdS/CFT-correspondence \cite{Witten1998, Maldacena:ads_cft}, which relates the asymptotic wavefunction of quantum gravity in the bulk to the partition function of a CFT on the boundary, has generated an incredible amount of interest. The correspondence applies strictly to special limits of Type IIB string theory and $\mathcal N = 4$ super Yang--Mills theory, but has lately been cast as an example of a more general gauge/gravity duality. In rough terms, the AdS/CFT dictionary relates radial evolution in AdS space with renormalization group flow of a conformal field theory  on the boundary.

The results of this section can be tentatively interpreted in terms of the correspondence. As it stands however, the connection is not direct and we will refrain from drawing too many parallels.  Heuristically, by considering a CMC trajectory that approaches a homogeneous spacetime at large volume so that the spatial volume becomes an asymptotic clock, the correspondence would arise from interpreting ``volume time'' as the ``Renormalization Group (RG) time'' of an Euclidean CFT partition function. The advantage of SD over GR in this setting is the following: in GR,  the implementation of all local constraints is complicated by the nonlinearity of the constraints \cite{Freidel2008}; whereas, in SD, one can determine the local physical degrees of freedom because the local constraints are \emph{linear} in the momenta.

\section{A practical way to calculate the global Hamiltonian.}

The defining characteristic of our global Hamiltonian is the non-local functional $\phi_0$, given by Proposition \ref{prop}. A more pragmatic approach to calculating the global Hamiltonian together with the functional $\phi_0$ is to use equation \eqref{equ:section_def}, which we reproduce here for convenience:
$$\frac{S({\mathcal T}_{\phi_0[g,\pi]}g_{ab},{\mathcal T}_{\phi_0[g,\pi]}\pi^{ab})}{\sqrt g}(x)=\frac{1}{V}\int d^3y S({\mathcal T}_{\phi_0[g,\pi]}g_{ab},{\mathcal T}_{\phi_0[g,\pi]}\pi^{ab})(y)N_0[{\mathcal T}_{\phi_0[g,\pi]}g_{ab},{\mathcal T}_{\phi_0[g,\pi]}\pi^{ab}](y).
$$
Then our aim is basically to simultaneously solve the two equations. We  assume a priori that there exists a unique solution (see theorem \ref{prop}):
\bea\label{equ:alt_def_H_gl} \frac{\mathcal{T}_\phi S}{\sqrt g}(x)&=&\hg\\
\label{equ:hat_cond}\mean{e^{6\phi}}&=&1
\eea
for $\phi$ and $\hg$.

 Proposition \ref{prop} allows us to assume that there exists a $\phi$ such that  \eqref{equ:alt_def_H_gl} implies $\frac{\mathcal{T}_\phi S}{\sqrt g}$ is a spatial constant. Thus we can take the mean without further consequences: $\mean{\mathcal{T}_\phi\frac{S}{\sqrt g}}=\frac{\mathcal{T}_\phi S}{\sqrt g}=\hg$. This will be our main tool in calculating the effective global Hamiltonian. Let us rewrite the conformally transformed scalar constraint \eqref{equ:transformedScalar} with a slightly more convenient notation here:
 \be\label{equ:transformedScalar2}\frac{\mathcal{T}_\phi S}{\sqrt g} = \frac{1}{g \Omega^{12}} \lf( \pi^{ab}\pi_{ab} - \frac{\pi^2}{2} - \frac{\mean{\pi}}{6} (1 -\Omega^6)^2 g + \frac{\mean{\pi}}{3} \pi (1 - \Omega^6) \sqrt g \rt) + 2\Lambda - \frac{R}{\Omega^4} + 8\frac{\nabla^2\Omega}{\Omega^5}
 \ee
  where $\Omega=e^\phi$ eliminates the divergence squared term from \eqref{equ:transformedScalar}.

  Before going on, we will consider a slightly different SD Hamiltonian, where the difference is however ``pure gauge". We do this as follows:
  as $\frac{\mathcal{T}_\phi S}{\sqrt g}$ is first-class wrt $\pi_\phi-D$, where $D=4(\pi-\mean{\pi}\sqrt g)$, we can choose a different basis of first class constraints made up of
  \be{\mathcal{T}_\phi S}'(x):={\mathcal{T}_\phi S}(x)-f[g,\pi,\phi, \pi_\phi; x)(\pi_\phi-D)(x)~, ~\mbox{and}~\pi_\phi(x)-D(x)
  \ee We can check that if we Poisson commute
  \be\{{\mathcal{T}_\phi S}',\pi_\phi\}\approx \{{\mathcal{T}_\phi S},\pi_\phi\}\ee
  which thus implies we can follow through the proof of Proposition \ref{prop} by merely substituting ${\mathcal{T}_\phi S}$ by ${\mathcal{T}_\phi S}'$  wherever it appears.

  Thus by setting $\pi_\phi$ strongly to zero, we see that instead of using \eqref{equ:transformedScalar2} to solve \eqref{equ:alt_def_H_gl} we can use
 \be\label{equ:ts'}  \frac{\mathcal{T}_\phi S'}{\sqrt g} = \frac{1}{g \Omega^{12}} \sigma^{ab} \sigma_{ab} - \frac{\mean{\pi}^2}{6} + 2\Lambda - \frac{R}{\Omega^4} + 8 \frac{\nabla^2\Omega}{\Omega^5} \approx 0
 \ee
  In the above expression, we have defined the traceless momenta $\sigma^{ab}$ as
\begin{equation}\label{equ:def:traceless_mom}
    \sigma^{ab} = \pi^{ab} - \frac 1 3 \mean{\pi} g^{ab} \sqrt g.
\end{equation}
For convenience we will also define:
\be\label{equ:def:P}
P=\frac{2}{3}\mean{\pi}.
\ee
From now on, we will use $\frac{\mathcal{T}_\phi S'}{\sqrt g}$ instead of $\frac{\mathcal{T}_\phi S}{\sqrt g}$ but drop the prime for convenience.

The last preparatory result is that, assuming again that a unique solution exists (Theorem \ref{prop}), we can exploit the conformal invariance and thus go to the Yamabe gauge (see section \ref{sec:Yamabe}).
This relies on the result that all closed manifolds are conformally constant curvature. For this, one needs to show that a constant $ R_0$ can be found such that
\begin{equation}
    R_0 = R(e^{4 \hat \lambda[g,x)} g)
\end{equation}
for some non--local functional $\hat \lambda[g,x)$ of $g$. The restriction $\mean{e^{6\hat\lambda}} = 1$ selects a unique value of $R_0$ and determines the metric representative (up to a  diffemomorphism).

\section{Large volume expansion}

To perform this expansion we write the SD Hamiltonian as a power series in $V^{-2/3}$, where $V$ denotes the total spatial volume.
To expand $\hg$ in powers of $V^{-2/3}$, the explicit $V$ dependence of $\hg$ must be isolated. This can be done using the change of variables $(g_{ab};\pi^{ab}) \to (V,\bar g_{ab}; P, \bar \sigma^{ab})$ given by
\begin{align}\label{equ:def:new_variables}
   \bar g_{ab} &= \lf(\frac{V}{V_0} \rt)^{-\frac{2}{3}} g_{ab}, \qquad \qquad V = \int d^3x \sqrt{g}, \\
   \bar \sigma^{ab} &= \lf(\frac{V}{V_0}\rt)^{\frac{2}{3}}\lf(\pi^{ab} - \frac 1 3 \mean{\pi} g^{ab} \sqrt{g} \rt),  ~ P= \frac 2 3 \mean{\pi}.
\end{align}
where $V_0 = \int d^3 x \sqrt{\bar g}$ is a fixed reference volume.
One can easily verify that $V$ and $P$, defined in \eqref{equ:def:P}, are canonically conjugate: $\pb{V}{P} = 1$. Furthermore  $\pb{P}{\bar g_{ab}} = \pb{P}{\bar \sigma^{ab}} = 0$, so that the barred variables are independent of $V$ (as they were designed to be). Note that $\bar g_{ab}$ and $\bar \sigma^{ab}$ are not canonically conjugate. Exploiting the fact that the unique solution for the conformal factor implies we can  use vpct invariance, we will consider all the variables to be taken in the Yamabe gauge \eqref{equ:YamabeGauge} from now on.

Then $\hg$ is found by simultaneously solving the equations
\begin{align}\label{equ:main hg}
   \hg^0 = \lf( 2\Lambda - \frac{3}{8}P^2 \rt) + \frac{\lf( 8 \bar{\nabla}_0^2 - \bar R_0 \rt) \Omega}{(V/V_0)^{2/3}\Omega^5}
       - \frac{\bar \sigma^{ab} \bar \sigma_{ab} }{(V/V_0)^2\Omega^{12}\bar g_0}
   \\
   \mean{\Omega^6} = 1, \label{equ:vpct_cond}
\end{align}
where barred quantities are calculated using $\bar g^0_{ab}$ and the (super)subscript $0$ in this section denotes the Yamabe gauge.

The large $V$ expansion is
\begin{equation}
   \hg = \sum_{n=0}^\infty  \left(\frac{V}{V_0} \right)^{-2n/3} \hn{n}, ~~ \Omega^6 = \sum_{n=0}^\infty  \left(\frac{V}{V_0} \right)^{-2n/3}  \wn{n} .
\end{equation}
The restriction \eqref{equ:vpct_cond} is trivially solved by $\mean{\wn{n}} = 0$ for $n \neq 0$ and $\mean{\wn{0}} = 1$. We can solve for the $\hn{n}$'s by inserting the expansion, taking the mean, and using the fact that $\bar R_0$ is constant.

The complete solution up to order $V^{-2}$is calculated in section \ref{sec:volume_expansion}. It is
\begin{equation}\label{equ:vol_expansion_fixed}
\mathcal H^0_{\text{gl}}=  2 \Lambda - \frac{3}{8} P^2  - \frac{{R_0}}{V^{2/3}} + \frac{1}{V^2} \left< \frac{\bar \sigma^{ab} \bar \sigma_{ab}}{ \bar g_0} \right> + \order{\lf(V/V_0\rt)^{-8/3}}~.
\end{equation}
A couple of comments are in order. First, we note that each term in the expansion is  diffeomorphism  invariant but vpct gauge dependent. Thus we conformally covariantize it, so that it coincides with the above equation over the Yamabe section. We get:
\be\label{equ:vol_expansion_general} \mathcal H_{\text{gl}}=  2 \Lambda - \frac{3}{8} P^2  - \frac{{R[e^{4\lambda[g]}g_{ab}]}}{V^{2/3}} + \frac{1}{V^2} \left< \frac{\bar \sigma^{ab} \bar \sigma_{ab}}{ \bar e^{12\lambda[g]}g} \right> + \order{\lf(V/V_0\rt)^{-8/3}}~.
\ee
Second, note that to first order in V the evolution generates just global conformal transformations. This does not mean of course that at asymptotic large volumes the Universe itself is homogeneous, which would indeed exclude all interesting asymptotic solutions of GR. It means only that evolution becomes homogeneous, or that in a sense evolution ``freezes out" at asymptotically late times.

\section{Hamilton-Jacobi equation for large volume.}

In the case of unconstrained systems, the Hamilton-Jacobi theory provides a bridge between classical and quantum mechanics. As we have all first class \emph{linear} local constraints here, we also have a classical Hamilton-Jacobi formulation of the theory for our case. With non-linear first class constraints, one is simply not able to implement the constraints at the level of Hamilton's principal function (see \cite{Henneaux&Teitelboim}, section 5.4.4). The principal function is of course very different from the large volume expansion studied in the previous section. First of all, we must choose an initial metric, then the Hamilton-Jacobi functional of a metric $g$ is considered to be the action of a solution connecting the initial metric and $g$. The value of the action can in principle change  if there are more than one solution between the two points. In the case of first class constraints, which generate gauge symmetries, this does not happen. That is the fundamental reason why it is not possible to implement second class constraints in the HJ approach as opposed to first class ones.

We can now solve the HJ equation for SD in the large volume limit. Again making use of our gauge principles, our point of departure will be the gauge fixed  \eqref{equ:vol_expansion_fixed}, as opposed to the gauge invariant \eqref{equ:vol_expansion_general}. After performing the necessary calculations in this particular gauge we will ``covariantize"  the results to a general one.

First we recall that $\bar g_{ab}$ and $\bar \sigma^{ab}$ are not canonically conjugate; the barred variables defined in \eqref{equ:def:new_variables} were only specifically designed to be independent of $V$. We show that the canonical Poisson brackets between the barred variables and $V$ and $P$ vanish in section \ref{sec:HJ_app}. Hence we can obtain the HJ equation  from  making the substitutions in our volume expanded Hamiltonian \eqref{equ:vol_expansion_fixed}:
\begin{align}
   P &\to \diby{S}{V} & \pi^{ab} &\to \diby{S}{ g^0_{ab}},
\end{align}
where $S= S(g^0_{ab}, \alpha^{ab})$ is the HJ functional that depends on the metric $g^0_{ab}$ and parametrically on  \emph{integration constants} $\alpha^{ab}$. These  constants $\alpha^{ab}$ are symmetric tensor densities of weight 1: they parametrize the non-gauge part of the initial conditions (or the initial point in the constraint manifold) in the Hamilton-Jacobi approach. We will fix our initial point to be given by asymptotic de Sitter, which translates into defining a homogeneous separation constant.  {These conditions are compatible with asymptotic (in time) dS space, which has maximally symmetric CMC slices.} The treatment of other separation constants is currently under investigation.
Unlike the usual (A)dS/CFT correspondence, this is not the most general case that can be considered within our framework (the generalization to the Euclidean AdS case is trivial, requiring the scalar constraint to be expressed as a radial, instead of a time, evolution operator).

We can express $\bar \sigma^{ab}$ in terms of $\diby{S}{ g_{ab}}$ and use the chain rule to write the result in terms of $\diby{S}{V}$ and $\diby{S}{\bar g_{ab}}$. We prove that the $V$ derivatives drop out of the final expression in \eqref{equ:traceless_dec}.  The final expression for $\sigma$ we  use is
\begin{equation}\label{equ:chain_rule_sigma}
   \bar \sigma^{ab} \to \diby{S}{\bar g_{ab}} - \frac 1 3 \mean{ \bar g_{ab} \diby{S}{\bar g_{ab}}} \bar g^{ab} \sqrt{\bar g}.
\end{equation}

Let us note that in the gauge fixed version,  the variations of $ R_0$ can be found using the standard variations of $R$. The $\sn{n}$'s can be found recursively using our solution for $\sn{0}$ and by collecting powers of $(V/V_0)^{-2/3}$.
The equation we are then trying to solve is:
\begin{multline}
2 \Lambda - \frac{3}{8} \left( \frac{\delta S}{\delta V}\right)^2  - \frac{{R_0}}{V^{2/3}}\\
 + \frac{1}{V^2} \mean{  \left(\diby{S}{{\bar g}^0_{ab}} - \frac 1 3 \mean{ {\bar g}^0_{ab} \diby{S}{{\bar g}^0_{ab}}} {\bar g}^0_{ab} \sqrt{{\bar g}^0}\right){\bar g}^0_{ac}{\bar g}^0_{cd} \left(\diby{S}{{\bar g}^0_{ab}} - \frac 1 3 \mean{ {\bar g}^0_{ab} \diby{S}{{\bar g}^0_{ab}}} {\bar g}^0_{ab} \sqrt{{\bar g}^0}\right)  } + \order{V^{-8/3}} = 0~.
\end{multline}
The expansion we are going to use, still of course in powers of $V^{-2/3}$, to solve this is
\begin{equation}
S = S_0 V + S_1 V^{1/3} + S_2 V^{-1/3} + \order{V^{-1}}
\end{equation}
We then insert this expansion into the HJ equation obtained using the substitutions above. as mentioned, to obtain a complete integral of the HJ equation, $\sn{0}$ can be taken of the form $\sn{0}=\int d^3 x \alpha^{ab}g^0_{ab}$. The linear constraints determine $\alpha^{ab}$ to be transverse and with covariantly constant trace. The leading order HJ equation determines the value of the trace of $\alpha^{ab}$, as we will see. This restricts the freely specifiable components of $\alpha^{ab}$ precisely to the freely specifiable momentum data in York's approach \cite{York:york_method_prl}. To encode an initial point that is asymptotically deSitter, we restrict ourselves to separation constants with vanishing transverse traceless part.

To be explicit , the first terms (reinstating $V_0$ from the calculations done in section \ref{sec:HJ_S_n}, are
\begin{align}
   \sn{0} &= \mp \sqrt{\frac {16} {3\Lambda} }V_0 \\
   \sn{1} &= \mp \sqrt{\frac 3 \Lambda }  R_0 \, V_0 = \mp \sqrt{\frac 3 \Lambda }  \int d^3x \sqrt { g_0}  R_0, \\
\sn{2} &= \pm \lf(\frac 3 \Lambda\rt)^{3/2} \int d^3 x \sqrt {\bar g} \lf( \frac 3 8 \tilde R^2 - \tilde R^{ab} \tilde R_{ab} \rt).\end{align}
Note that $\sn{0}$ and $\sn{1}$ are the only terms with positive dimension.  Gauge invariant solutions can be obtained by restoring the $\lambda[g,x)$ dependence of the tilded variables.  To draw a contrast of this result with that of \cite{Freidel2008, Boer2000}, we briefly note that these terms all solve the local HJ constraints of SD in asymptotic dS space. 

\part{Gauge Theory in Riem.}

\chapter{Riem as a principal fiber bundle}\label{sec:Riem}

In this Chapter we will build the technical tools to be used in Chapter \ref{chapter:connection_forms}. Most of the content of the present Chapter can be found in \cite{Ebin} and \cite{FiMa77}, albeit in slightly different language.
\section{Introduction}

Gauge theory, needless to say, has a long and rich history, and it is probably not an exaggeration to state it has by now permeated all areas of theoretical physics as an essential tool for existing frameworks and guide for future developments. It describes systems which possess some inherent symmetry in their parametrizations, and for classical fields over spacetime it has a well-developed geometrical understanding through the use of principal fiber bundles.

Geometrodynamics, as championed by Wheeler, is the study of gravitation through a primary focus on {\it space and
changes therein} rather than on space-time itself. Space-time is essentially `sliced-up' and described as an
evolution of the geometry of these spatial slices through time. It is fundamentally a dynamical view of GR,
 technically taking form as its constrained canonical, or  ADM formulation \cite{Arnowitt:1962hi}.

 Although widely regarded as a gauge theory (since all of its constraints are first class and thus interpreted as symmetry generating), there is no specific description of ADM as a gauge theory in the geometric, fiber bundle sense, making use of connection forms, sections and so forth. This is in part because a connection over configuration space seems to be far removed from reality. What would such a connection do? This is one of the questions we aim to answer in this part.

As is well known,  the unconstrained configuration space for General Relativity is defined as
$$\M:=\riem=~~~\mbox{the space of all 3-Riemannian metrics over }~~ M~$$
The Hamiltonian dynamics thus takes place on (a constraint submanifold of) $T^*\riem$. By a geometrical setting of gauge theory, mathematically we mean the existence of a principal fiber bundle and, most importantly, a connection form on it.
The first inkling of a connection form in $\riem $ arose in \cite{Giulini:1993ct,Giulini:2009np}, where the mention of horizontal and vertical components of metric velocities first appears. It is however our understanding that the concept was not fully explored, and one of the purposes of the present work, at least from the mathematical standpoint, is to investigate exactly what constitutes a connection over configuration space. That is,  what are the properties such a connection has to satisfy and how can we construct one both formally and explicitly. In doing so we would like to shed light on explicit infinite-dimensional geometrical gauge theories over configuration space, a point of view so far as we know original.
From the physical point of view, this may be connected with a richer history of relational ideas (see section \ref{sec:Barbour}).

A second reason why this approach has not been attempted before is because of the difficulties in interpreting the action of the Hamiltonian constraint as a group action, and other issues related to the infamous ``problem of time" \cite{Anderson:2010xm}. As in the first part of the thesis we have shown that it is possible to have a theory of gravity which no longer possesses the scalar (or Hamiltonian)  constraint, and thus no refoliation invariance. Unlike what is the case in ADM, the constraints of this dual theory then form subalgebras, reflecting the kind of group structure suitable for an exhaustive principal fiber bundle formulation. This points in a new direction for the development of  gauge theoretic tools for gravity and sets the stage for applying more standard methods for the quantization of gravity as a gauge theory.

 Motivated by the possibility of now describing the symmetry groups of general relativity in a full geometrical gauge-theoretic setting, we will attempt to make explicit the  gauge connections relating  to the action of these two groups; the group of three dimensional diffeomorphisms, which we denote by $\mathcal{D}$, and that of three-dimensional conformal transformations $\mathcal{C}$.  Both $\mathcal{C}$ and $\mathcal{D}$ groups have right actions on the natural configuration space $\M$.
 We will constrain our attention to the case of $M$ being compact and closed, which is of more interest to the relational approach for various reasons \cite{Barbour94}.

 \subsection{Principal fiber bundles and gauge theory.}

\begin{figure}\label{fig:PFB}
 \begin{center}
 \includegraphics[width=11cm]{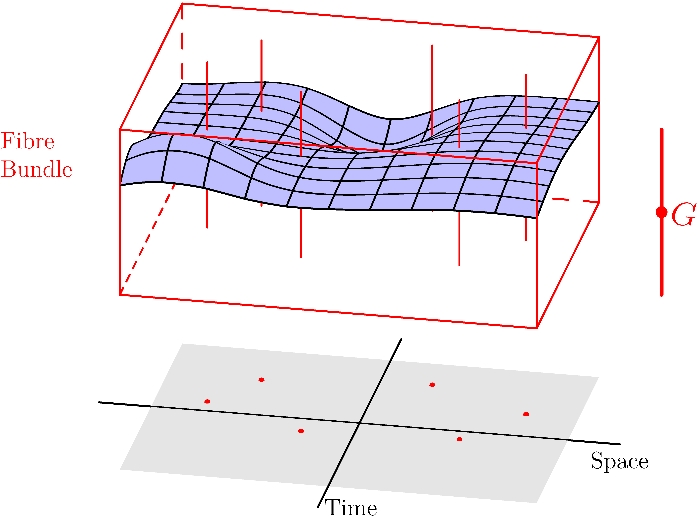}

\caption{A section in a principal fiber bundle over space-time.}

 \end{center}

\end{figure}

 Here we will briefly introduce the concept of a principal fiber-bundle, depicted in figure \ref{fig:PFB}. We will first present the formal definitions and

\begin{defi}\label{def:PFB}
A principal fiber bundle with smooth structural group $G$ is a smooth manifold $P$ on which $G$ acts $G\times P\rightarrow P$ and for which the action of $G$ is smooth and free. By a free action we mean that
$$G_p=\{h\in{G} ~|~ hp=p\}=\{\mbox{Id}\}$$
That is, the isotropy group of every point is the identity.
\end{defi}
One then constructs a projection
\be \mbox{pr}:P\rightarrow P/G=:B
\ee
where the base manifold $B$ is defined with the quotient topology with respect to the equivalence relation $p\simeq{q}\Leftrightarrow{p=h\cdot{q}}$, for some  $h\in G$. We call an orbit of $p\in P$ (or of $\mbox{pr}(p)=x\in B$) a \emph{fiber}, and also denote it by  $\mathcal{O}_x:=\mbox{pr}^{-1}(x)$.

For finite-dimensional manifolds, by the freedom of the group action, we can see that the orbits are isomorphic to the group $G$, but have no preferred identity element.\footnote{Such objects are in modern mathematical language called $G$-torsors. }\label{footn3} A smooth choice of identity element in each fiber coincides with the definition of a local section:
  \begin{defi}
Let $U$ be an open set in $B$. We define a \emph{local section} of $P$ over  $U$ as a submanifold
$\Sigma$ of $P$ such that for every $x\in U$,  $\Sigma$  is transversal to the orbits,~ $T_p\Sigma\oplus T_p\mathcal{O}_x=T_pP$,~ and $\Sigma$ intersects orbits over $U$ at a single point; i.e. for $p\in\Sigma$ then $\mathcal{O}_p\cap\Sigma=\{p\}$
  \end{defi}
 In rough terms, this means that a section $i)$ never has a component along the orbits and intersects every orbit (transversality) $ii)$ that it intersects each orbit once. These facts are enough to show that we can completely characterize an element of $P$ over $U$ by its location on the base and an element of the group which says where it is wrt the section. A choice of section is also called a \emph{choice of local gauge}. One can prove from this definition that in finite dimensions there always exists a \emph{slice} for our definition of a principal fiber bundle (definition \ref{def:PFB})  (see \cite{Palais}\footnote{Or see Theorem 19 in \cite{gomes-dissertation} for a detailed proof in the language used in section \ref{sec:slice}.}).  Indeed, it is using the same outlines of that proof that we are able to show that a section for the action of different groups on Riem exists (see section \ref{sec:slice}). A slice implies that $P$ has a \emph{local product structure} that we can patch  together to form an atlas of the manifold, and that all slices over the same open set are diffeomorphic. These \emph{sections} are then equivalent to the concept of {a gauge}, and transition maps from one gauge (or section) to the other can be shown to be functions $\Psi_{UU'}:U\cap U'\rightarrow G$.

 \subsubsection{Example: the bundle of bases.}

 The simplest and most telling example of a principal fiber bundle, is the one of all linear bases of $TM$, for a given manifold $M$. The group $\mbox{GL}(n)$ acts smoothly and has trivial isotropy, meaning it doesn't act trivially on any base. There is no preferred identity element (a preferred basis of each tangent space), and yet we can take every base to every other base by an action of $\mbox{GL}(n)$, making each fiber isomorphic to $\mbox{GL}(n)$. It is useful for us to already preview the concept of a \emph{connection} in this setting. Given a base $e$ over the point $x\in M$ and a vector $v\in T_xM$, a \emph{connection} will basically tell us which base corresponds to $e$ in that given direction, i.e., how to define parallel transport of the basis $e$ in each direction.

\section{The 3-diffeomorphism group.}\label{sec:3-diffeo}

Let $E=S^2T^*:=TM^*\otimes_STM^*$ denote the symmetric product of the cotangent bundle, and $\Gamma^\infty(S^2T^*)$ the
space of smooth sections over this bundle\footnote{It is a  Frech\'et space  (Metrizable Complete Locally Convex
Topological Vector space).}. The space of positive definite smooth sections of $S^2T^*$ is what we call $\M$. i.e.
$\M=\Gamma^\infty_+(S^2T^*)$, it is a positive  open cone over the vector space $S^2T^*$ (meaning that adding two metrics with positive coefficients is still a metric).

Let us also review the following general facts, which characterize the action of
what will play the role of a Lie algebra and Lie group \cite{Ebin}:
\begin{itemize}
\item The set $\DD:=\diff$ of smooth diffeomorphisms of $M$ is an infinite dimensional Lie group,  and it acts on $\M$
 on the right as a group of transformations by pulling back metrics:
\begin{eqnarray*}
\Psi:\M\times \DD &\ra& \M \\
~(g,f)&\mapsto & f^*g
\end{eqnarray*}
an action which is smooth with respect to the $C^\infty$-structures of  $\M$ and $\DD$\footnote{The natural
action is on the right since of course $(f_1f_2)^*g=f_2^*f_1^*g$.}. We call $\Psi_g:\DD\rightarrow\M$, the action for fixed $g\in\M$, the orbit map. It is clear that two metrics are isometric
if and only if they lie in the same orbit, $$g_1\sim g_2\Leftrightarrow g_1,g_2\in
\mathcal{O}_g:=\Psi_g(\DD)$$
\item The derivative of the orbit map
$\Psi_g: \DD \ra \M$ at the identity is\bea
\alpha_g:=T_{\Id}\Psi_g: \Gamma(TM) &\ra& T_g\M \nn\\
\label{jmath}X&\mapsto & L_Xg \eea where $X$ is the infinitesimal generator of a given curve of
diffeomorphisms of $M$.    The spaces $V_g$, tangent to the orbits will be called vertical and are defined as:
$$ V_g:=T_g(\mathcal{O}_g)=\{L_Xg~|~X\in \Gamma(TM)\}
$$ Since $M$ is compact, every $X\in\Gamma(TM) $ is complete and
$\Gamma(TM)$ forms an infinite dimensional Lie algebra under the usual commutator of vector fields,
$[X_1,X_2]\in\Gamma(TM)$.
\end{itemize}The quotient $\M/\DD$ is known to be a stratified manifold whose singular sets  correspond to the diffeomorphism classes of metrics with non-discrete isometry groups:
$$I_g(M):=\{f\in\DD~|~f^*g=g\}\subset \DD
$$
which are always groups of dimension at most 6. The singular sets are nested according to the dimension of $I_g(M)$.

When dealing with the space of metrics with no symmetries $\M'$, the space $\super'=\M'/\DD$  is indeed a manifold
and the existence of a section \cite{Ebin} allows us to construct
 its local product structure
 $\pi^{-1}(\mathcal{U}_\alpha)\simeq \mathcal{U}_\alpha\times \DD$
through bundle charts for $\mathcal{U}_\alpha$ and open set of the quotient and properly {\it define $\M'$ as a principal fiber bundle} (PFB). With the PFB  $\DD\hookrightarrow \M' \overset{\pi}{\ra}\M'/\DD=\super'$  we have
the usual constructions of gauge theory working properly, as we will see.

  There are other ways to resolve the singularities in the stratified structure of $\M/\DD$ than the one adopted here, which has the disadvantage of excising metrics with high degrees of symmetry such as the ones used to find explicit solutions of the Einstein equations. To excuse ourselves from that obvious criticism, we remark that only a meagre set of initial data will reach such boundaries, that our arguments are of a generic nature and that we can always approximate as well as we like any of those symmetric states. One of the other ways to resolve the singularity involves assuming that the topology of the underlying manifold does not allow for any continuous symmetry group, so called wild topologies, which are infinite in number. Another involves slightly modifying the group $\DD$ one works with, to $\DD_{\{x\}}$ the diffeomorphism which leaves point $x$ fixed. But perhaps the most useful route is to consider not $\M$ but $\M\times F(M)$, where $F(M)$ is the bundle of oriented frames over $M$. Since the action of $\DD$ can be seen to be free over this space, the quotient is indeed a manifold, and it is also a principal fiber bundle over said space. Our view here though is to be minimal with respect to the structures we use.

\section{Gauge structures over Riem: Slice theorem.}\label{sec:slice}

The important result for a gauge theory in $\M'$  is the Ebin-Palais  slice theorem \cite{Ebin}. It is analogous
to the usual slice theorem, and it is that which reveals the principal fiber bundle structure in $\super'$. We describe necessary material for the construction of a principal connection in $\M'$, with the main aim
being achieved in  Theorem \ref{proj theo}. But to see why the analogy between the free action of the $\DD$ group on
$\M'$ and finite-dimensional principal fiber bundles is more than an analogy we refer the reader to \cite{Michorbook}.

To a certain degree the material in the first section follows  \cite{Ebin}, but for the reader's convenience we
give a description in our language of the material that we need, i.e., the material necessary for the rigorous
definition  and construction of the connection through the use of a metric in $\M$.

\subsection{Constructing the vertical projection operator for the PFB-structure of $\M'$}\label{sec:constructingV}
The constructions here include technicalities needed in order to define the spaces we work with as proper Hilbert
manifolds, in order that we can use certain theorems only applicable in that domain. If the reader is happy that
we can make certain restrictions in $\M$ and $\DD$ so that we have Hilbert manifolds, on which  a
Riemannian metric is defined, she can skip the first two subsections. We use these Hilbert spaces and the
Riemannian metric in the third subsection, to define the structure of the $\DD$ orbits in $\M$. It is here that we
define and use the Fredholm alternative most intensely. The bundle normal
to the orbits (called the horizontal bundle in the main text) and the orthogonal projection with respect to such
a decomposition is constructed.
Hence this is the section used in the following chapter in the construction of the principal connection $\omega$ on $\M'$,
based on the existence of a metric on the Hilbert completion $\M^s$ (see below). Although the constructions here
are based in $\M^s$, it can be shown that they can be later transported to our merely $C^\infty$ setting of $\M'$
\cite{Ebin}. Lastly we state and sketch the remaining steps in the proof of the slice theorem, on which the whole
gauge apparatus is based.

\subsubsection{$H^s$-manifolds. Sobolev Lemma and all that}

Suppose that $E$ is a vector bundle over a smooth closed manifold $M$;  $\pi_{\mbox{\tiny E}}:E\ra M$. \begin{itemize}
\item Let $\Gamma^k(E)$ be the space of $k$-differentiable sections of $E$, this is a Banach space with topology of uniform
convergence up to $k$ derivatives.
\item Let $J^s(E)$ be the $s$-th jet bundle of $E$, which we endow with (for now) any Riemannian structure
$\langle\cdot,\cdot\rangle_s$. For a fixed volume element of $M$, let us call it $d^3x$, we get the inner product
on the space of sections  $\Gamma^\infty(J^s(E))$ by $$(a,b)_s=\int_M\langle a,b\rangle_s d^3x$$ Since there is a
natural linear map from $\Gamma^\infty(E)$ to $\Gamma^\infty(J^s(E))$ (basically given by successive
linearizations), this also defines an inner product on $\Gamma^\infty(E)$. Now we define
$$ H^s(E)\mbox{~~~is the completion of~~}~\Gamma^\infty(E) \mbox{~~~with respect to~~~} (\cdot,\cdot)_s
$$ As such it is a Hilbert space whose norm depends on the choices of inner product and volume form,
but whose topology does not. In local coordinates, this is the space of sections of $E$ which in local coordinates
have partial derivatives up to order $s$ square integrable, i.e. for $f\in H^s(E)$ the norm is given in local
coordinates by
$$ ||f||_{s}=\sum_{0\leq\alpha\leq s}||\partial^\alpha
f||_{L^2}=\sum_{0\leq\alpha\leq s}\sqrt{\int_M|\partial^\alpha f |^2d^3x}
$$ We note in passing that for $p\neq2$ the above is not a Hilbert space for the $L_p$ norm.
 \end{itemize}

Now to construct the appropriate manifolds, we will need the following
\begin{lem}[Sobolev Lemma]For $n=\mbox{dim}(M)$, if $s>k+n/2$ we have that $H^s(E)\subset\Gamma^k(E)$ and the
inclusion is a {\it linear continuous} map.
\end{lem} Note that the lemma is very far from trivial, since, of course we always have
$\Gamma^{k+1}(E)\subset\Gamma^k(E)$, but the $s$-th completion of the $\Gamma^\infty(E)$ sections could have
elements that were not smooth.

\subsection{Defining $\M^s$, a Riemmanian structure for $\M^s$, and an $\exp$ map.}

Let $E=S^2T^*:=T^*M\otimes_ST^*M$, the symmetric product of the cotangent bundle. The space of positive definite
smooth sections of $S^2T^*$ is what we call $\M$. i.e. $\M=\Gamma^\infty_+(S^2T^*)$. Abusing notation, let
$\Gamma_0(\M):=\Gamma_+^0(S^2T^*)\subset\Gamma^0(S^2T^*)$ be the space of merely continuous metrics on $M$, which
is an open subset of $\Gamma^0(S^2T^*)$. The set $\Gamma_0(\M)$ still is only endowed with a topology. To make it
into the appropriate Hilbert manifold, we define
$$\M^s:= H^s(S^2T^*)\cap\Gamma_0(\M)
$$ Now, by the Sobolev lemma,  the inclusion $\iota:H^s(S^2T^*)\hookrightarrow \Gamma^0(S^2T^*)$ is continuous for $s>1$ in $n=3$.
Since $\Gamma_0(\M)$ is an open subset of $\Gamma^0(S^2T^*)$, we have that $\M^s=\iota^{-1}(\Gamma^0(\M))$ is an
open set in $H^s(S^2T^*)$ and hence a Hilbert manifold. A similar construction is available to transform the group
of diffeomorphisms $\DD$ into a Hilbert manifold $\DD^s$, but as we will not get into the intricacies of the last
part of the proof of the Ebin-Palais section theorem, we will not need it, and hence just use the generic
$\Gamma(TM)$ as the tangent space to the identity of $\DD$.

For each point of the Hilbert manifold $\gamma\in\M^s$ we have that $\gamma$, being an inner product on $TM$,
induces an inner product in all product bundles over $TM$, and hence we have an induced inner product on $S^2T^*$,
which we call $\langle\cdot,\cdot\rangle_\gamma$. It furthermore induces a volume form, and thus we have the
induced inner product on each $T_\gamma\M^s\simeq H^s(S^2T^*)\ni \alpha, \beta$. \be\label{section
ip}(\alpha,\beta)_\gamma=\int_M \langle\alpha,\beta\rangle_\gamma d\mu_\gamma \ee Since $\M^s\subset\Gamma^0(\M)$,
$(\cdot,\cdot)_\gamma$ induces the $H^0$ topology on $H^s(S^2T^*)$,  there might be sequences in
$H^s(S^2T^*)$ that converge with respect to $(\cdot,\cdot)_\gamma$ but not to an element $H^s(S^2T^*)$. This is what we mean when we say that
$(\cdot,\cdot)_\gamma$ is  merely a weak Riemannian metric on $\M^s$.\footnote{ This sort of lack of metric convergence in $H^s$, poses certain issues when objects are only implicitly defined by the metric.}

For $f\in\DD$, as extensively used in the main text, $f^*:\M^s\rightarrow \M^s$ acts linearly, so furthermore
$T_\gamma f^*=(f^*)_{*}=f^*:H^s(S^2T^*)\rightarrow H^s(S^2T^*)$. From the properties $\langle
Tf^*\alpha,Tf^*\beta\rangle_{f^*\gamma}=\langle\alpha,\beta\rangle_\gamma\circ f$ and $d\mu_{f^*g}=f^*d\mu_g$ it
is straightforward to show that $(\cdot,\cdot)_\gamma$ is $\DD$-invariant.

\subsection{The orbit manifold and splittings}\label{orbit manifold section}

Consider now the map
\begin{eqnarray*}
\Psi:\M^s\times \DD &\ra& \M^s \\
~(g,f)&\mapsto & f^*g
\end{eqnarray*}
As in the previous section,  the image of $\Psi_g$, $\mathcal{O}_g=\Psi_g(\DD)$ is called {\it the orbit of $\DD$ through $g$}.
We have that the derivative of the orbit map $\Psi_g: \DD \ra \M$ at the identity, which we will call $\alpha_g$:
 \be\label{equ:def:alpha}\alpha_g:=T_{\Id}\Psi_g: X\mapsto
L_Xg=\imath_X(L_\cdot g)\ee where $X\in \Gamma(TM)$ is the infinitesimal generator of a given curve of
diffeomorphisms of $M$. We may also write $\alpha_g$ as $\alpha_g(X)=(T_{(g,\Id)}\Psi)\cdot(0,X)$, which may make the meaning of the map more clear. It takes each element of the Lie algebra into its fundamental vector field, i.e. it gives directions along the orbits corresponding to certain directions along the group.

 We want to calculate what $T_{f}\Psi$ is with respect to
$T_{\Id}\Psi$. For $\eta, f\in \DD$ and $r_f$  the right action of diffeomorphisms (for which $T
(r_f)=(r_f)_{*}:\Gamma(TM)\rightarrow \Gamma(TM)$), we have, $$ f^*\circ\Psi(g,r_{f^{-1}}(\eta))=f^*\circ\Psi(g,\eta\circ
f^{-1})=f^*(\eta\circ f^{-1})^*(g)=\Psi(g,\eta) $$ therefore \be\Psi=f^*\circ\Psi\circ r_{f^{-1}}\ee and thus
\be\label{orbit transf} T_f\Psi=T_gf^*\circ T_{\Id}\Psi\circ (r_{f^{-1}})_*=f^*\circ\alpha\circ
(r_{f^{-1}})_*:T_f\DD\rightarrow H^s(S^2T^*)\ee This equation is of course equivalent to saying that at $f$
$$T_{f}\Psi_g(X\circ f)=f^*\alpha_g(X)
$$

Since the maps above are isomorphisms, we conclude from \eqref{orbit transf} that $ T_f\Psi(T_f\DD)$ is isomorphic to $
T_{\Id}\Psi(T_{\Id}\DD)= T_{\Id}\Psi(\Gamma(TM))$, and thus all tangent spaces to the orbits are isomorphic.

For a  finite dimensional vector space $E$, we can always algebraically split a subspace $F_1$ from its complement $F_2=F_1^C$. For infinite-dimensional vector spaces, a closed finite-dimensional subspace also always has a closed complement subspace. In the general case of closed infinite-dimensional subspaces though, the complement $F_1^C$ of $F_1$ is not necessarily  closed, and upon closure it might not be in the complement (see Section \ref{sec:Closed_graph}).

Now we have to show that the tangent space to the orbits splits. I.e. that not only is the image of
$T_{\Id}\Psi=\alpha$ a closed linear subspace of $H^s(S^2T^*)$, but also that it has a closed complement $\mbox{Im}\alpha)^{\mbox{\tiny C}}$ and thus
 $H^s(S^2T^*)\simeq\mbox{Im}\alpha\oplus(\mbox{Im}\alpha)^{\mbox{\tiny C}}$. We will do this in the following
detour through functional analysis.

\subsubsection{Splitting of $T\M$ by $T\mathcal{O}$.}

In local charts of $E$ and $F$,  for $E$ and $F$ vector bundles over $M$, a $k$-th order differential operator
$D:\Gamma^\infty(E)\ra \Gamma^\infty(F)$, acting on $f\in \Gamma^\infty(E)$  can be written as \footnote{Here we
use $f$ to make the analogy with vector-valued functions in local charts more transparent}:
$$ D(f)=\sum_{0\leq|i|<ka_i}a^i\frac{\partial^{|i|}f}{\partial x^{i_{\mbox{\tiny 1}}}\cdots \partial x^{i_{\mbox{\tiny
n}}}}
$$ where $i=(i_{\mbox{\tiny 1}},\cdots,i_{\mbox{\tiny n}})$, $n=$dim$M$ and $|i|=\sum i_{\mbox{\tiny n}}$
and $a^i(x)\in L(E_x,F_x)$.

For each $x\in M$ and for ${\bf p}\in T_x^*M$, the symbol of an operator $D$ is a linear map $\sigma_p(D):E_x\ra
F^*_x$. Basically what one does, in local coordinates, is to replace the highest order partial derivatives by the
components of $p$: $\partial/\partial x^i\ra p_i$.  The symbol of a differential operator will be said to be
injective if the resulting linear operator is injective.

The $k$-th order differential operator $D:\Gamma^\infty(E)\ra \Gamma^\infty(F)$ trivially extends uniquely to a
continuous linear map between the Hilbert spaces $D:H^s(E)\ra H^{s-k}(F)$. If inner products
$\langle\cdot,\cdot\rangle_E, \langle\cdot,\cdot\rangle_F$ in $E$ and $F$ respectively, together with a measure
for $M$ are given, we call $(\cdot,\cdot)_E, (\cdot,\cdot)_F$ the inner products induced in $H^s(E)$ and
$H^{s-k}(F)$ respectively. By the Riesz representation theorem, there then exists a unique adjoint for any such
$D$: \be\label{adjoint D}(a,Db)_E= (D^*a,b)_F \mbox{~~~for~~~}a\in H^s(E), b\in H^{s-k}(F)\ee

Now, a well-known theorem in functional analysis tells us that, if a differential operator is elliptic  it
possesses the splitting property :
\begin{theorem}[Fredholm Alternative]\label{theo:Fredholm}Let $D$ be an elliptic differential operator of $k$-th order,\footnote{
There are subtleties here regarding the order $s$ of the Sobolev spaces in each side \cite{Ebin}, but these do not
concern us here. For the avid reader, the order of the spaces can be worked out by the Regularity theorem , which
states that for an elliptic operator of order $k$, and $f\in L_2(E)$, $D(f)\in H^{s-k}$ implies $f\in H^s$. The
Weyl lemma, stating that if the Laplacian (which is an elliptic operator) of an $L_2$ function is zero (and zero
is in $H^s$ for any $s$) then the function $f$ is $C^{\infty}$, is an immediate corollary of the regularity
theorem.} then
 \be\label{Splitting}
H^{s-k}(F)=\mbox{Im}(D)\oplus \mbox{Ker}D^*\ee \end{theorem}This stems from the more general fact, that for any linear densely-defined (i.e., having a domain of definition that is dense in H)
operator $A$, not necessarily bounded, we have the splitting property:
\be\label{equ:weaker_splitting}A=\overline{\mbox{Im}(A)}\oplus \mbox{Ker}(A^*)
\ee where the overline denotes closure.
  We will not dwell on the proof, we merely mention that
the necessary ingredients are norm bounds in the presence of elliptic operators to show that $\mbox{Im}(D)$ is
closed, and that $\mbox{Im}(D)\bot_{L_2} \mbox{Ker}D^*$ implies an $H^s$ splitting. For a different take on the Fredholm alternative, see \cite{Ramm}.

 The operator $\alpha_g:\Gamma(TM)\ra H^s(S^2T^*): X\mapsto X_{(i;j)}$
 can easily be shown to have injective symbol, since for $p\in T_x^*M$, $v\in T_xM$
 such that $\xi=g(v,\cdot)$ we have $$ \sigma_p(\alpha)(v)=\xi\otimes_{\mbox{\tiny S}} p
 $$ where again the subscript $S$ stands for the symmetrized tensor product.
  Furthermore, since
$\sigma(D^*\circ D)=\sigma(D)^*\circ\sigma(D)$, it follows that if $\sigma(D)$ is injective, then for
positive definite inner product we automatically have $\sigma(D^*)$ surjective and
$\Ker(\sigma(D^*))\cap\mbox{Im}(\sigma(D))=0$ (trivial, see proof of Proposition \ref{propo}).
Then $\sigma(D^*\circ
D)$ is an isomorphism, which by definition  makes $D^*\circ D$, or  in our case, $\alpha^*\circ\alpha$
an elliptic operator. Applying the above equation
\eqref{Splitting} to $\alpha^*\circ\alpha$ we  arrive at
$$\Gamma(TM)=\mbox{Im}(\alpha^*\circ\alpha)\oplus \mbox{Ker}(\alpha^*\circ\alpha)$$
from which we conclude that $\alpha^*\circ\alpha: \mbox{Im}(\alpha^*\circ\alpha)\ra \mbox{Im}(\alpha^*\circ\alpha)$ is
an isomorphism.

We will now sketch how under the present conditions, using ellipticity of $\alpha^*\circ\alpha$, a similar splitting
automatically applies
for $D=\alpha$.
\begin{prop}\label{propo}If $D^*\circ D$ elliptic and  the restricted inner products $(\cdot,\cdot)_{E_{| \mbox{\tiny Im}(D)}}$ and $(\cdot,\cdot)_{F_{| \mbox{\tiny Im}(D^*)}}$ are non-degenerate, then  $\mbox{Ker}(D^*\circ D)=\mbox{Ker}D$ and $\mbox{Im}(D^*\circ
D)=\mbox{Im}D^*$ which implies
 \be H^s(S^2T^*)=\mbox{Im}(D)\oplus
\mbox{Ker}(D^*)\ee for $D=\alpha$.\end{prop}
 {\rm Proof.}
That $\mbox{Ker}(D^*\circ D)\supset \mbox{Ker}D$ is clear. Now suppose $a\in H^s(E), c\in H^{s-k}(F)$, then if
$D^*\circ Da=0$, we have $(Da,Db)_F=0$ for all $b\in  H^s(E)$ which implies $Da=0$ if the inner product restricted to Im$(D)$ is non-degenerate. This shows $\mbox{Im}(D)\cap
\mbox{Ker}D^*=0$.

Also $\mbox{Im}(D^*\circ D)\subset\mbox{Im}D^*$ from the
outset. To show $\mbox{Im}(D^*\circ D)\supset\mbox{Im}D^*$,  since $H^{s-k}(E)=\mbox{Im}(D^*\circ D)\oplus
\mbox{Ker}(D^*\circ D)$ and $\mbox{Ker}(D^*\circ D)=\mbox{Ker}D$ we have merely to show that  $\mbox{Im}(D^*)\cap
\mbox{Ker}D=0$. Suppose $b\in \mbox{Im}(D^*)\cap \mbox{Ker}D$, i.e. $b=D^*c$ and $Db=0$, then  $(D^*c,D^*d)_E=0$ for all $d\in H^{s-k}(F)$.
Then if the inner product in $E$ restricted to $\mbox{Im}(D^*)$ is non-degenerate,  $D^*a=b=0$. Thus we have proved the first part of the proposition.

Now for the second part we already have  $\mbox{Im}(D)\cap \mbox{Ker}D^*=0$; to show that $H^s(S^2T^*)$
is generated by $\mbox{Im}(\alpha)+\mbox{Ker}(\alpha^*)$ we write:\footnote{Even though we have only shown the above
direct sum exists in the linear algebraic sense, the closed graph theorem guarantees it extends to
the topological domain (see section \ref{sec:Closed_graph}).},
$$H^s(S^2T^*)=(D^*)^{-1}(\mbox{Im}(D^*))=(D^*)^{-1}(\mbox{Im}(D^*\circ D))=(D^*)^{-1}(D^*\circ D(H^s(TM))$$ Since
$\mbox{Im}(D)\cap \mbox{Ker}D^*=0$  $$(D^*)^{-1}(D^*\circ D(H^s(TM)))=H^s(S^2T^*)=\mbox{Im}(D)\oplus
\mbox{Ker}(D^*)~~~~~~~~\square$$

Note  that, in the first part of the proposition,  $\mbox{Im}(D^*\circ D)\supset\mbox{Im}D^*$ is equivalent to
$\mbox{Ker}(D^*\circ D)=\mbox{Ker}D$, if the inner product in $H^s(E)$ is positive-definite. And of course, if $D$
is injective, this is equivalent to $\Ker(D^*)\cap \mbox{Im}(D)=0$, which is the usual equation to define the
orthogonality relation (but not a projection).

Now it is relatively straightforward to show that \eqref{Splitting} is valid for $D=\alpha$, which shows that for
$\M'$, the map orbits are injective immersions. To show that they are also embeddings requires more work, which
again we will not go through since it does not contribute anything to our constructions. Thus omitting the prof we shall, for $g\in{\M'}^s$, take $\Psi_g:\mathcal{O}_g\ra{\M'}^s$ to be an embedding.

\subsection{The normal bundle to the orbits and construction of the vertical projection operator.}\label{projection section}
The  bundle orthogonal to $\mathcal{O}_g$ is defined as \be\label{ortho bundle}\nu(\mathcal{O}_g):=\{n\in
T\M^s_{|\mathcal{O}_g}~|~(n,v)=0, \mbox{~~for~~}v\in T\mathcal{O}_g\}\ee Given a Riemannian structure on $\M^s$,
the bundle orthogonal  with respect to it would  automatically be a smooth subbundle, however we possess so far
merely a weak Riemannian metric, and so must put in a little more effort.

From the previous subsection we have seen that for any $g\in \M'$,  there exists an isomorphism
\be\label{Splitting alpha}T_g\M\simeq H^s(S^2T^*)\simeq\mbox{Im}\alpha\oplus \mbox{Ker}(\alpha^*)\ee Hence, since
for $v\in\mbox{Ker}(\alpha^*)$ it follows that $(\alpha(X),v)=0$ and $\mbox{Im}(\alpha_g)\simeq T_g(\mathcal{O}_g)$
we have that $\nu(\mathcal{O}_g)_h=\mbox{Ker}(\alpha_h^*)$.

 We shall thus define a smooth, surjective
map: $P: T\M^s_{|\mathcal{O}_g}\ra T\mathcal{O}_g$, such that $\Ker(P)=\mbox{Ker}(\alpha^*)=\nu(\mathcal{O})$
which will turn out to be exactly the vertical projection $\hat V$ we need for the definition of the principal
connection $\omega$. Before proceeding we note that in finite dimensions an orthogonal projection operator can be
easily defined from a basis, but that in the present case an orthogonality relation  does not automatically define a projection, even for a positive definite inner product.

From Proposition \ref{propo} we have  Im$(\alpha^*\circ\alpha)=\mbox{Im}(\alpha^*)$; hence for each point
$g\in\M$, $\alpha^*(H^s(S^2T^*))=\alpha^*\circ\alpha(\Gamma(TM))$. From the above consideration we can regard
$\alpha^*\circ\alpha_{|\mbox{Im}(\alpha^*)}$ as a map from $\mbox{Im}(\alpha^*\alpha)$ to itself, which,  from
self-adjointness and ellipticity, means it is in fact an isomorphism, having thus a smooth inverse.
 Hence we define: \be\label{P vertical}P:=\alpha\circ(\alpha^*\circ\alpha)^{-1}\circ\alpha^*:H^s(S^2T^*)\ra H^s(S^2T^*)\ee
It is clear that  $P^2=P$,  that $\nu(\mathcal{O}_g)_h=\mbox{Ker}(\alpha_h^*)=\mbox{Ker}P_h$, and that for a
vertical vector, i.e. $v=\alpha(X)$, we get
$P(v)=\alpha\circ(\alpha^*\circ\alpha)^{-1}\circ\alpha^*\alpha(X)=\alpha(X)$, hence the projection acts as the identity on the
vertical space. Thus the following decomposition holds\footnote{ Again, to go from merely
algebraic decomposition to topological decomposition, one must use the closed graph theorem, which says that for
Banach spaces $A, C$ then for a continuous linear operator $f$ such that $f(A)$ is a closed subspace, then  there is a
closed complement $B$, such that $C=f(A)\oplus B$.}: $W=\mbox{Im}T\oplus\Ker T $.
 and thus:
 $$
 H^s(F)=\mbox{Ker}(P)\oplus \mbox{Im}(P)$$
  All that is left to do is check the transformation properties of
$P$.

Let us recall first of all that $\alpha=T_{\Id}\Psi$, and from \eqref{orbit transf}
$$ \alpha_f=f^*\circ\alpha\circ ((r_{f})_*)^{-1}\mbox{~~~and~~~}\alpha^*_f=(r_{f})_*\circ\alpha^*\circ(f^*)^{-1}
$$ thus we can prove the \emph{equivariance} of $P$:
\bea \alpha_f\circ(\alpha_f^*\circ\alpha_f)\circ \alpha_f^* &=&
 f^*\circ\alpha\circ ((r_{f})_*)^{-1}((r_{f})_*\alpha^*\circ\alpha\circ
  ((r_{f})_*)^{-1})(r_{f})_*\circ\alpha^*\circ(f^*)^{-1}\nn\\
  &=& f^*(\alpha\circ(\alpha^*\circ\alpha)^{-1}\circ\alpha^*)(f^*)^{-1}\label{equivariant V}\eea
Since $\alpha_f$ is automatically smooth, all that is left to check is that $\alpha^*_f$ is smooth, since
$\alpha^*_f\circ\alpha_f$ is an isomorphism and the  inverse map in the restricted Banach space is smooth. We
shall not perform this calculation, which stems directly from the construction of the adjoint. Thus we have proven
the following theorem\footnote{We have actually proven it for the Hilbert extension $\M^s$, but it is shown in
\cite{Ebin} how these constructions can be more or less straightforwardly translated to the $C^\infty$ setting.}
\begin{theorem}\label{proj theo}
Given a $\DD$ invariant positive definite metric in $\M'$,   the operator \be
P:=\alpha\circ(\alpha^*\circ\alpha)^{-1}\circ\alpha^*:\Gamma^\infty(S^2T^*)\ra \Gamma^\infty(S^2T^*)\ee has
the following properties: \begin{itemize}\item $\DD$-equivariant. \item $P^2=P$ and $H^s(F)=\mbox{Ker}(P)\oplus \mbox{Im}(P)$
\item $P(\alpha(X))=\alpha(X)$.
\item $\nu(\mathcal{O}_g)_h=\mbox{Ker}(\alpha_h^*)=\mbox{Ker}P_h$. \end{itemize}
\end{theorem} So we call it {\it the vertical projection
operator} for this metric.

Let us go through the exact structures that were needed for this theorem (and were implied by positive-defiteness
of the $H^s(E)$ and $H^s(F)$ inner products ).
\begin{itemize} \item The adjoint of the operator $\alpha$  exists and is
smooth (in the main text $\alpha=T_{\Id}\Psi$ where $\Psi:\DD\times\M'\ra\M'$ is the group multiplication
operator).
\item  $\alpha^*\circ\alpha$ is elliptic (which can be checked by its symbol). Then from self-adjointness and the
decomposition \eqref{Splitting} we concluded that $\alpha^*\circ\alpha_{|\mbox{Im}(\alpha^*\circ\alpha)}$ was an
isomorphism. \item Im$(\alpha^*\circ\alpha)=\mbox{Im}(\alpha^*)$, which allowed us to regard
$\alpha^*\circ\alpha_{|\mbox{Im}(\alpha^*)}$ as a map from $\mbox{Im}(\alpha^*\alpha)$ to itself, which meant
$\alpha^*\circ\alpha_{|\mbox{Im}(\alpha^*)}$ was in fact an isomorphism, having thus a smooth inverse. Note that
for this, from Proposition \ref{propo}, we needed only that $\Ker(\alpha^*)\cap\mbox{Im}(\alpha)=0$ and
$\langle\cdot,\cdot\rangle_E$ be positive definite. Thus the injectivity of $\alpha$, combined with the previous item says
$\mbox{Ker}(\alpha_h^*)=\mbox{Ker}P_h$.
\item The metric in $H^s(T^*M\otimes T^*M)$ is $\DD$ invariant. We used to derive the transformation
properties of $P$.\end{itemize}

\subsection{The Slice Theorem}\label{subsec:slice} Since we have come
this far into the constructions in a reasonable degree of detail,
we now state
\begin{theorem}[Slice for $\M/\DD$, \cite{Ebin}]\label{theo:slice}
For each $g\in\M$ there exists a contractible submanifold $\Sigma$ of $\M$ containing $g$ such that
\begin{enumerate}
\item $f\in I_gM\Rightarrow f^*\Sigma=\Sigma$
\item $f\notin I_gM\Rightarrow f^*\Sigma\cap \Sigma=\emptyset$
\item There exists a local cross section $\tau:Q\subset \DD/I_g(M)\ra\DD$ where $Q$ is an open neighborhood of the
identity, such that \bea F: Q\times \Sigma &\ra& U_g\\
(f,s)&\mapsto& \tau(f)^*s \eea where $U_g$ is an open neighborhood of $g\in\M$, is a diffeomorphism.
\end{enumerate}\end{theorem}
For $\M'$ the space of metrics with no symmetries, the space $\super'=\M'/\DD$  is indeed a manifold
and the existence of a section above allows us to construct
 its local product structure
 $\pi^{-1}(\mathcal{U}_\alpha)\simeq \mathcal{U}_\alpha\times \DD$
through bundle charts and properly {\it define $\M'$ as a PFB}.\footnote{In the  MK sense. See \cite{Michorbook} for an appropriate way to formulate the usual theorems of calculus in this infinite-dimensional setting.}
With this in hand, the usual properties of a principal fiber bundle are proved as in finite dimensions.

\subsubsection*{Remaining gaps in the proof}
For the convenience of the reader we point out the leftover gaps in the proof of the slice theorem. The
steps that we have omitted are $i)$ to take better care of the isotropy group, which we have largely ignored by restricting our attention to the subset of metrics without symmetries (see the notes \cite{gomes-2009} for
a more thorough topological treatment of this subset); $ii)$ the actual construction of a tubular neighborhood for
each fiber using the properties of the exponential map. However, since we have indeed  addressed the major issues
that separate the finite-dimensional case to the present infinite dimensional one, these remaining steps are
closely analogous to the usual finite-dimensional proofs.

 Regarding $i)$, the isotropy group at $g\in \M$ is defined as $I_g:=\{f\in \DD~|~f^*g=g\}$. As $I_g$ is a finite-dimensional, and hence splitting, subspace of $\DD$, all major infinite-dimensional difficulties are more or less
easily dissolved. Since the Lie bracket of vector fields over $M$ commutes with the pull-back by diffeomorphisms,
the distribution of the spaces tangent to $\{I_{f^*g}~|~f\in\DD\}\subset\DD$ is involutive. Hence, using the Frobenius
theorem, we can construct the quotient manifold $\DD/I_g$ and a section for $\pi_{\DD}:\DD\ra \DD/I_g$ on a
neighborhood of the identity, $\chi:U\subset \DD/I_g \ra \DD$. Now define $\Phi_g:\DD/I_g\ra \M$ by
$\Phi_g(I_g\circ f)=f^*g$. Basically we must now replace our results about orbit embeddings for $\Psi$ by the
same results for the effective action, $\Phi$, which is the embedding.

Regarding $ii)$, given a Riemannian metric on a Hilbert manifold, there exists a unique Levi-Civita connection
(which respects both metric compatibility and the no-torsion condition). As we mentioned before, existence of certain objects implicitly defined by a weak metric is not guaranteed, for these objects might lie in the Sobolev completion of the $\mathcal{H}^s$. Thus
uniqueness, but not existence is guaranteed for the Levi-Civita connection. From the two usual coordinate-free
Levi-Civita conditions, using the Jacobi identity one gets for $X,Y,Z$ vector fields on $\M^s$:
$$ (\nabla_XY,Z)_\gamma=\frac{1}{2}\left(X(Y,Z)_\gamma-Z(X,Y)_\gamma+Y(X,Z)_\gamma\right)
$$
We then explicitly calculate the formula above
for three arbitrary vector fields, and upon isolation of the $Z$ vector field on the right hand side find an
explicit expression for the Levi-Civita connection. We will not perform this calculation, which can be checked in \cite{Michor_article}. After this construction we have a smooth exponential map $\exp:T\M^s\ra\M^s$
that is furthermore a local diffeomorphism around the zero section (for fixed base points). Combining this with
the invariance of the metric we get  \bea
f^*(\nabla_XY_{|\gamma})=\nabla_{f^*(X_\gamma)}f^*(Y_\gamma)\\
\exp\circ Tf^*=f^*\circ\exp \eea These relations are instrumental in the building of a section for the action of
$\DD/I_\gamma$.

 We have thus constructed an exponential map for a Hilbert manifold, and we call  the  map $\mbox{Exp}:=\exp_{|\nu(\mathcal{O}_g)}$ the {\it normal
exponential}. It can be seen to be a diffeomorphism
onto a neighborhood of the zero section as follows: the tangent space at a zero normal vector over any point can
be given the  direct sum
 decomposition
 $T_{(g,0)}(\nu\mathcal{O}_g)\simeq T_g\mathcal{O}_g\oplus \nu_g\mathcal{O}_g
$. Over a fixed fiber of $T\M$, i.e. for $v\in{T_g\M}$ , $\mbox{Exp}(g,v)=\exp_g(v)$. We have, taking
$(w,u)=\xi\in T_g\M$,
\begin{eqnarray*}
T_{(g,0)}\mbox{Exp}(\xi)&=&\frac{d}{dt}{|_{t=o}}
\mbox{Exp}(\gamma(t),0)+\frac{d}{dt}{|_{t=o}}\mbox{Exp}(g,tu)=
\frac{d}{dt}{|_{t=o}}(\gamma(t),0)+\frac{d}{dt}{|_{t=o}}\exp_g(tu)\\
~&=&(w,0)+(0,u)=\xi
\end{eqnarray*}
So we have shown that $ T_{(g,0)}(\mbox{Exp})=Id_{|T_g\M}$ which by  the inverse function theorem for Hilbert
manifolds makes the normal exponential a local diffeomorphism that respects the normal decomposition.

Thus  all we now have to do is find a small enough neighborhoods of the zero section of the normal bundle  such
that the $\DD$-transported exponential of some neighborhood of zero on $\nu_h(\mathcal{O}_g)$ satisfies the first
and second item of Theorem \ref{theo:slice}. Finding an appropriate section $\chi:U\subset \DD/I_g \ra \DD$ of the isotropy
group such that the last property is satisfied requires only a small amount of extra work, but it is enough to
make it too much of a detour on the purpose of this section. We refer the reader to \cite{Ebin} for the remaining
details.

It is important to notice that by exponentiating the horizontal subspace at $T_g\M$ one does not necessarily
obtain a horizontal submanifold even for the simple positive definite $\DD$-invariant metric we have used here.
If this were so, there would exist a section in which the connection could be set to zero and thus the curvature of
the connection would automatically vanish. We can see this is not the case since the tool in maintaining
orthogonality through the push-forward of exponential map, the Gauss exponential lemma, only works if one of the
vectors is radial. In other words, in general notation $\langle T_v\exp_pv,T_v\exp_p w\rangle_{\exp_p(v)}=\langle
v,w\rangle_p$ but $\langle T_v\exp_pu,T_v\exp_p w\rangle_{\exp_p(v)}\neq\langle u,w\rangle$. Thus suppose $w$ is
vertical; then it will keep its orthogonality with the radial vector along the exponential, but it will not necessarily
keep orthogonal to the exponential push-forward of another horizontal vector $u\in \HH_g$. It will do this only if
the connection has no curvature.

\section{The conformal bundle.}
We now give a brief description of the action of the conformal group $\mathcal{C}$, since it has much nicer
mathematical properties and seems to be given a new importance in recent dual approaches to general relativity
described in the first part of this thesis.

{\flushleft{ \bf Basic results.}}\smallskip

Let $\mathcal{P}$ be the multiplicative group of positive smooth functions on $M$. We denote by
$$\mathcal{C}:=\DD\times\mathcal{P}~~~\mbox{the space of conformal transformations of}~~ M~$$
with group structure $(f_1,p_1)\cdot(f_2,p_2)=(f_1\circ f_2,p_2(p_1(f_2)))$ where $p_2(p_1(f_2))$ just means
scalar multiplication at each $x\in M$ as $p_2(x)(p_1(f_2(x)))$. As with $\DD$, $\mathcal{C}$ is an
infinite-dimensional regular Lie group and it acts on $\M$
 on the right as a group of transformations by:
\begin{eqnarray*}
\xi: \mathcal{C}\times\M &\ra& \M \\
~((f,p),g)&\mapsto & pf^*g
\end{eqnarray*}
For more information on the mathematical properties of conformal superspace and the analogous constructions of
Ebin \cite{Ebin}, see \cite{FiMa77}. For instance, it is fairly easy to prove
in the same fashion as done in section \ref{sec:slice} that a slice theorem exists also for
$\mathcal{C}$ (see  theorem 1.6 in \cite{FiMa77}). Here however, if we want to form properly a principal fiber bundle, we would have to
regard the manifold $\M''$ consisting of metrics with no non-trivial conformal isometries. It can be shown that
this restriction does not have serious topological implications \cite{gomes-2009}.

The derivative of the orbit map $\xi_g: \mathcal{C} \ra \M$ at the identity is given by \bea
\tau_g:=T_{(\Id,1)}\xi_g: \Gamma(TM)\times C^\infty(M) &\ra& T_g\M \nn\\
(X,p')&\mapsto & L_Xg +p'g\eea where $p'\in C^\infty(M)$ and which can be easily evaluated from
$\frac{d}{dt}_{t=0}(tp'+1)(f_t)f_t^*g$, where $f_t=\exp(tX)$.

\subsection{York splitting.}\label{sec:York_splitting}

As a last auxiliary result for the main text we here state and sketch the York splitting theorem for the action
of the conformal group of transformations.

From the demonstrated good behavior of the orthogonal projection operators in section \ref{sec:conformal_diff_group}, in all
cases of interest (i.e. for all $\beta$), we have a well defined normal bundle (see \eqref{Splitting alpha}) and
thus a slice for the action of $\mathcal{CD}$. This means that we can write
\be\label{integrable2}
T_g\M=\mbox{Ker}\tau^*\oplus \mbox{Im}\tau_g=T_g\Sigma_g\oplus T_g(\mathcal{CD}\cdot g)
 \ee
where $\Sigma_g$ is the section of $\mathcal{CD}$, given, for the canonical supermetric (DeWitt for $\lambda=0$),
by the exponentiation at $g\in\M$  of the  kernel of $\tau^*$ \eqref{Kernel}. This kernel is composed of
divergenceless (transverse) traceless tensors (TT tensors), which we denote by $S_2^{\mbox{\tiny TT}}\subset S_2T^*$ and can be by
definition decomposed further into
$$S_2^{TT}=S_2^{\DD}\cap S_2^T,
$$
where $S_2^{\DD}\in S_2T^*$ are the transverse (or divergenceless) tensors and $S_T^2\in S_2T^*$ are the traceless
tensors. Exponentiated these spaces respectively form the section $\Sigma_g^{\DD}$ for $\DD$ we used in the subsection \ref{subsec:slice}, and the space of constant volume forms
$$\mathcal{N}_{d\mu}:=\{g\in\M~|~d\mu(g)=d\mu\},$$ where $d\mu(g)$ is the volume form associated to $g$.

Equation \eqref{integrable2} is said to be an \emph{integrable} decomposition in the sense that the tangent space at any point $g$ on the lhs, is the direct sum of tangents to two submanifolds on the rhs.
What is more interesting to us though, is that the second factor in \eqref{integrable2} admits two sets of
integrable decompositions. One of the sets of suborbits, is the natural one given already by the action of the
group written as $\DD\times\mathcal{C}$, as mentioned above. As we are excluding the conformal Killing metrics,
i.e. $\Ker ~\tau=0$, the action of the algebras also clearly splits:
$$ \mbox{Im}(\Gamma(TM))\oplus\mbox{Im}(C^\infty(M))=T_g(\mathcal{CD}\cdot g)
$$
and it is easy to see that the orbit $\mathcal{C}\cdot g$ is a submanifold \cite{FiMa77}, and thus as we already
know this is also true for the $\DD$ group (see section \ref{orbit manifold section}) we have an
integrable decomposition.

The other can be seen by splitting the image of $\tau$ in its traceless and trace part:
\bea\label{York decomposition2} h&=&h_{TT}+L_Xg+Ng=h_{TT}+\frac{1}{3}(2 X^a_{~;a}+3N)g+(L_Xg-\frac{2}{3} X^a_{~;a}g))\\
T_g\M&=&T_g(\Sigma_g)\oplus T_g(\mathcal{CD}\cdot g)=T_g(\Sigma_g)\oplus T_g(\mathcal{C}\cdot g)\oplus
T_g(\mathcal{CD}\cdot g\cap \mathcal{N}_{d\mu(g)})\label{equ:integrable_decomp_conf}
 \eea
For the preceding decomposition to be true in the integrable sense written above \eqref{equ:integrable_decomp_conf}, we only need to check wether
$\mathcal{K}_g:=\mathcal{CD}\cdot g\cap \mathcal{N}_{d\mu(g)}$ is actually a manifold. One can see this since
$$\mathcal{K}_g=d\mu^{-1}(g)\cap \mathcal{CD}\cdot g$$
 where $d\mu:\M\ra \mathcal{V}$ is just the operator that
assigns volume forms to metrics (and $\mathcal{V}$ is the space of volume forms). Thus $\mathcal{K}_g$ is a
manifold, since it is given by an inverse regular value, as can be deduced from $T_gd\mu\cdot h=\text{tr}(h)d\mu(g)$,
which, for $h=L_Xg+Ng$, is surjective. I.e. $\mathcal{K}_g=d\tilde\mu^{-1}(g)$ where
$d\tilde\mu:\M_{\mathcal{CD}\cdot g}=\mathcal{CD}\cdot g\ra \mathcal{V}$. In the end we have the
decomposition: \be\label{York decomposition} T_g\M=T_g(\Sigma^{\DD}_g\cap \mathcal{N}_{d\mu(g)})\oplus
T_g(\mathcal{C}\cdot g)\oplus T_g(\mathcal{CD}\cdot g\cap \mathcal{N}_{d\mu(g)})\ee or in words, we can decompose
the space into ``the volume--form--preserving--divergenceless directions + the scaling--of--the--metric direction + the
diffeomorphisms--that preserve--the--volume--form directions".

\chapter{Connection Forms}\label{chapter:connection_forms}

In the following we have provided a natural extension of the construction of a principal fiber bundle structure to Riem. That next step is of course, the construction and interpretation of connections for infinite-dimensional groups acting on Riem, and that is the topic of this chapter.

\section{Introduction}

It is a much repeated story that in a diffeomorphism invariant theory points lose their meaning, their individuality  becoming
dissolved by the active interpretation of these global ``coordinate changes'' \cite{rovelli}.\footnote{In a sense then our notion of space is nothing but  a $\DD$-torsor (see footnote \ref{footn3}).} In fact, since we will be dealing exclusively with global, and thus active, diffeomorphisms, we will use the expression "\emph{change of labeling}" to distinguish the nomenclature from that of the passive, local ``coordinate changes".

It is the case that in pure gravity
only the metrics over the manifolds attribute any real significance to the spatial points of $M$. We indicate this dependence
by $M_g$, a family of diffeomorphic manifolds parametrized by $g$.

In the canonical analysis, the 3+1 decomposition of the four dimensional metric involves a `shift' vector field and a lapse scalar, which
parametrize the diffeomorphism from a globally hyperbolic space-time to $M\times\R$. This entails in our notation a one-parameter family of diffeomorphic manifolds $M_{g(t)}$.

The  lapse encodes the temporal distance
 element in the embedding of the one parameter family of hypersurfaces.
The 'shift' vector field effectively already requires some
 identification between the points of ``neighboring" $M_{g(t)}$'s.  The shift
 itself is an infinitesimal deviation from the background identification of $M_{g(t)}$ and $M_{g(t+\delta t)}$ by
 vectors orthogonal to $M_{g(t)}$ with respect to the ambient Lorentzian metric. If we propose here to at least
  momentarily disregard four dimensional embedding, specially in view of the first part of this thesis, then the shift vector field loses its meaning, and we must
  find a new way to string together the $M_{g(t)}$'s along time.

The need to somehow identify points of our manifolds along time, naturally brings us to the concept of
best-matching \cite{Barbour94}, and forces us to introduce a form of ``parallel transport" of point labels. From this
concept it is a small step to see this parallel transport as taking place in the gauge setting, where the
structural group should be  $\DD$, since it is this group that parametrizes the ways with which we can
connect $M_{g_0}$ and $M_{g_1}$. It is this concept and its generalization that we will study in this chapter. We regard this construction as an elaboration of the concept of best-matching, as spelled out in section \ref{sec:Barbour}.

Our aim is not just to construct the decomposition $T_g\mathcal{M}=H_g\oplus V_g$ of the principal fiber bundle, but to explicitly construct the Lie algebra valued one form $\omega$. This does not in fact require the introduction of new mathematical apparatus, but it certainly implies a shift in the way one views gauge theory over these configuration spaces towards indeed an original perspective. The connection form then has a very physical interpretation, as something that acts on metric velocities and yields vector fields. Its interpretation is that it yields a preferred infinitesimal ``label change" from each infinitesimal metric change. It is the general study of such connections and what they imply (as for example wrt locality) that this chapter is devoted. In principle, a connection could be derived from a more general action over the whole of Riem(M), including curvature forms. As this requires more work to be made sense of, we start with a certain type of metric induced connections, yielding equation  \eqref{connection from vertical} as our main result.

\begin{figure}
 \begin{center}
 \includegraphics[width=11cm]{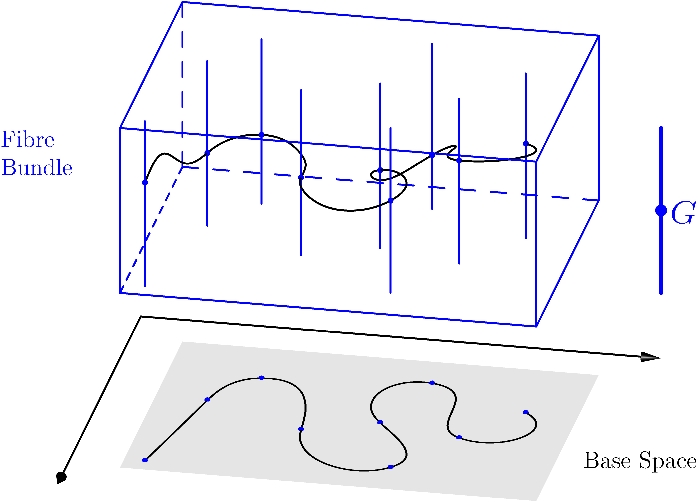}

\caption{A given evolution of the Universe making a trajectory in shape-space, and a given lift to a representation by 3-metrics.  A connection will act by extracting the vertical component of each infinitesimal segment of the line, but it exists on the whole bundle. We'll take it as orthogonally induced by a metric on Riem, but in principle it could be given by other means.}

 \end{center}

\end{figure}

\section{Connection forms: basic properties}

Let us first of all define a connection form for a finite-dimensional principal fiber bundle $P$:
\begin{defi}
 In finite-dimensions a connection form is defined as a Lie-algebra-valued one form $\omega\in\Gamma(T^*P\otimes\Lg)$ which acts on vertical vectors as: $\omega(T_{\Id}\mu(X))=X$, where $X\in \Lg$ and $\mu:P\times G\rightarrow P$ is the group action. The connection must also transform as
  \end{defi}
 $$ R_h\omega=\Ad(h^{-1})\omega
 $$
In the above $R_h$ is the push-forward right action: $R_h\omega(v)=\omega((\mu_h)_*v)$ where $v\in T_pP$ and $\mu_h:P\rightarrow P$ is the right action by the group element $h$. The adjoint acts on the algebra as
$$\Ad(h)X=\frac{d}{dt}_{|t=0}\left(h\circ \exp(tX) h^{-1}\right)
$$ where $\exp:\Lg\rightarrow G$ is the group exponential. Let us also define \emph{equivariance} as a term that encompasses both the covariant and contravariant denominations. Different spaces might have different equivariance properties. For instance the usual Yang-Mills curvature form transforms equivariantly in the adjoint representation, but the local expression for the connection does not, as it does not transform homogeneously.

The connection form in our present infinite-dimensional  setting will then be a Lie-algebra valued linear functional on $T\M$ (metric velocities).
 Since technically  the Lie algebra here is just the space of infinitesimal diffeomorphisms of $M$, the
connection form turns out to be a vector-field-valued distribution, taking metric velocities as test functions. We
are thus led to the meaning of a gauge connection over $\M'$ as representing a Machian notion of relational space,
since it relates spatial points along time in a manner depending on the dynamics of the entire
Universe (depending strictly and globally on the metric velocities). The connection form is not an empty mathematical construct. Its interpretation as yielding parallel transportation of spatial points\footnote{The connection form for the conformal group would then yield a notion of parallel transport of local scale.} is suggestive and interesting (especially from a relational point of view).

\subsection{ An equivariant splitting.}

  A choice of connection form in $P$ amounts as usual to choosing an
 equivariant decomposition
 \be\label{equivariant decomposition}T_g\M'=H_g\oplus V_g.\ee
 where by equivariant we mean that the decomposition is maintained by the group action.

   In the infinite-dimensional case, we have to separate the requirement of the direct sum in three separate conditions:
 \be\label{connection conditions} \begin{array}{c} V_g\mbox{~~and~~}H_g\mbox{~~are closed~~}\\
 V_g\cap H_g=\{0\}\\
 \overline{\mbox{span}\{V_g\cup H_g\}}=T_g\M'
 \end{array}\ee
 We will shortly discuss the mathematical instruments and conditions required for  these conditions.
  The physical purpose of the decomposition is so that we can distinguish what is a label change, which would be given by projection of  metric change along the orbits, and the remnant of the metric change, which could be identified with pure geometrical change.

 This amounts to defining
 an equivariant projection on the vertical space, which we call $\hat \V:TP\ra V$,
 having the properties $\hat \V\circ\hat \V=\hat \V$ and equivariance: $f^*\circ \hat\V=\hat \V\circ f^*$. But  to be able to construct the actual connection form we need more than just the equivariant direct sum decomposition, but also conditions on the representation of the algebra (that it forms a fundamental vector field with the correct equivariance properties). This is necessary to derive the correct transformation properties of the connection itself $R_f\omega=\Ad(f^{-1}) \omega$. We also include this criterion and show that indeed such a connection has these properties.

 \subsection{ The connection form obtained from the vertical projection}

The vertical sub-bundle, the bundle tangent to the orbits, is given by
$V_g:=\{L_Xg~|~X\in \Gamma(TM)\}$, where $L_X$ is the Lie derivative.
The
canonical  representation of  the diffeomorphism group on $\Gamma(TM)$ is the adjoint representation:
$$
 \Ad(f)(X)= \frac{d}{dt}(f\circ \exp(tX)\circ f^{-1})=f_*(X\circ f^{-1}).
$$
 Now, for the representation of the Lie algebra  on itself, we have
 $\ad_X=[X,\cdot]$ since $[X,Y]=\frac{d}{dt}_{|t=0}\Ad(\exp(tX))Y~~\mbox{which is an element of}~~\Gamma(TM).$
  Hence the Lie algebra bracket is just the vector field commutator.

  If we have a right action of a Lie group on any manifold $\mathcal{N}$ ($\mathcal{N}$ is allowed to be infinite-dimensional) $\Psi:\mathcal{N}\times G\ra \mathcal{N}$,
  for $X\in\Lg$ we define the {\it fundamental vector field} $\zeta_X\in\Gamma(T\mathcal{N})$
  by
  \be\label{equ:fundamental_vf}\zeta_X(x)=(T_{(x,\Id)}\Psi)\cdot(0_x,X)=:T_{\Id}\Psi_x(X),\ee
  where we redundantly keep the subscript $x$ in $O_x$, to remind ourselves that this is the zero vector at the point $x\in P$.
  \begin{lem}\label{integrable}
  A fundamental vector field must satisfy the following properties for $f\in G$:
  \begin{enumerate}
\item $\zeta:\Lg\ra\Gamma(T\mathcal{N})$ is linear.
\item $T_x(\Psi_f)(\zeta_X(x))=\zeta_{\Ad(f^{-1})X}(\Psi(x,f))$
\item $[\zeta_X,\zeta_Y]=\zeta_{[X,Y]}$
\end{enumerate}\end{lem}
where we have denoted, for fixed $f$, in the same way as in \eqref{equ:fundamental_vf}:
\be\label{equ:push_forward_at_f} T_x(\Psi_f):=(T_{(x,f)}\Psi)\cdot(\cdot, 0_{f}): T_x\mathcal{N}\rightarrow T_{\Psi(x,f)}\mathcal{N}
\ee

Let us explicitly check property 2 for the $\DD$ action on $\M$. To make the
actions clearer we expand $f^*g=\Psi(f, g)$ and so we have that, according to \eqref{equ:push_forward_at_f}, for $f\in\DD$ and $g\in\M$,
 \be\label{equ:fundamental_vf_diff}T_g(\Psi_{f}):=(T_{(g,f)}\Psi)\cdot (\cdot,0_f)
 :T_g\mathcal{\M}\ra
T_{\Psi({f},g)}\mathcal{\M}
\ee and
\be\label{equ:push_forward_at_f_diff} T_{\Id}\Psi_g:=(T_{(g,\Id)}\Psi)(0_g,\cdot ):\Gamma(TM)\ra T_g\M.
\ee
Using $r_f$ for the right action of the group on itself we also have:
 \be\label{equ:conjugate_action}\Psi(g,  r_f\circ h)=\Psi(f^*g,\mathrm{conj}_{f^{-1}}(h))\ee for any $h\in\DD$,
 where $\mathrm{conj}_f$
 is the conjugate action of the group on itself: $\mathrm{conj}_f(h)=f\circ h\circ f^{-1}$, and again $\circ$ is composition of maps (in this case diffeomorphisms).

Using \eqref{equ:push_forward_at_f_diff} and \eqref{equ:fundamental_vf_diff} we have: \bea
 T_g(\Psi_f)(L_X(g))&=& (T_{(g,f)}\Psi)\cdot ((T_{(g,\Id)}\Psi)\cdot(0_g,X ),0_f)\nn\\
 &=&(T_{(g,f)}\Psi)\cdot \left((T_{(g,\Id)}\Psi)\cdot(0_g,\frac{d}{dt}_{|t=0}\exp (tX) ),0_f\right)\nn\\
 &=&(T_{(g,f)}\Psi)\cdot \left(\frac{d}{dt}_{|t=0}\Psi(g,\exp(tX)),0_f\right)\nn\\
 &=&\frac{d}{dt}_{|t=0}\left(\Psi\left(\Psi(g,\exp(tX)),f\right)\right)=
 \frac{d}{dt}_{|t=0}\left(\Psi(g,\exp(tX)\circ f)\right)\nn\\
  &=&\frac{d}{dt}_{|t=0}\left(\Psi(f^*g,\mathrm{conj}_{f^{-1}}\exp(tX))\right)=
  \frac{d}{dt}_{|t=0}\left(\Psi(f^*g,\exp(t\Ad_{f^{-1}}X))\right)\nn\\
\label{omegaprop} ~&=&L_{\Ad(f^{-1})X}(f^*g),
 \eea where we used \eqref{equ:conjugate_action} to go from the fifth to the sixth line.

  We can then identify $\zeta_X$ in {\bf Lemma \ref{integrable}}
  with $L_X$ and verify that it automatically satisfies the first and third
  identities required for  {\it a fundamental vector field}.
 So the key properties of the action of the
  Lie algebra on the bundle are satisfied by the Lie derivative of vector fields, a fact that is of utmost importance for our treatment and that allows us to take the ``gauge analogy" to be not merely an analogy.

   We then
  define the Lie-algebra valued connection form:
  \begin{defi}
  Given a tangential decomposition as in \eqref{connection conditions}, if we can construct a vertical projection $\hat \V:TP\ra V$ satisfying $\hat \V\circ\hat \V=\hat \V$ and\footnote{ We pause to note that for usual push-forward and pull-back maps for ``constant" diffeomorphisms, we could have used
extra ``$*$"'s: $(f^*)_*:T_g\M\ra T_{f^*g} \M$. This is the exact analogous of the tangent left translation
${l_g}_*$ in the usual action of Lie groups. In this setting this is superfluous since due to the vector space
structure of $S_2M$, we have  ${f^*}_*=f^*$. From now on we will omit the double star notation.} $f^*\circ \hat\V=\hat \V\circ f^*$, we then define the vector field valued connection form as:
\be\omega_g=(T_{\Id}\Psi)^{-1}_g\circ \hat V_g:T\M'\ra\Gamma(TM)\ee
\end{defi}
Since
$\alpha_g:=T_{\Id}\Psi_g$ is an isomorphism over its image, i.e. over the vertical space, we can take
inverses.

Clearly we then have that
\be\label{vertical connection}\hat\V_g[\dot g]=\alpha_g\circ\omega_g[\dot g]=L_{\omega_g[\dot g]}g\ee where
$\dot g\in T_g\M$. Thus vertical projection of a velocity is equal to the Lie derivative of the metric along the preferred direction $\omega[\dot g]$.

By the transformation properties of the Lie derivative (see \eqref{omegaprop}) and equivariance of the vertical projection we have the usual
transformation property: \be\label{connectionprop} f^*\omega=\Ad(f^{-1})\omega \ee
confirming that this can indeed be interpreted as a connection form.

Because $\DD$  admits the exponential map, we have not only uniqueness, but also
existence of a $\DD$-equivariant smooth parallel transport. This means that we can integrate forward from some initial labeling, finding some relative preferred identification of spatial point along time.\footnote{We do note for completeness that it is well known that the exponential map of $\DD$ is not surjective on any neighborhood of the identity. We however abstain from speculating on the relevance of this fact for our approach.} To be more explicit, let us consider a heuristic example. Let us suppose we describe (supposing we could observe the entire Universe) a time evolution of the metric, i.e. a one-parameter family $g(t)$. Given an initial labeling of the manifold at $t=0$, we can then integrate the connection along time to find the preferred point that is ``equilocal"  to $x(0)$:
\be \label{equ:integral_bm}x(0)\mapsto \int_{[0,1]}\exp{\omega[\dot g(t)]}( x) dt.
\ee

 \subsubsection{Interlude: interpretation of non-trivial holonomy.}
 So what would it mean to have two paths in shape space $[g_1(t)]$ and $[g_2(t)]$ (this hypothetical situation then by definition already falls outside of the domain of classical physics), starting from the same shape $[g_0]$, such that their horizontal lifts $g^H_1(t)$ and $g^H_2(t)$, although arriving at the same shape $[g^H_1(1)]=[g^H_2(1)]$, fall on different places on the orbit $g^H_1(1)\neq g^H_2(1)$? For the diffeomorphism group, it would mean that best matched observers who agree on the ``location" of points initially would, through \eqref{equ:integral_bm}, disagree on their location in the end. We leave a thought her for the reader: what would it mean for the conformal group?

\subsection{ Locality of connection forms.}

Another factor of extreme importance  is the question of local representability of the connection form. That
is, $\omega$ at each $g$ is an element of $T^*_g\M\otimes\Gamma(TM)$. However, since we are dealing with
infinite-dimensional spaces, we cannot a priori identify the space of linear functionals acting on
$T_g\M'\simeq\Gamma(S_2T^*)$, which we call $T^*_g\M'$,  with $\Gamma(TM\otimes_STM)=\Gamma(S_2T)$.

As an initial attempt to construct such a local representation, we could choose a partition of unity of $M$,
defined by the characteristic functions $\{\chi_\alpha\}$ of the open sets $\{U_\alpha\in M\}$. Then for an
element $\lambda_g\in T_g\M'$ by linearity we have:
$$\lambda_g[\dot g]=\sum_\alpha{\lambda_g}_{|_{U_\alpha}}[\dot g_{|_{U_\alpha}}]
$$
where in this section we have denoted functional dependence by square brackets. In the limit,  this would come to: \be\label{dense representation}
\lambda_g[\dot g]=\int_M\lambda_g^{ab}(x)\dot g_{ab}(x)d\mu_g \ee for $\lambda_g\in\Gamma(TM\otimes_STM)$.

In fact, what we have is that elements of $T_g\M$ are tensors with compact support, which can thus be considered
as a space of {\it test functions} (or more precisely, test tensors\footnote{Since $M$ is compact, we can take the
components of an element of $T_g\M$ as the test functions.}). We have that the space $T^*_g\M'$ is of course {\it a space of
distributions on $T_g\M$}, and the space defined above by elements of the simple form of \eqref{dense representation} are dense inside $T^*_g\M$.

We will ignore these subtleties and now express $\omega$  (up to discrepancies on sets with vanishing measure) as the two-point tensor:
$\tilde\omega\in\Gamma(TM)\otimes\Gamma(TM\otimes_S TM)$. Pointwise:
 \begin{eqnarray}
\omega_g(x,x') ~
 \in (T_{x'}M\otimes_S T_{x'}M)\otimes T_xM\simeq L(T^*_{x'}M\otimes_S T^*_{x'}M, T_xM) \nn\\
  \label{connection} \int \omega^{ab'c'}(\dot
g_{b'c'}(x')) \sqrt{g}d^3x'=\omega^a_g[\dot g](x)\in T_xM\end{eqnarray} where we have used DeWitt's notation,
denoting tensorial character at $x'$ by primed indices. In the examples we will find, the connection form will always be given by the simple form of \eqref{dense representation}.

 The geometrical interpretation of the connection form viewed in this way is that, for each metric $g$, a
given metric velocity $\dot g(x)$ at a point $x\in M$ will contribute for the ``best-matching'' vector field at
each other point $x\in M$. In this way then, we get a non-local contribution to the best-matching vector field at
each point of $M$. These contributions however may  come from metric velocities at that and every other point of
$M$. This goes in line with relational arguments, since this implies that the stringing of points throughout time (equilocality) are determined by the \emph{kinematics of the entire universe}.

\section{Construction of connection forms in $\M$ through orthogonality}\label{Explicit construction}

Now that we have written down the basic structures that allow a gauge treatment of labelings using the $\DD$
group, we will derive explicit formulae for the connection forms through the use of the orbit maps and their adjoints. Of course in this
case a supermetric fixes the connection, if it exists, once and for all. It is not, as it is in Yang-Mills, determined by an action principle. So when we consider actions involving such fixed connections,
 our system works in analogy to a particle in a fixed electromagnetic potential.

 To let the connection be determined through a variational principle, one would require a term like $F[\Omega]$ in the action, where $\Omega$ is the curvature form of $\omega$. However, as $\Omega[\dot g,\dot g]=0$, it
 cannot appear in the `classical trajectory in $\M$' action we are considering.
 Nonetheless this can be done for a field theory in $\M$, a treatment which will show up in future work.

 Now we shall address the three items in \eqref{connection conditions} and present a direct formula for the connection form and the pre-requisite conditions on its components. We emphasize that no mathematical breakthrough is needed in  the construction of the vertical projection The only thing that is mathematically novel in the present work is the use of a connection, and  this is a very direct consequence of $i)$ $\Gamma(TM)$ forming a Lie algebra and $ii)$ the Lie derivative of the metric along these fields forming a fundamental vector field on $T\M$.

 Nonetheless, the emphasis here is completely different than that of usual treatments, as it introduces the explicit use of a connection form and focuses on the conditions necessary for its definition  from the basic ingredients. We will  generalize the statements in this section to the case of the conformal group.

\subsection{Construction through orthogonality and the momentum constraint}\label{sec:orthogonal_momenta_diff}

We could initially attempt to define an equivariant direct sum decomposition \eqref{equivariant decomposition} implicitly, through an equivariant inner product in $\M'$, by orthogonality with respect to the vertical bundle.

In other words, by  defining the horizontal subspace $\HH$ by orthogonality to the canonical fibers with respect to some
$\DD$-invariant supermetric $\langle\cdot,\cdot\rangle$: \be\label{ortho}
\mathcal{G}[\hat H[\dot g], L_Zg]=\int_M\langle \dot g-\hat V_g[\dot g], L_Zg\rangle dx^3=\int_M\langle\dot
g-L_{\omega[\dot g]}g, L_Zg\rangle_gdx^3=0 \ee
For instance, writing the canonical momentum as:
$$\pi^{ab}=\frac{1}{2N}G^{abcd}(\dot g_{cd}-L_\omega g_{cd})
$$ we obtain that the momentum constraint is written as
\be\label{momentum constraint}\mathcal{H}^a=\left(\frac{1}{2N}G^{abcd}(\dot g_{cd}-L_\omega g_{cd})\right)_{;b}=0\ee
 For $\omega=\omega_g[\dot g]$ given by \eqref{vertical connection}, we have that \eqref{momentum constraint} is implicitly exactly of the form of \eqref{ortho}, with $\langle\cdot,\cdot\rangle=\frac{G^{abcd}}{N}$:
$$\int_M\frac{1}{2N}G^{abcd}(\dot
g_{ab}-L_{\omega[\dot g]}g_{ab} )L_Zg_{cd}dx^3=0
$$ which has to be valid for all $Z\in\Gamma(TM)$. This is how the momentum constraint can be shown to be merely a statement to the effect that  the connection is induced by orthogonality to the fibers.

 If a horizontal space  is well defined with respect to such an invariant
supermetric, the projections should  themselves be equivariant, e.g. $(f^*)^*\hat \V_g=\hat\V_{f^*g}\circ{(f^*)}_*$
 However, the fact  that one is dealing with (completions in) function spaces obstructs such a direct approach. In the infinite-dimensional setting one cannot know if for instance $H$ and $V$ are closed, the first requirement of \eqref{connection conditions}. Furthermore, this procedure does not provide us with an explicit formula for our connection form. Even the vertical projection operator, which would in the finite dimensional case be defined as $P(w)=\sum_i\langle v_i, w\rangle v_i$, for $v_i$ an orthonormal basis for $V$, requires modification in the present infinite-dimensional case.

 We shall proceed differently, and find that there exists a more comprehensive way to define
a vertical projection operator and valid connection explicitly. This includes the orthogonality criterion. For this we need to
use the Fredholm alternative.

\subsection{Using the Fredholm Alternative}

In Subsection \ref{projection section} we have shown that  if the horizontal bundle is defined as the space
orthogonal to the orbits, i.e. orthogonal to $\mbox{Im}(\alpha)$ (where we remind the reader that $\alpha$ is the tangent to the orbit map at the identity), $H$ is given by $\mbox{Ker}(\alpha^*)$, (since $(\alpha(X),v)=0$
if $v\in\mbox{Ker}(\alpha^*)$).

 Taking any supermetric, without further assumptions
 (see alternative formulation of Theorem \ref{proj theo}):
 \begin{itemize}
 \item  The operator $\alpha$ and also its symbol $\sigma(\alpha)$ are injective. The first of these requirements is equivalent to a restriction to configurations that do not possess symmetry wrt the relevant group. For $\DD$ for example, it amounts to restricting our attention to $\M'$;
 \item A smooth adjoint of $\alpha$ exists
 with respect to the fiber metrics in $TM$ and $T^*M\otimes_S T^*M$,
 such that $\Ker(\alpha^*)\cap\mbox{Im}(\alpha)=0$ and $\Ker(\sigma(\alpha^*))\cap\mbox{Im}\sigma(\alpha)=0$. The condition $\Ker(\alpha^*)\cap\mbox{Im}(\alpha)=0$ can be seen in fact to be equivalent to requiring the supermetric on $\mbox{Im}(\alpha)$ to be non-degenerate (see  Proposition \ref{propo} in section \ref{sec:constructingV}). This is a condition automatically implemented in the  Fredholm alternative that encodes the criterion for $H_g\cap V_g=0$;
 \item The supermetric is $\DD$-equivariant. \end{itemize}
 Then the  operator defined by \eqref{P vertical}: \be \hat
V:=\alpha\circ(\alpha^*\circ\alpha)^{-1}\circ\alpha^*:T\M'\ra T\M'\ee is well-defined and satisfies all required
properties of a vertical projection operator: \begin{itemize}\item $\hat V$ is $\DD$-equivariant. \item It is idempotent,
$\hat V^2=\hat V$.
\item $\hat V(\alpha(X))=\alpha(X)$ for $X\in\Gamma(TM)$.
\item The space orthogonal to the orbits (or horizontal) satisfies:
$H:=\nu(\mathcal{O}_g)_h=\mbox{Ker}(\alpha_h^*)=\mbox{Ker}\hat V_h$ and $V=\mbox{Im}(\hat V)$ and thus
$T_g\M=H\oplus V$.
\end{itemize}
 In fact, the invariance of the supermetric
 is only used in the construction of the $\hat V$ operator in order to find  the necessary
transformation properties of $\alpha^*$. It ensures that the adjoint of $(r_f^{-1})_*$, where $r_f$ is right
translation by $f$,  is indeed $(r_f)_*$ and so $\alpha^*\circ\alpha$ transforms in the appropriate way. It is also worth noticing that the appearance of the inverse differential operator $(\alpha^*\circ\alpha)^{-1}$ in the definition of the vertical projection operator confirms the non-locality of the connection form explicit in \eqref{connection}.

From the vertical projection operator we obtain the connection form in the usual way:
\be \label{connection from vertical}
\omega:=\alpha^{-1}\circ\hat V=(\alpha^*\circ\alpha)^{-1}\circ\alpha^*\ee
Note that if the vertical operator is well-defined, so is $\alpha^{-1}_{|V_g}$.

\subsection{Equivariant metrics}

 We here first list a wide range of  inner products in $\M$ which are
$\DD$-invariant. We are able to prove equivariance for any supermetric of the form $FG_\beta$ where
$G^{abcd}_\beta=g^{ac}g^{bd}-\beta g^{ab}g^{cd}$  is a one-parameter family of supermetrics,
 weighted by a functional $F:\M\ra C^\infty(M)$, which we call {\it lapse
potential} and define as: \begin{defi}\label{lapse potential} A lapse potential is any functional $F:\M\ra
C^\infty(M)$ formed from $g$ and its curvature tensor by means of tensor product, index raising or lowering,
contraction and covariant differentiation.\end{defi}

To prove that the above mentioned class of supermetrics indeed induces an invariant inner product,  one must simply apply
 a theorem (see for instance {Theorem 9.12.13} of \cite{Bleecker}) which establishes  that, for such a lapse potential $F$,
$F(f^*g)=F(g)\circ f$. Furthermore it is easy to show that $L_Zg$, for any $Z\in\Gamma(TM)$, is a Killing vector
for the generalized supermetric \eqref{supermetric} \cite{Giulini:1993ct}. Combining these facts we have:
\be \int_M \frac{1}{F(f^*g)}G_\beta(f^*u,f^*v)_{f^*g}d\mu_{f^*g}=\int_M (\frac{1}{F(g)}G_\beta(u,v)_{g}\circ
f)f^*d\mu_g.\ee

\subsection{ Ellipticity of $\alpha^*\circ\alpha$}

We have already shown  that $\alpha$ has injective symbol in subsection \ref{orbit manifold section}, furthermore, by the very definition of $\M'$, it obviously true that
$\alpha$  is injective over $\M'$.
\begin{prop}\label{prop:injective_symbol} For each $g\in \M'$, for the inner products $g$ and
$G_{\beta}/N$ in $TM$ and $T^*M\otimes_{\mbox{\tiny S}}T^*M$ respectively, for $\beta\neq 1$ and $N$ any lapse
potential, $\Ker(\sigma(\alpha^*))\cap\mbox{Im}(\sigma(\alpha))=\emptyset$.
\end{prop}
{\rm Proof.~ } We first calculate the symbol of $\alpha^*$ (see subsection \ref{orbit manifold section}).
 For $\lambda\in T_x^*M$ and $v\in T_xM$
 such that $\xi=g(v,\cdot)$, we have $\sigma_\lambda(\alpha):T_xM\ra T_x^*M\otimes_{\mbox{\tiny S}}T_x^*M$ given by
 \be\label{equ:symbol1} \sigma_\lambda(\alpha)(v)=\xi\otimes_{\mbox{\tiny S}} \lambda=2v_{(a}\lambda_{b)}
 \ee From now on we omit the $\alpha$ in the notation. For $u_{ab}\in T_x^*M\otimes_{\mbox{\tiny S}}T_x^*M$
   the adjoint symbol can be directly defined by:
  $$\frac{G^{abcd}_\beta}{N} u_{ab}(\sigma_\lambda(v))_{cd}=(\sigma^*_\lambda(u))^cv_c$$  From this one easily
  calculates (we also omit the $\beta$ dependence to avoid cumbersome notation)
  $\sigma^*_\lambda: T_x^*M\otimes_{\mbox{\tiny S}}T_x^*M\ra T_xM$:
  \be(\sigma^*_\lambda(u))^a= \frac{2}{N(x)}(u^{(ab)}\lambda_b-\beta u^a_{~a}\lambda^a)\label{adjoint symbol}
  \ee
Now inserting $u_{ab}=\sigma_\lambda(v)=2v_{(a}\lambda_{b)}$ for some $v$, and assuming $u_{ab}\in\Ker(\sigma^*_\lambda)$,
we have \bea u^{(ab)}\lambda_b&=&\beta u^a_{~a}\lambda^a\nn\\
\frac{1}{2}||\lambda||^2v^a&=&(\beta-\frac{1}{2})( v^b\lambda_b)\lambda^a\nn
 \eea and thus $\lambda^a=c v^a$, which fed back into the equations can easily be seen to only
 have a solution for $\beta=1$.
$\square$.

So if we are not approaching $\beta=1$ this part of the requirements for the vertical projection operator for such supermetrics is satisfied. However, the value $\beta=1$ is the one present in the canonical 3+1 decomposition of general relativity, and is the one value for which one retains foliation invariance. However, there exist different approaches to gravity that encode fixed foliations, such as \cite{Horava:lif_point} and the theory of Shape Dynamics presented in the first part of this thesis. In particular, SD yields a theory even classically dynamically equivalent to GR which does include a preferred foliation through the use of the conformal group as a symmetry group.

As we will see, a trivial consequence of using the conformal group is that  the equivalent of
$\alpha^*\circ\alpha$ is indeed elliptic for all $\beta$ and lapse potentials.


\subsection{The intersection $\Ker(\alpha^*)\cap\mbox{Im}(\alpha)$}\label{Intersection section}

 There is a  potential problem even in the simple implicit
  orthogonality view, which stems from the non-definiteness  of the deWitt supermetric.
 If the direct sum decomposition is to be
determined by an orthogonality relation with respect to a metric that is not definite (it has signature
$-+++++$), we could run the risk of having elements of the vertical space that are orthogonal to the vertical
space, i.e. $v\perp\V_g$ such that $v=L_Xg$ for some $X\in\Gamma(TM)$, hence $v\in\V_g$ as well.

The adjoint $\alpha^*$ is given by:
\be\int_M
\frac{1}{N}G^{abcd}_\beta u_{ab}X_{(c;d)} d\mu_g=\left(\frac{1}{N}(u^{cd}-\beta g^{cd}u^a_{\phantom{a};a})\right)_{;c}X_d
\ee
Thus
\be \alpha^*(u_{cd})=\left(\frac{1}{N}(u^{cd}-\beta g^{cd} u^a_{\phantom{a};a})\right)_{;c}
\ee
However, even if we simplify the treatment to the case where the functional
$N(x;g]=N[g]$ is spatially constant, we can already glimpse severe obstructions to $\Ker(\alpha^*)\cap\mbox{Im}(\alpha)$ having zero  intersection. First note that
\begin{eqnarray*}
g^{ac}{X}_{(a;b);c}&=&\frac{1}{2}g^{ac}({X}_{a;bc}+{X}_{b;ac})=\frac{1}{2}g^{ac}(R^d_{~abc}{X}_d+{X}_{a;cb}+{X}_{b;ac})\\
~&=&\frac{1}{2}(R^d_{~b}{X}_d+({X}^d_{~;d})_{;b}+\nabla^2{X}_b)
\end{eqnarray*}
where
$\nabla^2{X}_b:=g^{ac}({X}_b)_{;ac}$ is the Riemannian Laplacian. Then
\be\label{one}\alpha^*\circ\alpha(X)= (g^{ac}g^{bd}-\beta g^{ab}g^{cd})({X}_{(a;b)})_{;c}=
\frac{1}{2}(R^{db}{X}_d+(1-2\beta)({X}^d_{~;d})^{;b}+\nabla^2{X}^b) \ee
 If one assumes $\beta\neq 1$, then the operator is elliptic. However, even for $\beta=1$, it can be shown  that   non-trivial \footnote{Since we have excluded Killing fields from our considerations, trivial
solutions to these equations are the ones for which $X_a=0$.} sets of solutions (or lack
thereof) of \eqref{one} (which is equivalent in this case to $\Ker(\alpha^*)\cap\mbox{Im}(\alpha)=0$) depend on the metrics $g$ \cite{Giulini:1993ct}.
There are a number of solutions and domains of validity for the condition
$\HH_g\cap\V_g=\{0\}$ even for $\beta=1$. For example,  for all Ricci-negative geometries (which always exist for
closed $M$ \cite{1986InMat..85..637G}) the condition holds, as well as for non-flat Einstein metrics. For a more
extensive study of these matters see \cite{Giulini:1993ct}. We remark though that as this would still not give a splitting of $\M'$, it would not count as a connection in the sense applied here, which requires it to exist on the whole principal fiber bundle.

\section{The conformal diffeomorphism group.}\label{sec:conformal_diff_group}

Now we apply the same reasoning as in the previous section to the case of the conformal group.

The symbol of $\tau_g$, for $\lambda\in T_x^*M, v\in T_xM$ and $c\in\R$ can be seen to be
 \be\label{symbol tau}
\sigma_\lambda(v,c)=cg_{ab}+\lambda_{(a}v_{b)} \ee Now,  take the metric $\langle\cdot,\cdot\rangle$ in
$T_x^*M\otimes_{\mbox{\tiny S}}T_x^*M$ to be $NG_\beta$. \footnote{Note that in our notation the lapse potential
here appears multiplying the metric, as opposed to the usual lapse in ADM which for the kinetic term appears
dividing it. This will make it easier to deal with powers and negative signs. } The inner product in
$T_xM\times\R$ is taken to be $g(v_1,v_2)+c_1c_2$. Then for $u_{ab}\in T_x^*M\otimes_{\mbox{\tiny S}}T_x^*M$
  from the definition of the adjoint symbol:
  $$NG^{abde}_\beta u_{ab}(\sigma_\lambda(v,c))_{de}=\langle(\sigma^*_\lambda(u)),(v,c)\rangle$$ we easily
  find
  $\sigma^*_\lambda: T_x^*M\otimes_{\mbox{\tiny S}}T_x^*M\ra T_xM$
  \be(\sigma^*_\lambda(u))=
  \left(2(u^{(ab)}\lambda_b-\beta u^a_{~a}\lambda^a),-(1-3\beta)u^a_{~a}\right).\label{adjoint symbol conformal}
  \ee
{\flushleft{ \bf Ellipticity of $\tau^*\circ\tau$}}\smallskip

 Now suppose $u_{(ab)}=\sigma_\lambda(v,c))_{ab}$
and $(\sigma^*_\lambda(u))=(0,0)$. Then we have that
\be\label{equ:symbol_conf}\begin{array}{rcl} c\lambda^a+||\lambda||^2v^a+\lambda^bv_b\lambda^a&=&0\\
3c+2\lambda^av_a&=&0\\
\Rightarrow ||\lambda||^2v^a-\frac{c}{2}\lambda^a&=&0\end{array}\ee
 Contracting the last equation with  $\lambda_a$ and
substituting \eqref{2} in the result yields $-2||\lambda^2||c=0$ which only has solution for $c=0$, in which case
$v^a$ is also obligatorily zero as well. Thus we have proven that
\begin{prop}
For the given action of $\mathcal{C}$ on $\M$, $\alpha$ is an elliptic  operator and
$\Ker(\sigma(\tau^*))\cap\mbox{Im}(\sigma(\tau))=0$. Thus $\tau^*\tau$ is an elliptic operator. \end{prop}

{\flushleft{ \bf The intersection $\Ker(\tau^*)\cap\mbox{Im}(\tau)$}}\smallskip

Since we have gone directly to the calculation of the symbol $\sigma^*(\tau)$, we now write down the actual
operator, for $v_{ab}\in \Gamma(S_2T^*)$. First of all, we check that the supermetric defined by
\eqref{supermetric} is equivariant with respect to the action of $\xi$ (global gauge transformations). One must
merely see that $\xi(f,p)$ acts on the covariant metric tensor $g^{ab}$, as
$$ \xi((f,p),g^{ab})=p^{-1}f^{-1}_*g^{ab}
$$
Thus for global transformations we have:
\be\mathcal{G}[u,v]=\int_Md^3x
\sqrt{pf^*g}N_{pf^*g}G^{abcd}_{pf^*g}(pf^*u)_{ab}(pf^*v)_{ab}=\int_Md^3x\sqrt{g}N_gG^{abcd}u_{ab}v_{ab} \ee  where
for the supermetric to be conformally invariant, $N$ must now not only be a {\it lapse potential}, but also must be further constrained:
 \begin{defi}\label{defi:conf_lapse}
A conformal lapse functional is one for which  $N_g(x)>0$ and
 \be\label{lapse potential
2}N_{pf^*g}(x)=p^{-3/2}N_g(f(x))\ee
\end{defi} We will give an example of
such a lapse potential below.

Calculating the adjoint operator we get: \be\label{Kernel} \tau^*(v)=\left(-2Nv^a_{~a}~,-(NG^{abde}_\beta
v_{de})_{;b}\right) \ee Since for the kernel of $\tau_\beta^*$ the trace part of $v_{ab}$ is zero from the first
component of \eqref{Kernel}, we immediately see that  (inputting the $\beta$ back into the notation for the
adjoint) $G^{abde}_\beta v_{de}=G^{abde}_0v_{de}$. Thus $\mbox{Ker}\tau^*_{~\beta}=\mbox{Ker}\tau^*_{~0}$. Hence
$\Ker(\tau^*_{~\beta}\circ\tau)= \Ker(\tau^*_{~0}\circ \tau )$. Thus, under the supposition that the lapse is a
strictly positive function, exactly as the result $\Ker(\alpha^*_{~0})\cap\mbox{Im}(\alpha)=0$ contained in Section \ref{sec:slice} (see also \cite{FiMa77}), a result dependent on a positive definite inner product on both the target and
domain spaces, one can show that $\Ker(\tau^*_{~0})\cap\mbox{Im}(\tau)=0$. For completeness, the specific
equations for elements of $\Ker(\tau^*_{~0})\cap\mbox{Im}(\tau)$, are, for $v_{ab}=X_{(a;b)}+\lambda g_{ab}$
\be\label{conformal symbol}
\begin{array}{c}
X^a_{~;a}+3\lambda=0\\
\left({N}(X^{(c;d)}+\lambda g^{cd})\right)_{;d}=0\end{array}\ee

 {
\flushleft{ \bf Equivariance of $\tau^*\circ\tau$.}}\smallskip

 Now all that is left to prove that indeed we
have a well-defined connection form for the conformal group (given implicitly by the generalized metrics in $\M$)
is to check wether $\hat V$ transforms equivariantly, or in other words, that  $\tau^*\circ\tau$ is extended to be
right invariant. As one can see from  equation \eqref{equivariant V}, this is dependent strictly on equation
\eqref{orbit transf}, which we now compute for this action. This computation is equivalent to finding out if the action of the group produces a fundamental vector field, as we did for the diffeomorphism group.

The left hand side of \eqref{orbit transf} gives, for $f_t$ the integral
diffeomorphism of $X$: \be T_{(f_0,p_0)}\xi_g(X\circ f_0,p'\circ
f_0)=\frac{d}{dt}_{|t=0}\xi((f_t,p_t),g)=f_0^*(p'g+p_0L_Xg) \ee where $f_t$ produces the integral curves of the
field $X(x)=\frac{d}{dt}_{|t=0}f_t(f_0(x))$. In its turn the right hand side  gives:
 \bea (l_{(f_0,p_0)})_*\circ\tau\circ (r_{(f_0^{-1},1/(p_0(f_0^{-1})))})_*(X,p')&=&
 (l_{(f_0,p_0)})_*\circ(\frac{d}{dt}_{|t=0}\xi_g(f_t,\frac{p't+p_0}{p_0}(f_0^{-1}))\nn\\
 &=& f_0^*(p'g+p_0L_Xg)\eea
 This is equivalent to the following:
\be\label{equivariant transf}T_{(f_0,p_0)}\xi_g(X\circ f_0,(p'\circ f_0)p_0)=p_0f_0^*\tau_g(X,p') \ee

 Hence we find that for the conformal group every structure works nicely and we have a metric-induced connection
 in $\mathcal{C}\hookrightarrow\M''\ra \M''/\mathcal{C} $ for every choice of $\beta$ in the supermetric
and positive lapse potential. The actual equations, writing $\omega[\dot g]=((\omega_{\DD}[\dot g])^a,
  \omega_{\mathcal{P}}[\dot g])$
  take the  form
\bea\label{conformal connection 1} (\omega_{\DD}[\dot g])^a_{~;a}+3\omega_{\mathcal{P}}[\dot g])=\dot g^a_{~a}\\
\label{conformal connection 2}\left(N((\omega_{\DD}[\dot g])^{(a;b)}+\omega_{\mathcal{P}}[\dot g]g^{ab}-\dot
g^{ab})\right)_{;b}=0
 \eea
meaning the corrected velocities are both traceless and transverse {\it with respect to the positive definite
(ultra-local) metric in $\M$} given by $G^{abcd}=Ng^{ac}g^{bd}$.

Thus we have guaranteed existence and uniqueness of solutions for the connection form for $\mathcal{C}$.

\subsection{A fully conformally invariant action with two metric degrees of freedom.}

\subsubsection{Horava gravity with detailed balance.}

Let us briefly examine one recent gravity theory which breaks foliation invariance and possesses powerful indications towards conformal invariance. The so-called Horava-Lifshitz gravity has recently received a great deal of attention, we present its ``detailed balance" formulation \cite{Horava:lif_point}:
\be\label{equ:Horava_action}S=\int dt\int d^3 x\sqrt g N\left(\frac{2}{\kappa^2}G_\lambda^{abcd}K_{ab}K_{cd}-\frac{\kappa^2}{2w^4}C^{ab}C_{ab}\right)
\ee
where $w$ and $\kappa$ are coupling constants. The cotton tensor used here is defined as
\be\label{equ:cotton_tensor} C^{ab}:=\epsilon^{acd}\nabla_c\left({R^b}_d-\frac{1}{4}\delta^b_d R\right)
\ee
The Cotton tensor is symmetric, transverse and traceless:
\be\label{equ:cotton_tensor_props} C^{ab}=C^{(ab)}~~,~~ {C^{ab}}_{;b}=0~~,~~{C^a}_a=0
\ee
It also homogeneously scales conformally with weight $-5/2$. Thus under $g_{ab}\rightarrow e^{4\phi}g_{ab}$ we get $C^{ab}\rightarrow e^{-10\phi}C^{ab}$.

  We shall not explain the more interesting aspects of \emph{why} the action \eqref{equ:Horava_action} was introduced in the first place. At this level, it suffices to say that it possesses different numbers of spatial and time derivatives, making it space-time anisotropic and power-counting renormalizable.
   By the previous work on
  this section we need not check that the constraints associated to the action of $\DD$ and $\mathcal{C}$ (the ``diffeomorphism" and ``conformal" constraints) propagate, have the right transformation properties, or any
  of a multitude of laborious computations; if the system is consistent, we have designed a well-defined gauge system, in every possible sense.
  Thus all that is required for conformal invariance is to impose equations \eqref{conformal connection 1}
  and \eqref{conformal connection 2} for our fixed connection form.
  The remaining constraint, the scalar constraint of theory \eqref{equ:Horava_action} is of the form:
  \be\label{Hamiltonian constraint1} \frac{2}{\kappa^2}K_{TT}^{ab}K^{TT}_{ab}-\frac{\kappa^2}{w^4}C^{ab}C_{ab}=0
  \ee
 where $NK_{TT}=\dot g_{TT}$ and
 \be\label{corrected}\dot g^{TT}_{ab}=(\dot g_{ab}+(\omega_{\DD}[\dot g])_{(a;b)}+\omega_{\mathcal{P}}[\dot
g]g_{ab})\ee are the traverse traceless metric velocities.

Let us proceed to count the degrees of freedom of the  conformally invariant system we expect to obtain (if the action is consistent).
  First of all, since we no longer have full Lorentz invariance, the modified
  version of the Hamiltonian constraint \eqref{Hamiltonian constraint1},
  does not automatically propagate, yielding one further constraint on our
   variables. Thus we have: $6+6+1=13$ (degrees of freedom in $g_{ab}, \dot g_{ab}$ and $N$) minus `the equations of
    motion for N (or `Hamiltonian constraint') and its propagation equation,
    which are (at least) 2 in number, the `conformal constraints on corrected
    velocities' \eqref{conformal connection 1} and
  \eqref{conformal connection 2} (which gives $4$ more), and finally the additional $4$ coming from the choice of
  section for $\mathcal{C}$. Thus we have $3$ remaining degrees of freedom, which is one too little.

{\flushleft{ \bf A model with the right number of degrees of freedom.}}\smallskip

First of all, we see that to make a fully $\mathcal{C}$-invariant theory, we must have a local lapse, and if so,
to get the right number of degrees of freedom the lapse equations of motion have to be automatically propagated by
the other constraints and equations of motion, which is a very tall order.

An alternative, which we propose here, does not have a lapse, and thus has no Hamiltonian constraint
 and yields a theory with the right number of degrees of freedom (12-4-4=4). For this we simply
choose the following lapse potential, satisfying \eqref{lapse potential 2}:
\be\label{Cotton}N_g(x)=\sqrt{C^{ab}(x)C_{ab}(x)}\ee Its conformal weight is given by $\frac{-20+8}{2}=-6 $, and it clearly satisfied all the items of  Definition \ref{lapse potential}) and thus all our gauge constructions are valid. We thus have the action: \be\label{new action} S= \int dt\int_M
\sqrt{g}d^3x \frac{1}{w^2}\sqrt{C^{ab}C_{ab}}(\dot g^{cd}_{TT}\dot g^{TT}_{cd})
  \ee The Hamiltonian for this is given by 
  \be H[\xi,\rho]=\int_M
d^3x \frac{G_{abcd}\pi^{ab}\pi^{cd}}{\sqrt{C^{ab}C_{ab}}}+\pi^{ab}\mathcal{L}_\xi g_{ab}+\rho\pi
  \ee

To stress, this is a fully $\mathcal{C}$-invariant action with the same number of degrees of freedom as GR, but
which does {\it not} have a Hamiltonian constraint.  Equation  \eqref{new action} is furthermore a purely geodesic-type action in Riem, with just one global lapse and thus one global notion of time, as such it also possesses inherent value in a relationalist setting.

In the first part of this thesis we were able to see GR fully as conformally invariant theory. So if any, this formalism has hopes only of
recovering this dual formulation of GR, something which will be investigated further.

\chapter{Conclusions}
In this chapter we will briefly sum up our vision of what has so far been achieved and possible immediate future directions. We repeat some of the statements made in the introductory chapter, now with more technical detail. 

\section*{Results and directions for Shape Dynamics.}

\subsubsection{Brief statement of results.}

 We have found a theory of gravity with two physical degrees of freedom that possesses local scale invariance. Only in a certain conformal gauge it is identical to ADM in constant mean curvature gauge. The total Shape Dynamics Hamiltonian is given by
\begin{equation}\label{equ:prelim_H_SD}
  H_{\text{SD}}=\alpha\mathcal{H}_{\mbox{\tiny{gl}}}+\int_\Sigma d^3 x \big(\rho(x) 4(\pi(x)-\mean{\pi}\sqrt g)+\xi^a(x)H_a(x)\big)
\end{equation}
in the ADM phase space $\Gamma$ parametrized by the usual coordinates $(g,\pi)$, where $\alpha\in\R$, $\rho(x)\in C^\infty(M)$ is an arbitrary Lagrange multiplier function,  and $\mathcal{H}_{\mbox{\tiny{gl}}}[g,\pi]$ is our unique global Hamiltonian, which is a \emph{non-local} functional of $g_{ab},\pi^{ab}$ which does \emph{not} depend on the point $x\in M$. Shape Dynamics possesses the local first class constraints
\begin{equation}
4(\pi(x)-\mean{\pi}\sqrt g)~,~{H}^a.
\end{equation} where $\mean{f}$ is  the global mean of the function $f$ over the 3-manifold $M$. These are the generators of volume-preserving conformal transformations and spatial diffeomorphisms, respectively.
The non-zero part of the constraint algebra is given solely by:
\begin{eqnarray}
\{H^a(\eta_a),H^b(\xi_b)\}&=&H^a([\vec\xi,\vec\eta]_a)\nonumber\\
\{H^a(\xi_a),\pi(\rho))\}&=&\pi\mathcal{L}_\xi(\rho)\nonumber
\end{eqnarray}
For the asymptotically flat case we have a similar result, where the conformal generator is given by $\pi(x)$ only, and its Lagrange multiplier respects certain asymptotic boundary conditions.

The way to achieve this was to enlarge the original theory by a process akin to the Stueckelberg mechanism \cite{Stueckelberg}, thereby obtaining a theory with more constraints then the original one, but still without conformal symmetry. By then performing two distinct preferred gauge fixings on the enlarged gauge theory, we regain either GR in ADM form, or the theory of Shape Dynamics (SD) outlined above. The technical crux of the matter is that the gauge fixing leading to SD leaves a single scalar constraint unfixed, and that the resulting second class constraints can be fully solved in terms of the formerly introduced extra fields.

We have furthermore found how we can extend the treatment that leads to \eqref{equ:prelim_H_SD} to the electromagnetic, massive and massless scalar fields (see chapter \ref{chapter:coupling}), with their usual Hamiltonians. The only thing that changes above is that one has a different $H^a$ (which nonetheless still generates diffeomorphisms), a different global Hamiltonian $\mathcal{H}_{\mbox{\tiny{gl}}}$, which is now also a functional of the added field variables, and, in the case of electromagnetism, the Gauss constraint is also added to the mix. It should also be said that for the massive scalar field the algorithm works only up to a certain field density. Here, the only necessary requirement was that the fields only scale through their coupling to the metric, i.e. the fields themselves do not scale, only the metric does. Once matter couplings are achieved, it becomes very simple to regain a causal (and indeed metric) structure of a space and time.

We also show here that different approximation schemes are available for the global Hamiltonian, and in this thesis we perform a large volume expansion for it (chapter \ref{sec:HJ}), obtaining the first three terms. We use this expansion to find the Hamilton--Jacobi version of the global Hamiltonian, a first step towards quantization. We have found that this bears strong resemblance to certain holographic dualities between gravity and traditional conformal field theories \cite{Boer2000}.

Other interesting results that are not yet ready for print and will not be included in this thesis are: the 2+1 formulation and quantization of the shape dynamics Hamiltonian, a second order expansion of the global Hamiltonian around the De Sitter vacuum, and the formulation of shape dynamics using Ashtekar variables.

\subsubsection{Why is this an interesting result?}
Here we present the significance of Shape Dynamics in light of previous problems and research programs, and then point to some promising directions of research in SD.

 Shape Dynamics provides the theory that fulfills the requirements of a complete theory of the gravitational field on conformal superspace.
Our results  justify York's intuitive remarks regarding the configuration space of gravity: conformal superspace is not the reduced configuration space of general relativity but that of  Shape Dynamics.
Shape Dynamics also meets Barbour's relational arguments for a truly relational theory of the Universe, encapsulated by the aphorism: ``size and motion are relative, and time is given by change".

It is also true, although unseen by us at the time of its conception, that SD is the completion and formalization of Dirac's 1958 paper \cite{Dirac:CMC_fixing}. Although his idea was put in a less developed form, was only valid for asymptotically flat spaces,  did not perceive the role of conformal invariance or symmetry trading, and suffers from a few other drawbacks (as made explicit in section \ref{sec:comparison:Dirac}), the idea behind the mathematical algorithm is basically the same. Made explicit and put into context however, it gains significance way beyond that of a mere ``fixation of the coordinates", providing truly an alternative description of gravity.

The local constraints are all linear in momenta, being easily implementable in configuration space. The true gravitational degrees of freedom are easily found. The constraint algebra is incredibly simple, making it possible that the attempts at the quantization of gravity that encounter the obstacle posed by structure functions being present in the algebra of constraints (as opposed to structure constants) might be more successful in Shape Dynamics.

As we can now couple matter and have a second order perturbation theory around DeSitter space, it also becomes possible to try our hand in perturbative cosmology through the prism of shape dynamics. The possibility of doing cosmology brings about another interesting possibility, this time concerning the uniqueness theorems of general relativity.

For 4-dimensionally covariant theories, there exist tight restrictions on the form of the action. In fact, for an action that is 4-dimensionally covariant, divergenceless, and only depends on the metric up to second derivatives, Lovelock's theorem (theorem \ref{theo:Lovelock}\footnote{There exists a result which can be said to be a 3+1 dimensional version of Lovelock's theorem (although it in fact uses Lovelock's theorem in its proof), \cite{HKT}.}) tells us that the only action available is indeed the Einstein--Hilbert one. But now we have a different set of symmetry principles with which to guide us, and thus a different theory space. We can construct other actions that only match SD (hence ADM in CMC) in certain limits, whilst still respecting the same symmetry principles. We already have two natural candidates, one of which is just taking the first three terms in the volume expansion mentioned above. This has an explicit form, is completely tractable, matches ADM in CMC for large volumes of the Universe, and by our results in coupling with matter and regaining a metric structure, can be tested classically against ADM in CMC (and thus GR). In my personal opinion this is now the most promising new area of research opened up by Shape Dynamics.

This brings us to another way to gain insight into GR through Shape Dynamics. Suppose we do not impose the gauge fixing $S=0$ in Shape Dynamics, but find solutions to some other gauge fixing condition an epsilon away from $S=0$. We would then have a modified gravity theory.  But, any solution of such a modified theory must have a cousin which is a solution to GR, a dual. We can then identify what this solution represents by returning to the linking theory and performing the appropriate gauge transformations.

Summing up, Shape Dynamics  definitely provides a completely different view on classical general relativity, and thus in prospects for its quantization. More than that, it provides interesting new ways to deform general relativity, breaking general covariance but not the 3-dimensional conformal covariance.

\appendix
\part*{Appendix}
\chapter{Variational Formulae}

First we establish some preliminary results, being as extensive as possible. We first note that:
\be\label{equ:inv_metric_var}\delta g^{cd}=-g^{ic}g^{jd}\delta g_{ij}
\ee
We need the variation of the Christoffel symbols. These can be derived from functional differentiation  of both
the the no torsion law: $\nabla_a\nabla_bf=\nabla_b\nabla_af$ (where $f$ is a smooth function) and compatibility
with the metric: $\nabla_ag_{cd}=g_{cd,a}-\Gamma^e_{ac}g_{ed}-\Gamma^e_{ad}g_{ec}=0$. The variation of this
last equation is given by: $$\nabla_a\delta g_{cd}-\delta\Gamma^e_{ac}g_{ed}-\delta\Gamma^e_{ad}g_{ec}=0$$ which can
then be used together with the symmetry given by the no torsion to yield: \be\label{christ
variation}\delta\Gamma^e_{ab}=\frac{1}{2}g^{ec}(\delta g_{bc;a}+\delta g_{ac_;b}-\delta g_{ab;c}) \ee Where we  already
input the semi-colon notation for the covariant derivative, which we utilize from now on, and denote
$\nabla_b\delta g_{ac}=:\delta g_{ac;b}$ to distinguish it from $\delta{g_{ac;b}}$. From these equations we can derive
the following equation for the variation of the Ricci scalar:
\be\label{equ:deltaR}\delta R=-R^{ab}\delta g_{ab}-\nabla^2\delta g+\delta
g_{ab}^{\phantom{ab};ab} \ee
 where $\delta g=g^{ab}\delta g_{ab}$ and $\nabla^2=g^{ab}\nabla_a\nabla_b$. The variation of the metric determinant can be seen to be $\delta g=gg^{ab}\delta g_{ab}$. Thus:
\be \delta\sqrt g= \frac{1}{2}\sqrt g g^{ab}\delta g_{ab}
\ee

\section{Poisson brackets of constraints.}\label{sec:importantApp}
\subsection{Pure gravity.}
The scalar constraint is given by
\be S(x)=\frac{G_{abcd}\pi^{ab}\pi^{cd}(x)}{\sqrt g}-R(x)\sqrt g(x)
\ee
Its variation is thus: \begin{multline}\label{equ:delta_gS}\frac{\delta S(x)}{\delta g_{ef}(y)}=
\frac{\delta}{\delta g_{ef}(y)}\left(\frac{1}{\sqrt g(x)}\right)G_{abcd}\pi^{ab}\pi^{cd}(x)+\frac{\pi^{ab}\pi^{cd}}{\sqrt g}(x)\frac{\delta G_{abcd}(x)}{\delta g_{ef}(y)}-\frac{\delta (R\sqrt g(x))}{\delta g_{ef}(y)}\\
= \left(-\frac{1}{2\sqrt g} g^{ef}G_{abcd}\pi^{ab}\pi^{cd}+\frac{2}{\sqrt g}(\pi^{eb}g_{bd}\pi^{fd}-\frac{\pi^{ef}\pi}{2})\right)(x)\delta(x,y)\\-\left(\frac{1}{2}\sqrt g g^{ef} R(x)\delta(x,y)+\sqrt g(x)\frac{\delta R(x)}{\delta g_{ef}(y)}\right)\\
= \left(-\frac{1}{2\sqrt g} g^{ef}G_{abcd}\pi^{ab}\pi^{cd}+\frac{2}{\sqrt g}(\pi^{eb}g_{bd}\pi^{fd}-\frac{\pi^{ef}\pi}{2})\right)(x)\delta(x,y)-\Big(\frac{1}{2}\sqrt g g^{ef} R(x)\delta(x,y)+\\
\sqrt g(x)(-R^{ef}(x)\delta(x,y)-g^{ef}(x)\nabla^2\delta(x,y)+\delta
(x,y)^{;ef} )\Big)
\end{multline}
The smeared version is
\begin{multline}\label{equ:delta_gSN}\int d^3x N(x)\frac{\delta S(x)}{\delta g_{ef}(y)}=
\left(-\frac{1}{2\sqrt g} g^{ef}G_{abcd}\pi^{ab}\pi^{cd}+\frac{2}{\sqrt g}(\pi^{eb}g_{bd}\pi^{fd}-\frac{\pi^{ef}\pi}{2})\right)N(y)-\Big(\frac{1}{2}\sqrt g g^{ef} RN(y)+\\
\sqrt g(y)(-R^{ef}N(y)-g^{ef}(y)\nabla^2N(y)+
N^{;ef}(y) )\Big)
\end{multline}

Now for the momentum variation:
\be\label{equ:delta_piS}\frac{\delta S(x)}{\delta \pi^{ef}(y)}=
2\frac{G_{efcd}\pi^{cd}}{\sqrt g}(x)\delta(x,y)=\frac{2g_{ec}g_{fd}\pi^{cd}-g_{ef}\pi}{\sqrt g}(x)\delta(x,y)
\ee
The only non-trivial Poisson bracket for ADM, since as we saw the momentum constraint generate only 3-diffeomorphisms, is $\{S(x),S(y)\}$. Using \eqref{equ:delta_gSN} and \eqref{equ:delta_piS} this is very easily calculated. For the smeared version, all the terms that are both linear in the smearings will cancel out upon anti-symmetrization, so in the end we only have to calculate:
\be \int d^3(x)N_2\left(2g_{ec}g_{fd}\pi^{cd}-g_{ef}\pi\right)\left(-g^{ef}\nabla^2N_1+
N_1^{;ef} \right)=2\int d^3x N_2\pi^{cd}{N_1}_{;cd}
\ee
And thus
\begin{multline}\label{equ:Dirac_ctraint_alg}\{S(N_1),S(N_2)\}= \int d^3 x\left(\pi^{cd}(N_1-N_2)_{;cd}\right)=\int d^3 x\left(\pi^{cd}({N_1}_{(;c}-{N_2}_{(;c})_{;d)}\right)\\=H^a(N_1\nabla_a N_2-N_2\nabla_a N_1)
\end{multline}
where the anti-symmetrization cancelled the mixed derivatives.
\subsubsection{Important variations for SD.}

Another calculation which will prove to be useful is the following one arising from $\delta_g\mean{f}$:
\be\label{equ:delta_gMean}\frac{\delta }{\delta g_{ef}(y)}\left(\frac{\sqrt g(x)}{V}\right)=\frac{1}{2V}\left((g^{ef}\sqrt g)(x)\delta(x,y)-\frac{(g^{ef}\sqrt g)(y)\sqrt g(x)}{V}\right)
\ee
We remark that it can easily be seen that
$$\int d^3y F_{ef}(y)\frac{\delta }{\delta g_{ef}(y)}\left(\frac{\sqrt g(x)}{V}\right)=\frac{1}{2V}(F-\mean{F})\sqrt g(x)
$$ where $F$ is the trace of the tensor $F_{ab}$.

The last preparatory results for us are the following:
\begin{multline}\label{equ:delta_gD}\frac{\delta (\pi-\mean\pi \sqrt g)(z)}{\delta g_{ef}(y)}=\pi^{ef}(z)\delta(z,y)-\pi^{ef}(y)\frac{\sqrt g(x)}{V}-\mean{\pi}V\frac{\delta }{\delta g_{ef}(y)}\left(\frac{\sqrt g(z)}{V}\right)\\
=\pi^{ef}(z)\delta(z,y)-\pi^{ef}(y)\frac{\sqrt g(z)}{V}-\mean{\pi}\frac{1}{2}\left((g^{ef}\sqrt g)(z)\delta(z,y)-\frac{(g^{ef}\sqrt g)(y)\sqrt g(z)}{V}\right)
\end{multline}
and
\be\label{equ:delta_piD}\frac{\delta (\pi-\mean\pi \sqrt g)(z)}{\delta \pi^{ef}(y)}=g_{ef}(z)\delta(z,y)-g_{ef}(y)\frac{\sqrt g(z)}{V}
\ee
Now we calculate the first part of the most important Poisson bracket for our results, that is $\{S(N),\pi(z)-\mean{\pi}\sqrt g(z)\}$:
\begin{multline}\label{equ:firstPbSpi}\int d^3 y\int d^3 x \frac{\delta S(x)}{\delta g_{ef}(y)}N(x)\frac{\delta (\pi-\mean\pi \sqrt g)(z)}{\delta \pi^{ef}(y)}
\\
=\int d^3 y\Big[\left(-\frac{1}{2\sqrt g} g^{ef}G_{abcd}\pi^{ab}\pi^{cd}+\frac{2}{\sqrt g}(\pi^{eb}g_{bd}\pi^{fd}-\frac{\pi^{ef}\pi}{2})\right)N(y)-\Big(\frac{1}{2}\sqrt g g^{ef} RN(y)+\\
\sqrt g(y)(-R^{ef}N(y)-g^{ef}(y)\nabla^2N(y)+\delta
N^{;ef}(y) )\Big)\Big]
\left(g_{ef}(z)\delta(z,y)-g_{ef}(y)\frac{\sqrt g(z)}{V}\right)\\
=\left(-\frac{3}{2}S+\frac{2}{\sqrt g} G_{abcd}\pi^{ab}\pi^{cd}+\sqrt g2(-R+\nabla^2)\right)N(z) -\mean{\mathcal A}
\end{multline}
where we completed $-\frac{3}{2\sqrt g} G_{abcd}\pi^{ab}\pi^{cd} $ by adding $\frac{3}{2}R-\frac{3}{2}R$, and again we use the notation that $\mathcal{A}$ denotes whatever comes before it in an equation (in this case $\mathcal{A}=(-\frac{3}{2}S(z)+\frac{2}{\sqrt g} G_{abcd}\pi^{ab}\pi^{cd}(z)+\sqrt g(z)(-R+\nabla^2))N(z)$).
Now
\begin{multline}\label{equ:secPbSpi}-\int d^3 y\int d^3 x \frac{\delta S(x)}{\delta \pi^{ef}(y)}N(x)\frac{\delta (\pi-\mean\pi \sqrt g)(z)}{\delta g_{ef}(y)}
\\
=-\int d^3 y \left[\frac{2g_{ec}g_{fd}\pi^{cd}-g_{ef}\pi}{\sqrt g}N(y)\right]\times\\
\left[\pi^{ef}(z)\delta(z,y)-\pi^{ef}(y)\frac{\sqrt g(z)}{V}-\mean{\pi}\frac{1}{2}\left((g^{ef}\sqrt g)(z)\delta(z,y)-\frac{(g^{ef}\sqrt g)(y)\sqrt g(z)}{V}\right)
\right]\\
=-\frac{2}{\sqrt g} G_{abcd}\pi^{ab}\pi^{cd}N(z)-\frac{1}{2}\pi(z)\mean{\pi}-\mean{\mathcal{A}}
\end{multline}
Combining \eqref{equ:firstPbSpi} and \eqref{equ:secPbSpi} we get:
\begin{multline}\label{equ:PbSpi}\{S(N),D(z)\}=\left(-\frac{3}{2}S+2\sqrt g(-R+\nabla^2)-\frac{1}{2}\pi\mean{\pi}\right)N(z)-\mean{\mathcal{A}}\\
\approx\left( 2\sqrt g(-R-\frac{1}{4\sqrt g}\pi\mean{\pi}+\nabla^2)\right)N(z)-\mean{\mathcal{A}}
\end{multline}
which we can rewrite if we so choose by using the scalar constraint as
\be\label{equ:PbSpi2}\{S(N),D(z)\}\approx2(-\frac{G_{abcd}\pi^{ab}\pi^{cd}}{\sqrt g}-\frac{1}{4}\pi\mean{\pi}+\sqrt g\nabla^2)N(z)-\mean{\mathcal{A}}
\ee

\section{Scalar and Electromagnetic fields.}
\subsection{Scalar.}
We start with the total scalar constraint, which is now:
\be S(x)=\frac{\pi^{ab}\pi_{ab}-\frac{1}{2}\pi^2+\pi_\psi^2}{\sqrt g}-\sqrt g (R-g^{ab}\nabla_a\psi\nabla\psi_b)
\ee
where we have simply added the scalar  field Hamiltonian:
\be H_{\mbox{\tiny Scal}}=\frac{\pi_\psi^2}{\sqrt g}+\sqrt g g^{ab}\nabla_a\psi\nabla\psi_b
\ee
To find the contribution this extra term will have to the lapse fixing equation \eqref{equ:PbSpi}, due to the absence of any terms containing the metric momenta, we must merely calculate:
\be \frac{\delta H_{\mbox{\tiny Scal}}(N)}{\delta g_{ab}(x)}g_{ab}(x)=\left(-\frac{3}{2}\frac{\pi^2_\psi}{\sqrt g}(x)+\frac{1}{2}\sqrt g g^{ab}\nabla_a\psi\nabla\psi_b\right)N(x)
\ee
where we used \eqref{equ:inv_metric_var}. But going back to \eqref{equ:PbSpi}, we must still complete the $-\frac{3}{2}S_g$ appearing there, with $-\frac{3}{2}H_{\mbox{\tiny Scal}}$, so that we can discard this term as weakly vanishing. To do so we add and subtract $2\sqrt g g^{ab}\nabla_a\psi\nabla\psi_b $ obtaining:
\be\label{equ:PbSpiScalar}
\left(-\frac{3}{2}S+2\sqrt g(-R-\frac{1}{4\sqrt g}\pi\mean{\pi}+g^{ab}\nabla_a\psi\nabla\psi_b+\nabla^2)\right)N(z)-\mean{\mathcal{A}}
\ee
which is the equation we need in the main text, in section \ref{sec:coupling:scalar}.
\subsection{Electromagnetic.}
Now we add to the gravitational Hamiltonian the electromagnetic Hamiltonian:
\be H_{\mbox{{\tiny EM}}} =-A_{[a,b]}A_{[c,d]}(x)g^{ac}(x)g^{bd}(x)\sqrt{g}(x)+
\frac{E^a(x)E^b(x)g_{ab}(x)}{\sqrt g}(x)\ee
Again we must calculate solely:
\be \frac{\delta H_{\mbox{\tiny EM}}(N)}{\delta g_{ab}(x)}g_{ab}(x)=\left(\frac{1}{2}g^{ac}g^{bd}A_{[a,b]}A_{[c,d]}\sqrt g-\frac{1}{2}\frac{E^aE^bg_{ab}}{\sqrt g}(x)\right)N(x)=-\frac{1}{2}H_{\mbox{\tiny EM}}N(x)
\ee
Again from \eqref{equ:PbSpi}, we must still complete the $-\frac{3}{2}S_g$ appearing there, with $-\frac{3}{2}H_{\mbox{\tiny EM}}$, so that we can discard this term as weakly vanishing. To do so we add and subtract $H_{\mbox{\tiny EM}}$ obtaining:
\be\label{equ:PbSpiEM}
\left(-\frac{3}{2}S+2\sqrt g(-R-\frac{1}{4\sqrt g}\pi\mean{\pi}+\frac{H_{\mbox{\tiny EM}}}{2\sqrt g}+\nabla^2)\right)N(z)-\mean{\mathcal{A}}
\ee
which is the equation we use in section \ref{sec:coupling:EM}.

\chapter{Relevant formulae for volume-preserving-conformal transformations. }\label{sec:importantAppII}

\section{Basic variations.}
We start from the definition of the surjection map given by \eqref{equ:vpct_def}:
\be
\hat \phi(x):=\phi(x)-\frac 1 6 \ln\langle e^{6\phi}\rangle_g\ee where we use the mean $\langle f\rangle_g:=\frac 1 V \int d^3x\sqrt{|g|} f(x)$ and 3-volume $V_g:=\int d^3x\sqrt{g}$.
We will mainly work with the simplified exponentiated version:
\be\label{equ:exp_phi_hat}e^{6\hat\phi}=\frac{e^{6\phi}V}{\int d^3 xe^{6\phi}\sqrt g}
\ee
Then we have:
\begin{eqnarray}\frac{\delta}{\delta g_{ij}(y)}\left(\frac{e^{6\phi(x)}V}{\int d^3 xe^{6\phi}\sqrt g}\right)&=&\frac{e^{6\phi(x)}V}{2\int d^3 xe^{6\phi}\sqrt g}\frac{\sqrt g(y)g^{ij}(y)}{V}-\frac{e^{6\phi(x)}V}{2\int d^3 xe^{6\phi}\sqrt g}\frac{\sqrt g(y)g^{ij}(y)}{V}\frac{e^{6\phi(y)}V}{\int d^3 xe^{6\phi}\sqrt g}\nonumber\\
&=&\frac{1}{2V}\sqrt g(y)g^{ij}(y)e^{6\hat\phi(x)}(1-e^{6\hat\phi(y)})\label{equ:del_gvpct}
\end{eqnarray}
Generalizing to
\be\label{equ:ndel_gvpct}\frac{\delta(e^{n\hat\phi})}{\delta g_{ij}(y)}=\frac{n}{12V}\sqrt g(y)g^{ij}(y)e^{n\hat\phi(x)}(1-e^{6\hat\phi(y)})
\ee
In the same way:
\begin{eqnarray}\frac{\delta}{\delta \phi(y)}\left(\frac{e^{6\phi(x)}V}{\int d^3 xe^{6\phi}\sqrt g}\right)&=&6\left(e^{6\hat\phi(x)}\delta(x,y)-
\frac{e^{6\phi(x)}V}{{\int d^3 xe^{6\phi}\sqrt g}^2}e^{6\hat\phi(y)}\sqrt g(y)\right)\nonumber\\
&=&6e^{6\hat\phi(x)}\left(\delta(x,y)-
\frac{e^{6\hat\phi(y)}\sqrt g(y)}{V}\right)\label{equ:del_phivpct}
\end{eqnarray}
Generalizing to
\be\label{equ:ndel_phivpct}\frac{\delta(e^{n\hat\phi})}{\delta \phi(y)}=ne^{n\hat\phi(x)}\left(\delta(x,y)-
\frac{e^{6\hat\phi(y)}\sqrt g(y)}{V}\right)
\ee

\section{Canonical transformation properties}\label{sec:canonical_transfs_app}

With the basic variations calculated in the last section  we can, as a consistency check, verify explicitly that the transformation is canonical:
\begin{eqnarray}\{\mathcal{T}_\phi g_{ab},\pi_\phi\}&=&4\mathcal{T}_\phi \{g_{ab},\pi-\mean{\pi}\sqrt g\}\\
\{\mathcal{T}_\phi \pi^{ab},\pi_\phi\}&=&4\mathcal{T}_\phi \{\pi^{ab},\pi-\mean{\pi}\sqrt g\}\\
\{\mathcal{T}_\phi g_{ab},\mathcal{T}_\phi \pi^{ab}\}&=&\mathcal{T}_\phi \{g_{ab},\pi^{ab}\}\label{equ:explicitCanonical}
\end{eqnarray}
The first two also verify that the constraint $\mathcal{Q}$ commutes with any function $\mathcal{T}_\phi f(g,\pi)$.

 Let us start with a direct proof of the last identity \eqref{equ:explicitCanonical}. As $\pi_\phi$ does not appear in the equation, and $\mathcal{T}_\phi g_{ab}$ does not contain $\pi^{ab}$, the calculation is made a lot simpler.
\be
\frac{\delta \mathcal{T}_\phi \pi^{ab}(x)}{\delta \pi^{ef}(y)}=e^{-4\hat\phi(x)}\left(\delta^{ab}_{ef}\delta(x,y)-\frac{1}{3V}g_{ef}(y)(\sqrt g g^{ab}(1-e^{6\hat\phi}))(x)\right)\ee
The second element we need is (using \eqref{equ:ndel_gvpct})
\be \frac{\delta \mathcal{T}_\phi g_{cd}(x)}{\delta g_{ef}(y)}=\frac{1}{3V}e^{4\hat\phi(x)}(1-e^{6\hat\phi(y)})\sqrt g(y)g^{ef}(y)g_{cd}(x)+e^{4\hat\phi(x)}\delta^{ef}_{cd}\delta(x,y)
\ee Upon multiplying and integrating over $y$ we get:
\be \delta^{ab}_{cd}\delta(x,y)-\frac{1}{3V^2}\int d^3 y\left(\sqrt g g_{cd}g^{ab}(1-e^{6\hat\phi})\right)(x)\left((1-e^{6\hat\phi(y)})\sqrt g\right)(y)=\delta^{ab}_{cd}\delta(x,y)
\ee
where we already discarded the terms that come in with opposite signs, and used the fact that the integral present runs only over the $y$ dependent terms, and $\int d^3 y (1-e^{6\hat\phi(y)})\sqrt g(y)=0$.

To finish this explicit verification  that we are indeed deling with a canonical transformation, we separate the remaining steps into three: $i)$ $\phi$ does not change under the transformations, $\phi\rightarrow \phi$, and the change in $\pi_\phi$ is just the conserved charge, which is $\phi$ independent, and thus the Poisson brackets $\{\phi,\pi_\phi\}$, $\{\phi,\phi\}$ and $\{\pi_\phi,\pi_\phi\}$ are conserved. $ii)$  $\{\phi,g_{ab}\}$ clearly stays the same as does $\{\phi,\pi^{ab}\}$, as none of the original canonical variables transforms to something containing $\pi_\phi$. $iii)$ This step is the most difficult one. Here we must explicitly compute that $\{\pi_\phi,\mathcal{T}_\phi g_{ab}\}=4\{D,\mathcal{T}_\phi g_{ab}\}$. To do so, we use  the preceding proof of \eqref{equ:explicitCanonical} and the fact that $D$ is invariant under  $\mathcal{T}_\phi$, to resort to the equivalent calculation of $4\mathcal{T}_\phi\{D, g_{ab}\}$.

By \eqref{equ:ndel_phivpct}, we have that
\be \{\mathcal{T}_\phi g_{ab},\pi_\phi\}=\frac{\delta(e^{4\hat\phi})}{\delta \phi(y)}g_{ab}(y)=4e^{4\hat\phi(x)}\left(\delta(x,y)-
\frac{e^{6\hat\phi(y)}\sqrt g(y)}{V}\right)g_{ab}(y).
\ee
By \eqref{equ:delta_piD}, we have that
\be \{g_{ab}(y),D(x)\}=\frac{\delta (\pi-\mean\pi \sqrt g)(x)}{\delta \pi^{ef}(y)}=g_{ab}(x)\delta(x,y)-g_{ab}(y)\frac{\sqrt g(x)}{V}
\ee which upon acting with $\mathcal{T}_\phi$ and multiplying by 4 clearly gives us the sought for equality.

Now, using the form of $\mathcal{T}_\phi\pi^{ab}(x)$ given in \eqref{equ:canonicalTransformation_grav}, and again \eqref{equ:ndel_phivpct}:
\begin{multline}\label{equ:check}
\{\mathcal{T}_\phi \pi^{ab},\pi_\phi\}=\diby{e^{-4\hat\phi(x)}}{\phi(y)}\left(\pi^{ab}(x) -\frac{g^{ab}(x)}{3}\sqrt {g(x)}\langle \pi\rangle (1-e^{6\hat\phi(x)})\right)+\diby{e^{6\hat\phi(x)}}{\phi(y)}\frac{g^{ab}(x)}{3}\sqrt {g(x)}\langle \pi\rangle e^{-4\hat\phi(x)}\\
=-4 e^{-4\hat\phi(x)}\left(\delta(x,y)-e^{6\hat\phi(y)}\frac{\sqrt {g(y)}}{V}\right)\left(\pi^{ab}(y)-\frac{1}{3}\mean{\pi}g^{ab}(y)\sqrt{g(y)}+\frac{1}{6}\mean{\pi}g^{ab}(y)\sqrt{g(y)}e^{6\hat\phi(y)}\right).
\end{multline}
On the other hand we have by \eqref{equ:delta_gD}
\begin{multline} \{D(x),\pi^{ab}(y)\}=\frac{\delta (\pi-\mean\pi \sqrt g)(x)}{\delta g_{ab}(y)}\\
=\pi^{ef}(x)\delta(x,y)-\pi^{ef}(y)\frac{\sqrt g(x)}{V}-\mean{\pi}\frac{1}{2}\left((g^{ef}\sqrt g)(x)\delta(x,y)-\frac{(g^{ef}\sqrt g)(y)\sqrt g(x)}{V}\right)
\end{multline}
which upon acting with $\mathcal{T}_\phi$ and multiplying by 4, after a little manipulation using the fact that $\mean{\pi}$ transforms trivially, yields \eqref{equ:check}.

\section{Tangent space to the space of volume-preserving conformal transformations.}\label{sec:tangent_space_vpct}

Thus if we fix the metric, it is easy to see what constitutes the space of tangent functions to the volume-preserving ones (at $\phi=0$):
\be{\frac{d}{dt}}_{|t=0|}e^{6\hat\phi_t(x)}=6{\frac{d}{dt}}_{|t=0|}\hat\phi_t
\ee
for a one parameter family of $\hat\phi$, such that $\hat\phi_0=0$. Let us call the function $\phi':=f\in C^\infty(M)$. Since the metric is fixed we get from \eqref{equ:del_phivpct}:
\be\label{equ:tangent_hatphi} {\frac{d}{dt}}_{|t=0}e^{6\hat\phi_t(x)}=\frac{\delta}{\delta \phi(y)}e^{6\hat\phi(x)}\cdot f(y)=6(f(x)-\mean{f})
\ee
where we have used the $\cdot$ notation employed in section \ref{sec:PbsLinear}. And we have thus proven the assertion needed in section \ref{sec:Const.Surf.SD}.

\section{Group structure.}\label{sec:group_structure}

Now we check that indeed volume preserving conformal transformations form a groupoid. For each metric $g$ the action of $\mathcal{C}/\mathcal{V}$ will form a subgroup.  Let $\phi_1,\phi_2$ be the generators of two vpcts.
From \eqref{equ:canonicalTransformation_grav} we have
\be
\begin{array}{rcl}
\mathcal{T}_{\phi_1} g_{ab}&=&e^{4\hat{\phi_1}(x)}g_{ab}(x)\\
\mathcal{T}_{\phi_1}\pi^{ab}&=&e^{-4\hat{\phi_1}(x)}\left(\pi^{ab}(x) -\frac{g^{ab}}{3}\sqrt {g}\langle \pi\rangle (1-e^{6\hat{\phi_1}})\right)\end{array}\ee
Now we see what happens when we iterate the transformation with $\phi_2$. For ease of manipulation, we will just call the transformed variables above by $\bar g_{ab}$ and $\bar\pi^{ab}$. All we have to do now is replace all occurrences of $\phi_1$ by $\phi_2$, and all occurrences of the unbarred variables by the barred variables. It turns out to be more convenient to express
\be e^{4\hat\phi}=\left(\frac{e^{6\phi}V}{\int e^{6\phi}\sqrt g}\right)^{2/3}\ee
Now
\be  \int e^{6\phi_2}\sqrt{\bar g}=\frac{V\int e^{6\phi_2}e^{6\phi_1}\sqrt  g}{\int e^{6\phi_1}\sqrt g}
\ee
and thus
\be\label{equ:vpct_comp} \frac{e^{6\phi_2}V}{\int e^{6\phi_2}\sqrt {\bar g}}=\frac{e^{6\phi_2}\int e^{6\phi_1}\sqrt { g}}{\int e^{6\phi_1}e^{6\phi_2}\sqrt { g}}.
\ee
Finally:
\be
\mathcal{T}_{\phi_2} \bar g_{ab}=\left(\frac{e^{6\phi_2}V}{\int e^{6\phi_2}\sqrt{\bar g}}\right)^{2/3}\left(\frac{e^{6\phi_1}V}{\int e^{6\phi_1}\sqrt g}\right)^{2/3}g_{ab}=\left(\frac{Ve^{6(\phi_2+\phi_1)}}{\int e^{6(\phi+\phi_2)}\sqrt g}\right)^{2/3}g_{ab}.
\ee

For the momenta the equations are more involved. Let us first right down $\mathcal{T}_{\phi_1}\bar\pi^{ab}$ in terms of $\phi_2$ and barred variables (except for the last term, where we already input \eqref{equ:vpct_comp}):

\begin{multline}
 \left(\frac{e^{6\phi_2}\int e^{6\phi_1}\sqrt { g}}{\int e^{6\phi_1}e^{6\phi_2}\sqrt { g}}
\right)^{-2/3}\left(\bar\pi^{ab}(x) -\frac{\bar g^{ab}}{3}\sqrt {\bar g}\langle \pi\rangle (1-\left(\frac{e^{6\phi_2}\int e^{6\phi_1}\sqrt { g}}{\int e^{6\phi_1}e^{6\phi_2}\sqrt { g}}
\right))\right)\\
= \Big(\frac{e^{6\phi_2}\overbrace{\int e^{6\phi_1}\sqrt { g}}^A}{\int e^{6\phi_1}e^{6\phi_2}\sqrt { g}}
\Big)^{-2/3}\Big\{\Big[\Big(\frac{e^{6\phi_1}V}{\underbrace{\int e^{6\phi_1}\sqrt g}_A}\Big)^{-2/3}\left(\pi^{ab}(x) -\frac{g^{ab}}{3}\sqrt {g}\langle \pi\rangle (1-\overbrace{\left(\frac{e^{6\phi_1}V}{\int e^{6\phi_1}\sqrt g}\right)}^B)\right)\Big]\\
 -\frac{1}{3}g^{ab}\sqrt {g}\Big[\frac{e^{6\phi_1}V}{\underbrace{\int e^{6\phi_1}\sqrt g}_A}\Big]^{-2/3}\Big[\frac{e^{6\phi_1}V}{\underbrace{\int e^{6\phi_1}\sqrt g}_C}\Big]\langle \pi\rangle \big[\underbrace{1}_B-\big(\frac{e^{6\phi_2}\overbrace{\int e^{6\phi_1}\sqrt { g}}^C}{\int e^{6\phi_1}e^{6\phi_2}\sqrt { g}}
\big)\big]\Big\}\\
=\Big(\frac{e^{6(\phi_2+\phi_1)}}{\int e^{6(\phi_1+\phi_2)}\sqrt { g}}
\Big)^{-2/3}\left(\pi^{ab}(x) -\frac{g^{ab}}{3}\sqrt {g}\langle \pi\rangle \Big[1-\frac{e^{6(\phi_1+\phi_2)}V}{\int e^{6(\phi_1+\phi_2)}\sqrt g}\Big]\right)
\end{multline}
where in the above equation we use the letters $A,B,C$ to mean operations in that order. $A$ pairs multiplicative inverses, $B$ cancels terms, and $C$ (which can be used only after $B$) also pairs multiplicative inverses.  This finishes the proof that volume-preserving conformal transformations acts as a commutative groupoid on phase space. We note that for commutative groupoids, the structure constants of the algebra are zero, and thus for the algebra there is no leftover dependence on the base point.

\chapter{Hamilton-Jacobi auxiliary calculations.}\label{sec:HJ_app}
\section{Volume decoupling for $\sigma^{ab}$.}
Here we show that the canonical Poisson brackets between the  barred variables
 \begin{align}
   \bar g_{ab} &= \lf(\frac{V}{V_0} \rt)^{-\frac{2}{3}} g_{ab}, \qquad \qquad V = \int d^3x \sqrt{g}, \\
   \bar \sigma^{ab} &= \lf(\frac{V}{V_0}\rt)^{\frac{2}{3}}\lf(\pi^{ab} - \frac 1 3 \mean{\pi} g^{ab} \sqrt{g} \rt),  ~ P= \frac 2 3 \mean{\pi}.
\end{align}
 in section \ref{sec:HJ} are trivial. Furthermore we will show how to use the chain rule to prove the important identity \eqref{equ:chain_rule_sigma}. Let us start with the relations:
\begin{eqnarray}\diby{\left(\frac{V}{V_0}\right)^{-2/3}}{g_{ab}(x)}&=&-\frac{1}{3}\left(\frac{V}{V_0}\right)^{-4/3}\frac{\sqrt{\bar g}\bar g^{ab}(x)}{V_0}\\
 \diby{\bar g_{cd}(y)}{g_{ab}(x)}&=&\diby{\left(\frac{V}{V_0}\right)^{-2/3}}{g_{ab}(x)}g_{cd}(y)+\left(\frac{V}{V_0}\right)^{-2/3}\diby{ g_{cd}(y)}{g_{ab}(x)}\nn\\
 &=&\left(\frac{V}{V_0}\right)^{-2/3}\left(\delta^{ab}_{cd}(x,y)-\frac{1}{3}\frac{\bar g_{cd}(y)}{V_0}\bar g^{ab}(x)\sqrt {\bar g} (x)\right)
\end{eqnarray}
thus
\be \int d^3y \diby{S}{\bar g_{cd}(y)}\diby{\bar g_{cd}(y)}{g_{ab}(x)}=\left(\frac{V}{V_0}\right)^{-2/3}\left[\diby{S}{\bar g_{ab}(y)}-\frac{1}{3}\frac{\bar g^{ab}(x)}{V_0}\sqrt {\bar g} (x)\int d^3 y \diby{S}{\bar g_{cd}(y)}\bar g_{cd}(y)\right]
\ee
Thus
\be \diby{S}{V}\diby{V}{g_{ab}(x)}=\frac{1}{2}\diby{S}{V}\left(\frac{V}{V_0}\right)^{1/3}\sqrt {\bar g}\bar g^{ab}(x)
\ee
We then have
\bea \diby{S}{ g_{ab}(x)}&=&\diby{S}{V}\diby{V}{g_{ab}(x)}+\int d^3y \diby{S}{\bar g_{cd}(y)}\diby{\bar g_{cd}(y)}{g_{ab}(x)}\nn\\
 &=&\frac{1}{2}\diby{S}{V}\left(\frac{V}{V_0}\right)^{1/3}\sqrt {\bar g}\bar g^{ab}(x)+\left(\frac{V}{V_0}\right)^{-2/3}\left[\diby{S}{\bar g_{ab}(x)}-\frac{1}{3}\frac{\bar g^{ab}(x)}{V_0}\sqrt {\bar g} (x)\int d^3 y \diby{S}{\bar g_{cd}(y)}\bar g_{cd}(y)\right]\nn\\
\eea
which means
\be \int d^3x \diby{S}{ g_{ab}(x)}g_{ab}(x)= \frac{3}{2}\diby{S}{V}\frac{V}{V_0}V_0
\ee Finally,
 under $\pi^{ab}(x)\rightarrow \diby{S}{g_{ab}(x)}$, $\sigma^{ab}$ goes to:
 \begin{multline}\label{equ:traceless_dec} \bar\sigma^{ab}\rightarrow \lf(\frac{V}{V_0}\rt)^{\frac{2}{3}}\lf( \diby{S}{ g_{ab}(x)}- \frac {1}{3V} \int d^3 y\left[{\diby{S}{ g_{ab}(y)}} g^{ab}(y)\right] \sqrt{g(x)} g^{ab}(x)\rt)\\
 =\diby{S}{\bar g_{ab}(x)}-\frac{1}{3}\frac{\bar g^{ab}(x)}{V_0}\sqrt {\bar g} (x)\int d^3 y \diby{S}{\bar g_{cd}(y)}\bar g_{cd}(y)
 \end{multline}

 \section{Volume expansion for the SD Hamiltonian}\label{sec:volume_expansion}

 We find  $\hg$  by simultaneously solving the equations
\begin{align}
   \hg^0 = \lf( 2\Lambda - \frac{3}{8}P^2 \rt) + \frac{\lf( 8 \bar{\nabla}_0^2 - \bar R_0 \rt) \Omega}{(V/V_0)^{2/3}\Omega^5}
       - \frac{\bar \sigma^{ab} \bar \sigma_{ab} }{(V/V_0)^2\Omega^{12}\bar g_0}\label{equ:main hg2}
   \\
   \mean{\Omega^6} = 1, \label{equ:vpct_cond2}
\end{align}
where barred quantities are calculated using $\bar g^0_{ab}$ and the (super)subscript $o$ denotes the Yamabe gauge.

The large $V$ expansion is
\begin{equation}
   \hg = \sum_{n=0}^\infty  \left(\frac{V}{V_0} \right)^{-2n/3} \hn{n}, ~~ \Omega^6 = \sum_{n=0}^\infty  \left(\frac{V}{V_0} \right)^{-2n/3}  \wn{n} .
\end{equation}
Explicitly \begin{equation}
\mathcal H_{\text{gl}} = \mathcal H_0 + V^{-2/3} \mathcal H_1 + V^{-4/3} \mathcal H_2 + V^{-2} \mathcal H_3 + \dots
\end{equation}
\begin{equation}
\Omega^6 = \omega_0 + V^{-2/3} \omega_1 + V^{-4/3} \omega_2 + V^{-2} \omega_3 + \dots
\end{equation}
From \eqref{equ:vpct_cond2}, the restriction  is trivially solved by $\mean{\wn{n}} = 0$ for $n \neq 0$ and $\mean{\wn{0}} = 1$. We can solve for the $\hn{n}$'s by inserting the expansion, taking the mean, and using the fact that $\bar R_0$ is constant. That is
\begin{equation}
\left< \omega_0 \right> = 1~, \qquad \left< \omega_j \right> = 0 ~, ~~~ j>0 ~.
\end{equation}
For order zero we get thus:
\begin{equation}
\mathcal H_0 =  2 \Lambda - \frac{3}{8} P^2~,
\end{equation} as it does not depend on the volume expansion, this already cancels out separately in the expansion \eqref{equ:main hg2}.

To order $V^{-2/3}$ we have
\begin{equation}
\mathcal H_1 = - \frac{1}{\omega_0^{5/6} V^{2/3}} \left( 8 \bar \nabla_0^2  - {R_0} \right) \omega_0^{1/6} = - R \left( \omega_0^{2/3} \bar g_0 \right)
\end{equation}
where we used equation \eqref{equ:RicciConf2} on the last equality. Taking the mean
\begin{equation}
\mathcal H_1 = - \left< R \left( \omega_0^{2/3} \bar g_0 \right) \right> ~,
\end{equation}
the equation now reads
\begin{equation}
R \left( \omega_0^{2/3} \bar g_0 \right) =  \left< R \left( \omega_0^{2/3} \bar g_0 \right) \right>  ~,
\end{equation}
which tells us that $\omega_0$ would take the metric to the Yamabe gauge. But since
we're already working in the Yamabe gauge, and our Yamabe metric is unique (up to conformal diffeomorphisms),  this equation reduces to $\omega_0 =1$.
Then the solution for $\mathcal H_1$ is:
\begin{equation}
\mathcal H_1 =  -  R_0 ~,
\end{equation}

Now let us expand the conformal factor to second order:
\be\label{equ:omega_exp} \Omega^{-5}  = 1-\frac{5 }{6}  \omega_1 V^{-2/3}+ \left(\frac{55}{72}\omega_1^2-\frac{5}{6} \omega_2\right) V^{-4/3}
~~\mbox{and}~~ \Omega^1  = 1+\frac{1}{6}  \omega_1 V^{-2/3}+\left( -\frac{5}{72}  \omega_1^2 + \frac{1}{6} \omega_2\right) V^{-4/3}
\ee
Substituting, for order $V^{-4/3}$ we get:
\begin{equation}
\mathcal H_2 = -\frac{1}{6} \left( 8{\bar \nabla_0}^2 -{R_0}  \right)  \omega_1-\frac{5}{6}R_0\omega_1= -\frac{2}{3} \left( {R_0} + 2 \bar \nabla_0^2 \right)  \omega_1 ~,
\end{equation}
Taking the mean, and using $\mean{\omega_1}=0$, we get $\mathcal H_2 =0$. The conformal factor $\omega_1$ is given by
 $$-\frac{2}{3} \left( {R_0} + 2 \bar \nabla_0^2 \right)  \omega_1=0
 $$
 We will assume that the operator $(\nabla_0^2+aR_0)$ does have a unique inverse, and thus $\omega_1=0$ is the unique solution\footnote{We note that since the manifold is compact, $\nabla_0^2$ has discrete spectrum. Thus an operator of the form $(\nabla_0^2+aR_0)$ will generically have a unique Green's function, as generically $R_0\in \R$ will not fall into that spectrum. If the reader finds this argument insufficient, using more sophisticated analytical tools \cite{Yamabe}, one can show that if the metric is not the standard one on $S^3$, then the operator above has a unique Green's function. And if it is the standard one, we still generically have uniqueness. }.

Finally to order $V^{-2}$, using again \eqref{equ:omega_exp}
\begin{equation}
\mathcal H_3 =  - \frac{2}{3}  \left( {R_0} +  2 \nabla_0^2 \right) \omega_2
+\frac{\bar \sigma^{ab}\bar \sigma_{ab}}{ \bar g_0}
~,\end{equation}
whose solution is
\begin{equation}
\mathcal H_3 =- \frac{2}{3}R_0\mean{\omega_2}+ \left< \frac{\bar \sigma_{ab} \bar \sigma_{ab}}{ \bar g_0} \right> =\left< \frac{\bar \sigma^{ab} \bar \sigma_{ab}}{ \bar g} \right>~,
\end{equation} where we used that $\mean{\omega_i}=0$ for $i\neq 0$. Note that already at this stage the solution of $\omega_2$ becomes significantly more complex,  $$\left( {R_0} +  2 \nabla^2 \right)^{-1}\left(\frac{\bar \sigma_{ab} \bar \sigma_{ab}}{ \bar g_0}-\mean{\frac{\bar \sigma_{ab} \bar \sigma_{ab}}{ \bar g_0}}\right).$$

The complete solution, up to order $V^{-2}$ is:
\begin{equation}
\mathcal H^0_{\text{gl}}=  2 \Lambda - \frac{3}{8} P^2  - \frac{{R_0}}{V^{2/3}} + \frac{1}{V^2} \left< \frac{\bar \sigma^{ab} \bar \sigma_{ab}}{ \bar g_0} \right> + \order{\lf(V/V_0\rt)^{-8/3}}~.
\end{equation}
A couple of comments are in order. First, we note that each term in the expansion is  diffeomorphism  invariant but vpct gauge dependent. Thus we conformally covariantize it, so that it coincides with the above equation over the Yamabe section. We get:
\be \mathcal H_{\text{gl}}=  2 \Lambda - \frac{3}{8} P^2  - \frac{{R[e^{4\lambda[g]}g_{ab}]}}{V^{2/3}} + \frac{1}{V^2} \left< \frac{\bar \sigma^{ab} \bar \sigma_{ab}}{ \bar e^{12\lambda[g]}g} \right> + \order{\lf(V/V_0\rt)^{-8/3}}~.
\ee

\subsection{Explicit calculation of the first three terms in the Hamilton-Jacobi volume expansion.}\label{sec:HJ_S_n}

The equation we are then trying to solve is:
\begin{multline}
2 \Lambda - \frac{3}{8} \left( \frac{\delta S}{\delta V}\right)^2  - \frac{{R_0}}{V^{2/3}}\\
 + \frac{1}{V^2} \mean{  \left(\diby{S}{{\bar g}^0_{ab}} - \frac 1 3 \mean{ {\bar g}^0_{ab} \diby{S}{{\bar g}^0_{ab}}} {\bar g}^0_{ab} \sqrt{{\bar g}^0}\right){\bar g}^0_{ac}{\bar g}^0_{cd} \left(\diby{S}{{\bar g}^0_{ab}} - \frac 1 3 \mean{ {\bar g}^0_{ab} \diby{S}{{\bar g}^0_{ab}}} {\bar g}^0_{ab} \sqrt{{\bar g}^0}\right)  } + \order{V^{-8/3}} = 0~.
\end{multline}
The expansion we are going to use, still of course in steps of $V^{-2/3}$, to solve this is:
\begin{equation}
S = S_0 V + S_1 V^{1/3} + S_2 V^{-1/3} + \order{V^{-1}}.
\end{equation}

The 0-th order equation then becomes:
\be 2\Lambda- \frac{3}{8} \bar\alpha^2-\left(\alpha^{ab}-\frac{1}{3}g_0^{ab}\right)\left(\alpha_{ab}-\frac{1}{3}g^0_{ab}\right)=2\Lambda- \frac{3}{8} \bar\alpha^2=0~,
\ee
where we used  $\bar\alpha\sqrt g=\alpha$ and that $\alpha^{ab}$ is for our boundary conditions pure trace, and thus its traceless part vanishes. Thus we have
\be\label{equ:sol_S_0}S_0=\pm\sqrt{\frac{16\Lambda}{3}}
\ee
For the next order, the relevant terms are those that contribute with $V^{-2/3}$, coming from $- \frac{{R_0}}{V^{2/3}}$ and:
\be\label{equ:mid_cal} -\frac{3}{8}\left(\diby{S_0V+S_1V^{1/3}}{V}\right)^2=-\frac{3}{8}\left(S_0+\frac{1}{3}S_1V^{-2/3}\right)^2=\order{1}-\frac{1}{4}S_0S_1V^{-2/3}
-\frac{1}{24}S_1^2V^{-4/3}
\ee
Thus we get
\be\label{equ:sol_S_1} S_1= -4\frac{R_0}{S_0}
\ee

For the next order term we get the term $-\frac{2}{3} \frac{R_0^2}{S_0^2}$ coming from the last order of \eqref{equ:mid_cal}, the term coming from
$$ -\frac{3}{8}\left(\diby{S_0V+S_2V^{-1/3}}{V}\right)^2=-\frac{3}{8}\left(S_0-\frac{1}{3}S_2V^{-4/3}\right)^2=\order{1}+\frac{1}{4}S_0S_2V^{-4/3}
-\frac{1}{24}S_2^2V^{-8/3}
$$ which is just
$$ \frac{1}{4}S_0S_2
$$
and finally, the term:
$$\left< \frac{1}{ g_0} \left( \frac{\delta S_1}{\delta  g^0_{ab} }- \frac{1}{3}  g_0^{ab} \sqrt{ g_0} \left< \frac{\delta S_1}{\delta  g^0_{kl} } g^0_{kl} \right> \right)  \left( \frac{\delta S_1}{\delta  g^0_{cd} }- \frac{1}{3}  g_0^{cd} \sqrt{ g_0} \left< \frac{\delta S_1}{\delta  g^0_{ij} } g^0_{ij} \right> \right)  g^0_{ac} g^0_{bd}\right>~,
$$
yielding
\begin{equation}\label{equ:calc_S_2}
S_2 =  \frac{8}{3} \frac{{R_0}^2}{S_0^3} -
\frac{4}{S_0} \left< \frac{1}{ g_0} \left( \frac{\delta S_1}{\delta  g^0_{ab} }- \frac{1}{3}  g_0^{ab} \sqrt{ g_0} \left< \frac{\delta S_1}{\delta  g^0_{kl} } g^0_{kl} \right> \right)  \left( \frac{\delta S_1}{\delta  g^0_{cd} }- \frac{1}{3}  g_0^{cd} \sqrt{ g_0} \left< \frac{\delta S_1}{\delta  g^0_{ij} } g^0_{ij} \right> \right)  g^0_{ac} g^0_{bd}\right> ~,
\end{equation}
Now we use the fact that\footnote{Assuming for simplicity that in the barred variables $V_0=1$.} $R_0(x)=R_0=\int d^3 x R_0\sqrt {g_0}$, we can discard the boundary terms of the variation to get:
\begin{equation}
\frac{\delta S_1}{\delta  g^0_{ab} }
=  \frac{4}{S_0} \sqrt{ g_0} \left( R_0^{ab} - \frac{1}{2} R_0 \,  g_0^{ab} \right) ~,
\end{equation}
Contraction with $g^{ab}_0$ yields $-\frac{2R_0}{S_0} $ and thus
\begin{multline}
 \frac{\delta S_1}{\delta  g^0_{ab} }- \frac{1}{3}  g_0^{ab} \sqrt{ g_0} \left< \frac{\delta S_1}{\delta  g^0_{cd} } g^0_{cd} \right> =
 \frac{4}{S_0} \sqrt{ g_0} \left( R_0^{ab} - \frac{1}{2} R_0 \,  g_0^{ab} \right)\sqrt g_0-\frac{1}{3}  g_0^{ab} \sqrt{ g_0}\left(-\frac{2R_0}{S_0}\right)\\
 = \frac{4}{S_0}\left(R_0^{ab} - \frac{1}{3} R_0 \,  g_0^{ab} \right)\sqrt g_0.
 \end{multline}
As can easily be seen the term  $\mean{\sigma\cdot\sigma}$ yields to this order:
$$\mean{ \frac{16}{S^2_0}\left(R_0^{ab} - \frac{1}{3} R_0 \,  g_0^{ab} \right)\left(R^0_{ab} - \frac{1}{3} R_0 \,  g^0_{ab} \right)}
=\frac{16}{S_0^2}\mean {R_0^{ab}R^0_{ab}- \frac{1}{3} R_0^2}$$
Inputting this back into \eqref{equ:calc_S_2} yields:
\begin{equation}
S_2 =  \frac{8}{S_0^3}\left( \frac{{R_0}^2}{3} -{8}\mean {R_0^{ab}R^0_{ab}-\frac{1}{3}R_0^2}\right)=\frac{24}{S_0^3}\left( {R_0}^2 -\frac{8}{3}\mean {R_0^{ab}R^0_{ab}}\right)
\end{equation}

Complete solution:
\begin{equation}
S =    \pm \left(  4\sqrt{ \frac{\Lambda}{3}  } \, V  -  \sqrt{ \frac{3}{ \Lambda} } \,{R_0} \, V^{1/3} + \frac{9}{8\Lambda}\sqrt{\frac{3}{\Lambda}} \left( {R_0}^2 -\frac{8}{3}\mean {R_0^{ab}R^o_{ab}}\right)V^{-1/3} + \dots \right)~.
\end{equation}
where we separated the factors $\sqrt{\frac{3}{\Lambda}}$ to show that the $V^{2/3}$ step is accompanied by a $\frac{3}{\Lambda}$ one.
\chapter{Manifold structure for constraint set.}

\section{Poisson Brackets and Linear maps}\label{sec:PbsLinear}
To make what we mean more precise and evaluation more straightforward (specially the linear algebra part), I will employ the Fischer-Marsden notation. So we start by notation. Riem$(M)=:\mathcal{M}$.

We define the variation as a tangent map in these spaces (for example for $\mathcal{F}:T^*\mathcal{M}\rightarrow C^\infty(M)
$ which we are considering to be densities):
\begin{equation}
\delta\mathcal{F}_{(g,\pi)}\cdot(h,w)=\delta\mathcal{F}_{(g_0,\pi_0)}\cdot(h,w)=\delta_g\mathcal{F}_{(g_0,\pi_0)}\cdot h+\delta_\pi\mathcal{F}_{(g_0,\pi_0)}\cdot w\end{equation}
where
$$ \delta_g\mathcal{F}_{(g_0,\pi_0)}\cdot h=\int_M \left.{\frac{\delta\mathcal{F}}{\delta g_{ab}(x)}}\right|_{(g_0,\pi_0)}h_{ab}(x)
~~~{\mbox {and}} ~~~\delta_g\mathcal{F}_{(g_0,\pi_0)}=\left.{\frac{\delta\mathcal{F}}{\delta g_{ab}(x)}}\right|_{(g_0,\pi_0)}$$
and so on. Note we are omiting in the $\delta_g$ notation that this has both continuous $(x)$ and discrete $(ab)$ indices. Both types of indices are summed over through the dot notation. This mimics the action of matrices in linear algebra.

Now, we have  natural inner products on $C^\infty(M)$ and $T^*\mathcal{M}$:
\be \label{equ:innerprod}\begin{array}{rcl} \langle f,m\rangle_{C^\infty(M)}&:=&\int d^3x\sqrt g fm\\
\langle (h,\pi),(k,w)\rangle_{T^*\mathcal{M}}&:=&\int d^3x\sqrt g \left(g^{ac}g^{bd}h_{ab}k_{cd}+\frac{g_{ac}g_{bd}}{g}\pi^{ab}w^{cd}\right)\end{array}
\ee
In this way, what we usually mean for a smearing of a function $F\in C^\infty(M)$ is seen as an inner product:
$$F(N):=\langle F,N\rangle_{C^\infty(M)}.
$$
We will omit the subscript $C^\infty(M)$ from now on. Of course to be more precise we should be working not with the space of smooth functions (on the second entry of the inner product), but the space of square-integrable functions. However, as we can sidestep most of the difficulties arising from this simplification, we will merely make a side note whenever the difference becomes relevant.

We can write the Poisson bracket as:
 \be\label{Equ:Pb_Fischer}\{\mathcal{F}(N),\mathcal{S}(x)\}=\langle \delta_g\mathcal{F}\cdot\delta_\pi\mathcal{S}(x)-\delta_\pi\mathcal{F}\cdot\delta_g\mathcal{S}(x),N\rangle=
\langle \delta_{(g,\pi)}\mathcal{F}\cdot J\delta_{(g,\pi)}\mathcal{S}(x),N\rangle
\ee where one can simplify notation by using the $J$ map (the symplectic structure) which inverts the order of contraction (so to speak).

In much the same way as in linear algebra, we can define the adjoint of the tangent map linear operator (in this case for $\mathcal{F}$ taking values in $C^\infty(M)$):
$$\langle (\delta_g\mathcal{F})\cdot h,N\rangle_{C^\infty(M)}=\langle (\delta_g\mathcal{H})^*\cdot N,h\rangle_{\mathcal{M}}$$
It is exactly by finding the adjoint that we actually find the variational derivative; i.e. we have to isolate $h$, more usually denoted by $\delta g$. As an example, let us study both the scalar and the momentum constraints \eqref{equ:scalar constraint}, \eqref{equ:momentum constraint}.

 Let us start with the smeared version of the momentum constraint:
\be H^a(\xi_a)=-\int d^3 x\pi^{ab}\mathcal{L}_\xi g_{ab}=\int d^3 xg_{ab}\mathcal{L}_\xi \pi^{ab}.
\ee
where we have used integration by parts to transfer the Lie derivative. There is an easy way to check consistency of this formula by using the formula for the Lie derivative of the density:
 \be\label{Lie deriv}\mathcal{L}_\xi
\pi^{ab}=\xi^e\pi^{ab}_{\phantom{ab};e}-\xi^a_{\phantom{i};e}\pi^{eb}-\xi^b_{\phantom{i};e}\pi^{ea}+\xi^e_{\phantom{i};e}\pi^{ab}
\ee
Upon contraction with $g_{ab}$ and discarding the term  $\xi^e\pi^{ab}_{\phantom{ab};e}+\xi^e_{\phantom{i};e}\pi^{ab}$ as a total derivative, one obtains the desired relation. This makes it very easy to calculate any Poisson bracket between the smeared momentum constraint and functionals of $(g,\pi)$. The fundamental Poisson brackets are easily calculated to be
\begin{eqnarray*}
\{H^a(\xi_a),g_{ab}(x)\}=\mathcal{L}_\xi g_{ab}(x)\\
\{H^a(\xi_a),\pi^{ab}(x)\}=\mathcal{L}_\xi \pi^{ab}(x)\end{eqnarray*}
And thus by the chain rule we get that for any functional of phase space:
$$ \{H^a(\xi_a),f(g,\pi)(x)\}=\mathcal{L}_\xi f(g,\pi)(x)
.$$

Using the above notation, for the scalar constraint we have, from \eqref{equ:delta_gS} and \eqref{equ:delta_piS}:
\begin{multline}
\delta S_{(g,\pi)}\cdot(h,w)=\left(-\frac{1}{2\sqrt g} g^{ef}G_{abcd}\pi^{ab}\pi^{cd}+\frac{2}{\sqrt g}(\pi^{eb}g_{bd}\pi^{fd}-\frac{\pi^{ef}\pi}{2})\right)h_{ef}-\Big(\frac{1}{2}\sqrt g g^{ef} Rh_{ef}+\\
\sqrt g(-R^{ef}h_{ef}-g^{ef}\nabla^2h_{ef}+h_{ef}^{;ef} )\Big)+
+2\frac{G_{efcd}\pi^{cd}}{\sqrt g}w^{ef}
\end{multline}
The adjoint is given using \eqref{equ:innerprod}
\be\label{equ:delta*S} \delta^* S_{(g,\pi)}\cdot N= (A^{ab}, N(\frac{2g_{ac}g_{bd}\pi^{cd}-g_{ab}\pi}{\sqrt g}))
\ee where $A^{ab}$ is given by \eqref{equ:delta_gSN}.

For the momentum constraint we have (remembering that if we consider the momentum constraint as a map into the space of vector fields $C^\infty(TM)$, we need to lower indices with the metric):
\begin{eqnarray}
\delta {H_a}_{(g,\pi)}\cdot(h,w)&=&-2\left({{w_a}^b}_{;b}+h_{ac}\pi^{cb}_{;b}+\pi^{bc}(h_{ab;c}-\frac{1}{2}h_{bc;a})\right)\\
\delta^* H^a_{(g,\pi)}\cdot \xi_a&=&(-\mathcal{L}_\xi\pi^{ab},L_\xi g_{ab})
\end{eqnarray}

 The next issue is do we get such things as the lapse fixing equation, which is
 \begin{equation}\label{Equ:pi_phi_derivative}\{\mathcal{T}\mathcal{H}(N),\pi_\phi(\rho)\}=\langle (\delta_\phi \mathcal{T}\mathcal{H})\cdot\rho,N\rangle_{C^\infty(M)}=\langle (\delta_\phi \mathcal{T}\mathcal{H})^*\cdot N,\rho\rangle_{\mathcal{C}}
=0 \end{equation}thus $(\delta_\phi\mathcal{H})^*\cdot N=0$ is the lapse fixing equation and $(\delta_\phi\mathcal{H})^*$ is the lapse fixing operator. We can still use the canonical transformation properties and $\mathcal{D}_{({\mathcal T}_\phi g,{\mathcal T}_\phi\pi,\phi)}=\mathcal{D}_{( g,\pi)}$. We get:
$$\langle (\delta_\phi T\mathcal{H})\cdot\rho,N\rangle=\langle \delta_{(g,\pi)}\mathcal{H}\cdot J\delta_{(g,\pi)}\mathcal{D}(\rho),N\rangle.$$

\section{Constraint manifold for GR.}\label{sec:constraint_manifoldGR}

We now investigate under which circumstances the set of phase space points obeying both constraints form a manifold. I.e., under which conditions the intersection
\be \left(S^{-1}(0):=\{(g,\pi)\in T^*\M~|~S(g,\pi)=0\}\right)\cap \left({H_a}^{-1}(0):=\{(g,\pi)\in T^*\M~|~H^a(g,\pi)=0\}\right)
\ee
forms a manifold. From the form of the constraint algebra for the constraints \eqref{equ:Dirac_ctraint_alg} we already know that the constraint set is maintained by the evolution equations for all choices of lapse and shift.

Now suppose again that if $\pi^{ab}\equiv 0$ then $g_{ab}$ is not flat. We prove the following proposition:
\begin{prop}\label{prop:regularValue_S}
Under the assumption that whenever $\pi^{ab}\equiv 0$ then $g_{ab}$ is not flat, $S^{-1}(0)$ forms a manifold.
\end{prop}
To prove this we use the Fredholm alternative \eqref{Splitting} for the differential operator
\be \delta S_{(g,\pi)}:T_g\M\times T^*_g\M\rightarrow C^\infty(M)
\ee
where we remind the reader that $T^*_g\M$ is not the actual space of linear functionals on $T_g\M$, but the space of sections of $TM\otimes_STM$ the symmetric product of the tangent bundle. See section \ref{sec:Riem} for a more thorough explanation of these spaces.
The reasoning is quite familiar from section \ref{sec:constructingV} and again quite simple: if the operator is elliptic, we can use the Fredholm splitting
\be
C^\infty(M)=\mbox{Im}(\delta S)\oplus \mbox{Ker}\delta^*S,\ee
where $\oplus$ is an $L_2$ orthogonal splitting (i.e. in the positive definite metric used in this section \eqref{equ:innerprod}). Then we must merely show that  $\delta^*S$ is injective, which will imply that $\delta S$ is surjective. By the regular value theorem we then have that $S^{-1}(0)$ is (at least locally) a submanifold, with tangent space $\mbox{Ker}\delta S$. For $\sigma(D)$ denoting the principal symbol of the operator $D$, we have
$\sigma(D^*\circ D)=\sigma(D)^*\circ\sigma(D)$. This means that if $\sigma(D)$ is injective,  for
positive definite inner product, we automatically have $\sigma(D^*)$ surjective and thus that the operator is elliptic (see proposition \ref{propo}). And thus we are left to prove that the operator is injective and has injective symbol.

This calculation is typical of chapter \ref{chapter:connection_forms} (see \eqref{equ:symbol1}, \eqref{adjoint symbol conformal}, \eqref{equ:symbol_conf}, etc) and by now we can skip the preliminaries and assert that the symbol of $\delta^*S$, given in \eqref{equ:delta*S}:
\be \sigma_\xi(\delta^*S)=(-\xi_a\xi_b+g_{ab}\xi^c\xi_c, 0)
\ee
which by doing the usual trick of taking the trace, guarantees the operator is elliptic. Now from \eqref{equ:delta*S}, we have two equations that we must satisfy if $N\in\mbox{Ker}\delta^*S $. Taking the trace of the second one we get $N\pi=0$, which when input back into the same equation yields $N\pi^{ab}=0$. When input back into the first of these equations (denoted by $A^{ab}$) we get:
\be -\Big(\frac{1}{2}\sqrt g g^{ef} RN(y)+
\sqrt g(y)(-R^{ef}(y)-g^{ef}(y)\nabla^2N(y)+
N^{;ef}(y) )\Big)=0\ee
By taking the trace we arrive at:
\be -2\nabla^2 N+\frac{1}{2}RN=0
\ee
Now if we substitute the scalar constraint for $R$, and since $N\pi^{ab}=0$ we get $\nabla^2N=0$ which means $N$ is a constant. But we furthermore have that $\pi^{ab}\not\equiv 0$, which implies that $N=0$ everywhere (i.e $N\equiv 0$), as it is a constant. Thus $\delta^*S$ is injective. $\square$.

We will not prove the same for the diffeomorphism group, which goes through quite simply.
What changes when we try to prove that the intersection is a manifold? Now we have the mapping:
\be (\delta S_{(g,\pi)},\delta H^a_{(g,\pi)}):T_g\M\times T^*_g\M\rightarrow C^\infty(M)\times C^\infty(T^*M)
\ee
And the adjoint given in \eqref{equ:delta*S} differs from that by the term $(-\mathcal{L}_\xi\pi^{ab},\mathcal{L}_\xi g_{ab})$. We will not go through the calculations here (since the tricks used are basically the same as before), but the only condition under which this is still injective is if $g_{ab}\pi^{ab}=c$, a constant \cite{EinsteinCentenary-FM}. This is further circumstantial evidence that the domain which has good theoretical properties is not the entire one of general relativity, but that of shape dynamics.

\section{Propagation of the constraints.}\label{sec:prop_const_Lagrange}

  We now show how the Poisson brackets of the ADM constraints, propagate. As mentioned in section \ref{sec:geometric_Pb}, we will take this opportunity to illustrate how the  Lagrangian formalism,  as opposed to the Hamiltonian one, is cumbersome when dealing with canonical dynamical systems. In the Hamiltonian formalism the only  non-trivial bracket is the one given by $\{S(x),S(y)\}$, which was easily calculated in \eqref{equ:Dirac_ctraint_alg}. Not so in the Lagrangian formalism, where we will have to prove every propagation non-trivially. We will also take the opportunity to demonstrate two other points that arise in the text. First is the propagation of the constraints in BSW form (section \ref{sec:Barbour}). Second is the use of a more general supermetric in the propagation (section \ref{sec:Riem}) and the conditions it gives rise to. To emphasize then, in this section we will compute propagation of the BSW constraints with a generalized supermetric.

 \subsection{Momentum Constraint}

We shall start with the BSW action given by the Lagrangian density $ \mathcal{L}=\sqrt{gRT_\lambda}$, $L=\int_M
d^3x\mathcal{L}$, $M$ a closed manifold without boundary again, and where
$$T_\lambda=(g^{ac}g^{bd}-\lambda g^{ab}g^{cd})(\dot g_{ab}-2\xi_{(a;b)})(\dot g_{cd}-2\xi_{(c;d)}).$$
To distinguish the BSW momenta from the ADM one, we denote it by $p^{ij}$, as opposed to $\pi^{ij}$.

Now we are ready to start with: \be\label{momentum} p^{ij}=\frac{\delta\mathcal{L}}{\delta \dot
g_{ij}}=\sqrt{\frac{gR}{T_\lambda}}(g^{ic}g^{jd}-\lambda g^{ij}g^{cd})(\dot g_{cd}-2\xi_{(c;d)}) \ee
We also assume both the Hamiltonian and momentum constraints:
\bea -p^{lj}p_{lj}+\frac{\lambda}{3\lambda-1}p^2+gR=0\label{Hamiltonian ctraint}\\
p^{ij}_{\phantom{ij};j}=0
\eea

Now,\be \dot
p^{ij}=\frac{\delta \mathcal L}{\delta g_{ij}}=\int_Md^3x\left((\frac{\delta\sqrt{g}}{\delta
g_{ij}})\sqrt{RT_\lambda}+\sqrt{g}(\frac{\delta\sqrt{R}}{\delta
g_{ij}})\sqrt{T_\lambda}+\sqrt{g}\sqrt{R}(\frac{\delta\sqrt{T_\lambda}}{\delta g_{ij}})\right) \ee The first term
yields naturally: \be (\delta\sqrt{g})\sqrt{RT_\lambda}=\frac{1}{2}\sqrt{RT_\lambda}\sqrt{g}g^{ij}\delta g_{ij}
\ee The second term gives: \be\label{second term}
\sqrt{g}(\delta\sqrt{R})\sqrt{T_\lambda}=\frac{1}{2}\sqrt{\frac{gT_\lambda}{R}}(-R^{ij}\delta
g_{ij}-g^{ij}\Delta(\delta g_{ij})+(\delta g_{ij})_{;cd}g^{ic}g^{jd}) \ee Integrating by parts, and noting that
the covariant derivative of the density function $g$ is zero,  we have an equivalence of \eqref{second term} up to
a boundary term, with: \be
\frac{\sqrt{g}}{2}\left(-R^{ij}\sqrt{\frac{T_\lambda}{R}}-g^{ij}\Delta\left(\sqrt{\frac{T_\lambda}{R}}\right)+
\left(\sqrt{\frac{T_\lambda}{R}}\right)_{;dc}g^{ic}g^{jd}\right)\delta g_{ij} \ee

Now for the third term we have that the inverse of $(g^{ac}g^{bd}-\lambda g^{ab}g^{cd})$ is:
$g_{ae}g_{bf}-\frac{\lambda}{3\lambda-1}g_{ab}g_{cd}$. We must first calculate: \be \frac{\delta
(g^{ac}g^{bd}-\lambda g^{ab}g^{cd}))}{\delta
g_{ij}}=-g^{ai}g^{cj}g^{bd}-g^{ac}g^{bi}g^{dj}+\lambda(g^{ai}g^{bj}g^{cd}+g^{ab}g^{ci}g^{dj})\ee Inverting
\eqref{momentum} we have:
 \be\label{gdot}\dot g_{ij}=\frac{N}{\sqrt g}\left(p_{ij}-\frac{\lambda}{3\lambda-1}g_{ij}p\right)+2\xi_{(i;j)}
 \ee
and thus:
\begin{multline} (\delta(g^{ic}g^{jd}-\lambda g^{ij}g^{cd}))(\dot g_{cd}-2\xi_{(c;d)})(\dot g_{ij}-2\xi_{(i;j)})=
(\delta(g^{ic}g^{jd}-\lambda g^{ij}g^{cd}))\frac{T}{gR}p_{cd}p_{ij}\nn\\
=\frac{T}{gR}\left(-g^{ai}g^{cj}g^{bd}-g^{ac}g^{bi}g^{dj}+\lambda(g^{ai}g^{bj}g^{cd}+g^{ab}g^{ci}g^{dj})\right)p_{cd}p_{ij}\nn\end{multline}
\bea~&=&\frac{2T}{R}\left((p^{id}-\frac{\lambda}{3\lambda-1}g^{id}p)(p^j_{\phantom{j}d}-\frac{\lambda}{3\lambda-1}\delta^j_{\phantom{j}d}p)+\lambda p
(p^{ij}-\frac{\lambda}{3\lambda-1}g^{ij}p)(1-\frac{\lambda}{3\lambda-1})\right)\nn\\
~&=&\frac{2T}{R}\left(-(p^{id}p^j_{\phantom{j}d}-\frac{2\lambda}{3\lambda-1}pp^{ij}+\frac{\lambda}{3\lambda-1}^2p^2g^{ij})-
\frac{\lambda}{3\lambda-1}(pp^{ij}-\frac{\lambda}{3\lambda-1}p^2g^{ij})\right)\nn\\
~&=&\frac{2T}{R}(-p^{id}p^j_{\phantom{j}d}+\frac{\lambda}{3\lambda-1}pp^{ij})\eea

 Finally the third term yields:
\bea \sqrt{gR}({\delta\sqrt{T_\lambda}})&=& \frac{1}{2}{\sqrt\frac{gR}{T_\lambda}}
(\delta(g^{ic}g^{jd}-\lambda g^{ij}g^{cd}))(\dot g_{cd}-2\xi_{(c;d)})(\dot g_{ij}-2\xi_{(i;j)})\nn\\
&-&2{\sqrt\frac{gR}{T_\lambda}}(g^{ic}g^{jd}-\lambda g^{ij}g^{cd})(\dot g_{cd}-2\xi_{(c;d)})(\delta \xi_{(i;j)})\nn\\
&=& - {\sqrt\frac{T_\lambda}{g R}}(p^{ic}p_c^{\phantom{c}j}-\frac{\lambda}{3\lambda-1}pp^{ij})\delta
g_{ij}+2p^{ij}(\xi_e\delta\Gamma^e_{ij}) \eea Now from \eqref{christ variation} we get:
\bea 2p^{ab}(\xi_e\delta\Gamma^e_{(ab)})&=&p^{ab}\xi^{c}(\delta g_{bc;a}+\delta g_{ac_;b}-\delta g_{ab;c})\nn\\
&\hat=&-((\xi^cp^{ab})_{;a}\delta g_{bc}+(\xi^cp^{ab})_{;b}\delta g_{ac}-(\xi^cp^{ab})_{;c}\delta g_{ab})
\label{christ variation 2}\eea where $\hat=$ is equivalence up to boundary terms.  Now, the formula for the Lie
derivative of a density reads: \be\label{Lie deriv}\mathcal{L}_\xi
p^{ij}=\xi^ep^{ij}_{\phantom{ij};e}-\xi^i_{\phantom{i};e}p^{ej}-\xi^j_{\phantom{i};e}p^{ei}+\xi^e_{\phantom{i};e}p^{ij}
\ee and thus we get from \eqref{christ variation 2}, already discarding the terms that disappear due to  the
momentum constraint $p^{ab}_{\phantom{ab};b}=0$: \bea 2p^{ab}\left(\xi_e\frac{\delta\Gamma^e_{(ab)}}{\delta
g_{ij}}\right)&=& -
((\xi^cp^{ab})_{;a}\delta^{ij}_{bc}+(\xi^cp^{ab})_{;b}\delta^{ij}_{ca}-(\xi^cp^{ab})_{;c}\delta^{ij}_{ab})\nn\\
&=& -(\xi^j_{\phantom{j};a}p^{ai}+\xi^i_{\phantom{j};b}p^{jb} -\xi^c_{\phantom{j};c}p^{ij}-\xi^c
p^{ij}_{\phantom{ij};c})\nn\\
&=&\mathcal{L}_\xi p^{ij}\eea
 Putting it all together we have:
 \begin{multline} \dot p^{ij}=-\frac{1}{2}\sqrt{g}\left((R^{ij}-g^{ij}R)N+g^{ij}\Delta N-
 N_{;cd}g^{ic}g^{jd}\right)\\
 -{\frac{N}{\sqrt g}}(p^{ic}p_c^{\phantom{c}j}-\frac{\lambda}{3\lambda-1}pp^{ij})+   \mathcal{L}_\xi p^{ij} \label{pdot}
 \end{multline}
  where $N=\sqrt{\frac{T_\lambda}{R}}$.

   Now we move forward to check if the momentum constraint is propagated. We must thus calculate
   $\dot{(p^{ij}_{\phantom{ij};j})}$. We have that for a tensor density of weight one:
   \be p^{ij}_{\phantom{ij};j}=(p^{ij}_{\phantom{ij},j}+\Gamma^i_{\phantom{i}jl}p^{lj}+
   \Gamma^j_{\phantom{i}jl}p^{il})-\Gamma^j_{\phantom{i}jl}p^{il}
=p^{ij}_{\phantom{ij},j}+\Gamma^i_{\phantom{i}jl}p^{lj}
   \ee
 and thus,
 \be\label{mom cons}  \dot{(p^{ij}_{\phantom{ij};j})}=\dot p^{ij}_{\phantom{ij};j}+\dot\Gamma^i_{\phantom{i}jl}p^{lj}
 \ee

 Then using \eqref{christ variation} in \eqref{mom cons} we get for the second term:
  \bea  \dot\Gamma^i_{\phantom{i}jl}p^{lj}&=& (\dot g^i_{\phantom{i}j;l}-\frac{1}{2}\dot {g_{lj}}^{;i})p^{lj}\nn\\
  &=& \frac{1}{\sqrt{g}}\left(p^{lj}(N(p^i_{\phantom{i}j}-\frac{\lambda}{3\lambda-1}\delta^i_{\phantom{i}j}p))_{;l}-
  \frac{1}{2}p^{lj}(N(p_{lj}-\frac{\lambda}{3\lambda-1}pg_{lj}))^{;i}\right)
  +2g^{ic}(\xi_{(j;c);l}-\frac{1}{2}\xi_{j;lc})p^{lj}\nn\\
 &=&
\frac{1}{\sqrt{g}}\left((N(p^{lj}p^i_{\phantom{i}j}-\frac{\lambda}{3\lambda-1}p^{li}p))_{;l}-
 \frac{1}{4}gg^{ij}((NR)_{;j}+N_{;j}R)\right)+2g^{ic}(\xi_{(j;c);l}-\frac{1}{2}\xi_{j;lc})p^{lj}\nn\\
\label{Gamma1} \eea
Where on the first term we used the momentum constraint and on the second term we had:
\bea -\frac{1}{2}p^{lj}(N(p_{lj}-\frac{\lambda}{3\lambda-1}pg_{lj}))^{;i}&=&-\frac{1}{2}p^{lj}(N^{;i}(p_{lj}-\frac{\lambda}{3\lambda-1}pg_{lj}))
-\frac{1}{2}(p^{lj}N(p_{lj}-\frac{\lambda}{3\lambda-1}pg_{lj})^{;i})\nn\\
&=&2\left(-\frac{1}{4}(N^{;i}(p^{lj}p_{lj}-\frac{\lambda}{3\lambda-1}p^2))\right)
-\frac{1}{4}(N(p^{lj}p_{lj}-\frac{\lambda}{3\lambda-1}p^2)^{;i})\nn\\
&=&-\frac{1}{4}(N(p^{lj}p_{lj}-\frac{\lambda}{3\lambda-1}p^2))^{;i}-\frac{1}{4}(N^{;i}(p^{lj}p_{lj}-\frac{\lambda}{3\lambda-1}p^2))\nn\\
&=&-\frac{1}{4}gg^{ij}\left((NR)_{;j}+N_{;j}R\right)\label{Gamma3}
\eea
 where we used the Hamiltonian constraint on the last equality.

Finally, going back to \eqref{pdot}:
  \begin{multline}\label{cov p dot} \dot p^{ij}_{;j}=-\frac{1}{2}\sqrt{g}\left
  (\left((R^{ij}-g^{ij}R)N\right)_{;j}+\left(g^{ij}\Delta N-
  N_{;cd}g^{ic}g^{jd}\right)_{;j}\right)\\
   -\left(\frac{N}{\sqrt g}(p^{ic}p_c^{\phantom{c}j}-\frac{\lambda}{3\lambda-1}pp^{ij})\right)_{;j}+
  ( \mathcal{L}_\xi p^{ij})_{;j}
   \end{multline}
   Then from \eqref{mom cons}, the third term of \eqref{cov p dot} can be seen to cancel with the first term of  \eqref{Gamma1}.

    Combining the part $-\sqrt{g}\frac{1}{4}(Ng^{ij}R)_{;j}$ of the second term
   of \eqref{Gamma1} (or the first term of \eqref{Gamma3}) with the first term of \eqref{cov p dot} we obtain:
\begin{multline} -\frac{1}{2}\sqrt{g}\left((R^{ij}-g^{ij}R)N\right)_{;j}-\sqrt{g}\frac{1}{4}(Ng^{ij}R)_{;j}=
-\frac{1}{2}\sqrt{g}\left((R^{ij}-\frac{1}{2}g^{ij}R)N\right)_{;j}\\
=-\frac{1}{2}\sqrt{g}\left((R^{ij}-\frac{1}{2}g^{ij}R)N_{;j}\right)\label{R}
\end{multline}
where we used the Bianchi identity in 3D to set $(R^{ij}-\frac{1}{2}g^{ij}R)_{;j}=0$.
 The second term of \eqref{cov p dot} is given by:
   \begin{multline} -\frac{1}{2}\sqrt{g}\left(g^{ij}\Delta N-
  N_{;cd}g^{ic}g^{jd}\right)_{;j}=\frac{1}{2}\sqrt{g}(N^{;ik}_{\phantom{;ik}k}-N_{;k}^{\phantom{k}ki})=
  \frac{1}{2}\sqrt{g}(
N_{;k}^{\phantom{;k}ik}-N_{;k}^{\phantom{k}ki})=
\frac{1}{2}\sqrt{g}N_{;j}R^{j\phantom{k}ik}_{\phantom{j}k}\\
=\frac{1}{2}\sqrt{g}N_{;j}R^{ij}\label{NR}
   \end{multline}
Now combining \eqref{R} and \eqref{NR} we obtain: \be \frac{1}{4}\sqrt{g}g^{ij}RN_{;j} \ee which cancels with the
second term of \eqref{Gamma3}. Thus, in \eqref{pdot}, we have already used  up all the terms that don't explicitly
involve $\xi$, being left with the third term of \eqref{Gamma1}, the fourth term of \eqref{cov p dot}.

We start by writing out $(\mathcal{L}_\xi p^{ij})_{;j}$:
\bea
(\mathcal{L}_\xi p^{ij})_{;j}&=& \xi^c_{\phantom{c};j}p^{ij}_{\phantom{ij};c}+\xi^cp^{ij}_{\phantom{ij};cj}
-\xi^c_{\phantom{c};jc}p^{ij}-
\xi^i_{\phantom{c};cj}p^{cj}-
{\xi^j_{\phantom{c};c}p^{ic}_{\phantom{ij};j}}+\xi^c_{\phantom{c};cj}p^{ij}\nn\\
&=&\xi^cp^{ij}_{\phantom{ij};cj}-\xi^c_{\phantom{c};jc}p^{ij}-\xi^i_{\phantom{c};cj}p^{cj}+\xi^c_{\phantom{c};cj}p^{ij}
\nn\\
&=& -\xi^c(R_{cj\phantom{i}k}^{\phantom{cj}i}p^{jk}+R_{c\phantom{k}kj}^{\phantom{c}k}p^{ij})+
\xi^kR_{k\phantom{k}cj}^{\phantom{c}c}p^{ij}
-\xi^i_{\phantom{c};cj}p^{cj}\nn\\
&=& -\xi^cR_{cj\phantom{i}k}^{\phantom{cj}i}p^{jk}
-\xi^i_{\phantom{c};cj}p^{cj}
\eea
where on the last line we have used the momentum constraint, implying $p^{ij}_{\phantom{ij};cj}-p^{ij}_{\phantom{ij};jc}
=p^{ij}_{\phantom{ij};cj}$ and the Riemman curvature formula as applied to a type $(0,2)$ tensor.
Now combining the third (last) term of \eqref{Gamma1} and the fourth (last) term of \eqref{cov p dot}. We get:
\bea(\mathcal{L}_\xi p^{ij})_{;j}+(\xi_{j\phantom{;i}l}^{\phantom{j};i}+
\xi_{\phantom{i};jl}^{i}-\xi_{j;l}^{\phantom{j;l}i})p^{lj}
=-\xi^cR_{cj\phantom{i}k}^{\phantom{cj}i}p^{jk}
+(\xi_{j\phantom{;i}l}^{\phantom{j};i}-\xi_{j;l}^{\phantom{j;l}i})p^{lj}= 0\nn
\eea
And so finally we have shown that \eqref{mom cons} is indeed zero, and thus the momentum constraint propagates.

\subsection{Hamiltonian Constraint}

We rewrite equation \eqref{Hamiltonian ctraint}, for convenience:
$$H= -g_{kl}g_{ij}p^{lj}p^{ki}+\frac{\lambda}{3\lambda-1}(g_{ij}p^{ij})^2+gR=0
$$
Now we dot it:
\begin{multline}
\dot H=g\left((g^{ij}R-R^{ij})\dot g_{ij}-\Delta(g^{ij}\dot g_{ij})+\dot g_{ij}^{\phantom{ij};ij}\right)\\
+\left(-2\dot g_{ij}p^i_{\phantom{i}l}p^{jl}+\frac{2\lambda}{3\lambda-1}\dot
g_{ij}p^{ij}p\right)+\left(-2p_{ij}\dot p^{ij}+\frac{2\lambda}{3\lambda-1}\dot p^{ij}g_{ij}p\right) \label{Hdot
0}\end{multline} Substituting for now only the undifferentiated $\dot g_{ij}$ through the use of \eqref{gdot}:
\begin{multline}
\dot H=\sqrt g\left(N(g^{ij}R-R^{ij})(p_{ij}-\frac{\lambda}{3\lambda-1}pg_{ij})\right)+g\left(\dot g_{ij}^{\phantom{ij};ij}-\Delta(g^{ij}\dot g_{ij})\right)\\
+\frac{N}{\sqrt g}\left(-2(p_{ij}-\frac{\lambda}{3\lambda-1}pg_{ij})p^i_{\phantom{i}l}p^{jl}+
\frac{2\lambda}{3\lambda-1}(p_{ij}-\frac{\lambda}{3\lambda-1}pg_{ij})p^{ij}p\right)+\left(-2p_{ij}\dot p^{ij}+\frac{2\lambda}{3\lambda-1}\dot p^{ij}g_{ij}p)\right)+\nn\\
2\xi_{(i;j)}\left({g}(g^{ij}R-R^{ij})-2\left(p^i_{\phantom{i}l}p^{jl}-\frac{\lambda}{3\lambda-1}p^{ij}p\right)\right)
=\sqrt gN\left((p-\frac{3\lambda}{3\lambda-1}p)R-R^{ij}p_{ij}+\frac{\lambda}{3\lambda-1}pR)\right)+\\
\frac{2N}{\sqrt g}\left(-p_{ij}p^i_{\phantom{i}l}p^{jl}+\frac{2\lambda}{3\lambda-1}p_{ij}p^{ij}p
-\left(\frac{\lambda}{3\lambda-1}\right)^2p^3\right)+\left(-2p_{ij}\dot p^{ij}+\frac{2\lambda}{3\lambda-1}\dot p^{ij}g_{ij}p)\right)\nn\\
+g\left(\dot g_{ij}^{\phantom{ij};ij}-\Delta(g^{ij}\dot g_{ij})\right)+2\xi_{(i;j)}\left({g}(g^{ij}R-R^{ij})-2\left(p^i_{\phantom{i}l}p^{jl}-\frac{\lambda}{3\lambda-1}p^{ij}p\right)\right)\end{multline}
  where we left on the last line the terms we will deal with shortly. Let us re-write the last equation, enumerating some of the terms so that they are more easily manipulable:
\begin{multline}\label{Hdot}
\dot H=\sqrt gN\left(\overbrace{\frac{\lambda-1}{3\lambda-1}pR}^1-\overbrace{R^{ij}p_{ij}}^2\right)\\
+\frac{2N}{\sqrt g}\left(\overbrace{-p_{ij}p^i_{\phantom{i}l}p^{jl}}^3+\overbrace{\frac{2\lambda}{3\lambda-1}
p_{ij}p^{ij}p}^4
 -\overbrace{\left(\frac{\lambda}{3\lambda-1}\right)^2p^3}^5\right)+\left(-2p_{ij}\dot p^{ij}+
 \frac{2\lambda}{3\lambda-1}\dot p^{ij}g_{ij}p)\right)\\
+g\left(\dot g_{ij}^{\phantom{ij};ij}-\Delta(g^{ij}\dot g_{ij})\right)+2\xi_{(i;j)}\left({g}(g^{ij}R-R^{ij})-
2\left(p^i_{\phantom{i}l}p^{jl}-\frac{\lambda}{3\lambda-1}p^{ij}p\right)\right)\end{multline}

  Now we have that for the $\dot p_{ij}$ terms, using \eqref{pdot}:
\begin{multline}2\left(-p_{ij}\dot p^{ij}+\frac{\lambda}{3\lambda-1}\dot p^{ij}g_{ij}p)\right)=\\
-p_{ij}\left(-\sqrt{g}\left((R^{ij}-g^{ij}R)N+g^{ij}\Delta N-
 N_{;cd}g^{ic}g^{jd}\right)
 -{\frac{2N}{\sqrt g}}(p^{ic}p_c^{\phantom{c}j}-\frac{\lambda}{3\lambda-1}pp^{ij})+ 2\mathcal{L}_\xi p^{ij}\right)+\\
 +\frac{\lambda}{3\lambda-1}p
 \left(-\sqrt{g}\left(-2RN+2\Delta N
 \right)
 -{\frac{2N}{\sqrt g}}(p^{ij}p_{ij}-\frac{\lambda}{3\lambda-1}p^2)+ 2g_{ij}\mathcal{L}_\xi p^{ij}\right)=\\
 \sqrt{g}\left(\overbrace{p_{ij}R^{ij}}^2+p(\overbrace{RN}^1-\Delta N)(\frac{2\lambda}{3\lambda-1}-1)-
 N_{;ij}p^{ij}\right)
 +\overbrace{{\frac{2N}{\sqrt g}}p_{ij}p^{ic}p_c^{\phantom{c}j}}^3\\
 -\overbrace{{\frac{4N}{\sqrt g}}\frac{\lambda}{3\lambda-1}pp_{ij}p^{ij}}^4
 +
 \overbrace{{\frac{2N}{\sqrt g}}\left(\frac{\lambda}{3\lambda-1}\right)^2p^3}^5
- 2\left(p_{ij}-\frac{\lambda}{3\lambda-1}pg_{ij}\right)\mathcal{L}_\xi p^{ij}
 \nn
\end{multline}
So from \eqref{Hdot} we are left with:
\begin{multline}
\dot H=g\left(\dot g_{ij}^{\phantom{ij};ij}-\Delta(g^{ij}\dot g_{ij})\right)+2\xi_{(i;j)}\left({g}(g^{ij}R-R^{ij})-2\left(p^i_{\phantom{i}l}p^{jl}-\frac{\lambda}{3\lambda-1}p^{ij}p\right)\right)\\
+\sqrt{g}\left(\underbrace{p\Delta N}_a(1-\frac{2\lambda}{3\lambda-1})-
 \underbrace{N_{;ij}p^{ij}}_b\right)- 2\left(p_{ij}-\frac{\lambda}{3\lambda-1}pg_{ij}\right)\mathcal{L}_\xi p^{ij}
\end{multline} Where we already denoted the terms that will be cancelled by the next set of equations with underbraces.

 Now, to substitute the $g\left(\dot g_{ij}^{\phantom{ij};ij}-\Delta(g^{ij}\dot g_{ij})\right)$ term, using \eqref{gdot}:
 \bea g\dot g_{ij}^{\phantom{ij};ij}&=&g\left(\frac{N}{\sqrt g}\left(p_{ij}-\frac{\lambda}{3\lambda-1}g_{ij}p\right)+2\xi_{(i;j)}\right)^{;ij}\nn\\
&=& {\sqrt g}\left({N}^{;ij}\left(p_{ij}-\frac{\lambda}{3\lambda-1}g_{ij}p\right)-\frac{\lambda}{3\lambda-1}\left(g_{ij}{N}^{;i}p^{;j}+N\Delta p\right)\right)+2g\xi_{(i;j)}^{\phantom{(i;j)}ij}\nn\\
~\\
-gg_{ij}(\dot g)^{;ij}&=&-gg_{ij}\left(\frac{N}{\sqrt g}\left(-\frac{1}{3\lambda-1}p\right)+2\xi^k_{\phantom{k};k}\right)^{;ij}\nn\\
~&=&\label{cov g dot}\sqrt{g} \frac{1}{3\lambda-1}\left(p\Delta N+g^{ij}N_{;i}p_{;j}+N\Delta p\right)-2g{\xi^k_{\phantom{k};k}}^l_{\phantom{l}l}
\eea And thus:
\be g\dot g_{ij}^{\phantom{ij};ij}-gg_{ij}(\dot g)^{;ij}=\sqrt{g} \left(\frac{1-\lambda}{3\lambda-1}\left(\overbrace{p\Delta N}^a+g^{ij}N_{;i}p_{;j}+N\Delta p\right)
+\overbrace{{N}^{;ij}p_{ij}}^b\right)+2g(\xi_{(i;j)}^{\phantom{(i;j)}ij}-{\xi^k_{\phantom{k};k}}^l_{\phantom{l}l})
\ee

Thus finally we are left with:
\begin{multline}\label{Hdot 2}\dot H=2\xi_{(i;j)}\left({g}(\overbrace{g^{ij}R}^1-\overbrace{R^{ij}}^2)-2\left(\overbrace{p^i_{\phantom{i}l}p^{jl}-\frac{\lambda}{3\lambda-1}p^{ij}p}^3\right)\right)
+\overbrace{2g(\xi_{(i;j)}^{\phantom{(i;j)}ij}-{\xi^k_{\phantom{k};k}}^l_{\phantom{l}l})}^4\\
\overbrace{- 2\left(p_{ij}-\frac{\lambda}{3\lambda-1}pg_{ij}\right)\mathcal{L}_\xi p^{ij}}^5+
\frac{1-\lambda}{3\lambda-1}\sqrt{g}\left(g^{ij}N_{;i}p_{;j}+N\Delta p\right)\end{multline} Where once again we
group the terms for easier future manipulation.

Now, using \eqref{Lie deriv}, we expand the remaining $\xi$ term, term 5:
\begin{multline}- 2\left(p_{ij}-\frac{\lambda}{3\lambda-1}pg_{ij}\right)\mathcal{L}_\xi p^{ij}=- 2\left(p_{ij}-\frac{\lambda}{3\lambda-1}pg_{ij}\right)
\left(\xi^ep^{ij}_{\phantom{ij};e}-\xi^i_{\phantom{i};e}p^{ej}-\xi^j_{\phantom{i};e}p^{ei}+\xi^e_{\phantom{i};e}p^{ij}\right)\\
=-\xi^k\left(p^{ij}p_{ij}-\frac{\lambda}{3\lambda-1}p^2\right)_{;k}+\overbrace{4\xi_{(i;j)}\left(
p^i_{\phantom{i}l}p^{jl}-\frac{\lambda}{3\lambda-1}p^{ij}p\right)}^3-
\overbrace{2\xi^k_{\phantom{k};k}\left(p^{ij}p_{ij}-\frac{\lambda}{3\lambda-1}p^2\right)}^1
\end{multline}Where
\be\label{2}-\xi^k\left(p^{ij}p_{ij}-\frac{\lambda}{3\lambda-1}p^2\right)_{;k}=-g\xi^kR_{;k}=-2\xi^kgR_{ik}^{\phantom{ik};i}
\ee where we used the Bianchi identity. Combining \eqref{2} with element 2 of \eqref{Hdot 2}:
\begin{multline} -2g\left(\xi_kR^{ik}_{\phantom{ik};i}+R^{ij}\xi_{i;j}\right)=-2g(\xi_kR^{ik})_{;i}=
-2g\left(\xi_kR_l^{\phantom{l}ilk}\right)_{;i}=-2g\left(\xi_kR^{{m}ilk}g_{lm}\right)_{;i}=2g\left(\xi_kR^{klmi}g_{lm}\right)_{;i}\\
=2gg_{lm}\left(\xi^{l;mi}-\xi^{l;im}\right)_{;i}=2g\left({\xi^l_{\phantom{l};l}}^i_{\phantom{i}i}-\xi^{l;i}_{\phantom{l;i}li}\right)
\end{multline}
This cancels with term 4 of \eqref{Hdot 2}.
Thus we are left with:
\be\label{Hdot 3}\dot H=\frac{1-\lambda}{3\lambda-1}\sqrt{g}\left(g^{ij}N_{;i}p_{;j}+N\Delta p\right)
\ee

This vanishes for $\lambda=1$, but for different $\lambda$ it generates a further constraint, $p=\mbox{cte}$. Taking the mean we get $p=\mean{p}\sqrt g$, but this does \emph{not} mean that the constraint \emph{is}   $p=\mean{p}\sqrt g$. It could be given by any constant functional $p(x)=F[g,p]$ that this would still propagate the scalar constraint. We will not calculate propagation of the new constraint or follow the Dirac analysis in the Lagrangian formalism, as it is obviously too unwieldily. If one just assumes that $p$ is any constant functional, i.e. $p=\mbox{cte}$, then we know from section \ref{sec:importantApp} that this implies the familiar equation:
\be (\nabla^2_R)N=0
\ee
which can be solved for uniquely only on asymptotically flat manifolds. This case $F[g,\pi]=\mbox{cte}$, is the one analyzed in \cite{Bellorin:2010je}, where they reach the same conclusion. Other works by Barbour et al, such as \cite{barbour-2002-19}, had already reached the same conclusion using the Lagrangian framework some years ago. This also points to an interesting connection between the present theory and Horava theory, which is what is being studied in \cite{Bellorin:2010je}.

\chapter{Some useful mathematical theorems.}

Except for Lovelock's theorem and the Yamabe conjecture, for which we give the specific references, all theorems can be found in the form in which we present them here in \cite{Lang}. In the category of topological vector spaces, the appropriate name of isomorphisms are called \emph{topological linear isomorphims} (toplinear).
\subsection{Lovelock's theorem}
Lovelock's theorem was proved in a series of papers \cite{Lovelock1, Lovelock2, Lovelock3}, and its statement is the following:
\begin{theo}\label{theo:Lovelock}
In 4 dimensions, if the tensor $A^\mu_\nu$ depends exclusively on the metric tensor $^4g_{\mu\nu}$ and on its first and second partial derivatives, and if it also satisfies the continuity equation ${A^\mu_\nu}_{;\mu}=0$ then necessarily
\be A^\mu_\nu=\alpha\delta^\mu_\nu +\beta G^\mu_\nu
\ee
where $\alpha,\beta$ are constants and
$$ G_{\mu\nu}=R_{\mu\nu}-\frac{R}{2}g_{\mu\nu}
$$ is the Einstein tensor.
\end{theo}
It is highly restrictively on the possible actions for General Relativity, as the second order condition is required if we would like to keep initial data that just depend on positions and velocities.
\subsection{Closed graph theorem.}\label{sec:Closed_graph}
We use two aspects of the closed graph theorem:
\begin{prop} Every continuous bijective linear between Banach spaces $E$ and $F$ is a toplinear isomorphim.
\end{prop}
\begin{prop}If $E$ is a Banach space and $F_1, F_2$ are two closed subspaces which are complementary ($E=F_1+F_2$ and $F_1\cap F_2=0$) then the map of $F_1\times F_2\rightarrow E$ given by the sum is a toplinear isomorphism.
\end{prop}
We will say that a \emph{closed} subspace $F$ of a Banach space $E$ is such that there exists a \emph{closed} complement $F_1$ such that $E$ is isomorphic to the product $F\times F_1$ in the above manner, that $F$ \emph{splits} $E$.
\subsection{Regular value theorem.}
The regular value theorem is of the same family as the implicit function theorem. It gives us a local description of a submanifold as always being given by the regular values of some function in the ambient manifold.
\begin{theo}\label{theo:regular_value}
Let $U$ be an open subset of a Banach space $E$ and $f:U\rightarrow F$ a map into a Banach space $F$. Let $x_0\in U$ and assume that the tangent map $Tf_{x_0}$ is surjective and that its Kernel splits. Then there exists an open subset $U'$ of $U$ containing $x_0$ and an isomorphism:
$$
h:V_1\times V_2\rightarrow U'$$ such that the map $f\circ h$ is a projection:
$$V_1\times V_2\rightarrow V_1\rightarrow F
$$ where the second map is an isomorphism.
\end{theo}
\subsection{Implicit function theorem.}
\begin{theo}\label{theo:implicit_function}
Let $U, V$ be open sets of Banach spaces $E$ and $F$, respectively, and let
$$
f:U\times V\rightarrow H$$
be a $C^r$ mapping. Let $(a,b)\in U\times V$, and assume that $T_2f_{(a,b)}\rightarrow H$ is a toplinear isomorphism. Let $f(a,b)=0$. Then for a sufficiently small neighborhood $U_0$ of $a$ there exists a unique continuous map $h:U_0\rightarrow V$ defined on an open neighborhood $U_0$ of $a$ such that $h(a)=b$ and such that
$$ f(x,h(x))=0
$$for all $x\in U_0$.
\end{theo}
\subsection{Yamabe problem}\label{sec:Yamabe}
The Yamabe problem, which was proven in different dimensions by different people (see \cite{Yamabe} for a review), can be simply stated as
\begin{theo} Given a compact closed metric manifold $(M,g)$ of dimension $\geq 3$,  there exists a conformal transformation of $g$, let us call it $\tilde g$, such that $(M,\tilde g)$ has constant scalar curvature. Furthermore $\tilde g$ is unique up to global scaling.
\end{theo}
For us, this means that we can implicitly go uniquely to the Yamabe gauge: \be\label{equ:YamabeGauge}\mathcal{T}_\phi g_{ab}\rightarrow \mathcal{T}_\phi {g}^{(\lambda)}_{ab}:=\mathcal{T}_\phi(\mathcal{T}_{\lambda[g,x)}g_{ab}).
\ee


\end{document}